\documentclass[twocolumn]{aastex631}

\usepackage{graphicx}	
\usepackage{amsmath}	
\usepackage{booktabs}  
\usepackage{natbib}
\usepackage{float}
\usepackage{hyperref}
\usepackage{rotating}

\begin{document}
\title{Investigating the young stellar populations and hierarchies in nearby galaxies with the UVIT. II. Presenting the properties of $\sim$25,000 UV-detected star-forming clumps}

\shortauthors{Shashank et al.}

\correspondingauthor{Shashank Gairola}
\email{shashank.gairola@iiap.res.in, shashankgairola17@gmail.com}

\author[0009-0008-1250-6128]{Gairola Shashank}
\affiliation{Indian Institute of Astrophysics, Koramangala II Block, Bangalore-560034, India}
\affiliation{ Pondicherry University, R.V. Nagar, Kalapet, 605014, Puducherry, India}

\author[0000-0002-5331-6098]{Smitha Subramanian}
\affiliation{Indian Institute of Astrophysics, Koramangala II Block, Bangalore-560034, India}
\affiliation{ Pondicherry University, R.V. Nagar, Kalapet, 605014, Puducherry, India}
\affiliation{Leibniz-Institut für Astrophysik Potsdam, An der Sternwarte 16, D-14482 Potsdam, Germany}

\author[0000-0003-4531-0945]{Chayan Mondal}
\affiliation{S. N. Bose National Centre for Basic Sciences Block-JD, Sector-III, Salt Lake, Kolkata-700106, India}
\affiliation{Academia Sinica Institute of Astronomy and Astrophysics (ASIAA), No. 1, Section 4, Roosevelt Road, Taipei 10617, Taiwan}

\author[0000-0001-5944-291X]{Shyam H. Menon}
\affiliation{Center for Computational Astrophysics, Flatiron Institute, 162 5th Avenue, New York, NY 10010, USA}
\affiliation{Department of Physics and Astronomy, Rutgers University, 136 Frelinghuysen Road, Piscataway, NJ 08854, USA}

\author[0000-0002-5331-6098]{Annapurni Subramaniam}
\affiliation{Indian Institute of Astrophysics, Koramangala II Block, Bangalore-560034, India}

\begin{abstract}

Studying young stellar populations within galaxies can help refine our understanding of recent star formation in galaxies and their evolution. With this motivation, we present a catalog of $\sim$25,000 recently formed (within 400 Myr) star-forming clumps (SFCs) in 17 morphologically diverse nearby galaxies, including 8 massive, classic spirals, 6 intermediate-mass, flocculent spirals, and 3 dwarf irregulars. We used far- and near-UV observations from the UltraViolet Imaging Telescope (UVIT), whose $\sim$1.5\arcsec~angular resolution and 28\arcmin~field-of-view allow us to probe SFCs at a mean physical scale of $\sim$54 parsec, within the full extent of our galaxies. We adopted a homogeneous SFC detection criterion, corrected for spatially varying dust attenuation (using 6\arcsec~resolution $A_V$ maps, made by combining FUV with archival infrared observations), and estimated the SFC ages by comparing the observed UV color-magnitude diagrams with Starburst99 simple stellar population models. Using our SFC catalog, we studied the age demographic of the recently formed stellar populations across different galaxy morphologies and observed age trends consistent with several well-known phenomena, such as the inside-out formation of disc galaxies, local gravitational instabilities leading to flocculent spiral arms, and the stochastic nature of star formation in dwarf galaxies. Leveraging full galaxy coverage and far-UV data, our catalog complements existing optically-identified star cluster catalogs in the literature towards improving our understanding of star formation across a wide range of galaxy morphologies, masses, and environments. We make the SFC catalog and $A_V$ maps of our 17 galaxies publicly available with this paper. 

\end{abstract}


\section{Introduction}
\label{sec_intro}

\begin{figure*}[t]
    \centering
    \includegraphics[width= 0.96\textwidth]{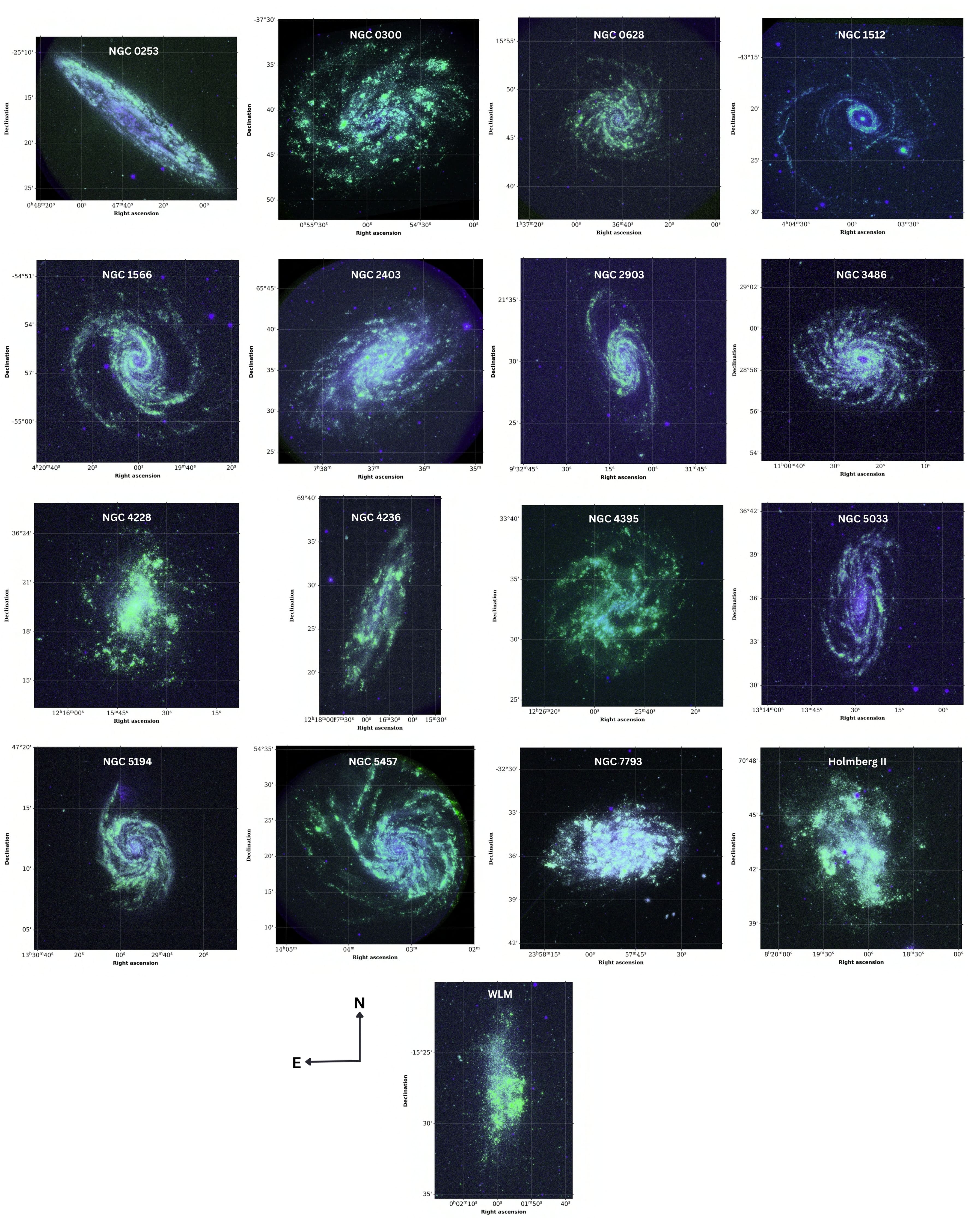}
    \caption{Color-composite images of the 17 galaxies studied in this paper. Color scheme chosen to generate these images: Red = (FUV + NUV)/2, Green = FUV$\times$1.5, Blue = NUV$\times$1.5.}
    \label{fig_galaxies}
\end{figure*}

Galaxies, giant agglomerations of stars, gas, dust, and dark matter, come in different masses, sizes, star formation rates (SFRs), colors, and morphologies (\citealt{1936rene.book.....H, 1991rc3..book.....D, 2007dvag.book.....B, 2007ApJS..173..267S}). They are the unitary building blocks of the universe, so it is important to understand how galaxies assemble their baryonic components and evolve over many gigayears (Gyrs) of their lifetimes, often transitioning across morphologies. Star formation is one of the important processes that governs galaxy growth and evolution. In galaxies of different masses and morphologies, the star formation process is influenced by a wide variety of common or mutually exclusive physical processes such as gas accretion, dynamical shocks, local or global instabilities, feedback and galaxy interactions (\citealt{2004ARA&A..42..603K, 2011ApJ...733...74L, 2013seg..book....1K, 2019ARA&A..57..227K, 2019ApJ...875...54H}). Given the context, nearby galaxies serve as ideal testing grounds for theories of star formation and galaxy evolution. Due to their proximity, high-resolution multi-wavelength observations of nearby galaxies can be used to study them in great detail - from individual stars to star clusters to any larger stellar structures. By analyzing the spatial distribution of different stellar populations in nearby galaxies, we can better understand how different internal or external processes affect the assembly and evolution of galaxies. 

Many studies have leveraged the Ultraviolet (UV) emission arising from young, massive stars to characterize the recently formed (within the past $\sim$400 million years (Myrs)) stellar populations in nearby galaxies. On one hand, far- and near-UV (FUV and NUV) observations from Galaxy Evolution Explorer (GALEX) were used to address topics such as the inside-out growth of galaxies, identification and characterization of extended-UV disc galaxies and the spatial distribution of stellar populations (\citealt{2005ApJ...619L..79T, 2005ApJ...619L..67T, 2007ApJS..173..538T, 2007ApJS..173..185G, 2007ApJ...658.1006M, 2011ApJ...733...74L}). GALEX's 1.5 degree diameter field-of-view (FoV) allowed the probing of complete star-forming discs of nearby galaxies, but its coarse $\sim$6\arcsec~angular resolution tends to miss out on the details of star formation on smaller scales, where interesting gas, dust, and feedback physics is involved. On the other hand, studies using the Hubble Space Telescope (HST) probed star-forming regions at an incredibly high, $\sim$0.1\arcsec~angular resolution with multi-band near ultraviolet (NUV) and optical observations. Legacy ExtraGalactic Ultraviolet Survey (LEGUS) and Physics and High ANGular resolution in Nearby Galaxies Survey (PHANGS) have produced extensive HST-based star cluster catalogs - (ages between 1 Myr and few Gyrs) (\citealt{2015AJ....149...51C, 2017ApJ...841..131A, 2019MNRAS.484.4897C, 2022ApJS..258...10L, 2024ApJS..273...14M, 2025ApJS..280....1T}). In conjunction with these star cluster catalogs, HST data have brought insights into stellar population modeling, star cluster evolution, hierarchical star formation, stellar feedback, and the association of spiral arms with star formation (\citealt{Radburn-Smith_2012, 2018MNRAS.478.3590S, 2019ApJ...878....1S, 2020A&A...644A.101R, 2021MNRAS.507.5542M, Linden_2022, 2025ApJ...987...33M}). However, due to HST's 3\arcmin$\times$3\arcmin~field-of-view (FoV), often multiple pointings are needed to cover even a single nearby galaxy. Usually, only the inner disc of galaxies is covered in these observations, which leaves the lower metallicity, low-density outskirts of the galaxies unexplored.

Therefore, the optimization between the angular resolution and FoV that the UltraViolet Imaging Telescope (UVIT) \citep{2006AdSpR..38.2989A, 2012SPIE.8443E..1NK, 2014SPIE.9144E..1SS} offers ($\sim$1.5\arcsec~angular resolution, 28\arcmin~diameter FoV) is a valuable asset that can be utilized to better understand the recent star formation activity in galaxies. UVIT has $\sim$4 times better angular resolution than GALEX (UVIT's angular resolution corresponds to $\sim$73 pc at 10 megaparsec distance), and it offers a significant FoV advantage over the HST. Moreover, UVIT's FUV coverage offers greater sensitivity in detecting the youngest stellar populations in galaxies as compared to the NUV filter (the bluest available waveband) used in the HST-based studies. Several studies in the past have leveraged these features of the UVIT to study different aspects of extragalactic star formation (\citealt{Mondal_2018, Mondal_2021, 2021ApJ...914...54Y, 2022MNRAS.516.2171U, 2024MNRAS.532..322H, 2024AJ....168..255H, Hassani_2024, 2024MNRAS.530.2199A, 2024ApJ...974..206W, 2025PASA...42...73S, 2025PASA...42...56A, 2025A&A...702A.222C, 2026arXiv260222860A}). These studies used diverse source detection techniques and dust attenuation correction methods as per the requirements of their science goals. However, most of these studies either focus on a small sample of galaxies or focus on galaxies of a specific morphology. Thus, systematic UVIT-based studies for a large galaxy sample, spanning a wide range in mass and morphology, are rare. Ideally, such studies should characterize stellar populations in different galaxies with a homogeneous methodology and carefully account for the spatially varying internal dust attenuation (\citealt{Hao_2011, 2016A&A...591A...6B, 2025arXiv250808451C}). 

It is against this background that we utilized UVIT observations of 17 nearby (distance $<$20 Mpc), morphologically diverse sample of galaxies to construct an extensive catalog of $\sim$25,000 recently formed (within 400 Myrs), UV-detected star-forming clumps (SFCs). Facilitated by the UVIT's 28\arcmin~ FoV, we covered all the 17 galaxies fully at $\lesssim$1.5\arcsec~angular resolution, adopted a homogeneous SFC detection and characterization approach (see Section \ref{sec_catalog_preparation}), and took into account spatially variable internal dust attenuation. Framed in the context of past surveys and studies, our catalog offers the following key advancements: 1) Compared to the past surveys (e.g. LEGUS and PHANGS surveys), which primarily target massive, star-forming main-sequence spiral galaxies, our sample includes many intermediate- and low-mass galaxies, with flocculent spiral structure or irregular morphology. This allowed us to study star formation in a representative sample of morphologically diverse, nearby galaxies. These high-resolution multi-wavelength studies (LEGUS/PHANGS) covered a few tens of percent of galaxy area, mostly located in the inner regions of galaxies, whereas our FoV advantage enables full galaxy coverage. This is advantageous for galaxies of large angular size ($>$20\arcmin), such as NGC 5457, NGC 300, and NGC 2403, which would have required multiple pointings with the HST/JWST-based studies. This enabled us to study the full star-forming extent of galaxies, including the low-density outer regions, which provide intriguing environments for star formation studies owing to their low stellar surface density, metal- and dust-poor nature, and abundance of HI \citep{2007ApJS..173..538T, 2021ApJ...914...54Y}. Furthermore, UVIT FUV observations allow the detection and characterization of the youngest star-forming regions, unlike the NUV- or V-band-based detection method used in LEGUS. 2) Compared to GALEX-based studies, our catalog has 3-4 times improved spatial resolution. 3) Relative to our past UVIT study \cite{2025A&A...693A.188S} (Paper I from here on), we have now added 13 more galaxies to our sample, nearly quadrupling its size. Moreover, the spatially resolved attenuation correction represents a significant improvement and the discussion of the age demographics is a major highlight (see Section \ref{sec_age_demographic}). 4) Finally, with our homogeneous analysis method and spatially varying attenuation correction, our catalog attempts to bridge the gap left by the plethora of past UVIT-based studies of star formation in nearby galaxies.

This SFC catalog can be used to study different topics such as the galaxy assembly process, secular or interaction-driven star formation and galaxy evolution, hierarchical star formation, and the UV luminosity function of star-forming regions in galaxies. This catalog can also be combined with kinematic data in the future to understand the symbiotic interplay between galactic dynamics and star formation. In this paper, we have utilized this catalog to investigate the recently formed stellar population demographic in galaxies of different morphologies. In our forthcoming paper (Shashank et al. (in prep.), Paper III from here on; current paper acts as Paper II), we will use our catalog to explore hierarchical star formation in these 17 galaxies. Recent studies suggest that the hierarchical star formation process occurring within galaxies exhibits a dependence on large-scale galaxy properties (\citealt{Elmegreen_2014, Grasha_2017_Spatial, 2020A&A...644A.101R, 2021MNRAS.507.5542M}). In Paper I, we had investigated this dependence in 4 spiral galaxies and demonstrated UVIT's capabilities in effectively probing hierarchical star formation in nearby galaxies - on scales between a few parsecs (pc) up to several kiloparsecs (kpc). Paper III will expand our investigation of hierarchical star formation to a larger sample of 17 galaxies, spanning a broader parameter space in stellar mass and morphology.

The remaining paper is structured as follows. In Section \ref{sec_data}, we describe our galaxy sample, along with the observational data we have used. Section \ref{sec_catalog_preparation} describes the various methods involved in the preparation of the extensive SFC catalog in our sample galaxies. In Section \ref{sec_age_demographic}, we present a discussion on the age demographics of SFCs characterized within our galaxies, and finally, in Section \ref{sec_summary}, we summarize our key findings and also outline our future goals.\

\begin{table*} 
\caption{Some important properties of the selected galaxies in this study.}
\resizebox{1\textwidth}{!}{
\begin{tabular}{cccccccccc}
\hline
Galaxy & R.A. & Dec. & Morphological & P.A. & Incl. & Distance &  Stellar mass & UVIT PSF & Spatial \\
 & (deg) & (deg) & subclass & (deg) & (deg) & (Mpc)  &  ($M_{\odot}$) & FWHM (\arcsec)  & resolution (pc) \\
(1) & (2) & (3) & (4) & (5) & (6) & (7) & (8) & (9) & (10) \\\hline
NGC 0253 & 11.8880 & -25.2880 & Massive, classic spiral              & 52.5$^{(a)}$ & 75.0$^{(a)}$ & 3.7$^{(a)}$   & 5.0 $\times$10$^{10}$ $^{(a)}$ & 1.5 &  27  \\
NGC 0300 & 13.7227 & -37.6842 & Intermediate-mass flocculent spiral  & 114.3$^{(a)}$ & 39.8$^{(a)}$ & 2.1$^{(a)}$ & 1.5 $\times$10$^{9}$ $^{(a)}$ & 1.3 &  12  \\
NGC 0628 & 24.1741 & +15.7837 & Massive, classic spiral              & 20.7$^{(b)}$ & 8.9$^{(b)}$ & 9.8$^{(b)}$   & 1.1 $\times$10$^{10}$ $^{(b)}$  & 1.5 &  71 \\
NGC 1512 & 60.9757 & -43.3488 & Massive, classic spiral              & 261.9$^{(c)}$ & 42.5$^{(c)}$ & 18.8$^{(c)}$   & 5.2 $\times$10$^{10}$ $^{(c)}$  & 1.5 &  137 \\
NGC 1566 & 65.0016 & -54.9379 & Massive, classic spiral              & 214.7$^{(b)}$ & 29.6$^{(b)}$ & 17.7$^{(b)}$   & 2.7 $\times$10$^{10}$ $^{(b)}$  & 1.2 &  103 \\
NGC 2403 & 114.2137 & +65.6027 & Intermediate-mass flocculent spiral & 124.0$^{(d)}$ & 63.0$^{(d)}$ & 3.2$^{(d)}$    & 5.0 $\times$10$^{9}$ $^{(d)}$  & 1.3 &  19 \\
NGC 2903 & 143.0421 & +21.5008 & Massive, classic spiral             & 203.7$^{(c)}$ & 66.8$^{(c)}$ & 10.0$^{(c)}$    & 4.4 $\times$10$^{10}$ $^{(c)}$  & 1.1 &  50 \\
NGC 3486 & 165.0996 & +28.9751 & Intermediate-mass flocculent spiral & 80.0$^{(e)}$ & 50.0$^{(e)}$ & 11.4$^{(f)}$   & 6.3 $\times$10$^{9}$ $^{(g)}$  & 1.5 &  83 \\
NGC 4228 & 183.9132 & +36.3268 & Dwarf irregular                     & 65.0$^{(d)}$ & 44.0$^{(d)}$ & 2.9$^{(d)}$    & 6.3 $\times$10$^{8}$ $^{(d)}$  & 1.4 &  20  \\
NGC 4236 & 184.1755 & +69.4626 & Intermediate-mass flocculent spiral & 162.0$^{(h)}$ & 75.0$^{(h)}$ & 4.5$^{(h)}$   & 9.2 $\times$10$^{8}$ $^{(i)}$ & 1.2 &  21  \\
NGC 4395 & 186.4536 & +33.5469 & Intermediate-mass flocculent spiral & 147.0$^{(j)}$ & 38.0$^{(j)}$ & 4.3$^{(j)}$    & 2.5 $\times$10$^{9}$ $^{(j)}$  & 1.1 &  23 \\
NGC 5033 & 198.3645 & +36.5939 & Massive, classic spiral             & 352.0$^{(k)}$ & 68.0$^{(k)}$ & 16.5$^{(k)}$   & 3.7 $\times$10$^{10}$ $^{(l)}$ & 1.5 &  120  \\
NGC 5194 & 202.4696 & +47.1952 & Massive, classic spiral             & 173.0$^{(b)}$ & 22.0$^{(b)}$ & 8.6$^{(b)}$   & 2.4 $\times$10$^{10}$ $^{(b)}$ & 1.1 &  46  \\
NGC 5457 & 210.8023 & +54.3489 & Massive, classic spiral             & 39.0$^{(b)}$ & 18.0$^{(b)}$ & 6.7$^{(b)}$  & 1.9 $\times$10$^{10}$ $^{(b)}$ & 1.5 &  49 \\
NGC 7793 & 359.4572 & -32.5910 & Intermediate-mass flocculent spiral & 98.0$^{(b)}$ & 55.0$^{(b)}$ & 3.6$^{(b)}$   & 3.2 $\times$10$^{9}$ $^{(b)}$  & 1.4 &  24 \\
Holmberg II  & 124.7701 & +70.7199 & Dwarf Irregular                     & 177.0$^{(d)}$ & 41.0$^{(d)}$  & 3.4$^{(d)}$ & 2.0 $\times$10$^{9}$ $^{(d)}$ & 1.3 &  21  \\
WLM      & 0.4924 & -15.4611 & Dwarf Irregular                       & 181.0$^{(m)}$ & 69.0$^{(m)}$ & 1.0$^{(m)}$ & 4.3 $\times$10$^{7}$ $^{(n)}$  & 1.4 &  6 \\\hline
\end{tabular}
}
\tablecomments{(1) Galaxy name, (2) right ascension, (3) declination, (4) morphological classification (described in Section \ref{subsec_sample}), (5) position angle, (6) inclination angle, (7) distance, (8) stellar mass, (9) UVIT point spread function (PSF) full-width half maximum (FWHM) and (10) corresponding spatial resolution for our sample galaxies, respectively. References for P.A., Incl., distance and stellar mass are indicated as bracketed superscripts where, (a) - \cite{Hassani_2024}, (b) - \cite{2021MNRAS.507.5542M}, (c) - \cite{2021ApJS..257...43L}, (d) - \cite{2008AJ....136.2782L}, (e) - \cite{2015AJ....149....1Z}, (f) - NED, (g) - \cite{2022MNRAS.515.3270S}, (h) - \cite{2007A&A...462..933C}, (i) - \cite{2019A&A...621A..51H}, (j) - \cite{2023ApJ...950...81N}, (k) - \cite{1997MNRAS.290...15T}, (l) - \cite{2019MNRAS.488.3826B}, (m) - \cite{Mondal_2018}, (n) - \cite{2019MNRAS.490..467Z}}
\label{table_galaxy_properties}
\end{table*}
\section{Data}
\label{sec_data}

\subsection{A morphologically diverse galaxy Sample}
\label{subsec_sample}

For this study, we compiled a UVIT-based, morphologically diverse sample of 17 star-forming galaxies so as to include grand design spirals, flocculent spirals, and dwarf irregulars, all located within a distance of 20 Mpc (see Table \ref{table_galaxy_properties} for distances and Figure \ref{fig_galaxies} for the UVIT images of the galaxies). The mean galaxy distance for our sample is $\sim$7.4 Mpc, which translates to a reasonably good spatial resolution of $\sim$54 pc. We used the stellar mass, visual inspection of the spiral structure, and the literature-based classifications to divide our galaxy sample into three categories as follows. Our galaxy sample includes 3 dwarf irregulars (dIrr) - with no prominent spiral structure, each having mass less than 10$^{9}$$M_{\odot}$, 6 intermediate mass, flocculent spirals - with fragmented spiral structure, each having mass ranging from $\sim$10$^{9}$$M_{\odot}$ to 10$^{10}$$M_{\odot}$ and 8 massive, classic spirals - with well-defined spiral structure, each having mass greater than 10$^{10}$$M_{\odot}$. 
Among the galaxies in our sample, the dIrr galaxy WLM has the lowest stellar mass ($M_{*}$ = 4.3 $\times$ 10$^{7}$ $M_{\odot}$), while the massive classic spiral NGC 1512 is the most massive ($M_{*}$ = 5.2 $\times$ 10$^{10}$ $M_{\odot}$. The physical characteristics of our sample galaxies are provided in Table \ref{table_galaxy_properties}.

These galaxies were selected based on the availability of archival, UVIT FUV and NUV observations, as the FUV$-$NUV color of the SFCs is crucial for our age-estimation method. To correct for dust attenuation in a spatially resolved manner, we used the method presented by \cite{2016A&A...591A...6B}, which requires near-infrared (NIR) and mid-infrared (MIR) observations (see Section \ref{subsec_attenuation} for dust attenuation correction). We took the Two Micron All-Sky Survey 2MASS J-band images from \cite{2003AJ....125..525J} as our NIR data and Multi-band Imaging Photometer for Spitzer (MIPS) 24$\mu$ images from \cite{2009ApJ...693.1821D} as our MIR data. Ultimately, the availability of the aforementioned FUV, NUV, NIR, and MIR data dictated the final number of 17 galaxies studied in this paper. \

\subsection{Observational data}
\label{subsec_data}

\begin{table*}
\caption{Description of the UVIT observations of the sample galaxies and the number of SFCs identified in each galaxy.}
\resizebox{0.995\textwidth}{!}{%
\centering
\begin{tabular}{cccccccccccccc}
\hline
Galaxy & FUV filter & FUV exposure & NUV filter & NUV exposure & $R_{FUV}$ & $R_{NUV}$ & Mag. error & $N_{SFC}$ & Completeness magnitude & Completeness age & Age remark \\
(1) & (2) & (3) & (4) & (5) & (6) & (7) & (8) & (9) & (10) & (11) & (12) \\\hline
N0253   & F169M & 11.29 ks & N245M &  11.98 ks & 2.55 & 2.43 & 0.10 & 1348 & 20.00 mag & 25 Myr & UL \\
N0300   & F148W & 8.07 ks & N242W & 11.71 ks & 2.69 &  2.50 & 0.10 & 4947 & 21.60 mag & 210 Myr & UL \\
N0628   & F154W & 4.42 ks & N263M & 1.83 ks & 2.61 & 2.11 & 0.25 & 883  & 20.80 mag & 10 Myr & OK \\
N1512   & F154W & 4.78 ks & N242W & 2.80 ks & 2.61 & 2.50 & 0.20  & 385 & 21.90 mag & 6 Myr & UL \\
N1566   & F148W & 2.94 ks & N263M & 2.96 ks & 2.69 & 2.11 & 0.20 & 1194 & 21.30 mag & 30 Myr & OK \\
N2403   & F148W & 5.38 ks & N242W & 5.99 ks & 2.69 & 2.50 & 0.10 & 2144 & 20.80 mag & 60 Myr & UL \\
N2903   & F148W & 3.35 ks & N263M & 3.35 ks & 2.69 & 2.11 & 0.20 & 1180 & 20.70 mag & 6 Myr & OK \\
N3486   & F148W & 1.91 ks & N263M & 2.20 ks & 2.69 & 2.11 & 0.25 & 864 & 21.60 mag & 15 Myr & OK \\
N4228   & F148W & 5.70 ks & N219M & 2.80 ks & 2.69 & 3.16 & 0.25 & 648  & 21.10 mag & 80 Myr & UL \\  
N4236   & F148W & 3.89 ks & N263M & 3.79 ks & 2.69 & 2.11 & 0.20 & 1194 & 22.20 mag & 100 Myr & OK \\
N4395   & F148W & 19.02 ks & N263M &  1.45 ks & 2.69 &  2.11 & 0.20 & 947  & 21.70 mag & 75 Myr & OK \\ 
N5033   & F148W & 2.93 ks & N263M & 2.95 ks & 2.69 & 2.11 & 0.25 & 647 & 22.30 mag & 15 Myr & OK \\
N5194   & F148W & 1.89 ks & N263M & 1.02 ks & 2.69 & 2.11 & 0.20 & 1694 & 20.40 mag & 10 Myr & OK \\
N5457   & F148W & 3.34 ks & N263M & 3.28 ks & 2.69 & 2.11 & 0.10 & 1507 & 20.40 mag & 14 Myr & OK \\
N7793   & F148W & 7.57 ks & N242W & 8.11 ks & 2.69 & 2.50 & 0.10 & 1775 & 21.20 mag & 70 Myr & UL \\
Holmberg II & F154W & 18.38 ks & N245M &  17.38 ks & 2.51 & 2.43 & 0.10 & 1181 & 22.50 mag & 190 Myr & UL \\
WLM     & F148W & 10.18 ks & N263M &  5.25 ks & 2.69 &  2.11 & 0.20 & 1357 & 23.20 mag & $>$400 Myr & OK \\\hline
\end{tabular}
}
\tablecomments{(1) Galaxy name, (2) FUV filter, (3) FUV exposure time, (4) NUV filter, (5) exposure time, (6) attenuation coefficient for the FUV and, (7) NUV filters derived using Cardelli's law \citep{1989ApJ...345..245C}, (8) finalized FUV and NUV magnitude error cut (refer Section \ref{subsec_SFC_ages}), (9) final number of SFCs characterized, (10) completeness limit FUV magnitude and (11) the corresponding age (for an SFC of 10$^{4}$$M_{\odot}$) (refer Section \ref{subsec_completeness}) and (12) remark about the age-estimate (refer Section \ref{subsec_age_caveats}), respectively. Abbreviations : OK - age estimates are reliable, UL - age estimates may be systematically overestimated and therefore represent upper limits.}
\label{table_UVIT_data}
\end{table*}

\subsubsection{UV imaging data and reduction}
\label{subsubsec_UVdata}
We used archival FUV and NUV observations, taken with the 37.5 cm aperture, UVIT telescope, which is one of the five science instruments onboard AstroSat satellite, operating in a low-earth orbit \citep{2006AdSpR..38.2989A, 2012SPIE.8443E..1NK, 2014SPIE.9144E..1SS}. UVIT is a twin-telescope with simultaneous imaging capability in FUV (1200 - 1800 {\AA}), NUV (1800 - 3200 {\AA}) and visible (VIS : 3200 - 5500 {\AA}) wavebands. VIS observations are not used for science purposes and are only used in the drift correction process (which corrects for the movement of the telescope's pointing between orbits of observation). UVIT offers angular resolutions better than 1.5\arcsec~in both FUV and NUV, and it has a suite of FUV and NUV filters, with varying central wavelengths and bandwidths. 

We downloaded the level 1 data from the AstroSat archive and reduced it using the dedicated UVIT data reduction software CCDLAB \citep{2017PASP..129k5002P}. The detailed data reduction process to create the final UVIT science-ready images is described in Section 2 of Paper I. It involves multiple steps such as drift correction, flat-fielding, aligning and merging observations taken in different orbits, point spread function (PSF) optimization, and establishing the world coordinate system in the images. Small differences in the accuracy with which these steps can be performed in CCDLAB affects the final PSF of the science ready images \citep{2017PASP..129k5002P}. This meant that our UVIT images of 17 galaxies have PSF FWHM values ranging from 1.1\arcsec~to 1.5\arcsec~(see Table \ref{table_galaxy_properties}). In order to maximize the exposure times of our FUV and NUV images, we also merged multi-epoch observations for some of the galaxies, taken by different observing proposals. We make the science-ready FUV and NUV images of our sample galaxies publicly available with this paper. The details of all the UVIT data used in this work and the total exposure times are summarized in Table \ref{table_UVIT_data}.

\subsubsection{Near- and mid- infrared (MIR and NIR) data}
\label{subsubsec_NIRMIR_data}

To correct for dust attenuation (see Section \ref{subsec_attenuation}), we used the archival NIR and MIR data. For NIR, we used archival J-band data from the 2MASS survey \citep{2003AJ....125..525J}. 2MASS was an all sky NIR survey offering angular resolutions of $\sim$2" in the J-band (1.2 micron ($\mu$)). In \cite{2003AJ....125..525J}, the global NIR properties along with the J-, H- and K-band images of the 100 brightest galaxies covered by 2MASS were presented. For MIR, we used archival MIPS 24$\mu$ data from the Spitzer Infrared Nearby Galaxies Survey (SINGS) (\citealt{2003PASP..115..928K, 2009ApJ...693.1821D}). There are 75 galaxies in the SINGS survey, imaged at $\sim$6\arcsec~angular resolution. The aim of the SINGS survey was to characterize the IR emission of galaxies across a broad range of galaxy properties and star formation environments. We obtained the science-ready NIR and MIR images directly from the NASA Extragalactic Database (NED) \footnote{\url{http://ned.ipac.caltech.edu}} for our analysis.

\section{SFC catalog preparation}
\label{sec_catalog_preparation}

The age-demographic of different stellar populations within galaxies can help us better understand the physical processes responsible for shaping the structure of galaxies. For this purpose, full galaxy coverage age maps of stellar populations are essential, which motivated us to create this catalog, consisting of $\sim$25,000 SFCs. Through UV-selection, we are only probing the stellar populations formed within the past 400 Myrs in these galaxies, which convey the picture of the recent star formation and galaxy assembly process.

This paper is the second in a series of papers, following up on \cite{2025A&A...693A.188S} (Paper I), investigating hierarchical-star formation in a statistically large sample of nearby galaxies using a homogeneous methodology. Most of the methods used in the current paper related to SFC characterization are the same as in Paper I. We provide a complete description of our methodology here. However, we refer interested readers to Paper I for a more nuanced description and clear motivations for the choices made in this paper. Each successive Paper in this series differs from its predecessors, as we try to implement more science and a larger number of galaxies into our investigation. For example, NGC 1566, NGC 5194, NGC 5457, and NGC 7793 were also part of Paper I. But in Paper I, we could not take into account spatially variable dust attenuation and instead used galaxy-averaged attenuation correction. This warrants the inclusion of these 4 galaxies in the galaxy sample selected for this paper.

\subsection{Identification of SFCs from the FUV data}
\label{subsec_SFC_detection}

UV emission arising from galaxies is a direct tracer of recent star formation and FUV emission traces the youngest stellar populations much more effectively as compared to NUV (\citealt{2007ApJS..173..538T, 2007ApJS..173..185G, Mondal_2018, Mondal_2021, 2021ApJ...914...54Y, 2022MNRAS.516.2171U, 2024A&A...681A...8S, 2024MNRAS.530.2199A, 2024ApJ...974..206W, 2025A&A...702A.222C}). So, in order to detect the SFCs in our sample galaxies, we applied the Astrodendro source detection algorithm on their FUV images. Astrodendro was developed to explore the hierarchical distribution of flux in astronomical images \citep{2008ApJ...679.1338R}. It establishes hierarchical connections between different regions in the two-dimensional (2D) flux maps (in the form of a dendrogram tree) by scanning from the peak flux point to the minimum flux floor. It classifies different flux regions into trunks, branches, and leaves - in increasing order of flux density and decreasing order of compactness. Leaves are the structures of the most interest to us, as these are the smallest structures identifiable in a given flux map by Astrodendro and which cannot be divided any further. For this reason, Astrodendro leaves are used as SFCs in our study.

Astrodendro uses three input parameters - 1) min\_value, which we set equal to three times the standard deviation ($\sigma$) of the sky background plus one times the sky background (bg) and it serves as the detection threshold below which no structures should be identified; 2) min\_delta, which we set equal to one times the standard deviation ($\sigma$) and it is used as a de-blending criteria to create segmentation between two closely located SFCs; and 3) min\_npix, which we set equal to 11 pixels (equivalent to a circular PSF of 1.5\arcsec diameter FWHM), and corresponds to the angular resolution of the UVIT - in pixel units. Both min\_value and min\_delta values are defined per pixel and all the leaf structures identified by the Astrodendro are a collection of contiguous pixels. In Appendix \ref{apdx_Astrodendro_sensitivity}, we present a sensitivity analysis which demonstrates that the number of detected SFCs and our scientific results are robust against the variation in astrodendro input parameters, at least by a factor of two. We used the index, position and the sizes of the leaf structures identified by Astrodendro in our subsequent analysis. 

\subsection{Estimation of SFC ages}
\label{subsec_SFC_ages}
Several studies in the literature have demonstrated that dust attenuation corrected UV color of a star-forming region can serve as an indicator of its age (\citealt{bianchi2005recent, 2005ApJ...619L..79T, 2005ApJ...619L..67T, 2007ApJ...658.1006M, Mondal_2021, 2022MNRAS.516.2171U}). So, we estimated ages of the detected SFCs by comparing their observed, attenuation corrected FUV$-$NUV color vs FUV magnitude color-magnitude diagram (termed UV CMD from here on) against the synthetic UV CMDs generated using Starburst99 (SB99) stellar population synthesis models \citep{1999ApJS..123....3L}. We have described our dust attenuation correction methodology in Section \ref{subsec_attenuation}. However, throughout this section, it should be assumed that each SFC was corrected for spatially varying dust attenuation.

\begin{figure*}[t]
    \centering
    \includegraphics[width=0.70\textwidth]{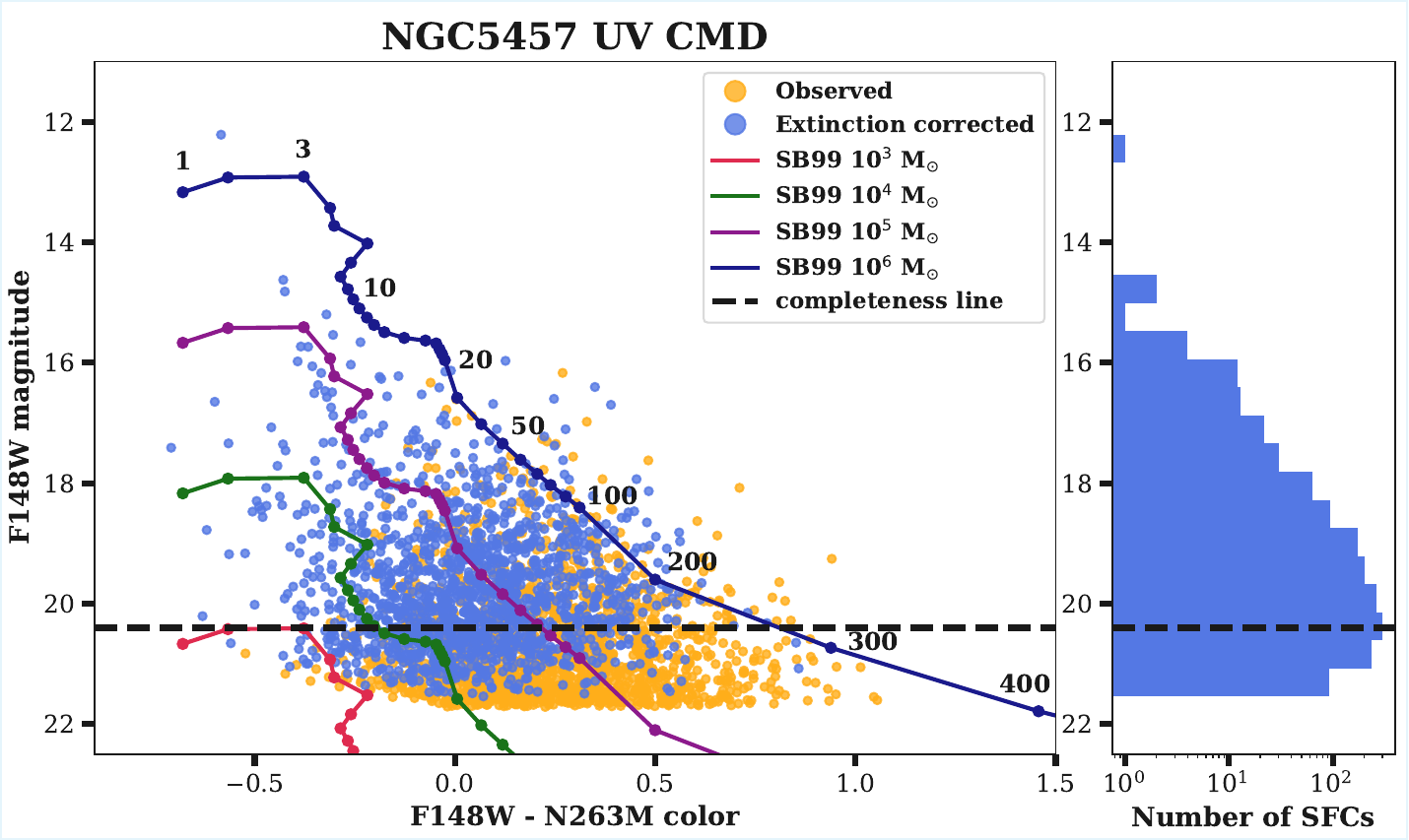}
    \caption{Synthetic Starburst99 v/s observed UV CMD for NGC 5457 which is used for the age estimation of SFCs (refer Section \ref{subsec_SFC_ages}). The synthetic colors and magnitudes derived for different ages and masses are represented by the four tracks. The numbers written adjacent to the 10$^{6}$$M_{\odot}$ track denote the synthetic SFC age in Myrs. The observed SFCs with their dust attenuation uncorrected/corrected colors and magnitudes are represented by the orange/blue points. The right hand side plot shows the FUV magnitude histogram for the SFCs in NGC 5457. The horizontal black dashed line denotes the completeness limits of SFCs and it corresponds to 20.40 FUV magnitude and $\sim$14 Myr for an SFC of 10$^{4}$$M_{\odot}$ (refer Section \ref{subsec_completeness}).}
    \label{fig_CMD_5457}
\end{figure*}

To produce the observed UV CMD, we estimated the FUV and NUV magnitudes for the SFCs by performing circular aperture photometry on the leaves/SFCs identified by Astrodendro and using the UVIT-specific counts to AB magnitude conversion provided in \cite{Tandon_2020}. For a given SFC, the area of the circular aperture is kept equal to the area of the irregularly shaped SFC detected by Astrodendro. We also measured the FUV and NUV magnitude errors associated with each SFC by applying error propagation on the counts to AB magnitude conversion formula, and assuming the error on the observed counts to be Poisson errors. The sky background (bg) contribution was also subtracted during the measurement of FUV and NUV magnitudes. Next, we corrected the observed FUV and NUV magnitudes and FUV$-$NUV color of the SFCs for dust attenuation using our spatially resolved $A_V$ maps (see Section \ref{subsec_attenuation} and Figures \ref{fig_AV_5457}, \ref{fig_AV_maps}). 

To produce the synthetic SB99 CMD, we generated synthetic spectra for star clusters of varying ages ranging from 1 to 400 Myrs, four discrete stellar mass values (10$^{3}$$M_{\odot}$, 10$^{4}$$M_{\odot}$, 10$^{5}$$M_{\odot}$ and 10$^{6}$$M_{\odot}$) and three metallicity values (z = 0.02, 0.008, 0.004). It was assumed that the stars would follow Kroupa initial mass function (IMF) \citep{2001MNRAS.322..231K} and Geneva high mass loss rate stellar evolutionary tracks. A summary of the important input parameters used to generate the SB99 synthetic spectra is provided in Table \ref{table_SB99}. We note that we chose only one value of metallicity per galaxy, and the choice of metallicity is informed by the past measurements from literature or the stellar mass-metallicity dependence of the star-forming galaxies \citep{2004ApJ...613..898T}. At the same time, we acknowledge that negative radial metallicity gradients exist in galaxies which may affect our age estimations (see Appendix \ref{apdx_effect_of_metallicity} for a discussion.) We convolved the SB99 spectra with the UVIT filter transmission curves \citep{Tandon_2020} to create the synthetic UV CMDs. We show the age evolution of the synthetic SB99 CMD tracks for different star cluster masses in Figure \ref{fig_CMD_5457}. It is evident from the CMD that, in general, the FUV$-$NUV color of the SFCs becomes redder with increasing age. This color evolution is fast between ages 0 to 20 Myrs, and slows down significantly beyond 20 Myrs as the most massive, FUV emitting stars reach the end of their main sequence lives. The stellar mass of the synthetic CMD track has no effect on this color-age relationship. Finally, a comparison between the dust-attenuation corrected observed UV CMD and the synthetic SB99 UV CMD allowed us to interpolate the observed FUV$-$NUV color against the synthetic SB99 color and estimate the ages of the observed SFCs.\

\begin{table}
\caption{Summary of the important input parameters used in the Starburst99 simple stellar population synthesis models}
\centering
\resizebox{85mm}{!}{
\begin{tabular}{cc}
\hline
Parameter & Value\\\hline
Star formation  type & Instantaneous\\
Initial mass function (IMF) & Kroupa [slope : 1.3, 2.3]\\
Stellar mass range & 0.1, 0.5, 120 $M_{\odot}$\\
Clump mass & $10^3$,$10^4$,$10^5$,$10^6M_{\odot}$\\
Metallicity & z = 0.004 (WLM \citep{Mondal_2018})\\
& Z=0.008 (NGC 4228, Holmberg II) \\
& Z=0.008 (intermediate-mass flocc. spirals)\\
& Z=0.02 (high-mass classic spirals)\\
Evolutionary track & Geneva (high mass loss rate)\\
Age range & 1-400 Myr\\\hline
\end{tabular}}
\label{table_SB99}
\end{table}

In order to estimate the age error associated with a given SFC, we used a pilot approach leveraging the fact that the evolution of FUV$-$NUV color is quite rapid in the first 20 Myrs, slower in 20 - 100 Myr and even slower in 100 - 400 Myr (as can be seen in the UV CMD presented in Figure \ref{fig_CMD_5457}). We assumed that the age and the FUV$-$NUV color roughly follows linear relationships in these age bins and therefore we can use error propagation to estimate the age errors. This implies that the error on the SFC age is equal to the product of the error on the observed FUV$-$NUV color and the slope of the assumed linear relationship between the SFC age and its FUV$-$NUV color. Here, the slope represents the change in the SFC age corresponding to 1 magnitude change in its FUV$-$NUV color. We used three different values of this slope in our age error estimation method i.e slope1 (between 1 - 20 Myrs), slope2 (between 20 - 100 Myr) and slope3 (between 100 - 400 Myr). In this paper, we assigned these age errors to the SFCs.

We had detected thousands of leaves/SFCs per galaxy in the initial run of Astrodendro on the FUV images. This amounts to a total number much larger than the $\sim$25,000 SFCs finalized within the 17 galaxies. However, several past studies had utilized magnitude error cuts to reliably constrain the ages of the SFCs (\citealt{Mondal_2018, Mondal_2021, 2022MNRAS.516.2171U}; Paper I). As the age errors are propagated from the FUV$-$NUV color errors, which in turn depend on both the FUV and NUV magnitude errors, introducing magnitude error cuts can allow us to keep the age errors to a minimum. So, we started by using the 0.10 magnitude error cuts for all galaxies. However, in some of our galaxies, we could only detect a few hundred SFCs at the 0.10 magnitude error limit - owing to their shallow exposure time observations. The aforementioned studies had also shown that a sufficiently large number of SFCs is required in order to investigate the stellar populations demographics in nearby galaxies which is one of the aims of this paper. Moreover, a sufficiently large number of SFCs within each galaxy ($>$300-400) will eventually aid our exploration of stellar hierarchies in Paper III. Several studies have shown that in order to effectively explore stellar hierarchies, a larger number of star-forming regions need to be taken into consideration \citep{Grasha_2017_Spatial, 2021MNRAS.507.5542M, 2025A&A...693A.188S, 2025ApJ...987...33M}. Therefore, we increased the magnitude error cuts to 0.20 and 0.25 magnitudes in some of our galaxies. A summary of the galaxy-specific magnitude error cuts is given in Table \ref{table_UVIT_data}. We note that the magnitude error cut provided in Table \ref{table_UVIT_data} corresponding to any galaxy represents the maximum allowed photometric uncertainty. The actual magnitude errors associated with individual SFCs are smaller than the magnitude error cut and are provided with our SFC catalog presented in Table \ref{table_SFC_catalog}.

We have produced a catalog of approximately 25,000 SFCs in our sample galaxies (see Appendix \ref{apdx_catalog_table} for the SFC catalog and Appendix \ref{apdx_radius_distribution} for the radius distribution of the SFCs within our galaxies). This catalog includes information about the host galaxy name, positions, sizes, FUV, NUV magnitudes, magnitude errors, FUV$-$NUV color before and after attenuation correction, $A_V$ values, derived ages of all the SFCs, and the associated age errors. Our SFC catalog, along with the complete 6\arcsec~resolution dust attenuation maps (Tables \ref{table_SFC_catalog}, \ref{table_AV_values} respectively), is made public with this paper and can be accessed with the online version of this paper.

\subsection{Nature of SFCs, resolution effects and implications}
\label{subsec_nature_of_SFCs}
\textbf{T}he galaxies in our sample lie between 0.9 Mpc to 20 Mpc distance. The UVIT resolution of $\sim$1.5\arcsec~ corresponds to a broad range of values - from 6 pc to 137 pc (see column 9 and 10 of Table \ref{table_galaxy_properties}) in our galaxies; this spatial scale determines the minimum size of any detected SFC in our galaxies. Owing to this large variation in the physical resolution, the identified SFCs likely correspond to different stellar structures across the sample. For instance, the 6 pc resolution for WLM (the nearest galaxy) is comparable to the typical sizes of massive star clusters. In NGC 7793, the UVIT resolution of 24 pc is equivalent to the sizes of OB associations. Lastly, in NGC 1512 (the farthest galaxy), UVIT resolution of 137 pc corresponds to the sizes of large unbound stellar complexes in spiral galaxies. Apart from WLM, the SFC detection scale for our galaxies is significantly larger than the sizes of single-aged star clusters. Therefore, homogeneously comparing the SFC properties across our galaxy sample should be reasonable.

By overplotting LEGUS star clusters on our SFCs, we observed that many SFCs represent locations where multiple star clusters are clustered together. This implies that modeling the SFCs as single-aged simple stellar populations (SSPs) may not be strictly applicable. The derived age of an SFC encompassing multiple star clusters should represent some form of an average of the ages of its constituting clusters. Or, the derived SFC age may be older than the average due to the contamination from older stars in the galactic disk. Our choice of an instantaneous burst of star formation in estimating the SFC ages is perhaps oversimplified as these SFCs could be forming stars continuously, over many Myrs. Ideally, these SFCs should be modelled as composite stellar populations with complex, continuous star formation history (SFH). Instantaneous and continuous star formation represent two extreme cases of modeling stellar populations, providing lower and upper limit on ages, respectively. Adopting a continuous star formation model would shift the SFC of a given FUV$-$NUV color towards older ages, owing to the larger star formation timescale \citep{1999ApJS..123....3L, Levesque_2013}. In our tests, we find that although this age shift is observed for both young (1-20 Myr) and old ($>$20 Myr) SFCs, it is significantly more for redder/older SFCs.

However, several studies in the past have modelled GALEX/UVIT-detected UV-bright star-forming complexes as well as LEGUS based OB associations as SSPs, born in instantaneous bursts \citep{Iglesias-Paramo:2004mnq, 2005ApJ...619L..79T, bianchi2005recent, 2007ApJ...658.1006M, Pasquali_2008, 2017ApJ...841..131A, 2019MNRAS.484.4897C}. As we are detecting SFCs in FUV, using UV color for age determination, and using the instantaneous burst model in this paper, the derived SFC ages may be lower limits. Since the main aim of this paper is to construct an SFC catalog to understand the hierarchical nature of star formation (in Paper III), which evolves with age and disperses within tens of Myr, the lower limits of ages given by the instantaneous burst are more suitable. Our derived ages may not be equivalent to the star cluster ages derived in LEGUS/PHANGS, but may reflect ages of stellar associations in these surveys. Overall, using a continuous SFH instead of instantaneous SFH would lead to older ages at a given FUV–NUV color value. However, because the shift toward older ages between the instantaneous and continuous SFH is systematic in nature, our conclusions regarding the overall age demographics and age gradients are unlikely to be significantly affected.

\subsection{Dust attenuation correction}
\label{subsec_attenuation}

Starlight from external galaxies is attenuated by the dust present along the line of sight, across the UV, optical, and IR wavelengths, with the strongest attenuation in the UV. This attenuation has two contributing components: 1) due to the Milky Way (MW/foreground) dust and 2) due to the dust present within the host galaxy itself (internal) - this component is usually much more dominant than the foreground component. We corrected for the MW/foreground dust towards our sample galaxies using the dust reddening maps accessed from the NASA dust and reddening calculator \citep{2011ApJ...737..103S}. In Paper I, we used a literature-based, galaxy-averaged value of internal dust attenuation correction for all the SFCs within the 4 sample galaxies. However, dust attenuation in galaxies is spatially variable and not constant, so in this paper, we aimed to correct for internal dust attenuation in a spatially resolved manner. However, the spatially resolved internal dust attenuation maps of our sample galaxies, which have complete galaxy coverage, were not available, which motivated us to create our own attenuation maps. 

One of the most widely used methods to correct for dust attenuation in UV consists of combining the UV with the emission from the dust itself (in the infrared (IR) wavelengths). This is based on the fact that UV photons emitted by stars are often absorbed by dust, which in turn raises the dust temperature. This heated dust then emits the reprocessed radiation in IR wavelengths. Past studies based on multi-wavelength (UV, optical, and IR) observations of nearby galaxies have produced empirical calibrations to determine the fraction of IR flux corresponding to dust reprocessed UV emission - parameterized by the IR scaling coefficients ($k_{i}$) for UV obscuration (\citealt{2008MNRAS.386.1157C, Calzetti_2007, Hao_2011, 2005MNRAS.360.1413B, Buat_2005, 2025arXiv250808451C}). These calibrations, with a form of equation (1), can be used to retrieve the true UV flux ($FUV_{total}$) arising from a star-forming region by combining the observed star-produced UV fluxes ($FUV_{obs}$) with the dust-produced IR fluxes ($24\mu_{obs}$ in our case - which corresponds to the heated dust emission in MIPS 24$\mu$ waveband). 

\begin{equation}
FUV_{total} = FUV_{obs} + k_{24\mu} * 24\mu_{obs}  
\end{equation}

The value of $k_{i}$ ($k_{24\mu}$ in our case) reported in the literature varies, depending on whether the dust heated IR emission not related to recent star formation has been subtracted (a constant value of $k_{24\mu}$ = 6 from the study of \cite{2011ApJ...735...63L}) or not (a constant value of $k_{24\mu}$ = 3.89 from the study of \cite{Hao_2011}). It is now well-known that a significant fraction of IR emission arising from heated dust is caused by old stellar populations (\citealt{1992ApJ...396L..69S, 2016A&A...591A...6B}). 

Based on the SED fitting of spatially resolved regions, \cite{2016A&A...591A...6B} (MB16 from here on) found that $k_{i}$ varies within galaxies and also varies from galaxy to galaxy. This variation is a function of the relative contribution of young and old stellar populations towards dust heating. MB16 observed that $k_{i}$ shares a strong linear correlation with star formation history or specific star formation rate, which can observationally be probed with FUV$-$NIR color. This allowed them to calibrate $k_{i}$ using the FUV$-$NIR color, which is an observable quantity. In this work, we used this calibration relation (as given below) to estimate $k_{24\mu}$ in a spatially resolved manner and use it to find the dust attenuation corrected UV emission. 
\begin{equation}
k_{24\mu} = a + b * (FUV_{AB} - NIR_{AB})\  
\end{equation}

Here, $FUV_{AB}$ and $NIR_{AB}$ are the AB magnitudes in FUV and NIR wavelengths and a,b are calibration coefficients. We used UVIT FUV, 2MASS J-band (for NIR), and MIPS 24$\mu$ (for FIR) data and adopted (a,b) = (16.43,-2.12) as given in Table 4 of MB16. It is recommended that the FUV$-$J color lies strictly in the range of $-$0.97 to $+$6.66 magnitude. Equations 1 and 2 can be used to estimate dust attenuation correction in FUV, NUV, and optical bands as follows.\

\begin{equation}
A_{FUV} = 2.5*log_{10} ( 1 +  \frac {k_{i} * 24\mu\ luminosity} {FUV\\\ luminosity})
\end{equation}

\begin{equation}
A_{V} = A_{FUV} / R_{FUV}, \quad A_{NUV} = A_{V} * R_{NUV}
\end{equation}
where $R_{FUV}$ and $R_{NUV}$ are attenuation coefficients for the FUV and NUV filters used, and are derived using Cardelli's law \citep{1989ApJ...345..245C}.

We used the above equations to create spatially resolved $A_{FUV}$ or $A_{V}$ maps of our sample galaxies. These maps are constituted by numerous square bins of 6\arcsec~angular resolution, spanning the entire extent of our galaxies. Only in the special case of WLM, we used a constant value of $k_{24\mu}$ = 3.89, as WLM was not covered by the 2MASS survey in \cite{2003AJ....125..525J}. The 6\arcsec~angular resolution limit is set by the MIPS 24$\mu$ data, which has the poorest resolution out of all the wavebands used in our $A_{FUV}$ measurement. We matched the PSF of our FUV and J-band images to $\sim$6\arcsec~before the attenuation maps are generated. We present the $A_{V}$ values at the SFC positions of NGC 5457 in Figure \ref{fig_AV_5457} and for all the remaining galaxies in Appendix \ref{apdx_AVmaps}. We also provide mathematical relations valid for different UVIT filters, which are used to convert source counts into various photometric units used (e.g AB magnitude, ergs/s, Jy) in Appendix \ref{apdx_formulae}.\

There are a few caveats associated with the attenuation correction calibration relations of \cite{2016A&A...591A...6B}, who cautioned that these relations should be used in the range of physical conditions in which they were derived. The relations were derived using early- and late-type spirals, so their use on irregular galaxies is slightly uncertain (our sample contains three irregular galaxies). They also cautioned the users against the presence of strong active galactic nuclei (AGN) which provide an additional dust heating mechanism, independent of stellar-driven heating. We note that the central few kpc bins in the k$_i$ maps of AGN hosting galaxies such as NGC 1566 and NGC 5033 were removed from our A$_V$ maps because they fall outside of the 0.97 $<$ FUV $-$ J $<$ 6.66 magnitude color cut. This takes care of the AGN effect. Finally, their relations were derived at a few hundred parsec scale and they advise against using the relations at much smaller scales, where the assumptions of fully sampled IMF and star formation history may break down. However, in the majority of our galaxies, the 6\arcsec~A$_V$ bins correspond to a physical scale greater than 100 pc.

\begin{figure}
    \centering
    \includegraphics[width=0.45\textwidth]{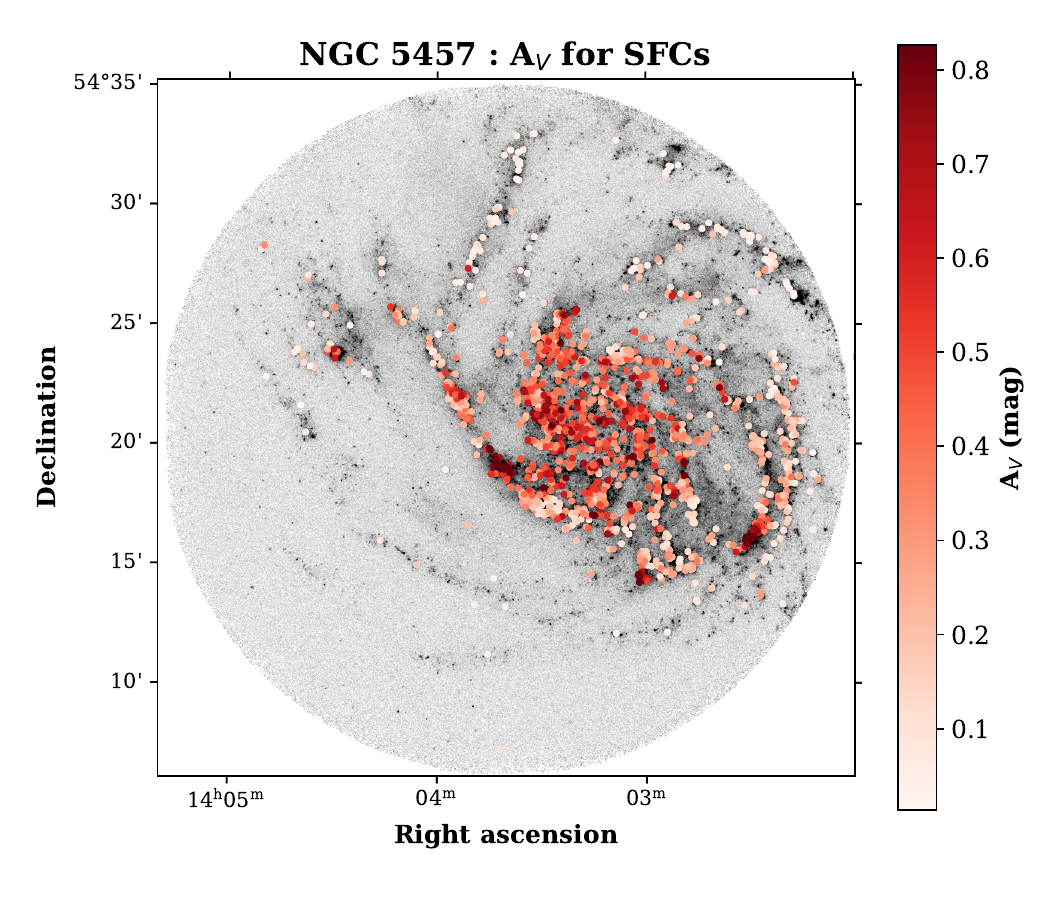}
    \caption{$A_V$ map for NGC 5457's SFCs positions. Similar $A_V$ maps for the remaining 16 galaxies are presented in Appendix \ref{apdx_AVmaps}.}
    \label{fig_AV_5457}
\end{figure}

To assign an $A_{FUV}$ value to a given SFC, we cross-matched the central SFC position with the 6\arcsec~bins in the $A_{V}$ maps. Due to the 1.5\arcsec~angular resolution of the UVIT as compared to that of the $A_{V}$ map, our SFC sizes are typically smaller than the 6\arcsec~angular resolution $A_V$ bin (6\arcsec~corresponds to $\sim$220 pc at the mean distance to the galaxies in our sample). In crowded regions, one 6\arcsec~$A_V$ bin may have more than one SFCs in it. In such cases, the same $A_V$ value is assigned to each SFC. Therefore, our dust attenuation correction relies on the assumption that the $A_{V}$ values do not vary significantly within our 6\arcsec~angular resolution limit. In Appendix \ref{apdx_AV_resolution_mismatch}, we demonstrate how the potential variation in A$_V$ within a resolution element does not significantly impact the SFC ages.

\begin{figure}
    \centering
    \includegraphics[width=0.45\textwidth]{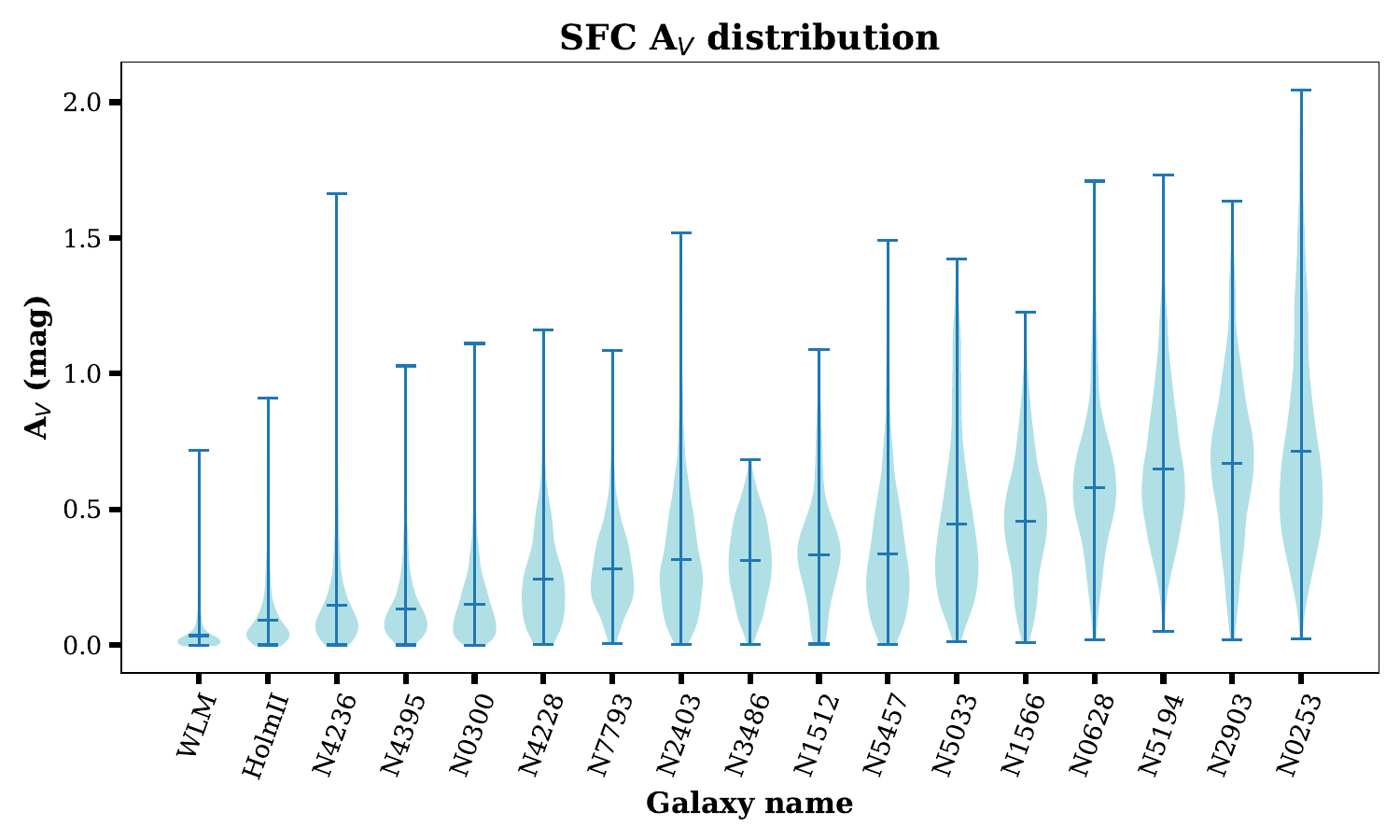}     
    \caption{Violin plots for the $A_V$ distribution of the observed SFCs in our 17 galaxies, arranged left to right in the increasing order of median $A_V$ values. The central, top, and bottom bars represent the median, maximum, and minimum $A_V$ values for the galaxy, respectively.}
  \label{fig_AV_distribution}
\end{figure}

\begin{figure*}
    \centering
    \includegraphics[width=0.90\textwidth]{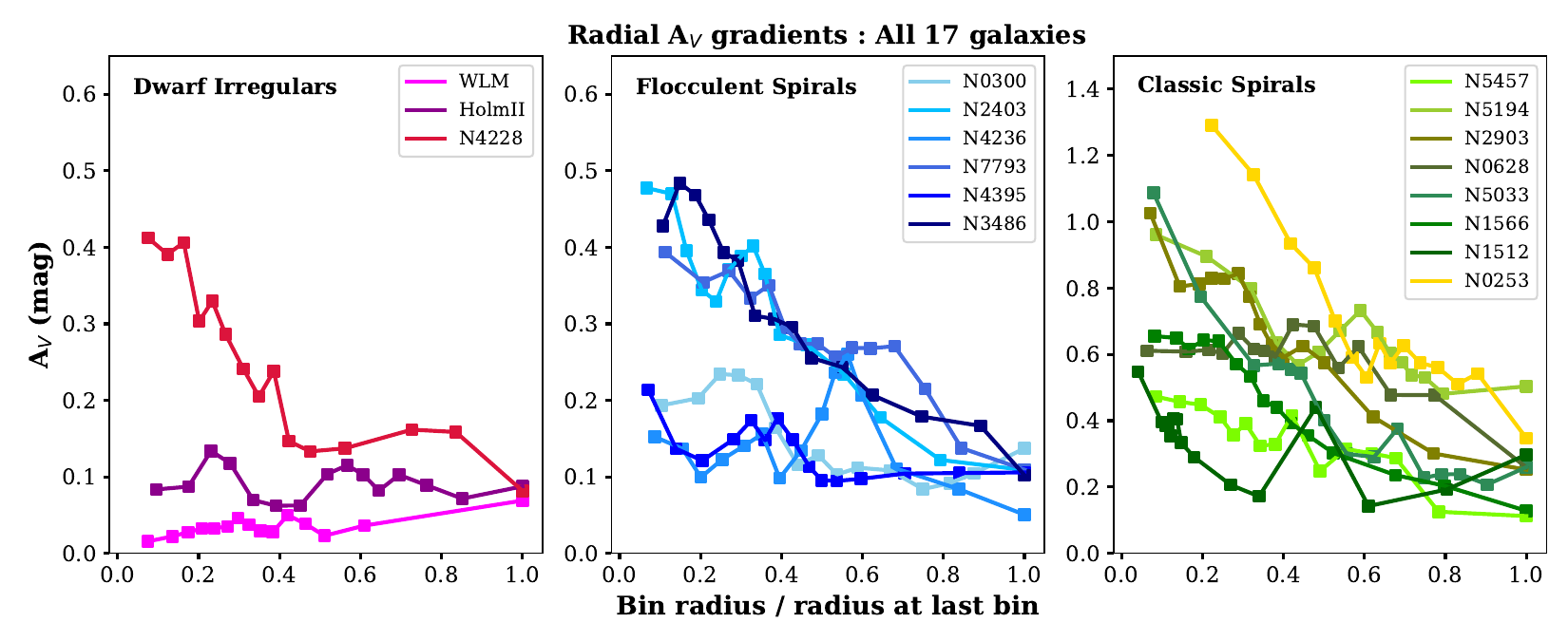}  
    \caption{Radial $A_V$ gradients with respect to the normalized galactocentric radius for our sample galaxies, separated by morphology.}
  \label{fig_AV_gradient}
\end{figure*}

\subsubsection{Distribution of $A_V$ values within our galaxies} 
\label{subsubsec_AV_distribution}

We created violin plots to describe the galaxy-wise $A_V$ values presented in our catalog corresponding to the SFC positions within our galaxies (see Figure \ref{fig_AV_distribution}). This plot demonstrates the spread in the $A_{V}$ values within our galaxies. The $A_{V}$ distribution is arranged in increasing order of the mean $A_{V}$ values. It shows that on average, dwarf irregular galaxies have the smallest values of $A_{V}$, followed by the intermediate mass flocculent spirals, and finally, the massive, classic spirals have the largest $A_{V}$ values. The two galaxies with high inclination angles i.e. NGC 2903 and NGC 0253 have the highest $A_{V}$ values.

\begin{figure*}
      \centering
		\includegraphics[width=0.75\linewidth]{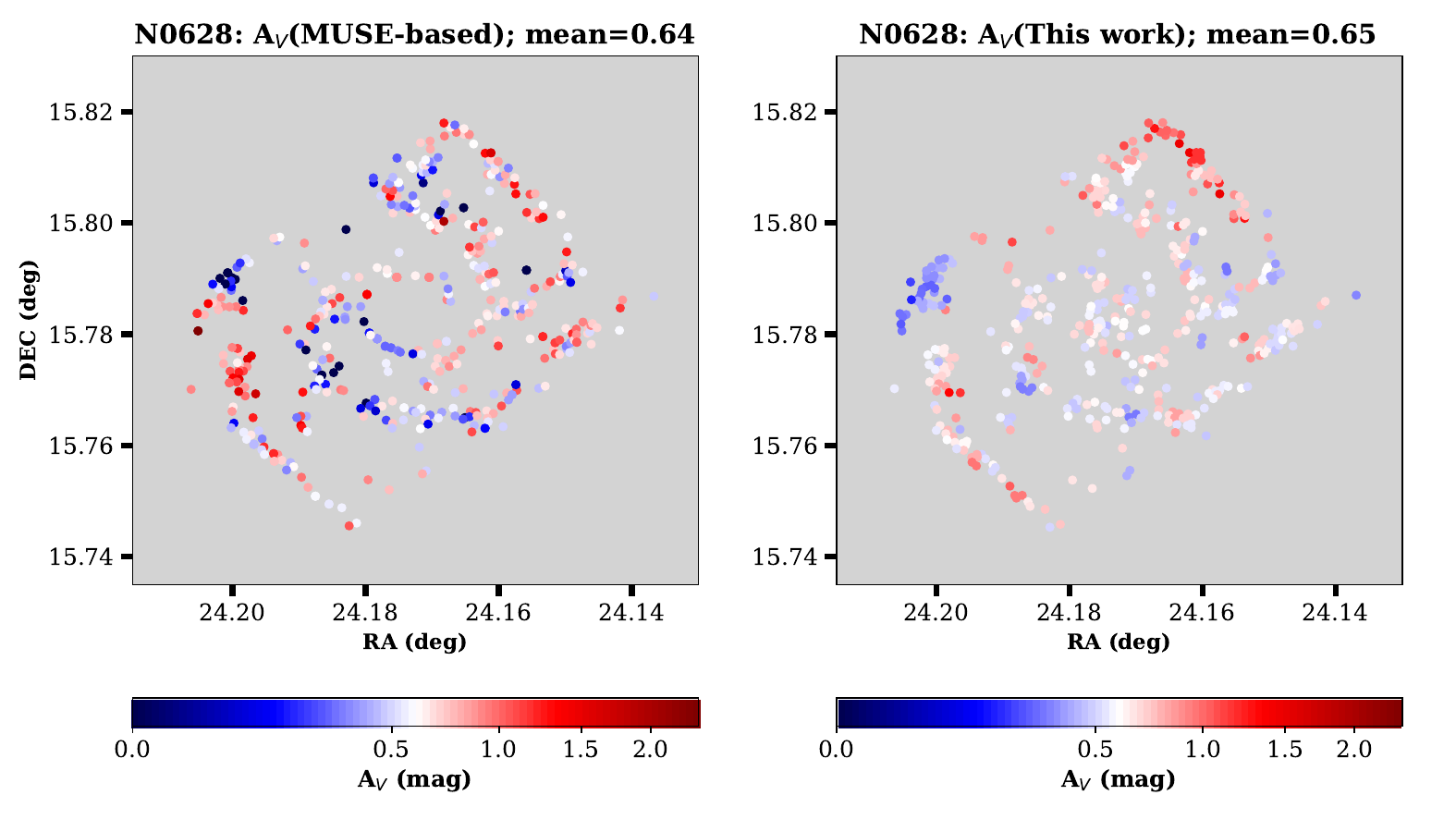}
        \vfill
        \includegraphics[width=0.75\linewidth]{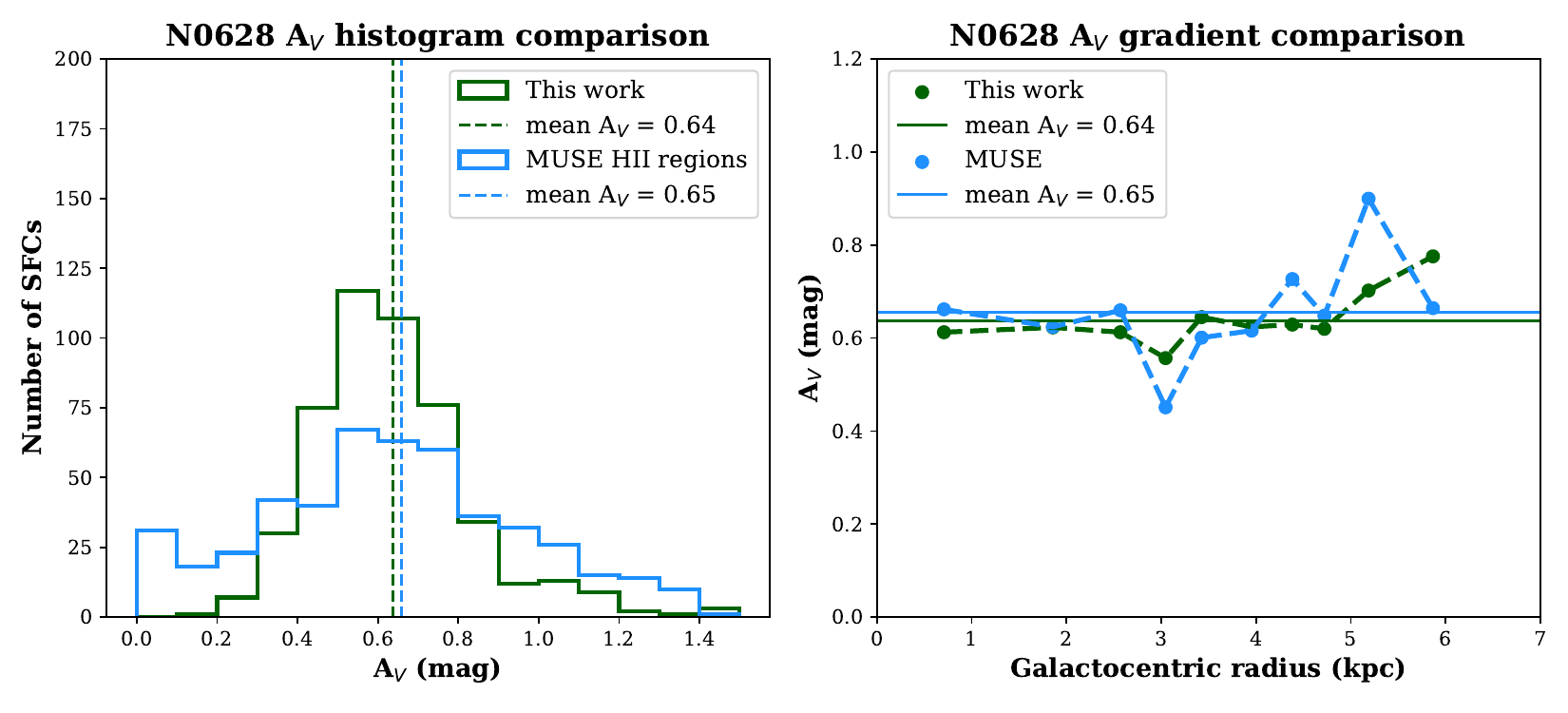}
    \caption{Illustrative figures showing the comparison of $A_V$ measurements in NGC 0628 using MUSE-based HII regions (derived using the Balmer decrement method) and the method outlined by MB16 and used in this paper. This analysis is performed only in those parts of NGC 0628 that were covered in MUSE observations. Top row: Spatial maps of $A_V$ values measured at the location of the SFCs according to MUSE (left) and the MB16 method (right, this work). Bottom left: Histogram of $A_V$ values measured with MUSE (in green) and the MB16 method (skyblue) - the mean and the range of $A_V$ values match quite well between the two methods. Bottom right: Radial profile of $A_V$ values measured with the two methods - we observe mutually consistent trends between the two methods.}
  \label{fig_AV_comparison_0628}
\end{figure*}

In Figure \ref{fig_AV_gradient}, we also present radial trends of the $A_V$ values, distinguished by the three galaxy morphology types (as described in Section \ref{subsec_sample}). These radial trends are generated by taking 15 bins, ranging from zero to the largest galactocentric radial distance at which an SFC is found - each bin having an equal number of SFCs. The x-axis of these plots is normalized by the largest galactocentric radial bin. Before the radial trends are measured, the SFC positions are de-projected using the position angle and inclination angle values of the galaxies as given in Table \ref{table_galaxy_properties}, following the method outlined in section 4.5 of Paper I. 

We observed negative $A_V$ gradients in nearly all of our spiral galaxies and the dwarf irregular galaxy NGC 4228. Similar negative $A_V$ gradients for disc galaxies have also been observed in many studies spanning a wide range of $A_V$ measurement techniques \citep{2013ApJ...771...62K, 2016A&A...591A...6B, 2019ApJ...884...21K,2023MNRAS.520.4902G}. The two dwarf irregular galaxies WLM and Holmberg II deviate notably from this trend of negative $A_V$ gradient - possibly owing to their unique irregular structure. The radial trends for $A_V$ illustrate the distribution of the attenuating dust content within galaxies of different morphological types. We had already observed in Figure \ref{fig_AV_distribution} that the mean $A_V$ values of different galaxy types exhibit significant variations. 

\begin{figure*}
      \centering
    	\includegraphics[width=0.75\linewidth]{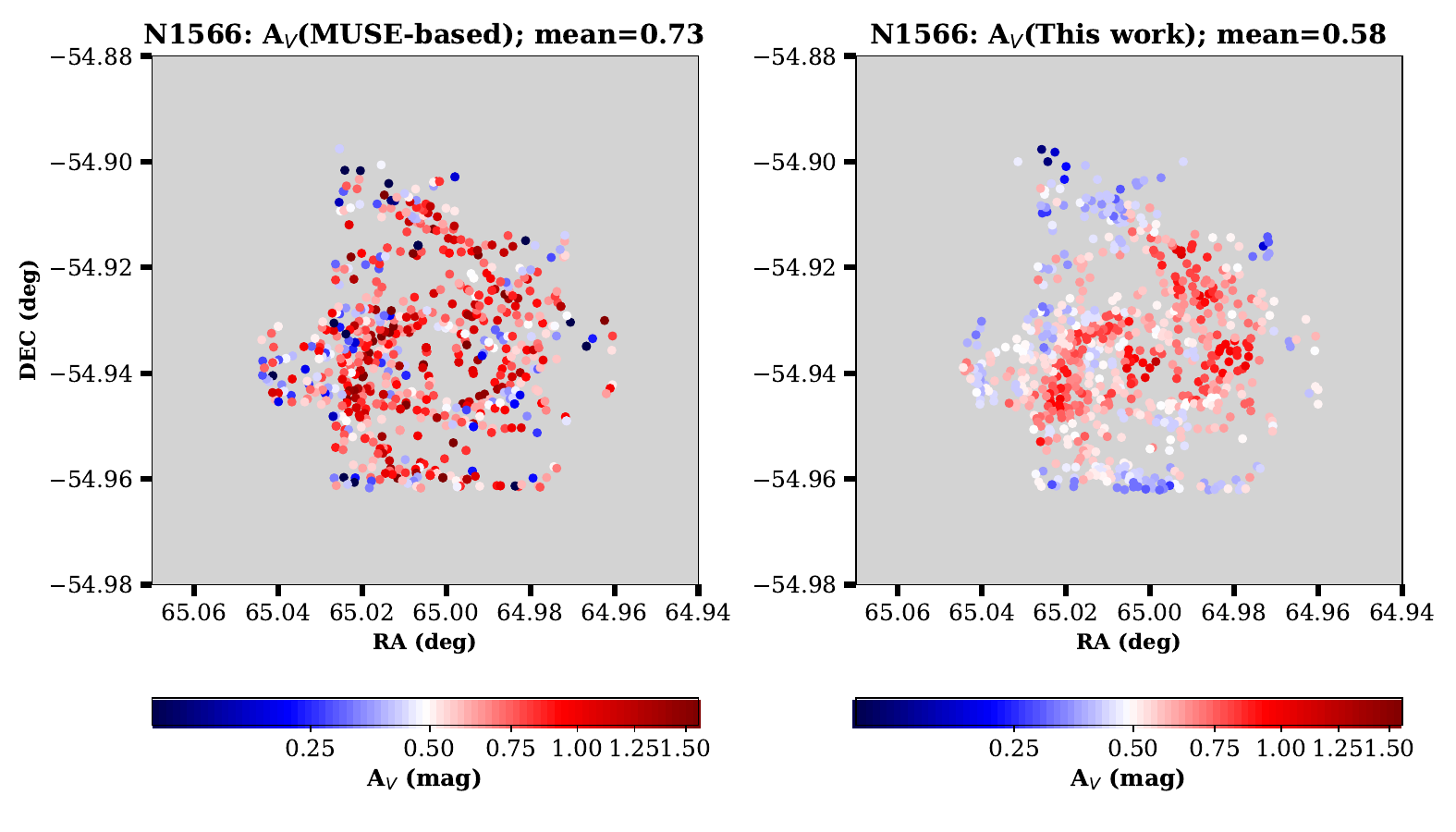}
        \vfill
		\includegraphics[width=0.75\linewidth]{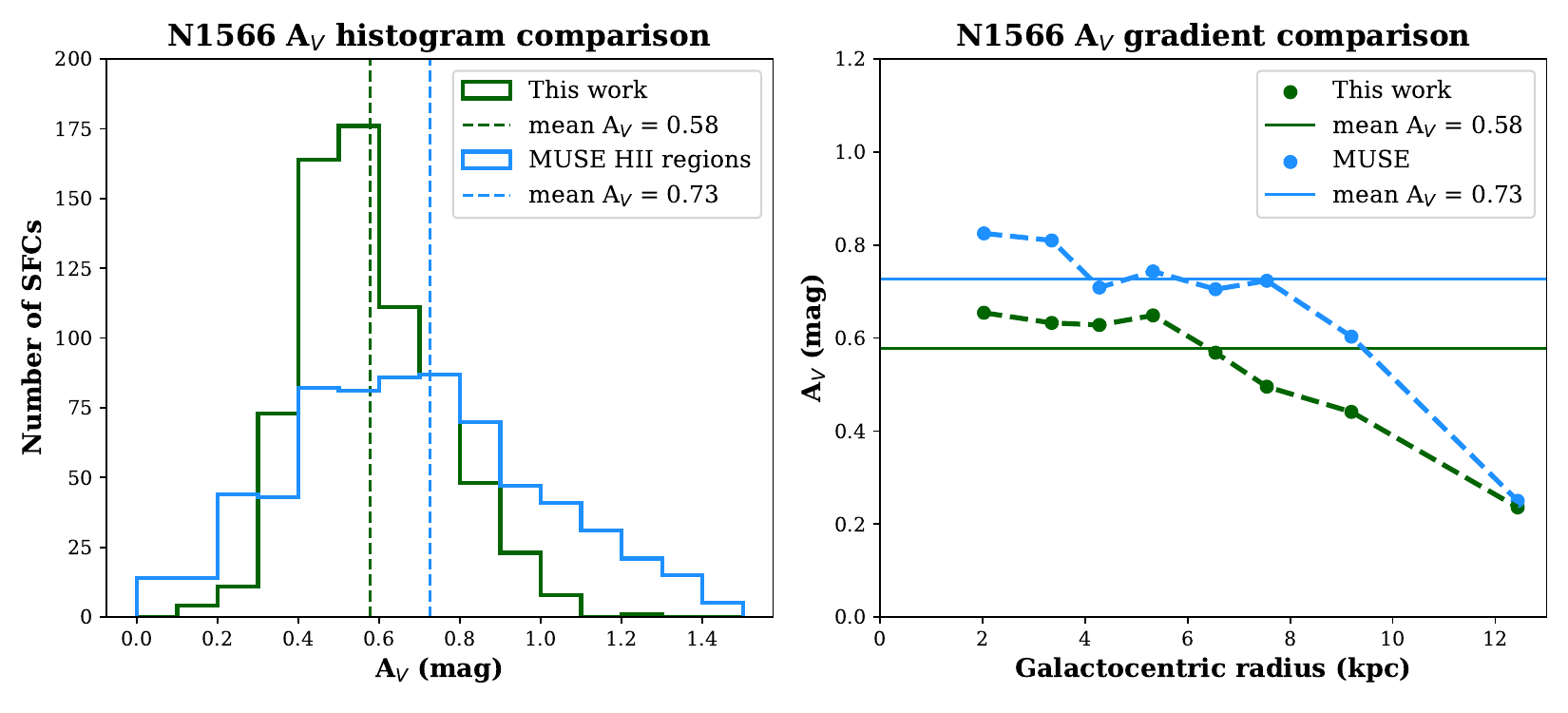}
    \caption{Same as Figure \ref{fig_AV_comparison_0628}, but for NGC 1566. All 4 plots (spatial maps, mean value, range of values and radial profile) indicate towards a reasonable match between the $A_V$ values derived using the MUSE-based and MB16-based method.}
  \label{fig_AV_comparison_1566}
\end{figure*}

\subsubsection{Effect of spatial resolution on the dust attenuation correction}
\label{subsubsec_AV_validation}

\indent One of the highest spatial resolution attenuation maps (to the order of the UVIT spatial resolution) available in the literature is the one provided by the PHANGS- Multi Unit Spectroscopic Explorer (MUSE) survey \citep{2022A&A...659A.191E, 2023MNRAS.520.4902G}. They provide $A_V$ values for a large number of ionized nebulae, including HII regions, in 19 PHANGS-MUSE galaxies, measured using the Balmer decrement method. Three galaxies in our sample (NGC 0628, NGC 1512, and NGC 1566) are also part of the PHANGS-MUSE sample, and their HII regions catalog was provided by \cite{2022A&A...658A.188S}. In order to check the effect of the spatial resolution of our attenuation maps on the SFC attenuation correction, we can compare our attenuation maps with those provided by the MUSE observations. However, due to the limited galaxy coverage of MUSE observations, the entire extent of these galaxies is not covered. So, we can only compare the $A_V$ values for the overlapping region between the UVIT and MUSE observations. We note that the attenuation maps compared here are derived using two different methods (one using UV and IR emission from the SFCs and the other using the Balmer decrement of HII regions), where the tracers and their lifetimes are different. 

MUSE observations are taken from the 8.1 meter Very Large Telescope and are assisted with adaptive-optics. Therefore, they achieve an average angular resolution of $\sim$0.7\arcsec~which is 1.5 to 2 times better than the UVIT's angular resolution. Additionally, these observations (suited to detect compact HII regions of luminosities in excess of a few times 10$^{36}$ ergs/sec) are much more sensitive than our UV observations. For reference, this allows PHANGS-MUSE to characterize $\sim$2000 HII regions as compared to our $\sim$500 UV SFCs in the galaxy area common between PHANGS-MUSE and UV observations of NGC 0628. Due to the smaller number of SFCs characterized in NGC 1512 and the fact that only the inner star-forming ring of the galaxy is covered by PHANGS-MUSE observation, we restrict our $A_V$ comparison analysis to only NGC 0628 and NGC 1566. 

We cross-matched the UV SFC positions in RA-DEC coordinates with the HII regions identified in the two galaxies, to an accuracy of $\sim$1.5\arcsec~which is closer to our UV angular resolution. This gives us two $A_V$ values corresponding to roughly the same SFC position - measured using two different methods. In Figures \ref{fig_AV_comparison_0628} and \ref{fig_AV_comparison_1566}, we demonstrate the results from our $A_V$ comparison. The top rows (left) and (right) show the spatial $A_V$ map derived using the Balmer decrement method for the HII regions and the measurement made in this paper for the SFCs following the MB16 method. We used the same color bar scaling in both images for an easier comparison, and we also quoted the mean $A_V$ values in the figure title. It can be observed that the MB16 based $A_V$ maps appear as slightly smoothed versions of the PHANGS-MUSE based $A_V$ maps. This is not surprising considering that PHANGS-MUSE maps are created at better than 1\arcsec~resolution, whereas the MB16-based maps are created at a poorer angular resolution of 6\arcsec. Overall, we observe quite a good agreement in the spatial $A_V$ maps as well as the mean values for both galaxies. This agreement is much better in NGC 0628. The bottom row (left) shows the histogram of the $A_V$ values and the respective mean values of the distribution as vertical dashed lines. The bottom row (right) shows the trend of the $A_V$ values with respect to the galactocentric radius. The spread of the $A_V$ distribution as well as the overall shape of the radial $A_V$-trend agree quite well between the two methods. Again, the agreement is much better in NGC 0628. Similar to the spatial maps, the bottom row plots too point towards good agreement between the $A_V$ values measured by the two methods. 

The observed small differences in the $A_V$ comparison between the two methods could be due to the difference in the tracers (and their lifetimes) used to measure attenuation and the spatial resolution of the two $A_V$ maps.  Previous studies have also found that nebular based attenuation gives higher values than the stellar based measurements \citep{2019ApJ...886...28Q}. The higher values of nebular-based attenuation than stellar-based values could be because Balmer decrement probes attenuation in dense, dusty, young star-forming regions whereas the stellar based attenuation combines the contributions from the dense, short-lived birth clouds and the diffuse dust permeating the ISM, getting heated by young and intermediate age stellar populations \citep{2019PASJ...71....8K}. Overall, these observed differences are small and unlikely to significantly affect the age demographics presented in this paper. 

Given the limited availability of high resolution, full galaxy coverage NIR, MIR data or dust attenuation maps, our current method of dust attenuation correction is the most that can be achieved within our current methodology, and it is clearly preferable over a constant or no dust attenuation correction. However, in the future, higher resolution $A_{V}$ maps of nearby galaxies can be created (at $\sim$1.5\arcsec~angular resolution) using UVIT FUV, full-galaxy coverage James Webb Space Telescope (JWST) Near-InfraRed CAMera (NIRCAM) F115W and JWST Mid-InfraRed Imager (MIRI) F2100W filter observations.

\subsection{Caveats with SFC ages} 
\label{subsec_age_caveats}
We note that the FUV and NUV observations used in this study for our 17 galaxies are not from the same UVIT filters. From the multi-filter, multi-epoch UVIT observations taken for any galaxy, our filter choice aims to maximize the number of detected SFCs. However, due to the existence of the 2175{\AA} bump in the Cardelli's Milky Way attenuation curve (which we have adopted in this study), the $R_{NUV}$ value for some of our galaxies in N219M, N242W and N245M filters is comparable to the $R_{FUV}$ value (see Table \ref{table_UVIT_data}). In such galaxies, the reddening correction for the FUV$-$NUV color is much smaller than the other galaxies where N263M filter is used. According to our tests based on the galaxies with multi-filter NUV observations, this implies that the derived SFC ages for such galaxies may be slightly overestimated and therefore should be treated as upper limits (UL). Moreover, the synthetic SB99 UV CMD tracks for our galaxies are dependent on our choice of metallicity (see Table \ref{table_SB99}). \cite{2021ApJ...914...54Y} and \cite{Mondal_2018} demonstrated that the synthetic FUV$-$NUV colors for a SFC of given age can vary if a different metallicity value is chosen in SB99. Although our adopted metallicity values for each galaxy are based on the latest reported values from the literature, we caution readers that if a different value of metallicity is found to be more suitable for any of our galaxies, it would result in a change in the derived SFC ages. For an observed value of FUV$-$NUV color, the SB99 CMD constructed with a higher metallicity value than the value assumed in this paper would result in younger SFC ages and vice versa \citep{2021ApJ...914...54Y}.

\begin{figure}[t]
    \centering
    \includegraphics[width=0.46\textwidth]{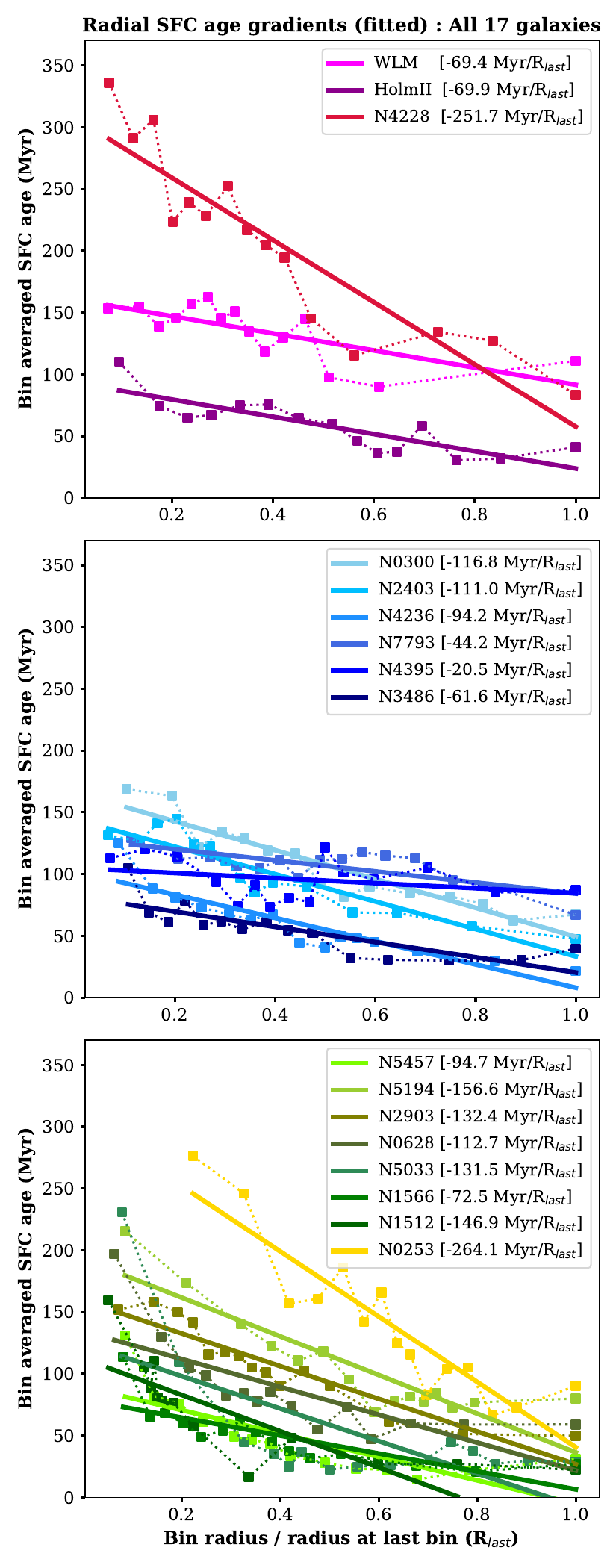}  
    \caption{Radial SFC age gradients with respect to the normalized galactocentric radius for our sample galaxies, separated by morphology. The gradients are fitted with a linear function and the best fit slopes are written in the legend.}
  \label{fig_age_gradients}
\end{figure}

\begin{figure}[t]
    \includegraphics[width=0.46\textwidth]{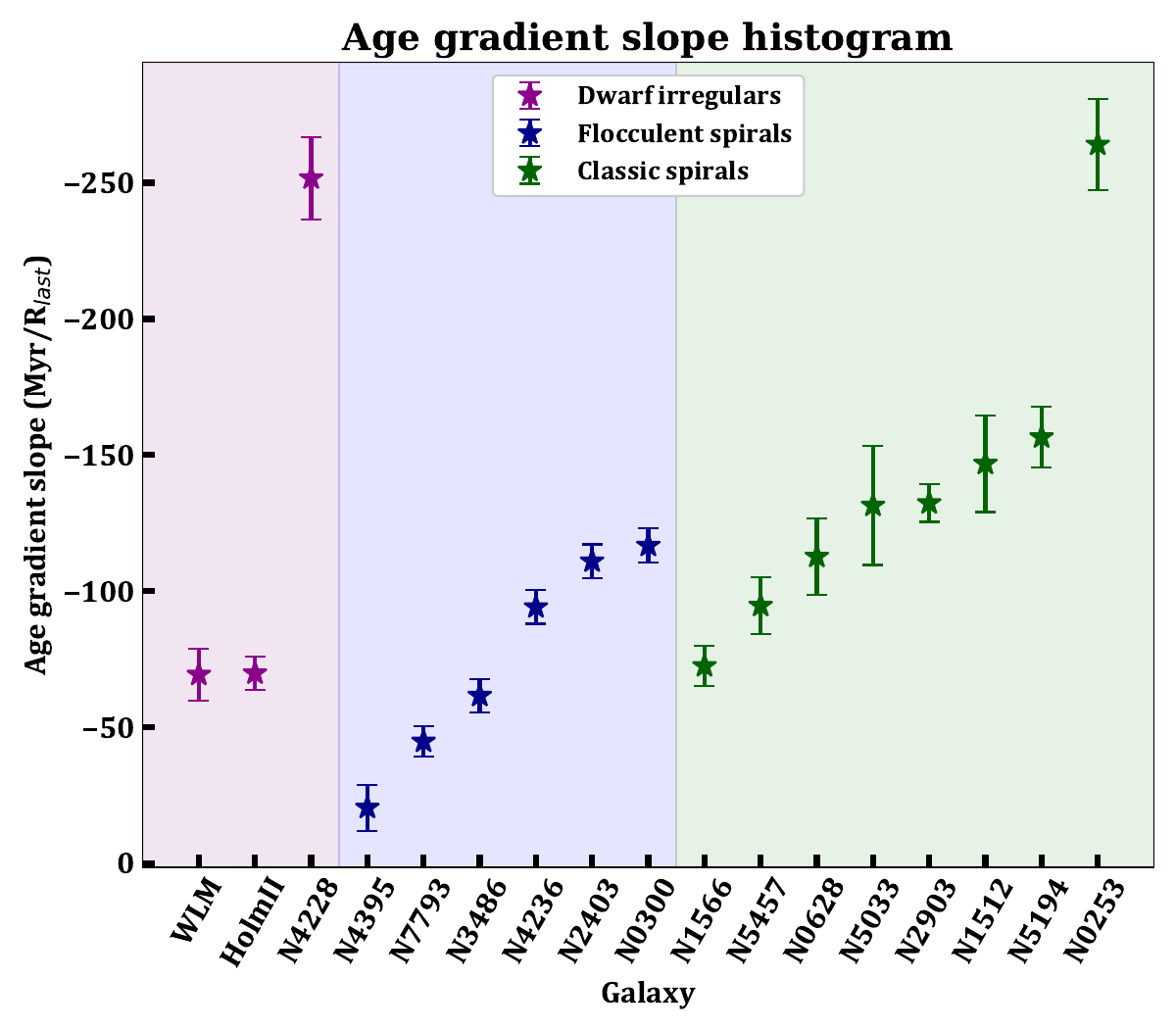}  
    \caption{Histogram of the radial SFC age gradient slopes for our galaxies. Colored vertical bands are added to distinguish among morphological subclasses. Galaxies belonging to a particular morphological subclass are arranged in the ascending order of the gradient slope.}
  \label{fig_gradient_morphology}
\end{figure}

\subsection{Completeness limit of SFCs}
\label{subsec_completeness}

The completeness limits for the SFCs identified in each galaxy depend on several factors, such as the distance of the galaxy, the exposure time of the FUV and NUV observations, and the magnitude error cuts used to characterize the SFCs. In the nearest galaxies, luminous SFCs of small physical size will be easily detectable. However, similar SFCs located in farther away galaxies will be smaller than the resolution element - thereby avoiding detection. Additionally, these SFCs may be significantly fainter at these distances, and therefore won’t survive the magnitude error cut. Only SFCs with sufficiently high signal-to-noise ratios survive the magnitude error cuts applied to each galaxy. Therefore, deeper exposure FUV and NUV images can reveal fainter SFCs, pushing the completeness limit to fainter magnitudes. This implies that the completeness limit for the SFCs identified in each galaxy will be different. 

We used the method outlined in Section 4.4 of Paper I to measure the SFC completeness. We had conducted tests in Paper I, by using UVIT images of NGC 7793 with different exposure times and plotting the attenuation corrected FUV magnitude histogram of the identified SFCs. We found that the peak of the histogram systematically corresponds to fainter magnitudes, if deeper exposure images are used to characterize the SFCs. Therefore, the magnitude corresponding to the histogram peak may be considered as the completeness limit \citep{2020ApJS..247...47L, 2023ApJ...946...65D, 2024AJ....168..255H} . This allowed us to measure the completeness limit FUV magnitude for each galaxy. Next, we measured the completeness limit age by measuring the age of a synthetic SB99 SFC of $10^4 M_{\odot}$ and FUV magnitude equal to the attenuation-corrected completeness limit. We applied this method to all 17 of our galaxies, and in Table \ref{table_UVIT_data}, we present the attenuation-corrected FUV magnitude value and the age (for an SFC of 10$^{4}$ $M_{\odot}$) corresponding to the completeness limit. Our SFCs are most complete in galaxies such as WLM, Holmberg II and NGC 0300 (a few 100 Myr), whereas the least complete in galaxies such as NGC 2903, NGC 0628, NGC 1512, and NGC 5194 (6-10 Myr). Overall, the combination of galaxy distance, exposure time and magnitude error cut are the primary drivers of the wide variation which we observe in the completeness limits of the galaxies.

\section{Age-demographic of SFCs within galaxies}
\label{sec_age_demographic}

In this section, we present and interpret the observed age distribution of SFCs within our galaxies with a special focus on the recent galaxy growth and assembly history across different morphologies. We reiterate that we have used UV observations for SFC identification, and our SFC ages are restricted to less than 400 Myrs. Therefore, all of our age trends describe the recent star formation activity in galaxies, rather than their entire star formation history spanning several Gyrs. Moreover, there exist partial/full galaxy coverage catalogs and age maps of stellar populations in the literature, for some of our galaxies. These catalogs differ from ours in terms of their spatial resolution, age range, age-estimation methodology and the fraction of galaxy area being covered. Therefore, a direct comparison of our SFC ages with the ages presented in such catalogs would not be possible. However, wherever we discuss the age trends in this section, we present a qualitative comparison between our age demographic and those revealed using similar catalogs in the literature.

In Figure \ref{fig_age_gradients}, we present the radial trends of the SFC ages within our galaxies, separated by the three galaxy morphology types (as described in Section \ref{subsec_sample}). We followed the same method as the one used to measure the radial $A_V$ trends in the Figure \ref{fig_AV_gradient}. The visual inspection of these gradients suggests negative age gradients in all of our sample galaxies. The observed slope of the age gradient is specific to the galaxy, and subtle variations in the shape of this gradient are attributed to the variation in the spiral structure of the galaxy at different galactocentric radii \citep{bianchi2005recent, 2005ApJ...619L..67T, 2007ApJ...658.1006M, 2011ApJ...743..137B, 2021MNRAS.502.5508P}.

We fitted the radial age gradient in each galaxy with a linear function. To perform the fitting, the change in SFC ages as a function of normalized galactocentric radius is measured. Since the UV extent of our galaxies can differ from R$_{25}$, the galactocentric radius was normalized with respect to the last radial bin (R$_{last}$) used in measuring the age gradient. In Table \ref{table_gradient_slopes}, we show that the R$_{25}$ and R$_{last}$ of our galaxies are quite similar, barring NGC 1512 which has a highly disturbed and extended morphology. As our study is based on UV emission, the use of R$_{last}$ over R$_{25}$ is justified. Therefore the age gradient slope is measured in units of Myr/R$_{last}$. Figure \ref{fig_age_gradients} shows the best-fit lines, and the corresponding slopes are provided in the legend. The best fit slopes, R$_{25}$, and R$_{last}$ are also tabulated in Table \ref{table_gradient_slopes}. 

The inside-out growth scenario of galaxies is supported by the fact that all measured gradients are negative. We note that because our galaxies span a wide range in size, the gradients presented are not absolute (i.e. in the Myr/kpc unit). Instead, the Myr/R$_{last}$ unit allows us to effectively compare the gradients across different galaxy morphologies, irrespective of galaxy size (interested readers can convert the Myr/R$_{last}$ unit to Myr/kpc unit by multiplying the gradient slopes with R$_{last}$). To compare the age gradient across morphological subclasses, we have plotted the individual gradient slopes in Figure \ref{fig_gradient_morphology}. The three different morphological subclasses have been arranged separately in this figure and individual galaxies are placed in the ascending order of the gradient slope. On average, the dwarf irregulars and intermediate-mass flocculent spirals exhibit shallower slopes (except the dwarf irregular NGC 4228, which exhibits a significantly steep negative gradient) as compared to the massive, classic spirals. This figure provides evidence for both galaxy-to-galaxy and morphology-dependent variations in age gradients. The powerful combination of UVIT’s 28\arcmin~ FoV and 1.5\arcsec~ angular resolution has enabled us to systematically characterize the SFCs and investigate the age gradients. Performing such an analysis on our sample of 17 galaxies would have been challenging with the HST due to its FoV limitation, or with GALEX due to its poorer, $\sim$5\arcsec~ angular resolution.

In the upcoming sections, we discuss the spatial age maps of galaxies, their connection with galaxy morphology, and the probable physical reasons for the observed age demographic. 
 
\begin{table}[htbp]
\centering
\caption{Galaxy properties and corresponding radial age gradient slopes.}
\begin{tabular}{lccc}
\hline
Galaxy & R$_{last}$ & $R_{25}$ & Slope \\
 & (kpc) & (kpc) & (Myr/R$_{last}$) \\\hline
WLM      & 2.08  & 1.8  & -69.4  \\
Holmberg II  & 4.02  & 3.7  & -69.9  \\
NGC 4228    & 4.25  & 2.9  & -251.7 \\
NGC 0300    & 6.72  & 5.9  & -116.8 \\
NGC 2403    & 10.44 & 7.3  & -111.0 \\
NGC 4236    & 12.8  & 15.2 & -94.2  \\
NGC 7793    & 5.34  & 4.9  & -44.2  \\
NGC 4395    & 7.85  & 8.2  & -20.5  \\
NGC 3486    & 11.02 & 11.7 & -61.6  \\
NGC 5457    & 22.74 & 27.9 & -94.7  \\
NGC 5194    & 9.21  & 13.9 & -156.6 \\
NGC 2903    & 16.02 & 17.4 & -132.4 \\
NGC 0628    & 13.14 & 15.0 & -112.7 \\
NGC 5033    & 27.17 & 29.7 & -131.5 \\
NGC 1566    & 22.64 & 21.4 & -72.5  \\
NGC 1512    & 49.2  & 23.1 & -146.9 \\
NGC 0253    & 11.78 & 14.4 & -264.1 \\\hline
\end{tabular}
\label{table_gradient_slopes}
\end{table}

\subsection{Spiral galaxies}
\label{subsec_spiral}

\begin{figure*}[b]
      \centering
		\includegraphics[width=0.32\linewidth]{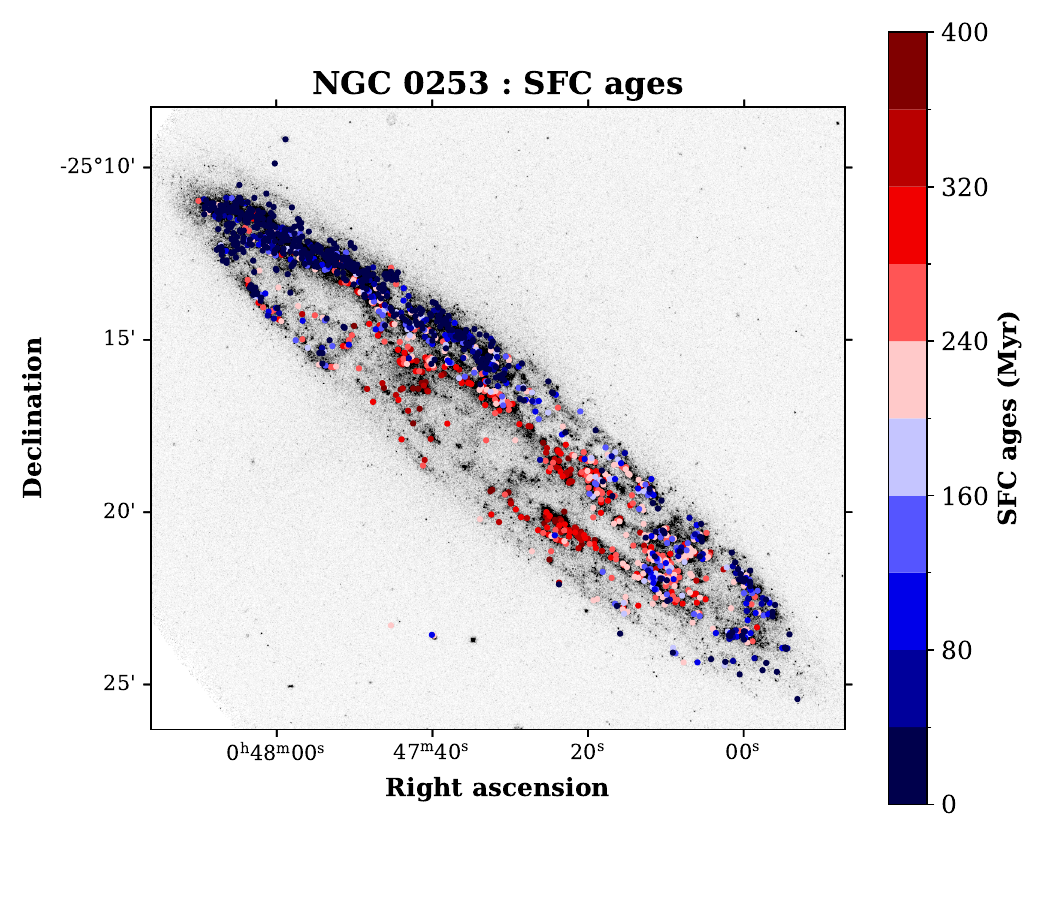}
        \hfill
		\includegraphics[width=0.32\linewidth]{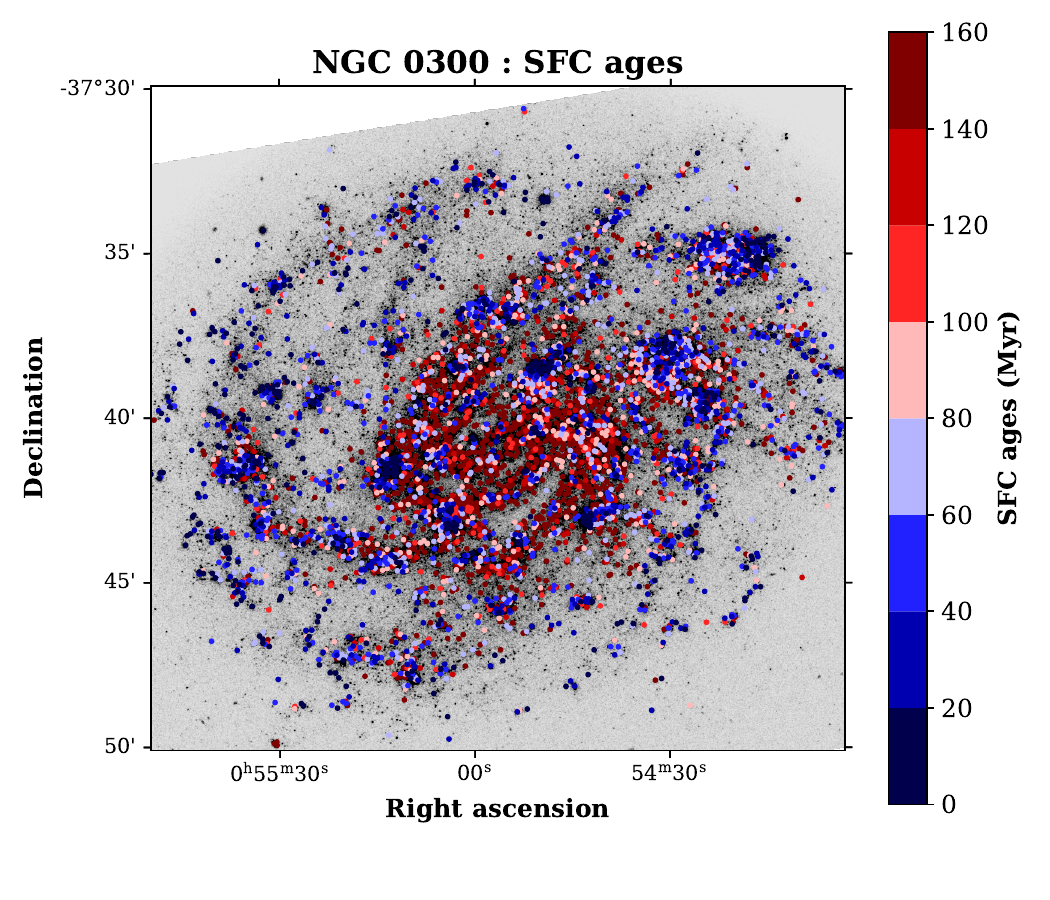}
        \hfill
		\includegraphics[width=0.32\linewidth]{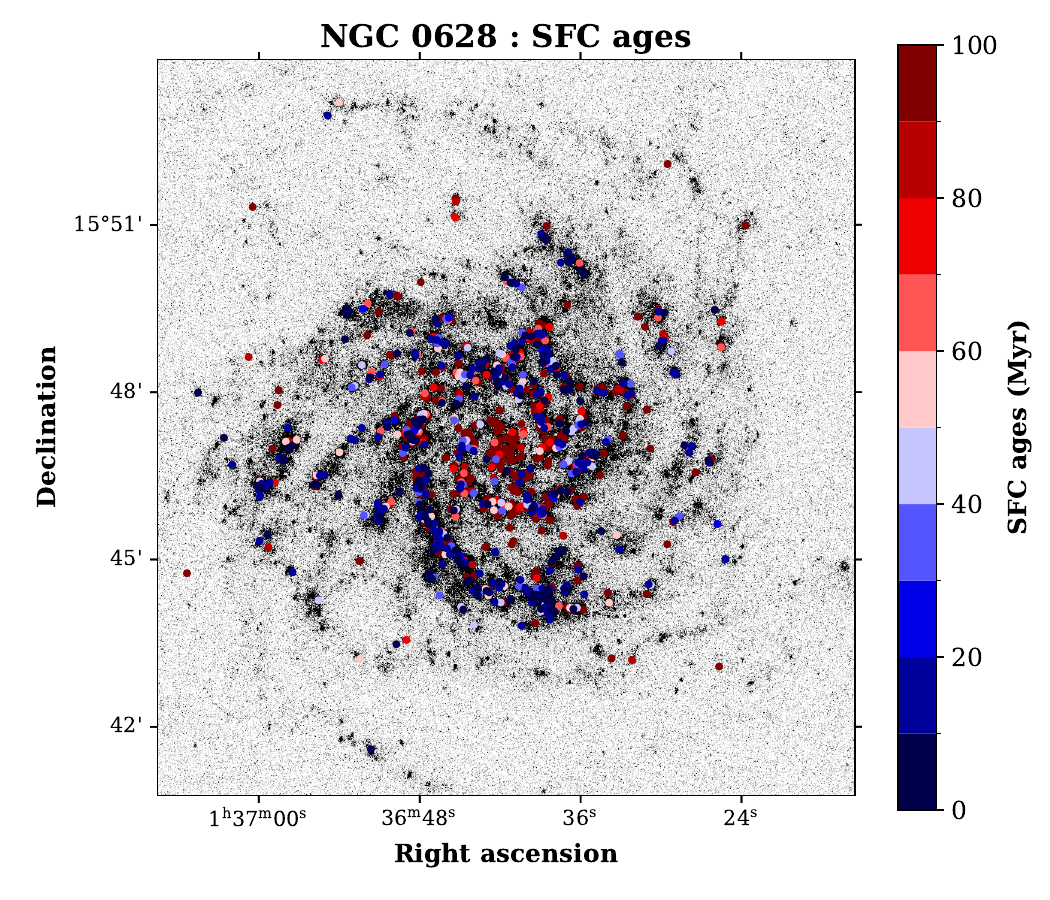}
	\vfill	
	\vfill	
		\includegraphics[width=0.32\linewidth]{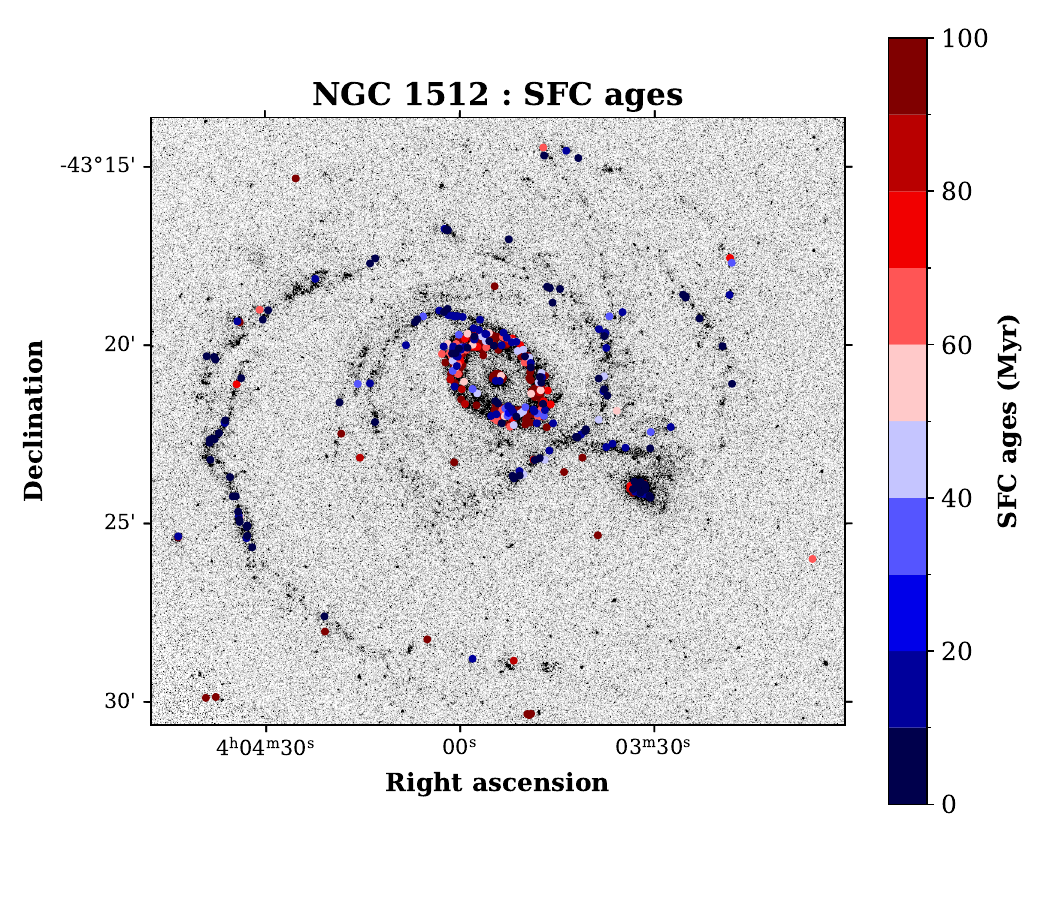}
	\hfill
		\includegraphics[width=0.32\linewidth]{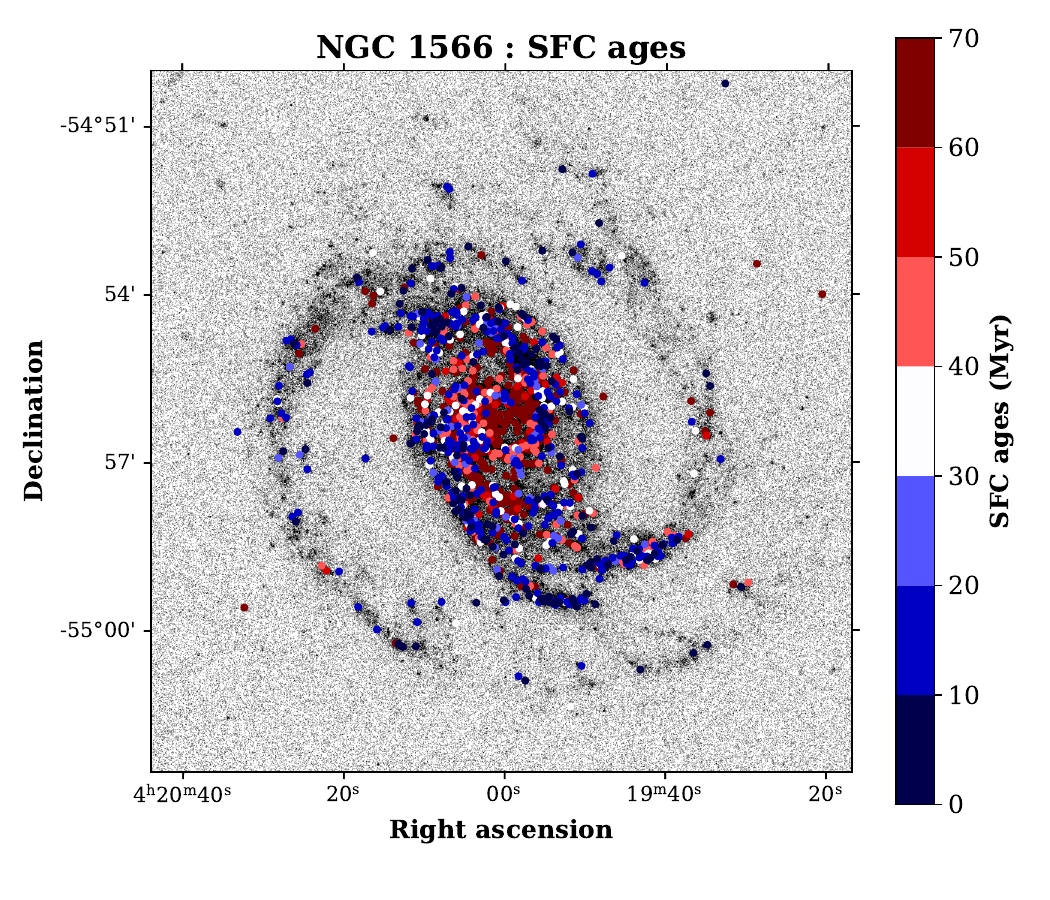}
        \hfill
		\includegraphics[width=0.32\linewidth]{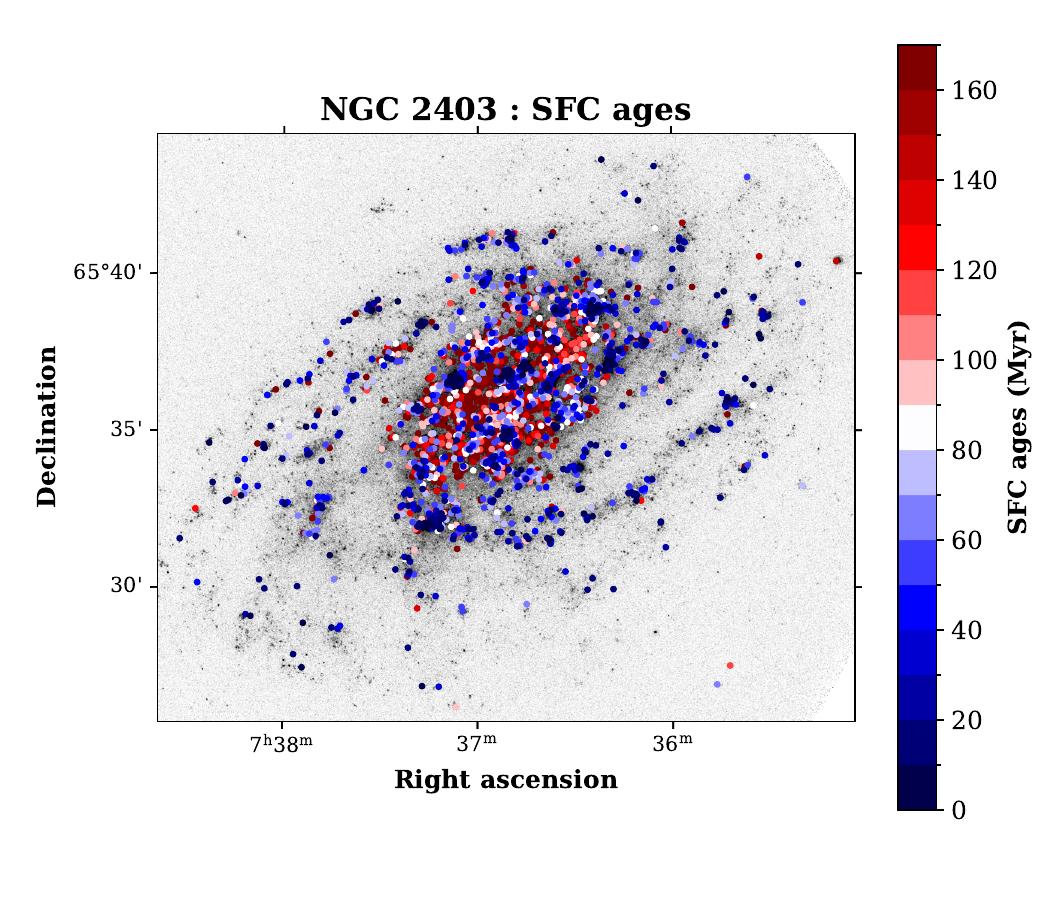}
        \vfill
        \vfill
		\includegraphics[width=0.32\linewidth]{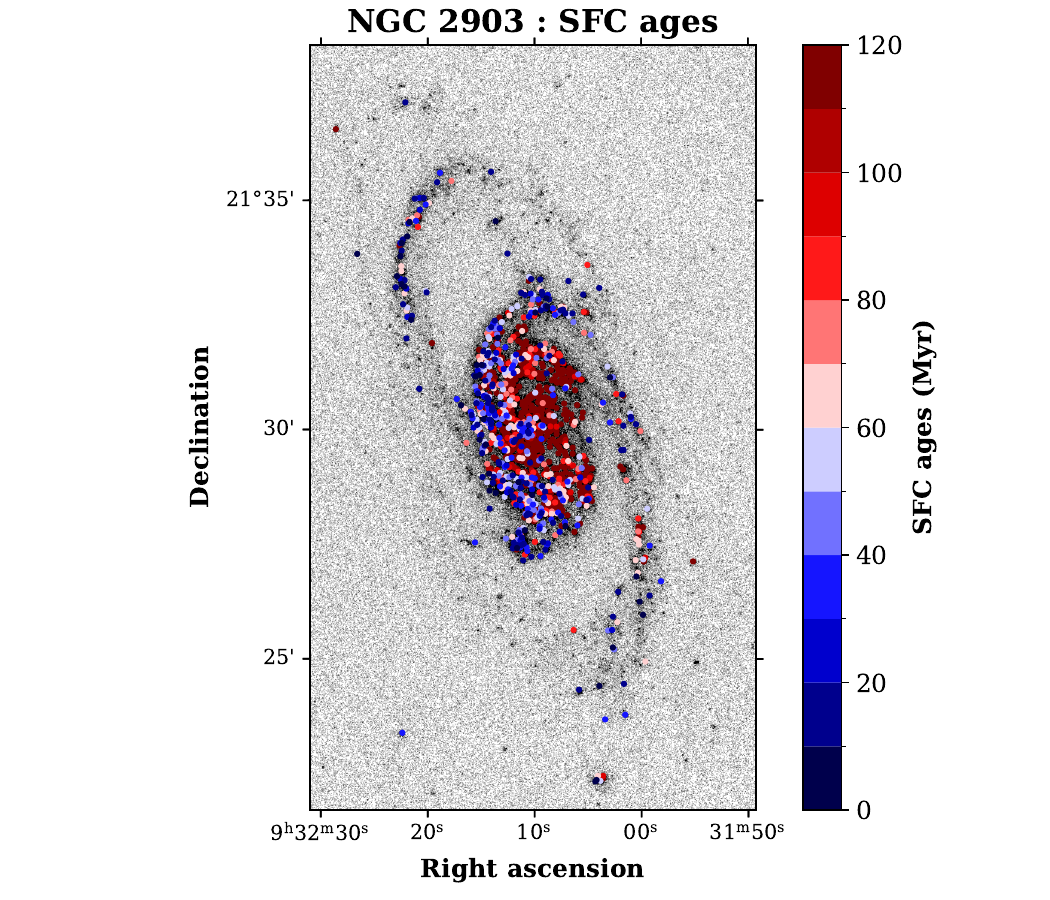}
	\hfill
		\includegraphics[width=0.32\linewidth]{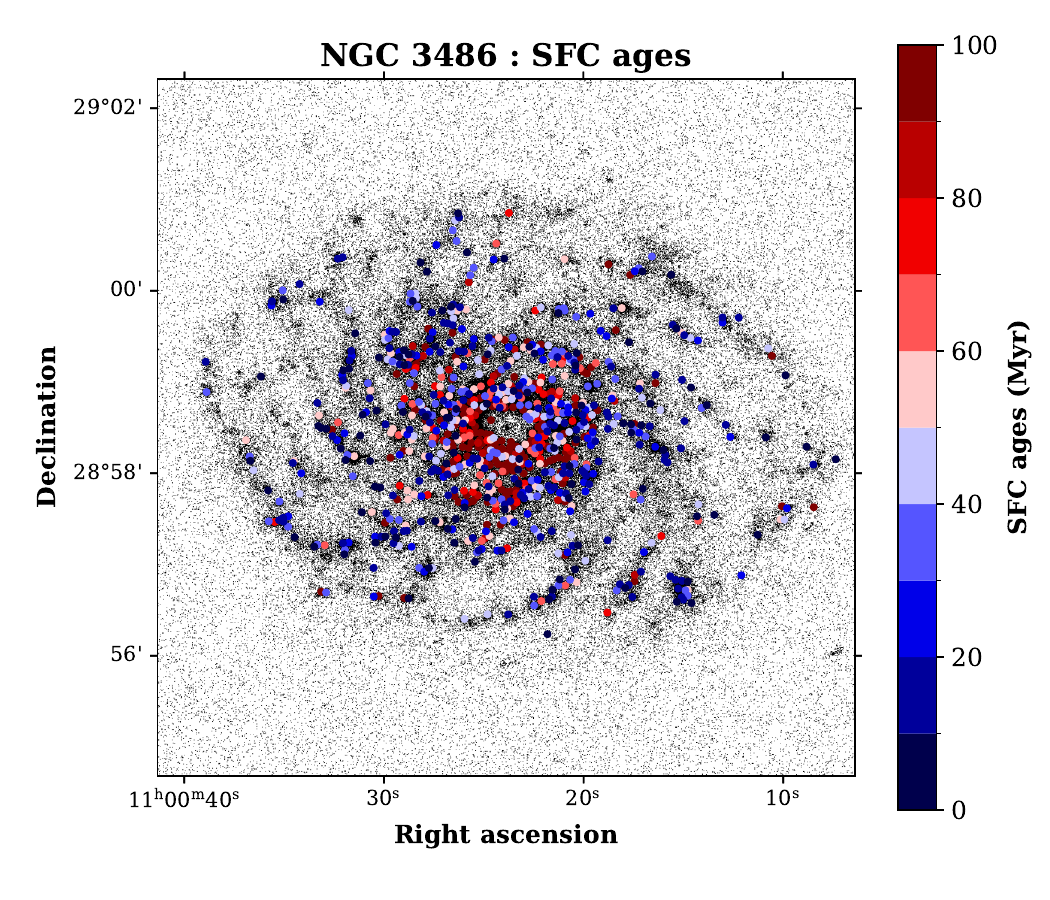}
        \hfill
		\includegraphics[width=0.32\linewidth]{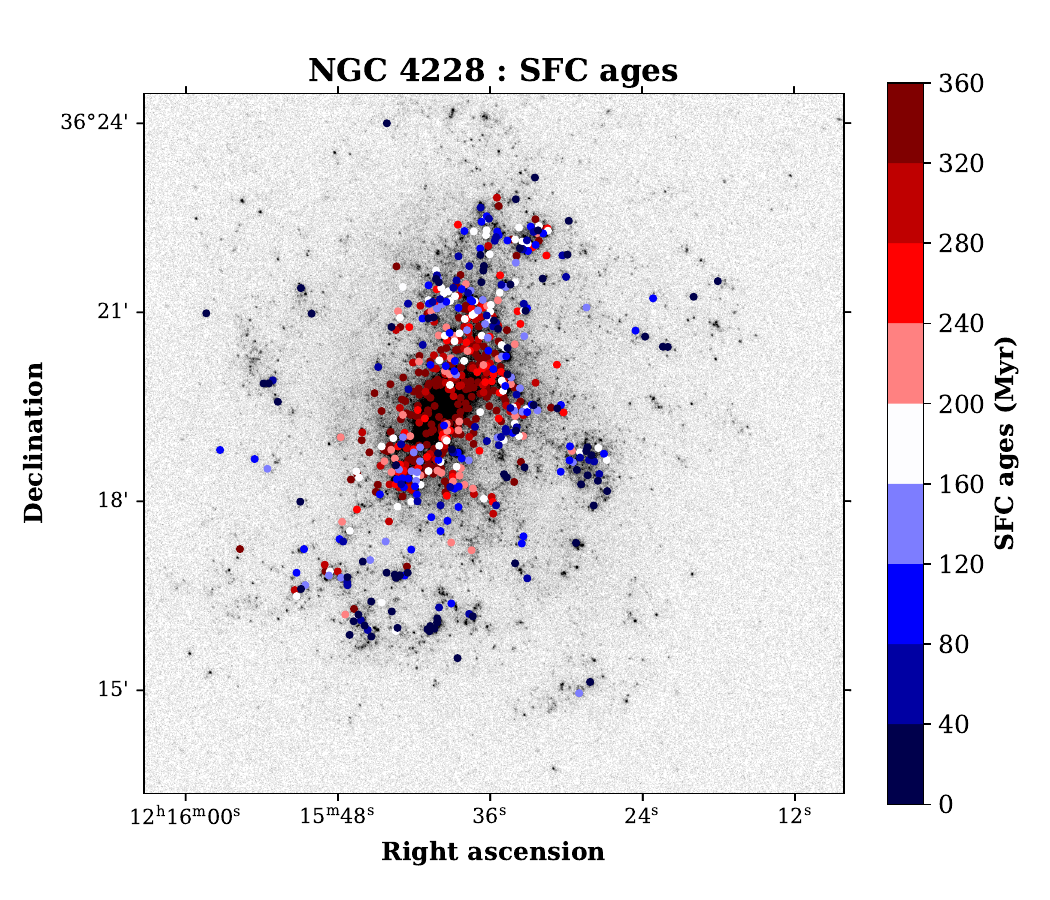}
        \vfill
        \vfill
		\includegraphics[width=0.32\linewidth]{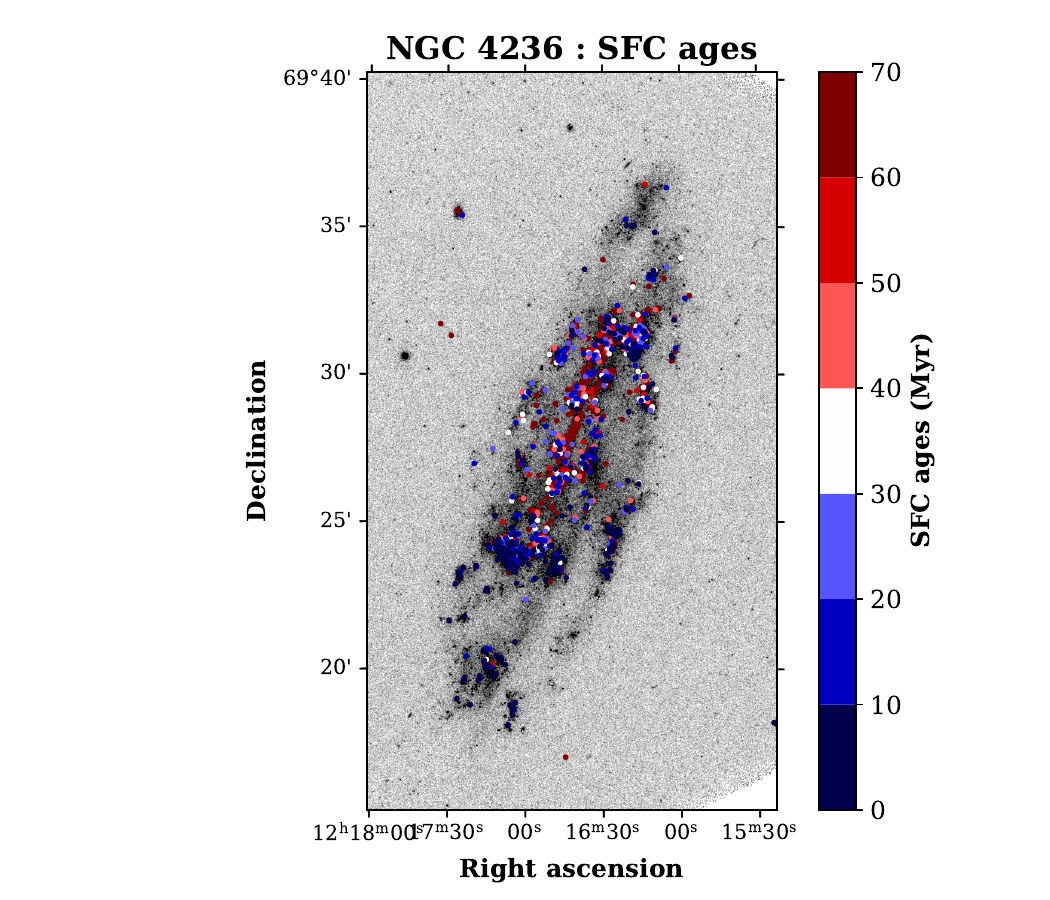}
        \hfill
            \includegraphics[width=0.32\linewidth]{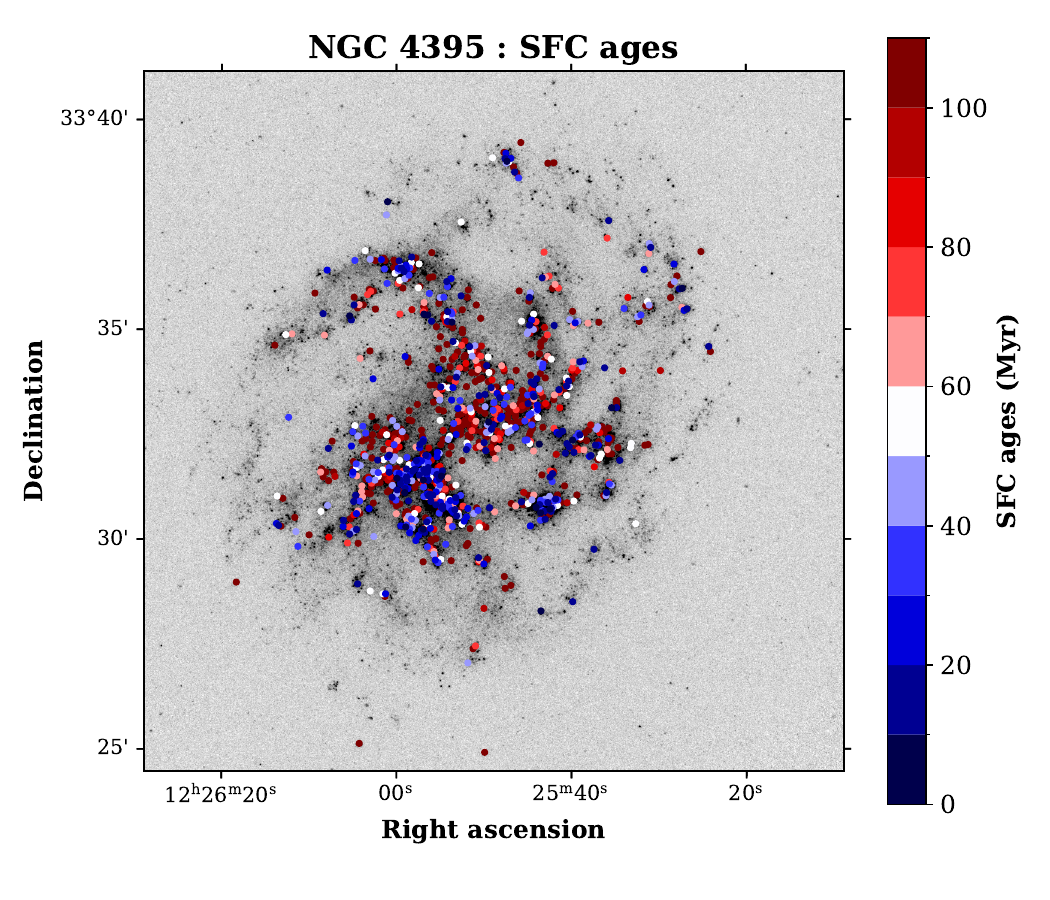}
        \hfill
		\includegraphics[width=0.32\linewidth]{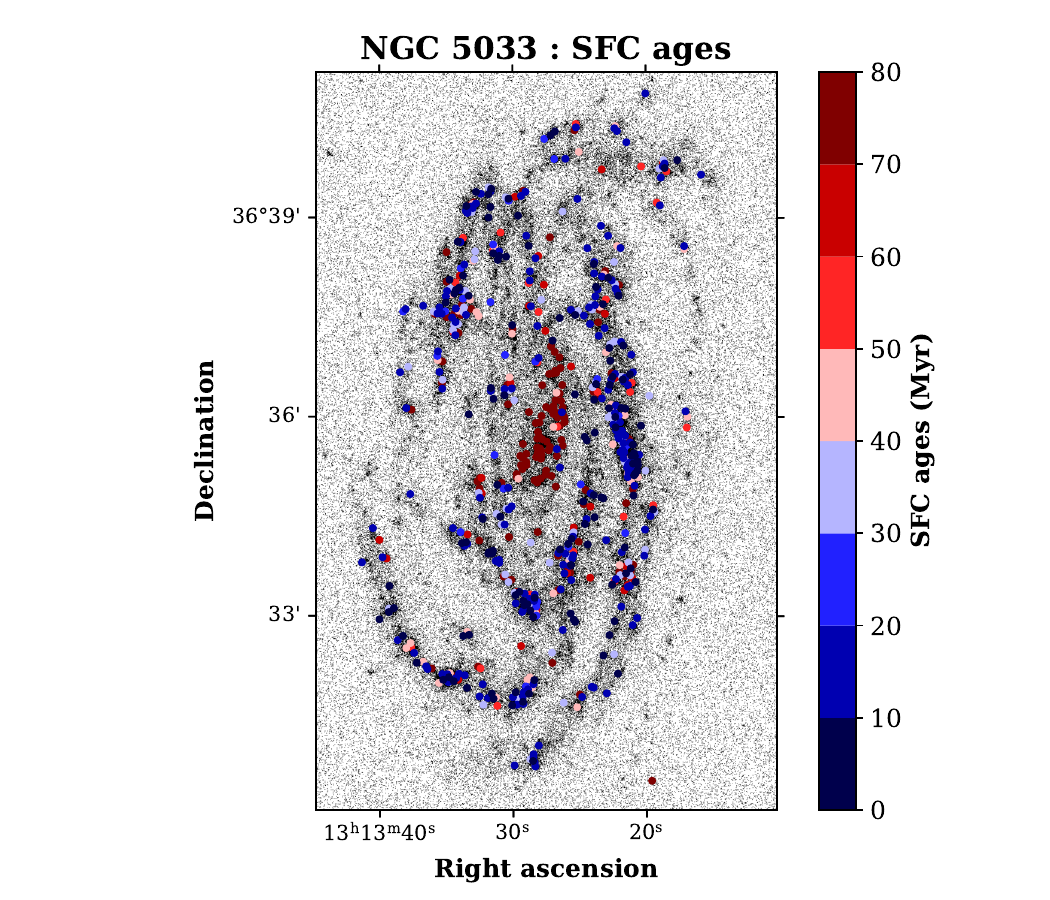}
    \caption{The derived age maps of the SFCs characterized in 12 out of the 17 galaxies. The colorbar represents SFC ages. We chose arbitrary age ranges in the colorbars in order to highlight the contrast between the positions of young and old SFCs, all formed within the past 400 Myrs. The complete list of photometric properties and ages of these SFCs is presented with this paper and it can be accessed from the supplementary material associated with this paper.}
  \label{fig_age_maps}
\end{figure*}

\begin{figure*}[hbt!]
      \centering
  	     \includegraphics[width=0.32\linewidth]{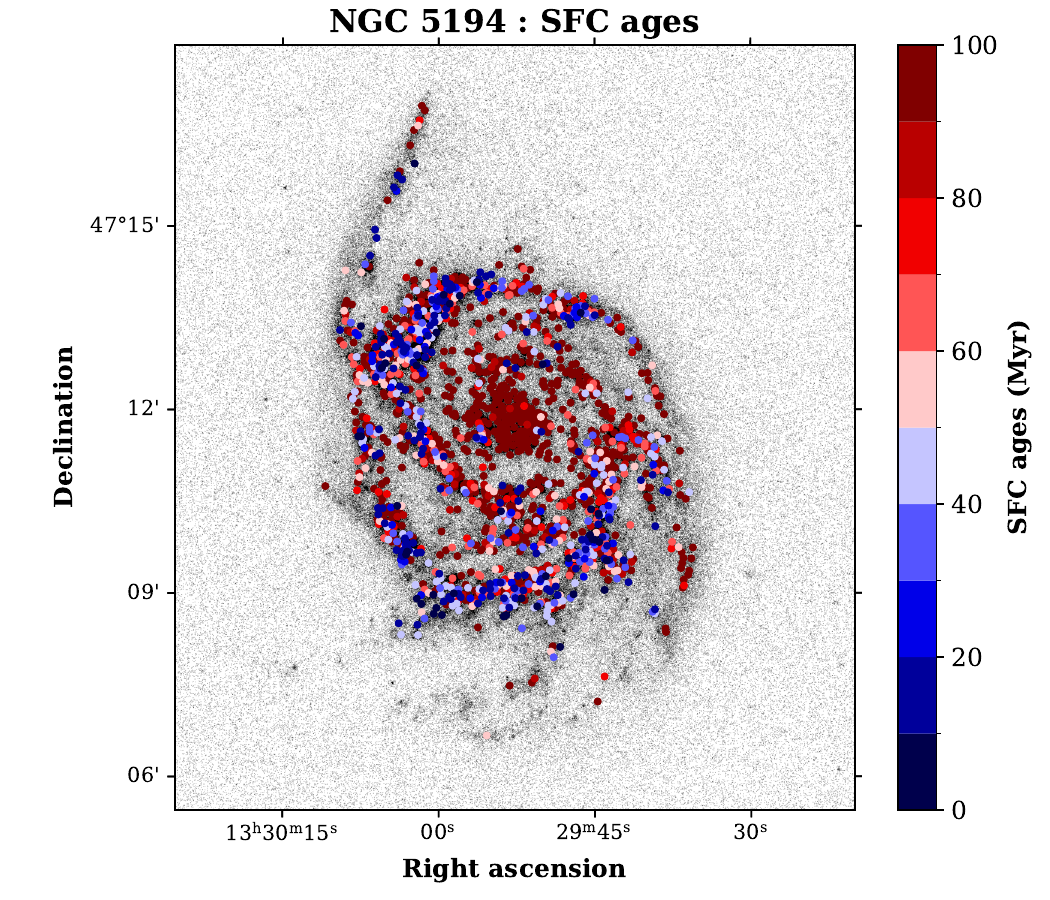}
	\hfill
		\includegraphics[width=0.32\linewidth]{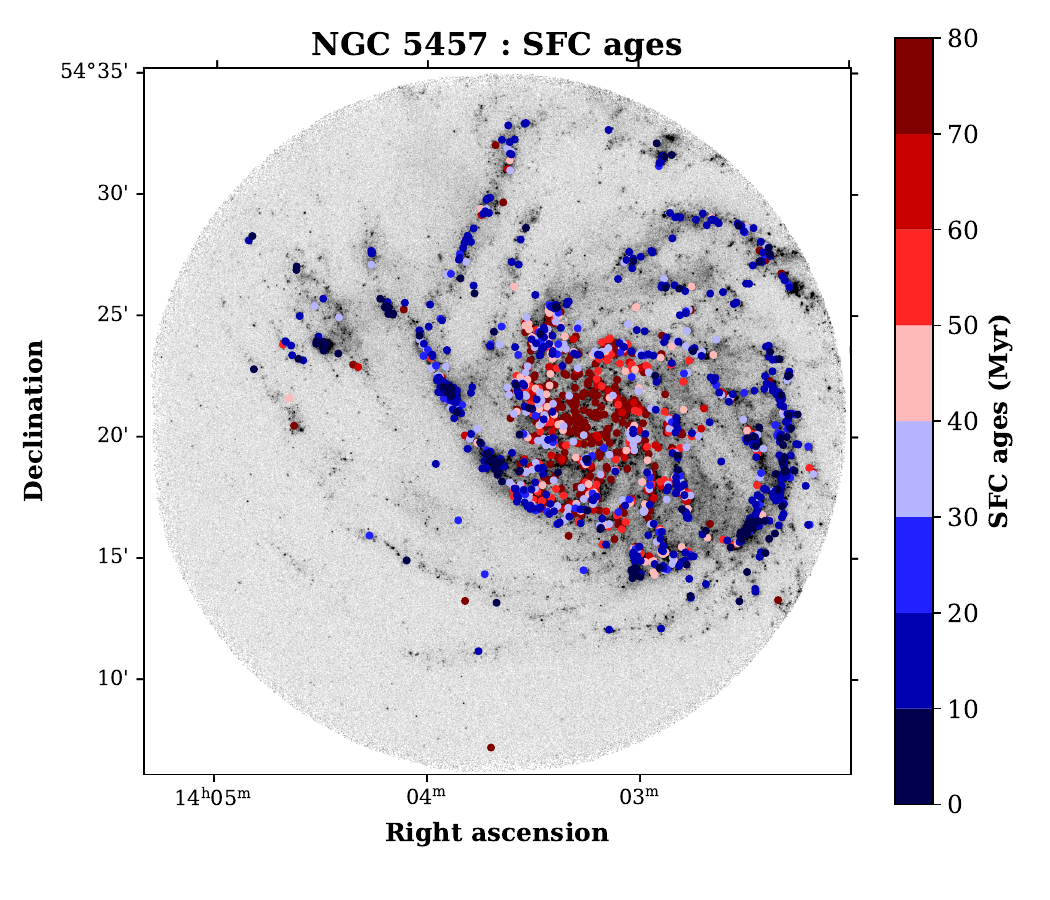}
        \hfill
		\includegraphics[width=0.32\linewidth]{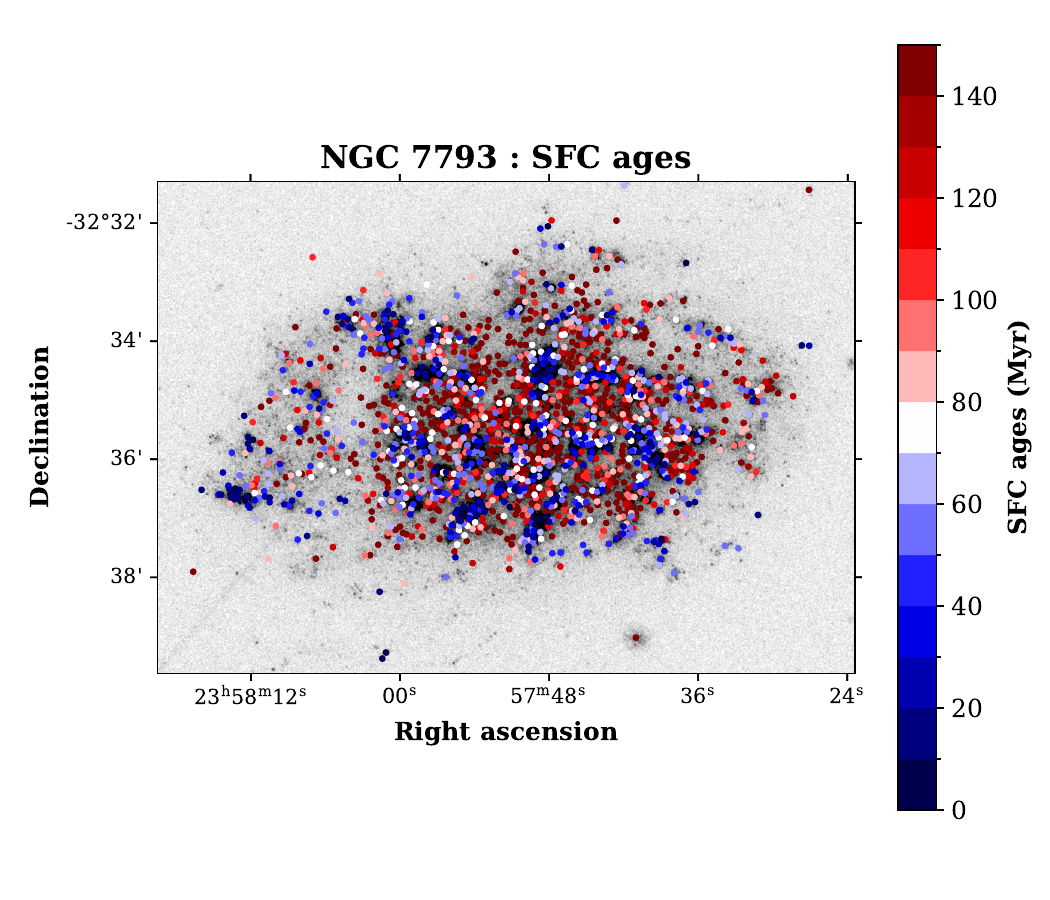}
        \vfill
        \vfill
		\includegraphics[width=0.32\linewidth]{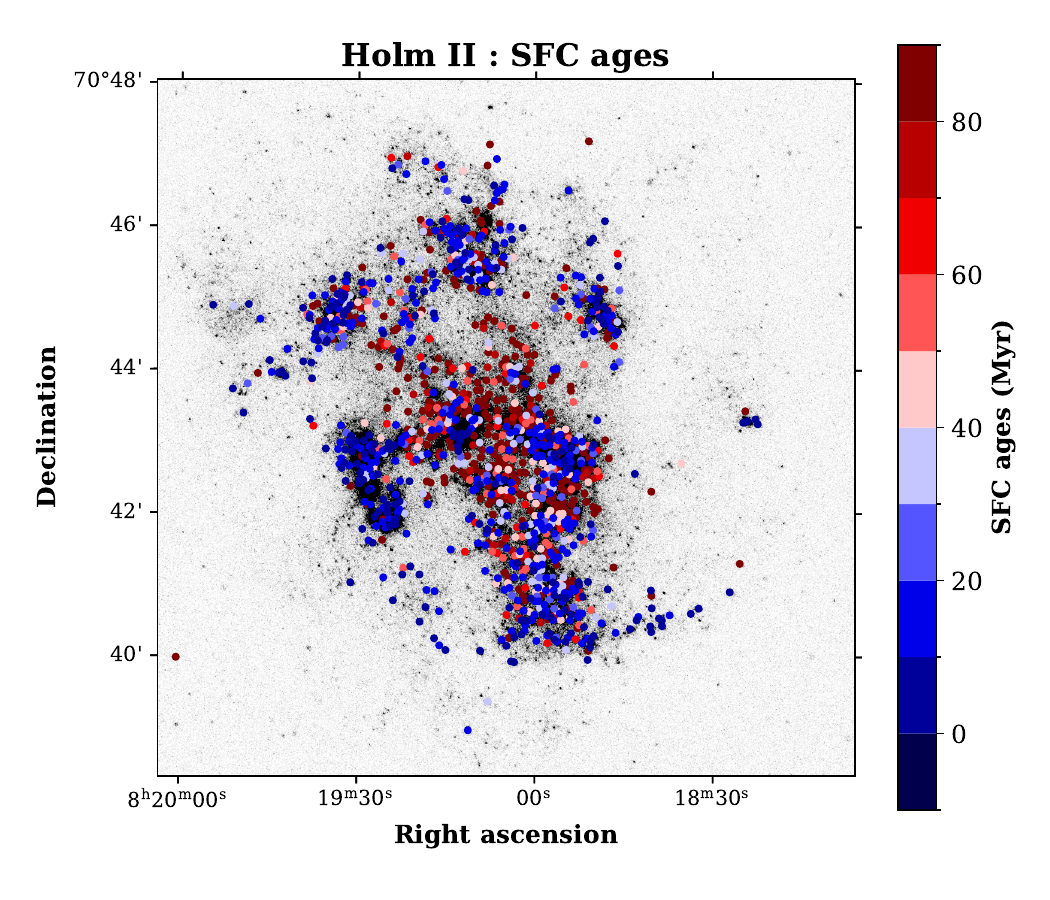}
	\hfill
		\includegraphics[width=0.32\linewidth]{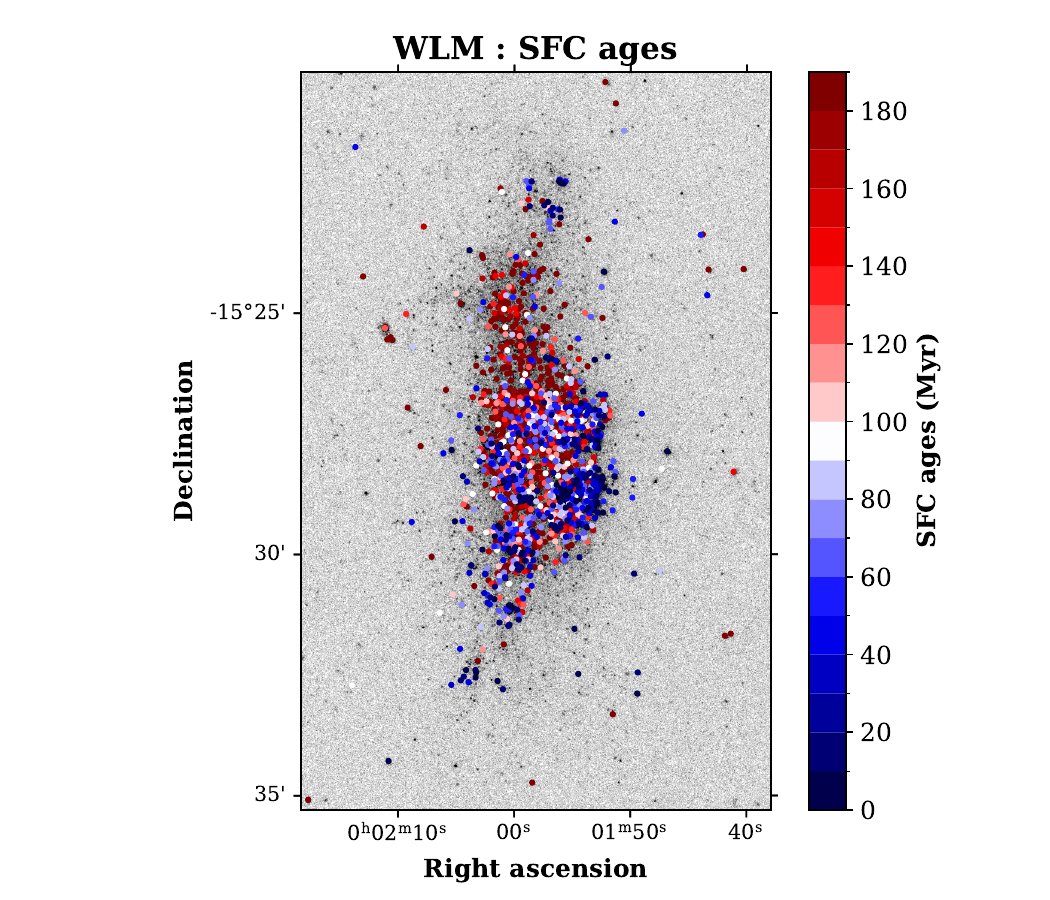}
    \caption{The derived age maps of the SFCs characterized in the remaining 5 out of the 17 galaxies.}
  \label{fig_age_maps_contd}
\end{figure*}  

The spatial age maps (Figures \ref{fig_age_maps}, \ref{fig_age_maps_contd}) for the majority of our spiral galaxies point towards the existence of a negative radial age gradient where the majority of the old SFCs (red colored points) are located at smaller galacto-centric radii and young SFCs (blue colored points) are predominantly found at larger radii (From this section onward, the classification of SFCs as "young” or “old” SFCs is made in a relative sense. We note that all these SFCs together represent recently formed stellar populations, within the past 400 Myr).

Clear negative age gradients were observed in all of our spiral galaxies (see the best fit lines and slope values in Figure \ref{fig_age_gradients}), which is a well-established signature of inside-out growth mechanism of disc galaxies - the galaxies acquire their inner regions quite early in their evolutionary history and their outer regions (that still harbor active star formation) are accumulated by continuous accretion of gas from the intergalactic medium in the later stages of the galaxy's evolution \citep{1976MNRAS.176...31L, 1991ApJ...379...52W, 1998MNRAS.295..319M,2006MNRAS.366..899N}. In the inside-out growth scenario, either the star-forming gas does not accumulate in the outer parts of galaxies until later times, or the star formation timescales in outer discs are longer. This scenario leads to predictions of negative age and metallicity gradients in galaxies, which have now been proven observationally using FUV, NUV data from GALEX \citep{bianchi2005recent, 2005ApJ...619L..67T, 2005ApJ...619L..79T, 2007ApJ...658.1006M, 2011ApJ...743..137B}, using HST observations \citep{2010ApJ...712..858G, Radburn-Smith_2012, 2019ApJ...878....1S}, and using integral field-unit observations \citep{2017A&A...608A..27G}. 

We observed that the youngest SFCs are often co-spatial with the position of spiral arms at all radii, in all spiral galaxies. This highlights the crucial role that spiral arm driven physical processes, such as spiral wave compression, shocks, and cloud collisions, play in the formation of the youngest stellar populations in a galaxy \citep{1964ApJ...140..646L, 2014ApJ...781...11E, 2018MNRAS.478.3590S, Mondal_2021, 2021ApJ...914...54Y, 2022MNRAS.516.2171U}. It can also be observed that the slightly older SFCs are located away from the spiral arms. This is consistent with the picture that it takes stars a few tens of Myrs to traverse the spiral arm of a galaxy. As the neutral Hydrogen (HI) discs of galaxies are much more extended than their stellar discs, the presence of young SFCs at large galactocentric radii may also be indicative of the cold gas accretion and its subsequent collapse to form young stars \citep{2013seg..book....1K, 2019ApJ...875...54H}. 

The flocculent spiral galaxy NGC 7793 shows a nearly uniform distribution of young and old SFCs as a function of radius and it exhibits one of the shallowest age gradients in our sample. \cite{Mondal_2021} had shown that the scale radius of NGC 7793 increases from optical to NUV to FUV wavebands, indicating inside-out growth. Similarly, \cite{2019ApJ...878....1S} found evidence for inside-out growth of NGC 7793 by analyzing resolved star formation history in different regions of the galaxy using the LEGUS-HST data. However, it appears that the negative age-gradient (inside-out growth indicator) in this flocculent spiral galaxy is not as prominent as the majority of the spiral galaxies in our sample. \cite{2005ApJ...619L..79T, 2005ApJ...619L..67T} had shown that typically, the stellar population in galaxies is found to be younger with increasing galactocentric radius but the steepness of this effect can vary from galaxy-to-galaxy. In another flocculent spiral galaxy, M33 (not studied in this paper), for instance, the negative age gradient was found to be small, and the FUV$-$NUV color remained roughly constant over a large range of galactocentric radius. This implied that the star formation activity in the galaxy over the past $\sim$200 Myr has been globally spread-out. NGC 7793 may be experiencing the same phenomenon. 

The young SFCs in NGC 7793 show good co-spatiality with its flocculent spiral arms, which agrees with similar observations made in \citep{Mondal_2018, 2018MNRAS.481.1016G, 2018MNRAS.478.3590S, 2019ApJ...878....1S}. Flocculent arms of a galaxy are believed to be not created by any coherent spiral wave, rather, gas collapses along the clumpy, disjoint, flocculent arms of the galaxy under the action of local gravitational instabilities \citep{Elmegreen_2006, 2014ApJ...781...11E, 2018PASP..130g2001G}. This leads to star formation along the flocculent arms of such galaxies, and the young SFC distribution of the flocculent spirals in our sample - NGC 0300, NGC 2403, NGC 3486, and NGC 7793 - clearly demonstrates this phenomenon.

For two of our most inclined galaxies NGC 2903 and NGC 0253, we observed a rough division of young and old SFCs in our age maps, where SFCs on one side of the major axis are predominantly old (e.g. south-eastern half of NGC 0253 and north-western half of NGC 2903) and SFCs on the other side of major axis are predominantly young (e.g. north-western half of NGC 0253 and south-eastern half of NGC 2903). These observations are unlikely to be physical in nature. We believe that such trends can be driven by insufficient dust attenuation correction due to the galaxy's high inclination. In such a scenario, diffuse dust located above or below the galaxy's thin stellar disc can cause the SFCs on the far side of the galaxy to suffer much more attenuation, so they consistently appear older (as the light path length traversing through the dust is longer), as compared to the SFCs on the near side which consistently appear younger. \cite{2013ApJ...771...62K} observed a similar contrast between the near- and far-side dust attenuation values for highly inclined galaxies NGC 2146 and NGC 7331, which supports our reasoning. The spatial age map of NGC 0253, in particular, appears to be the most affected by its high inclination induced attenuation problem. The young SFCs on the north-east side of the galaxy appear to be younger (darker shade of blue) than the young SFCs on the south-west side (lighter shades of blue). \cite{2010ApJ...725.1342D} had suggested that the high dust content of NGC 0253 strongly affects its appearance in UV wavelengths. Our $A_V$ maps (Figure \ref{fig_AV_maps}, first figure) and median $A_V$ values (Figures \ref{fig_AV_distribution}) prove that among our galaxy sample, NGC 0253 suffers the most strongly from dust attenuation. We posit that perhaps the unaccounted-for dust attenuation between the near- and far-side of our inclined systems NGC 2903 and NGC 0253 can give rise to the observed division between the SFC ages with respect to the galaxy's major/minor axis. This behavior of NGC 2903 and NGC 0253 points towards a limitation in our dust attenuation correction method in highly inclined galaxies. However, age determination for highly inclined galaxies can prove challenging even with the most sophisticated age-determination techniques (\citealt{2013ApJ...771...62K}; MB16). 

\subsection{Dwarf Irregular galaxies}
\label{subsec_dirr}
The three dwarf irregular galaxies in our sample show quite diverse age demographics (see spatial age maps in Figure \ref{fig_age_maps} and \ref{fig_age_maps_contd}). In NGC 4228, the old SFCs appear to be uniformly distributed in the main body of the galaxy, and most of the SFCs that are located in the patchy outskirts of the galaxy appear to be young. This is similar to the behavior shown by the spiral galaxies in our sample - consistent with the inside-out growth scenario. In Holmberg II, both the young and old SFCs can be observed at all galactocentric radii. However, its young SFCs appear to be clustered together into distinct groups, particularly in the patchy outskirts of the galaxy. Its old SFC population appears to be uniformly distributed within the galaxy and their distribution follows what appears to be the main body of the galaxy. For comparison, in another galaxy NGC 1313, which has an irregular, disturbed morphology in its outskirts quite similar to Holmberg II, \cite{2026arXiv260222860A} recently studied the age demographic using UV and optical data. They too found that, similar to Holmberg II, the irregular, patchy outskirts of NGC 1313 host the youngest stellar populations, whereas the old SFCs dominate the stellar population in the inner regions of the galaxy. Finally, in WLM, the old SFC population can be found all over its main body, and some isolated, old SFCs are scattered farther away from the galaxy, too. The young SFCs of WLM are clustered together in three to four regions in the southern half of the galaxy, and one clustering of young SFCs can be observed at the northern most part of the galaxy. 

\cite{2024ApJ...977...20R} (and references therein) suggest that dwarf galaxies, too, are formed inside out and are expected to exhibit negative age gradients. This is supported by our observed age gradient slopes. However, in dwarf galaxies, bursty star formation can induce strong stellar feedback that imparts outward momentum to star-forming gas. The stars formed from this gas will migrate towards outer regions owing to their inherently outward radial velocity. This can flatten or even erase the negative age gradient in dwarf galaxies, which may explain why the age-gradient slopes of Holmberg II and WLM are shallower than those of our massive spiral galaxies. However, this mechanism may not be strong enough to produce the same effect in large, spiral galaxies \citep{2019MNRAS.490.1186G}. Overall, the diversity in the distribution of different-aged stellar populations in our small sample of three dwarf irregular galaxies may be indicative of their more complex galaxy assembly history, as compared to the spiral galaxies in our sample. Our age maps for dwarf irregular galaxies also appear to be consistent with the stochastic nature of star formation in them - often triggered by internal processes such as local gas instabilities and stellar feedback \citep{2019MNRAS.484.4897C, Mondal_2018, 2024MNRAS.530.2199A}.

\section{Summary and Future Plans}
\label{sec_summary}
In this paper, we have presented a complete galaxy coverage, FUV and NUV based catalog of $\sim$25,000 recently formed ($\lesssim$400 MYrs) star-forming clumps (SFCs) in 17 nearby galaxies. The production of this catalog is motivated by a dearth of systematic UV-based studies of recent star formation at high spatial resolution in a large, diverse galaxy sample (spanning a wide range in morphology and mass) with a homogeneous methodology (including SFC detection, characterization, and a treatment of spatially varying dust attenuation correction). Such studies are needed to better understand different galaxy environments and physical processes responsible for galaxy evolution e.g. starbursts, feedback, interactions, and quenching.

Our sample of 17 galaxies, all located within 20 Mpc, includes 8 massive, classic spiral galaxies, 6 intermediate-mass, flocculent spiral galaxies,  and 3 dwarf irregular galaxies. In this work, we have utilized the 1.5\arcsec~angular resolution, archival FUV and NUV observations of our galaxies from the UltraViolet Imaging Telescope (UVIT) \citep{2012SPIE.8443E..1NK}. We used the Astrodendro package on the UVIT FUV images of our galaxies to identify SFCs using a homogeneous detection criteria. We performed photometry on the FUV and NUV images to estimate the corresponding magnitudes of the SFCs. Next, by comparing dust attenuation corrected, observed UV color-magnitude diagrams (UV CMD) against the synthetic Starburst99 simple stellar population model UV CMD \citep{1999ApJS..123....3L}, we estimated the ages of the identified SFCs. The dust attenuation correction was performed in a spatially resolved manner, all the while accounting for the age-dependence of the attenuation correction calibration coefficients. For this purpose, we followed the method outlined by \cite{2016A&A...591A...6B} and utilized a combination of UVIT FUV, 2MASS J-band and MIPS 24$\mu$ observations to create full galaxy coverage, 6\arcsec~angular resolution $A_V$ maps for our 17 galaxies. We applied FUV and NUV magnitude error cuts on the detected SFCs so as to choose only the brightest SFCs with reasonably small age-errors in the finalized SFC catalog. This ultimately led to an extensive, finalized catalog of $\sim$25000 SFCs. We make our SFC catalog and $A_V$ maps (as value-added products) for the 17 galaxies publicly available with this paper.

Along with presenting the methodology for constructing the SFC catalog, we also provided a discussion of the age demographics of different stellar populations in our galaxies. This discussion focused on the insights these age distributions can provide on galaxy growth, assembly history, the role of spiral arms in star formation, and the different drivers of star formation in galaxies of varied morphology and mass. With our age maps, we found trends consistent with several well-established phenomena spanning the entire galaxy mass and morphology range, such as the inside-out growth of disc galaxies, association of spiral arms with star formation, local gravitational instabilities causing star formation in the flocculent spiral arms, and the stochastic nature of star formation in dwarf galaxies. As our galaxy sample lies between 1-20 Mpc, the identified SFCs using UVIT's $\sim$1.5\arcsec~ resolution data likely correspond to different physical structures across the sample (e.g. individual clusters, stellar associations, exteneded unbound complexes). In Appendix \ref{apdx_robust_against_resolution}, we test the robustness of our results by degrading higher-resolution images to a common physical scale and demonstrate that our main trends are not driven by resolution effect.

In the future, this catalog can be coupled with the insights drawn from multi-wavelength data to improve the understanding of many other aspects of galaxy evolution, such as the impact of galactic dynamics on different stellar populations within galaxies, secular versus interaction driven galaxy evolution, enhancement/quenching of star formation and the UV luminosity function of star-forming regions in galaxies. Though the SFC catalog generated in this study is at $\sim$1.5\arcsec~angular resolution, due to our full galaxy coverage and our use of FUV data, this catalog can serve as a complementary resource to the excellent star cluster catalogs being created by large galaxy surveys such as LEGUS \citep{2015AJ....149...51C}, PHANGS \citep{2022ApJS..258...10L}, and Galaxy UV Legacy Project (GULP) (PI: Elena Sabbi; \citealt{2026arXiv260306510S}), to name a few.

In \cite{2025A&A...693A.188S} (Paper I), we had demonstrated with a pilot sample of 4 galaxies that UVIT's high resolution, FUV coverage, and large 28$^\prime$ field-of-view uniquely allows us to characterize the star formation hierarchies in galaxies - over scales ranging from tens of parsecs (pcs) up to tens of kiloparsecs (kpcs). The current paper (Paper II) aims to aid the expansion of the work done in Paper I onto this larger sample of 17 galaxies, with the advantage of a greater morphological diversity. In Paper III (Shashank et al, in prep.), we have utilized this SFC catalog to derive new insights about stellar hierarchies of these 17 galaxies, such as their limiting scale within galaxies, dispersal timescales, and how these are affected by host galaxy properties (stellar mass, size, shear, and nature of spiral arms). In the future, we will target a pre-selected sample of $\sim$30 more galaxies, which have archival UVIT FUV-only observations. We will combine UVIT FUV data with multi-band optical, narrow-band H$\alpha$, and multi-band IR observation and characterize the SFC populations in these galaxies using the spectral energy distribution (SED) fitting code Code Investigating GALaxy Emission (CIGALE; \citealt{2019A&A...622A.103B}). Our overarching goal with this larger sample will be to improve our understanding of the stellar population demographic and star formation hierarchies in nearby galaxies - spanning an even broader parameter space in morphology, mass, size, and metallicity. 

Significant improvements can be made in our methodology with a higher resolution ultraviolet imaging telescope with FUV and NUV imaging capabilities and a large field-of-view. With only a single pointing, such a telescope should allow the detection and characterization of thousands of parsec-scale, young star clusters within the full spatial extent of nearby galaxies. To this end, future missions such as the INdian Spectroscopic and Imaging Space Telescope (INSIST) \citep{2022JApA...43...80S} and Cosmological Advanced Survey Telescope for Optical and Ultraviolet Research (CASTOR)  \citep{10.1117/1.JATIS.11.4.042202} will allow us to observe a larger, untapped reservoir of star-forming galaxies in the local universe (which do not have existing high resolution UV observations) and further our knowledge of star formation related phenomena in galaxies. INSIST's proposed sub-arcsecond angular resolution will also facilitate the study of different stellar populations in galaxies beyond 20 Mpc and up to a few 100 Mpc.\

\section*{Data availability}
\label{sec_data_availability}

This publication uses data from the UVIT - one of the five science instruments onboard the AstroSat satellite of the Indian Space Research Organization (ISRO). The UVIT data is archived at the Indian Space Science Data Centre (ISSDC). The level 1 UVIT data for our galaxies can be extracted at (\href{https://astrobrowse.issdc.gov.in/astro_archive/archive/Home.jsp} {https://astrobrowse.issdc.gov.in/astro\_archive/archive/
Home.jsp)} using the galaxy name and RA, DEC information given in Table \ref{table_galaxy_properties}. The science-ready NIR and MIR images used in this paper were directly taken from the NASA Extragalactic Database (NED).

\section*{Acknowledgement}
\label{sec_acknowledgement}

GS and SS thank Dr. Alessandro Boselli and Dr. Médéric Boquien for their constructive suggestions regarding the dust attenuation corrections, which were crucial in constraining the SFC ages. GS also thanks D. Renu, Rakshit Chauhan and Sanal Ananthu for their help duing UVIT data analysis. GS and SS also acknowledge the PIs of the UVIT data used in this paper for targeting these majestic galaxies and making this work possible. SS acknowledges support from the Science and Engineering Research Board of India through the POWER research grant (SPG/2021/002672) and support from the Alexander von Humboldt Foundation. S.H.M. acknowledges the support of NASA grant No. 80NSSC20K0500, NSF grant AST-2009679, and the Simons Foundation. C.M. acknowledges support from the National Science and Technology Council, Taiwan (grant NSTC 112-2112-M-001-027-MY3) and the Academia Sinica Investigator award (grant AS-IA-112-M04). We acknowledge the use of Python (\citealt{python09}), ASTROML (\citealt{2012cidu.conf...47V}), scikit-learn (\citealt{JMLR:v12:pedregosa11a}), Matplotlib (\citealt{Hunter07}), NumPy (\citealt{NumPy20}), SciPy (\citealt{SciPy20}), AstroPy (\citealt{astropy_2018}), Astrodendro (\href{http://www.dendrograms.org/}{http://www.dendrograms.org/)}, photutils (\citealt{larry_bradley_2024_10967176}) and CCDLAB (\citealt{2017PASP..129k5002P}).

\bibliography{references2}

@BOOK{1936rene.book.....H,
       author = {{Hubble}, E.~P.},
        title = "{Realm of the Nebulae}",
         year = 1936,
       adsurl = {https://ui.adsabs.harvard.edu/abs/1936rene.book.....H}
}

@BOOK{1991rc3..book.....D,
       author = {{de Vaucouleurs}, Gerard and {de Vaucouleurs}, Antoinette and {Corwin}, Jr., Herold G. and {Buta}, Ronald J. and {Paturel}, Georges and {Fouque}, Pascal},
        title = "{Third Reference Catalogue of Bright Galaxies}",
         year = 1991,
       adsurl = {https://ui.adsabs.harvard.edu/abs/1991rc3..book.....D}
}

@BOOK{2007dvag.book.....B,
       author = {{Buta}, Ronald J. and {Corwin}, Harold G. and {Odewahn}, Stephen C.},
        title = "{The de Vaucouleurs Atlas of Galaxies}",
         year = 2007,
       adsurl = {https://ui.adsabs.harvard.edu/abs/2007dvag.book.....B}
}

@ARTICLE{2026arXiv260222860A,
       author = {{Ananthu}, Sanal and {Shashank}, Gairola and {Subramanian}, Smitha and {Jayanth}, Rao C. and {Menon}, Shyam H. and {Mondal}, Chayan and {Muraleedharan}, Sreedevi},
        title = "{A FUV - optical approach for studying hierarchical star formation in nearby galaxies with UVIT}",
      journal = {arXiv e-prints},
         year = 2026,
        month = feb,
          eid = {arXiv:2602.22860},
        pages = {arXiv:2602.22860},
          doi = {10.48550/arXiv.2602.22860},
archivePrefix = {arXiv},
       eprint = {2602.22860},
 primaryClass = {astro-ph.GA},
       adsurl = {https://ui.adsabs.harvard.edu/abs/2026arXiv260222860A}
}

@ARTICLE{2007ApJS..173..267S,
       author = {{Salim}, Samir and {Rich}, R. Michael and {Charlot}, St{\'e}phane and {Brinchmann}, Jarle and {Johnson}, Benjamin D. and {Schiminovich}, David and {Seibert}, Mark and {Mallery}, Ryan and {Heckman}, Timothy M. and {Forster}, Karl and {Friedman}, Peter G. and {Martin}, D. Christopher and {Morrissey}, Patrick and {Neff}, Susan G. and {Small}, Todd and {Wyder}, Ted K. and {Bianchi}, Luciana and {Donas}, Jos{\'e} and {Lee}, Young-Wook and {Madore}, Barry F. and {Milliard}, Bruno and {Szalay}, Alex S. and {Welsh}, Barry Y. and {Yi}, Sukyoung K.},
        title = "{UV Star Formation Rates in the Local Universe}",
      journal = {\apjs},
         year = 2007,
        month = dec,
       volume = {173},
       number = {2},
        pages = {267-292},
          doi = {10.1086/519218},
archivePrefix = {arXiv},
       eprint = {0704.3611},
 primaryClass = {astro-ph},
       adsurl = {https://ui.adsabs.harvard.edu/abs/2007ApJS..173..267S}
}

@ARTICLE{2011ApJ...733...74L,
       author = {{Lemonias}, Jenna J. and {Schiminovich}, David and {Thilker}, David and {Wyder}, Ted K. and {Martin}, D. Christopher and {Seibert}, Mark and {Treyer}, Marie A. and {Bianchi}, Luciana and {Heckman}, Timothy M. and {Madore}, Barry F. and {Rich}, R. Michael},
        title = "{The Space Density of Extended Ultraviolet (XUV) Disks in the Local Universe and Implications for Gas Accretion onto Galaxies}",
      journal = {\apj},
         year = 2011,
        month = jun,
       volume = {733},
       number = {2},
          eid = {74},
        pages = {74},
          doi = {10.1088/0004-637X/733/2/74},
archivePrefix = {arXiv},
       eprint = {1104.4501},
 primaryClass = {astro-ph.CO},
       adsurl = {https://ui.adsabs.harvard.edu/abs/2011ApJ...733...74L}
}

@ARTICLE{2004ARA&A..42..603K,
       author = {{Kormendy}, John and {Kennicutt}, Jr., Robert C.},
        title = "{Secular Evolution and the Formation of Pseudobulges in Disk Galaxies}",
      journal = {\araa},
         year = 2004,
        month = sep,
       volume = {42},
       number = {1},
        pages = {603-683},
          doi = {10.1146/annurev.astro.42.053102.134024},
archivePrefix = {arXiv},
       eprint = {astro-ph/0407343},
 primaryClass = {astro-ph},
       adsurl = {https://ui.adsabs.harvard.edu/abs/2004ARA&A..42..603K}
}

@ARTICLE{2019ARA&A..57..227K,
       author = {{Krumholz}, Mark R. and {McKee}, Christopher F. and {Bland-Hawthorn}, Joss},
        title = "{Star Clusters Across Cosmic Time}",
      journal = {\araa},
         year = 2019,
        month = aug,
       volume = {57},
        pages = {227-303},
          doi = {10.1146/annurev-astro-091918-104430},
archivePrefix = {arXiv},
       eprint = {1812.01615},
 primaryClass = {astro-ph.GA},
       adsurl = {https://ui.adsabs.harvard.edu/abs/2019ARA&A..57..227K}
}

@ARTICLE{2019ApJ...875...54H,
       author = {{Ho}, Stephanie H. and {Martin}, Crystal L. and {Turner}, Monica L.},
        title = "{How Gas Accretion Feeds Galactic Disks}",
      journal = {\apj},
         year = 2019,
        month = apr,
       volume = {875},
       number = {1},
          eid = {54},
        pages = {54},
          doi = {10.3847/1538-4357/ab0ec2},
archivePrefix = {arXiv},
       eprint = {1903.06840},
 primaryClass = {astro-ph.GA},
       adsurl = {https://ui.adsabs.harvard.edu/abs/2019ApJ...875...54H}
}

@INCOLLECTION{2013seg..book....1K,
       author = {{Kormendy}, John},
        title = "{Secular Evolution in Disk Galaxies}",
    booktitle = {Secular Evolution of Galaxies},
         year = 2013,
       editor = {{Falc{\'o}n-Barroso}, Jes{\'u}s and {Knapen}, Johan H.},
        pages = {1},
          doi = {10.48550/arXiv.1311.2609},
       adsurl = {https://ui.adsabs.harvard.edu/abs/2013seg..book....1K}
}

@ARTICLE{2018MNRAS.478.3590S,
       author = {{Shabani}, F. and {Grebel}, E.~K. and {Pasquali}, A. and {D'Onghia}, E. and {Gallagher}, J.~S. and {Adamo}, A. and {Messa}, M. and {Elmegreen}, B.~G. and {Dobbs}, C. and {Gouliermis}, D.~A. and {Calzetti}, D. and {Grasha}, K. and {Elmegreen}, D.~M. and {Cignoni}, M. and {Dale}, D.~A. and {Aloisi}, A. and {Smith}, L.~J. and {Tosi}, M. and {Thilker}, D.~A. and {Lee}, J.~C. and {Sabbi}, E. and {Kim}, H. and {Pellerin}, A.},
        title = "{Search for star cluster age gradients across spiral arms of three LEGUS disc galaxies}",
      journal = {\mnras},
         year = 2018,
        month = aug,
       volume = {478},
       number = {3},
        pages = {3590-3604},
          doi = {10.1093/mnras/sty1277},
archivePrefix = {arXiv},
       eprint = {1805.05643},
 primaryClass = {astro-ph.GA},
       adsurl = {https://ui.adsabs.harvard.edu/abs/2018MNRAS.478.3590S}
}

@ARTICLE{2015AJ....149...51C,
       author = {{Calzetti}, D. and {Lee}, J.~C. and {Sabbi}, E. and {Adamo}, A. and {Smith}, L.~J. and {Andrews}, J.~E. and {Ubeda}, L. and {Bright}, S.~N. and {Thilker}, D. and {Aloisi}, A. and {Brown}, T.~M. and {Chandar}, R. and {Christian}, C. and {Cignoni}, M. and {Clayton}, G.~C. and {da Silva}, R. and {de Mink}, S.~E. and {Dobbs}, C. and {Elmegreen}, B.~G. and {Elmegreen}, D.~M. and {Evans}, A.~S. and {Fumagalli}, M. and {Gallagher}, III, J.~S. and {Gouliermis}, D.~A. and {Grebel}, E.~K. and {Herrero}, A. and {Hunter}, D.~A. and {Johnson}, K.~E. and {Kennicutt}, R.~C. and {Kim}, H. and {Krumholz}, M.~R. and {Lennon}, D. and {Levay}, K. and {Martin}, C. and {Nair}, P. and {Nota}, A. and {{\"O}stlin}, G. and {Pellerin}, A. and {Prieto}, J. and {Regan}, M.~W. and {Ryon}, J.~E. and {Schaerer}, D. and {Schiminovich}, D. and {Tosi}, M. and {Van Dyk}, S.~D. and {Walterbos}, R. and {Whitmore}, B.~C. and {Wofford}, A.},
        title = "{Legacy Extragalactic UV Survey (LEGUS) With the Hubble Space Telescope. I. Survey Description}",
      journal = {\aj},
         year = 2015,
        month = feb,
       volume = {149},
       number = {2},
          eid = {51},
        pages = {51},
          doi = {10.1088/0004-6256/149/2/51},
archivePrefix = {arXiv},
       eprint = {1410.7456},
 primaryClass = {astro-ph.GA},
       adsurl = {https://ui.adsabs.harvard.edu/abs/2015AJ....149...51C}
}

@ARTICLE{2024ApJS..273...14M,
       author = {{Maschmann}, Daniel and {Lee}, Janice C. and {Thilker}, David A. and {Whitmore}, Bradley C. and {Deger}, Sinan and {Boquien}, M{\'e}d{\'e}ric and {Chandar}, Rupali and {Dale}, Daniel A. and {Wofford}, Aida and {Hannon}, Stephen and {Larson}, Kirsten L. and {Leroy}, Adam K. and {Schinnerer}, Eva and {Rosolowsky}, Erik and {{\'U}beda}, Leonardo and {Barnes}, Ashley T. and {Emsellem}, Eric and {Grasha}, Kathryn and {Groves}, Brent and {Indebetouw}, R{\'e}my and {Kim}, Hwihyun and {Klessen}, Ralf S. and {Kreckel}, Kathryn and {Levy}, Rebecca C. and {Pinna}, Francesca and {Rodr{\'\i}guez}, M. Jimena and {Tian}, Qiushi and {Williams}, Thomas G.},
        title = "{PHANGS-HST Catalogs for {\ensuremath{\sim}}100,000 Star Clusters and Compact Associations in 38 Galaxies. I. Observed Properties}",
      journal = {\apjs},
         year = 2024,
        month = jul,
       volume = {273},
       number = {1},
          eid = {14},
        pages = {14},
          doi = {10.3847/1538-4365/ad3cd3},
archivePrefix = {arXiv},
       eprint = {2403.04901},
 primaryClass = {astro-ph.GA},
       adsurl = {https://ui.adsabs.harvard.edu/abs/2024ApJS..273...14M}
}

@ARTICLE{2019MNRAS.484.4897C,
       author = {{Cook}, D.~O. and {Lee}, J.~C. and {Adamo}, A. and {Kim}, H. and {Chandar}, R. and {Whitmore}, B.~C. and {Mok}, A. and {Ryon}, J.~E. and {Dale}, D.~A. and {Calzetti}, D. and {Andrews}, J.~E. and {Aloisi}, A. and {Ashworth}, G. and {Bright}, S.~N. and {Brown}, T.~M. and {Christian}, C. and {Cignoni}, M. and {Clayton}, G.~C. and {da Silva}, R. and {de Mink}, S.~E. and {Dobbs}, C.~L. and {Elmegreen}, B.~G. and {Elmegreen}, D.~M. and {Evans}, A.~S. and {Fumagalli}, M. and {Gallagher}, J.~S. and {Gouliermis}, D.~A. and {Grasha}, K. and {Grebel}, E.~K. and {Herrero}, A. and {Hunter}, D.~A. and {Jensen}, E.~I. and {Johnson}, K.~E. and {Kahre}, L. and {Kennicutt}, R.~C. and {Krumholz}, M.~R. and {Lee}, N.~J. and {Lennon}, D. and {Linden}, S. and {Martin}, C. and {Messa}, M. and {Nair}, P. and {Nota}, A. and {{\"O}stlin}, G. and {Parziale}, R.~C. and {Pellerin}, A. and {Regan}, M.~W. and {Sabbi}, E. and {Sacchi}, E. and {Schaerer}, D. and {Schiminovich}, D. and {Shabani}, F. and {Slane}, F.~A. and {Small}, J. and {Smith}, C.~L. and {Smith}, L.~J. and {Taibi}, S. and {Thilker}, D.~A. and {de la Torre}, I.~C. and {Tosi}, M. and {Turner}, J.~A. and {Ubeda}, L. and {Van Dyk}, S.~D. and {Walterbos}, R. AM and {Wofford}, A.},
        title = "{Star cluster catalogues for the LEGUS dwarf galaxies}",
      journal = {\mnras},
         year = 2019,
        month = apr,
       volume = {484},
       number = {4},
        pages = {4897-4919},
          doi = {10.1093/mnras/stz331},
archivePrefix = {arXiv},
       eprint = {1902.00082},
 primaryClass = {astro-ph.GA},
       adsurl = {https://ui.adsabs.harvard.edu/abs/2019MNRAS.484.4897C}
}

@ARTICLE{2025ApJS..280....1T,
       author = {{Thilker}, David A. and {Lee}, Janice C. and {Whitmore}, Bradley C. and {Maschmann}, Daniel and {Henny}, Kiana and {Chandar}, Rupali and {Dale}, Daniel A. and {Deger}, Sinan and {Boquien}, M{\'e}d{\'e}ric and {Wofford}, Aida and {{\'U}beda}, Leonardo and {Razza}, Alessandro and {Barnes}, Ashley T. and {Belfiore}, Francesco and {Bigiel}, Frank and {Grasha}, Kathryn and {Groves}, Brent and {Kim}, Hwihyun and {Klessen}, Ralf S. and {Neumann}, Justus and {Pinna}, Francesca and {Rodr{\'\i}guez}, M. Jimena and {Rosolowsky}, Erik and {Schinnerer}, Eva and {Williams}, Thomas G.},
        title = "{PHANGS-HST Catalogs for {\ensuremath{\sim}}100,000 Star Clusters and Compact Associations in 38 Galaxies. II. Physical Properties from Decision-tree-based Spectral Energy Distribution Fitting of NUV-U-B-V-I Photometry with Categorical Priors Set by H{\ensuremath{\alpha}} Emission, Cluster Morphology, and Other Auxiliary Information}",
      journal = {\apjs},
         year = 2025,
        month = sep,
       volume = {280},
       number = {1},
          eid = {1},
        pages = {1},
          doi = {10.3847/1538-4365/addabb},
       adsurl = {https://ui.adsabs.harvard.edu/abs/2025ApJS..280....1T}
}

@ARTICLE{2022ApJS..258...10L,
       author = {{Lee}, Janice C. and {Whitmore}, Bradley C. and {Thilker}, David A. and {Deger}, Sinan and {Larson}, Kirsten L. and {Ubeda}, Leonardo and {Anand}, Gagandeep S. and {Boquien}, M{\'e}d{\'e}ric and {Chandar}, Rupali and {Dale}, Daniel A. and {Emsellem}, Eric and {Leroy}, Adam K. and {Rosolowsky}, Erik and {Schinnerer}, Eva and {Schmidt}, Judy and {Lilly}, James and {Turner}, Jordan and {Van Dyk}, Schuyler and {White}, Richard L. and {Barnes}, Ashley T. and {Belfiore}, Francesco and {Bigiel}, Frank and {Blanc}, Guillermo A. and {Cao}, Yixian and {Chevance}, Melanie and {Congiu}, Enrico and {Egorov}, Oleg V. and {Glover}, Simon C.~O. and {Grasha}, Kathryn and {Groves}, Brent and {Henshaw}, Jonathan D. and {Hughes}, Annie and {Klessen}, Ralf S. and {Koch}, Eric and {Kreckel}, Kathryn and {Kruijssen}, J.~M. Diederik and {Liu}, Daizhong and {Lopez}, Laura A. and {Mayker}, Ness and {Meidt}, Sharon E. and {Murphy}, Eric J. and {Pan}, Hsi-An and {Pety}, J{\'e}r{\^o}me and {Querejeta}, Miguel and {Razza}, Alessandro and {Saito}, Toshiki and {S{\'a}nchez-Bl{\'a}zquez}, Patricia and {Santoro}, Francesco and {Sardone}, Amy and {Scheuermann}, Fabian and {Schruba}, Andreas and {Sun}, Jiayi and {Usero}, Antonio and {Watkins}, E. and {Williams}, Thomas G.},
        title = "{The PHANGS-HST Survey: Physics at High Angular Resolution in Nearby Galaxies with the Hubble Space Telescope}",
      journal = {\apjs},
         year = 2022,
        month = jan,
       volume = {258},
       number = {1},
          eid = {10},
        pages = {10},
          doi = {10.3847/1538-4365/ac1fe5},
archivePrefix = {arXiv},
       eprint = {2101.02855},
 primaryClass = {astro-ph.GA},
       adsurl = {https://ui.adsabs.harvard.edu/abs/2022ApJS..258...10L}
}

@ARTICLE{2017ApJ...841..131A,
       author = {{Adamo}, A. and {Ryon}, J.~E. and {Messa}, M. and {Kim}, H. and {Grasha}, K. and {Cook}, D.~O. and {Calzetti}, D. and {Lee}, J.~C. and {Whitmore}, B.~C. and {Elmegreen}, B.~G. and {Ubeda}, L. and {Smith}, L.~J. and {Bright}, S.~N. and {Runnholm}, A. and {Andrews}, J.~E. and {Fumagalli}, M. and {Gouliermis}, D.~A. and {Kahre}, L. and {Nair}, P. and {Thilker}, D. and {Walterbos}, R. and {Wofford}, A. and {Aloisi}, A. and {Ashworth}, G. and {Brown}, T.~M. and {Chandar}, R. and {Christian}, C. and {Cignoni}, M. and {Clayton}, G.~C. and {Dale}, D.~A. and {de Mink}, S.~E. and {Dobbs}, C. and {Elmegreen}, D.~M. and {Evans}, A.~S. and {Gallagher}, III, J.~S. and {Grebel}, E.~K. and {Herrero}, A. and {Hunter}, D.~A. and {Johnson}, K.~E. and {Kennicutt}, R.~C. and {Krumholz}, M.~R. and {Lennon}, D. and {Levay}, K. and {Martin}, C. and {Nota}, A. and {{\"O}stlin}, G. and {Pellerin}, A. and {Prieto}, J. and {Regan}, M.~W. and {Sabbi}, E. and {Sacchi}, E. and {Schaerer}, D. and {Schiminovich}, D. and {Shabani}, F. and {Tosi}, M. and {Van Dyk}, S.~D. and {Zackrisson}, E.},
        title = "{Legacy ExtraGalactic UV Survey with The Hubble Space Telescope: Stellar Cluster Catalogs and First Insights Into Cluster Formation and Evolution in NGC 628}",
      journal = {\apj},
         year = 2017,
        month = jun,
       volume = {841},
       number = {2},
          eid = {131},
        pages = {131},
          doi = {10.3847/1538-4357/aa7132},
archivePrefix = {arXiv},
       eprint = {1705.01588},
 primaryClass = {astro-ph.GA},
       adsurl = {https://ui.adsabs.harvard.edu/abs/2017ApJ...841..131A}
}

@article{Radburn-Smith_2012,
doi = {10.1088/0004-637X/753/2/138},
url = {https://dx.doi.org/10.1088/0004-637X/753/2/138},
year = {2012},
month = {jun},
publisher = {The American Astronomical Society},
volume = {753},
number = {2},
pages = {138},
author = {Radburn-Smith, David J. and Roškar, Rok and Debattista, Victor P. and Dalcanton, Julianne J. and Streich, David and de Jong, Roelof S. and Vlajić, Marija and Holwerda, Benne W. and Purcell, Chris W. and Dolphin, Andrew E. and Zucker, Daniel B.},
title = {OUTER-DISK POPULATIONS IN NGC 7793: EVIDENCE FOR STELLAR RADIAL MIGRATION},
journal = {The Astrophysical Journal}
}

@ARTICLE{2024A&A...681A...8S,
       author = {{Subramanian}, Smitha and {Mondal}, Chayan and {Kalari}, Venu},
        title = "{Effect of low-mass galaxy interactions on their star formation}",
      journal = {\aap},
         year = 2024,
        month = jan,
       volume = {681},
          eid = {A8},
        pages = {A8},
          doi = {10.1051/0004-6361/202346536},
archivePrefix = {arXiv},
       eprint = {2310.02595},
 primaryClass = {astro-ph.GA},
       adsurl = {https://ui.adsabs.harvard.edu/abs/2024A&A...681A...8S}
}

@ARTICLE{2020A&A...644A.101R,
       author = {{Rodr{\'\i}guez}, M.~J. and {Baume}, G. and {Feinstein}, C.},
        title = "{Hierarchical star formation in nearby galaxies}",
      journal = {\aap},
         year = 2020,
        month = dec,
       volume = {644},
          eid = {A101},
        pages = {A101},
          doi = {10.1051/0004-6361/202038970},
archivePrefix = {arXiv},
       eprint = {2010.14419},
 primaryClass = {astro-ph.GA},
       adsurl = {https://ui.adsabs.harvard.edu/abs/2020A&A...644A.101R}
}

@ARTICLE{2019ApJ...878....1S,
       author = {{Sacchi}, E. and {Cignoni}, M. and {Aloisi}, A. and {Tosi}, M. and {Adamo}, A. and {Dale}, D.~A. and {Elmegreen}, B.~G. and {Elmegreen}, D.~M. and {Calzetti}, D. and {Gouliermis}, D.~A. and {Grasha}, K. and {Smith}, L.~J. and {Wofford}, A. and {Lee}, J.~C. and {Sabbi}, E. and {Ubeda}, L.},
        title = "{Star Formation Histories of the LEGUS Spiral Galaxies. I. The Flocculent Spiral NGC 7793}",
      journal = {\apj},
         year = 2019,
        month = jun,
       volume = {878},
       number = {1},
          eid = {1},
        pages = {1},
          doi = {10.3847/1538-4357/ab1de1},
archivePrefix = {arXiv},
       eprint = {1905.00020},
 primaryClass = {astro-ph.GA},
       adsurl = {https://ui.adsabs.harvard.edu/abs/2019ApJ...878....1S}
}

@article{Levesque_2013,
doi = {10.1088/0004-637X/779/2/170},
url = {https://doi.org/10.1088/0004-637X/779/2/170},
year = {2013},
month = {dec},
publisher = {The American Astronomical Society},
volume = {779},
number = {2},
pages = {170},
author = {Levesque, Emily M. and Leitherer, Claus},
title = {MODELING TRACERS OF YOUNG STELLAR POPULATION AGE IN STAR-FORMING GALAXIES},
journal = {The Astrophysical Journal}
}

@article{Iglesias-Paramo:2004mnq,
    author = "Iglesias-Paramo, Jorge and Boselli, A. and Gavazzi, G. and Zaccardo, A.",
    title = "{Tracing the star formation history of cluster galaxies using the H-alpha / UV flux ratio}",
    eprint = "astro-ph/0403620",
    archivePrefix = "arXiv",
    doi = "10.1051/0004-6361:20034572",
    journal = "Astron. Astrophys.",
    volume = "421",
    pages = "887--897",
    year = "2004"
}

@article{Pasquali_2008,
doi = {10.1086/591658},
url = {https://doi.org/10.1086/591658},
year = {2008},
month = {nov},
publisher = {},
volume = {687},
number = {2},
pages = {1004},
author = {Pasquali, A. and Leroy, A. and Rix, H.-W. and Walter, F. and Herbst, T. and Giallongo, E. and Ragazzoni, R. and Baruffolo, A. and Speziali, R. and Hill, J. and Beccari, G. and Bouché, N. and Buschkamp, P. and Kochanek, C. and Skillman, E. and Bechtold, J.},
title = {The Large Binocular Telescope Panoramic View of the Recent Star Formation Activity in IC 2574},
journal = {The Astrophysical Journal}
}

@ARTICLE{2006AdSpR..38.2989A,
       author = {{Agrawal}, P.~C.},
        title = "{A broad spectral band Indian Astronomy satellite {\textquoteleft}Astrosat{\textquoteright}}",
      journal = {Advances in Space Research},
         year = 2006,
        month = jan,
       volume = {38},
       number = {12},
        pages = {2989-2994},
          doi = {10.1016/j.asr.2006.03.038},
       adsurl = {https://ui.adsabs.harvard.edu/abs/2006AdSpR..38.2989A}
}

@INPROCEEDINGS{2014SPIE.9144E..1SS,
       author = {{Singh}, Kulinder Pal and {Tandon}, S.~N. and {Agrawal}, P.~C. and {Antia}, H.~M. and {Manchanda}, R.~K. and {Yadav}, J.~S. and {Seetha}, S. and {Ramadevi}, M.~C. and {Rao}, A.~R. and {Bhattacharya}, D. and {Paul}, B. and {Sreekumar}, P. and {Bhattacharyya}, S. and {Stewart}, G.~C. and {Hutchings}, J. and {Annapurni}, S.~A. and {Ghosh}, S.~K. and {Murthy}, J. and {Pati}, A. and {Rao}, N.~K. and {Stalin}, C.~S. and {Girish}, V. and {Sankarasubramanian}, K. and {Vadawale}, S. and {Bhalerao}, V.~B. and {Dewangan}, G.~C. and {Dedhia}, D.~K. and {Hingar}, M.~K. and {Katoch}, T.~B. and {Kothare}, A.~T. and {Mirza}, I. and {Mukerjee}, K. and {Shah}, H. and {Shah}, P. and {Mohan}, R. and {Sangal}, A.~K. and {Nagabhusana}, S. and {Sriram}, S. and {Malkar}, J.~P. and {Sreekumar}, S. and {Abbey}, A.~F. and {Hansford}, G.~M. and {Beardmore}, A.~P. and {Sharma}, M.~R. and {Murthy}, S. and {Kulkarni}, R. and {Meena}, G. and {Babu}, V.~C. and {Postma}, J.},
        title = "{ASTROSAT mission}",
    booktitle = {Space Telescopes and Instrumentation 2014: Ultraviolet to Gamma Ray},
         year = 2014,
       editor = {{Takahashi}, Tadayuki and {den Herder}, Jan-Willem A. and {Bautz}, Mark},
       series = {Society of Photo-Optical Instrumentation Engineers (SPIE) Conference Series},
       volume = {9144},
        month = jul,
          eid = {91441S},
        pages = {91441S},
          doi = {10.1117/12.2062667},
       adsurl = {https://ui.adsabs.harvard.edu/abs/2014SPIE.9144E..1SS}
}

@article{Calzetti_2025,
doi = {10.3847/1538-4357/adfbe0},
url = {https://doi.org/10.3847/1538-4357/adfbe0},
year = {2025},
month = {sep},
publisher = {The American Astronomical Society},
volume = {991},
number = {2},
pages = {198},
author = {Calzetti, Daniela and Kennicutt, Robert C. and Adamo, Angela and Sandstrom, Karin and Dale, Daniel A. and Elmegreen, Bruce and Gallagher, John S. and Gregg, Benjamin and Bajaj, Varun and Böker, Torsten and Bortolini, Giacomo and Boyer, Martha and Correnti, Matteo and De Looze, Ilse and Draine, Bruce T. and Duarte-Cabral, Ana and Faustino Vieira, Helena and Grasha, Kathryn and Hunt, L. K. and Johnson, Kelsey E. and Klessen, Ralf S. and Krumholz, Mark R. and Lai, Thomas S.-Y. and Lapeer, Drew and Linden, Sean T. and Messa, Matteo and Östlin, Göran and Pedrini, Alex and Relaño, Mònica and Sabbi, Elena and Schinnerer, Eva and Skillman, Evan and Smith, Linda J. and Tosi, Monica and Walter, Fabian and Weinbeck, Tony D.},
title = {Quantification of the Age Dependence of Mid-infrared Star Formation Rate Indicators},
journal = {The Astrophysical Journal}
}

@ARTICLE{2021A&A...650A.103D,
       author = {{Della Bruna}, Lorenza and {Adamo}, Angela and {Lee}, Janice C. and {Smith}, Linda J. and {Krumholz}, Mark and {Bik}, Arjan and {Calzetti}, Daniela and {Fox}, Anne and {Fumagalli}, Michele and {Grasha}, Kathryn and {Messa}, Matteo and {{\"O}stlin}, G{\"o}ran and {Walterbos}, Rene and {Wofford}, Aida},
        title = "{Studying the ISM at {\ensuremath{\sim}}10 pc scale in NGC 7793 with MUSE. II. Constraints on the oxygen abundance and ionising radiation escape}",
      journal = {\aap},
         year = 2021,
        month = jun,
       volume = {650},
          eid = {A103},
        pages = {A103},
          doi = {10.1051/0004-6361/202039402},
archivePrefix = {arXiv},
       eprint = {2104.08088},
 primaryClass = {astro-ph.GA},
       adsurl = {https://ui.adsabs.harvard.edu/abs/2021A&A...650A.103D}
}

@ARTICLE{2021MNRAS.502.1366T,
       author = {{Turner}, Jordan A. and {Dale}, Daniel A. and {Lee}, Janice C. and {Boquien}, M{\'e}d{\'e}ric and {Chandar}, Rupali and {Deger}, Sinan and {Larson}, Kirsten L. and {Mok}, Angus and {Thilker}, David A. and {Ubeda}, Leonardo and {Whitmore}, Bradley C. and {Belfiore}, Francesco and {Bigiel}, Frank and {Blanc}, Guillermo A. and {Emsellem}, Eric and {Grasha}, Kathryn and {Groves}, Brent and {Klessen}, Ralf S. and {Kreckel}, Kathryn and {Kruijssen}, J.~M. Diederik and {Leroy}, Adam K. and {Rosolowsky}, Erik and {Sanchez-Blazquez}, Patricia and {Schinnerer}, Eva and {Schruba}, Andreas and {Van Dyk}, Schuyler D. and {Williams}, Thomas G.},
        title = "{PHANGS-HST: star cluster spectral energy distribution fitting with CIGALE}",
      journal = {\mnras},
         year = 2021,
        month = mar,
       volume = {502},
       number = {1},
        pages = {1366-1385},
          doi = {10.1093/mnras/stab055},
archivePrefix = {arXiv},
       eprint = {2101.02134},
 primaryClass = {astro-ph.GA},
       adsurl = {https://ui.adsabs.harvard.edu/abs/2021MNRAS.502.1366T}
}

@article{Linden_2022,
doi = {10.3847/1538-4357/ac7c07},
url = {https://dx.doi.org/10.3847/1538-4357/ac7c07},
year = {2022},
month = {aug},
publisher = {The American Astronomical Society},
volume = {935},
number = {2},
pages = {166},
author = {Linden, S. T. and Perez, G. and Calzetti, D. and Maji, S. and Messa, M. and Whitmore, B. C. and Chandar, R. and Adamo, A. and Grasha, K. and Cook, D. O. and Elmegreen, B. G. and Dale, D. A. and Sacchi, E. and Sabbi, E. and Grebel, E. K. and Smith, L.},
title = {Star Cluster Formation and Evolution in M101: An Investigation with the Legacy Extragalactic UV Survey},
journal = {The Astrophysical Journal}
}

@ARTICLE{2005ApJ...619L..79T,
       author = {{Thilker}, David A. and {Bianchi}, Luciana and {Boissier}, Samuel and {Gil de Paz}, Armando and {Madore}, Barry F. and {Martin}, D. Christopher and {Meurer}, Gerhardt R. and {Neff}, Susan G. and {Rich}, R. Michael and {Schiminovich}, David and {Seibert}, Mark and {Wyder}, Ted K. and {Barlow}, Tom A. and {Byun}, Yong-Ik and {Donas}, Jose and {Forster}, Karl and {Friedman}, Peter G. and {Heckman}, Timothy M. and {Jelinsky}, Patrick N. and {Lee}, Young-Wook and {Malina}, Roger F. and {Milliard}, Bruno and {Morrissey}, Patrick and {Siegmund}, Oswald H.~W. and {Small}, Todd and {Szalay}, Alex S. and {Welsh}, Barry Y.},
        title = "{Recent Star Formation in the Extreme Outer Disk of M83}",
      journal = {\apjl},
         year = 2005,
        month = jan,
       volume = {619},
       number = {1},
        pages = {L79-L82},
          doi = {10.1086/425251},
archivePrefix = {arXiv},
       eprint = {astro-ph/0411306},
 primaryClass = {astro-ph},
       adsurl = {https://ui.adsabs.harvard.edu/abs/2005ApJ...619L..79T}
}

@ARTICLE{2005ApJ...619L..67T,
       author = {{Thilker}, David A. and {Hoopes}, Charles G. and {Bianchi}, Luciana and {Boissier}, Samuel and {Rich}, R. Michael and {Seibert}, Mark and {Friedman}, Peter G. and {Rey}, Soo-Chang and {Buat}, Veronique and {Barlow}, Tom A. and {Byun}, Yong-Ik and {Donas}, Jose and {Forster}, Karl and {Heckman}, Timothy M. and {Jelinsky}, Patrick N. and {Lee}, Young-Wook and {Madore}, Barry F. and {Malina}, Roger F. and {Martin}, D. Christopher and {Milliard}, Bruno and {Morrissey}, Patrick F. and {Neff}, Susan G. and {Schiminovich}, David and {Siegmund}, Oswald H.~W. and {Small}, Todd and {Szalay}, Alex S. and {Welsh}, Barry Y. and {Wyder}, Ted K.},
        title = "{Panoramic GALEX Far- and Near-Ultraviolet Imaging of M31 and M33}",
      journal = {\apjl},
         year = 2005,
        month = jan,
       volume = {619},
       number = {1},
        pages = {L67-L70},
          doi = {10.1086/424816},
archivePrefix = {arXiv},
       eprint = {astro-ph/0411311},
 primaryClass = {astro-ph},
       adsurl = {https://ui.adsabs.harvard.edu/abs/2005ApJ...619L..67T}
}

@ARTICLE{2007ApJS..173..185G,
       author = {{Gil de Paz}, Armando and {Boissier}, Samuel and {Madore}, Barry F. and {Seibert}, Mark and {Joe}, Young H. and {Boselli}, Alessandro and {Wyder}, Ted K. and {Thilker}, David and {Bianchi}, Luciana and {Rey}, Soo-Chang and {Rich}, R. Michael and {Barlow}, Tom A. and {Conrow}, Tim and {Forster}, Karl and {Friedman}, Peter G. and {Martin}, D. Christopher and {Morrissey}, Patrick and {Neff}, Susan G. and {Schiminovich}, David and {Small}, Todd and {Donas}, Jos{\'e} and {Heckman}, Timothy M. and {Lee}, Young-Wook and {Milliard}, Bruno and {Szalay}, Alex S. and {Yi}, Sukyoung},
        title = "{The GALEX Ultraviolet Atlas of Nearby Galaxies}",
      journal = {\apjs},
         year = 2007,
        month = dec,
       volume = {173},
       number = {2},
        pages = {185-255},
          doi = {10.1086/516636},
archivePrefix = {arXiv},
       eprint = {astro-ph/0606440},
 primaryClass = {astro-ph},
       adsurl = {https://ui.adsabs.harvard.edu/abs/2007ApJS..173..185G}
}

@ARTICLE{2007ApJS..173..538T,
       author = {{Thilker}, David A. and {Bianchi}, Luciana and {Meurer}, Gerhardt and {Gil de Paz}, Armando and {Boissier}, Samuel and {Madore}, Barry F. and {Boselli}, Alessandro and {Ferguson}, Annette M.~N. and {Mu{\~n}oz-Mateos}, Juan Carlos and {Madsen}, Greg J. and {Hameed}, Salman and {Overzier}, Roderik A. and {Forster}, Karl and {Friedman}, Peter G. and {Martin}, D. Christopher and {Morrissey}, Patrick and {Neff}, Susan G. and {Schiminovich}, David and {Seibert}, Mark and {Small}, Todd and {Wyder}, Ted K. and {Donas}, Jos{\'e} and {Heckman}, Timothy M. and {Lee}, Young-Wook and {Milliard}, Bruno and {Rich}, R. Michael and {Szalay}, Alex S. and {Welsh}, Barry Y. and {Yi}, Sukyoung K.},
        title = "{A Search for Extended Ultraviolet Disk (XUV-Disk) Galaxies in the Local Universe}",
      journal = {\apjs},
         year = 2007,
        month = dec,
       volume = {173},
       number = {2},
        pages = {538-571},
          doi = {10.1086/523853},
archivePrefix = {arXiv},
       eprint = {0712.3555},
 primaryClass = {astro-ph},
       adsurl = {https://ui.adsabs.harvard.edu/abs/2007ApJS..173..538T}
}

@ARTICLE{2007ApJ...658.1006M,
       author = {{Mu{\~n}oz-Mateos}, J.~C. and {Gil de Paz}, A. and {Boissier}, S. and {Zamorano}, J. and {Jarrett}, T. and {Gallego}, J. and {Madore}, B.~F.},
        title = "{Specific Star Formation Rate Profiles in Nearby Spiral Galaxies: Quantifying the Inside-Out Formation of Disks}",
      journal = {\apj},
         year = 2007,
        month = apr,
       volume = {658},
       number = {2},
        pages = {1006-1026},
          doi = {10.1086/511812},
archivePrefix = {arXiv},
       eprint = {astro-ph/0612017},
 primaryClass = {astro-ph},
       adsurl = {https://ui.adsabs.harvard.edu/abs/2007ApJ...658.1006M}
}

@ARTICLE{2021ApJ...914...54Y,
       author = {{Yadav}, Jyoti and {Das}, Mousumi and {Patra}, Narendra Nath and {Dwarakanath}, K.~S. and {Rahna}, P.~T. and {McGaugh}, Stacy S. and {Schombert}, James and {Murthy}, Jayant},
        title = "{Comparing the Inner and Outer Star-forming Complexes in the Nearby Spiral Galaxies NGC 628, NGC 5457, and NGC 6946 Using UVIT Observations}",
      journal = {\apj},
         year = 2021,
        month = jun,
       volume = {914},
       number = {1},
          eid = {54},
        pages = {54},
          doi = {10.3847/1538-4357/abf8c1},
archivePrefix = {arXiv},
       eprint = {2103.16819},
 primaryClass = {astro-ph.GA},
       adsurl = {https://ui.adsabs.harvard.edu/abs/2021ApJ...914...54Y}
}

@ARTICLE{2024MNRAS.530.2199A,
       author = {{Amrutha}, S. and {Das}, Mousumi and {Yadav}, Jyoti},
        title = "{A comparative study of star-forming dwarf galaxies using the UVIT}",
      journal = {\mnras},
         year = 2024,
        month = may,
       volume = {530},
       number = {2},
        pages = {2199-2231},
          doi = {10.1093/mnras/stae907},
archivePrefix = {arXiv},
       eprint = {2404.00537},
 primaryClass = {astro-ph.GA},
       adsurl = {https://ui.adsabs.harvard.edu/abs/2024MNRAS.530.2199A}
}

@ARTICLE{2025A&A...702A.222C,
       author = {{Chauhan}, Rakshit and {Subramanian}, Smitha and {Kudari}, Deepak A. and {Amrutha}, S. and {Das}, Mousumi},
        title = "{Dwarf-dwarf interactions and their influence on star formation: Insights from post-merger galaxies}",
      journal = {\aap},
         year = 2025,
        month = oct,
       volume = {702},
          eid = {A222},
        pages = {A222},
          doi = {10.1051/0004-6361/202554437},
archivePrefix = {arXiv},
       eprint = {2507.14695},
 primaryClass = {astro-ph.GA},
       adsurl = {https://ui.adsabs.harvard.edu/abs/2025A&A...702A.222C}
}

@ARTICLE{2022MNRAS.516.2171U,
       author = {{Ujjwal}, K. and {Kartha}, Sreeja S. and {Subramanian}, Smitha and {George}, Koshy and {Thomas}, Robin and {Mathew}, Blesson},
        title = "{Understanding the secular evolution of NGC 628 using UltraViolet Imaging Telescope}",
      journal = {\mnras},
         year = 2022,
        month = oct,
       volume = {516},
       number = {2},
        pages = {2171-2180},
          doi = {10.1093/mnras/stac2285},
archivePrefix = {arXiv},
       eprint = {2208.05999},
 primaryClass = {astro-ph.GA},
       adsurl = {https://ui.adsabs.harvard.edu/abs/2022MNRAS.516.2171U}
}

@article{Mondal_2018,
doi = {10.3847/1538-3881/aad4f6},
url = {https://dx.doi.org/10.3847/1538-3881/aad4f6},
year = {2018},
month = {aug},
publisher = {The American Astronomical Society},
volume = {156},
number = {3},
pages = {109},
author = {Mondal, Chayan and Subramaniam, Annapurni and George, Koshy},
title = {UVIT Imaging of WLM: Demographics of Star-forming Regions in the Nearby Dwarf Irregular Galaxy},
journal = {The Astronomical Journal}
}

@article{Mondal_2021,
doi = {10.3847/1538-4357/abe0b4},
url = {https://dx.doi.org/10.3847/1538-4357/abe0b4},
year = {2021},
month = {mar},
publisher = {The American Astronomical Society},
volume = {909},
number = {2},
pages = {203},
author = {Mondal, Chayan and Subramaniam, Annapurni and George, Koshy and Postma, Joseph E. and Subramanian, Smitha and Barway, Sudhanshu},
title = {Tracing Young Star-forming Clumps in the Nearby Flocculent Spiral Galaxy NGC 7793 with UVIT Imaging},
journal = {The Astrophysical Journal}
}

@ARTICLE{2025PASA...42...73S,
       author = {{Santhosh}, Geethika and {Rajalakshmi}, Rakhi and {George}, Koshy and {Subramanian}, Smitha and {Indulekha}, Kavila},
        title = "{Star formation in interacting galaxy systems: UVIT imaging of NGC 7252 and NGC 5291}",
      journal = {\pasa},
         year = 2025,
        month = apr,
       volume = {42},
          eid = {e073},
        pages = {e073},
          doi = {10.1017/pasa.2025.21},
archivePrefix = {arXiv},
       eprint = {2503.02777},
 primaryClass = {astro-ph.GA},
       adsurl = {https://ui.adsabs.harvard.edu/abs/2025PASA...42...73S}
}

@ARTICLE{2025PASA...42...56A,
       author = {{Akhil}, Krishna R. and {Kartha}, Sreeja S. and {Krishnan}, Ujjwal and {Mathew}, Blesson and {Robin}, Thomas and {Ray}, Shankar and {Devaraj}, Ashish},
        title = "{Connecting the dots: Tracing the evolutionary pathway of polar ring galaxies in the cases of NGC 3718, NGC 2685, and NGC 4262}",
      journal = {\pasa},
         year = 2025,
        month = apr,
       volume = {42},
          eid = {e056},
        pages = {e056},
          doi = {10.1017/pasa.2025.24},
archivePrefix = {arXiv},
       eprint = {2503.03709},
 primaryClass = {astro-ph.GA},
       adsurl = {https://ui.adsabs.harvard.edu/abs/2025PASA...42...56A}
}

@ARTICLE{2020ApJS..247...47L,
       author = {{Leahy}, D.~A. and {Postma}, J. and {Chen}, Y. and {Buick}, M.},
        title = "{AstroSat UVIT Survey of M31: Point-source Catalog}",
      journal = {\apjs},
         year = 2020,
        month = apr,
       volume = {247},
       number = {2},
          eid = {47},
        pages = {47},
          doi = {10.3847/1538-4365/ab77a9},
       adsurl = {https://ui.adsabs.harvard.edu/abs/2020ApJS..247...47L}
}

@ARTICLE{2023ApJ...946...65D,
       author = {{Devaraj}, A. and {Joseph}, P. and {Stalin}, C.~S. and {Tandon}, S.~N. and {Ghosh}, S.~K.},
        title = "{UVIT Observations of the Small Magellanic Cloud: Point-source Catalog}",
      journal = {\apj},
         year = 2023,
        month = apr,
       volume = {946},
       number = {2},
          eid = {65},
        pages = {65},
          doi = {10.3847/1538-4357/acba9c},
archivePrefix = {arXiv},
       eprint = {2302.01515},
 primaryClass = {astro-ph.GA},
       adsurl = {https://ui.adsabs.harvard.edu/abs/2023ApJ...946...65D}
}

@ARTICLE{2024MNRAS.532..322H,
       author = {{Hota}, Sipra and {Subramaniam}, Annapurni and {Dhanush}, S.~R. and {Cioni}, Maria-Rosa L. and {Subramanian}, Smitha},
        title = "{UVIT Study of the MAgellanic Clouds (U-SMAC) - I. Recent star formation history and kinematics of the Shell region in the north-eastern Small Magellanic Cloud}",
      journal = {\mnras},
         year = 2024,
        month = jul,
       volume = {532},
       number = {1},
        pages = {322-335},
          doi = {10.1093/mnras/stae1438},
archivePrefix = {arXiv},
       eprint = {2406.05093},
 primaryClass = {astro-ph.GA},
       adsurl = {https://ui.adsabs.harvard.edu/abs/2024MNRAS.532..322H}
}

@ARTICLE{2024AJ....168..255H,
       author = {{Hota}, Sipra and {Subramaniam}, Annapurni and {Nayak}, Prasanta K. and {Subramanian}, Smitha},
        title = "{UVIT Study of the MAgellanic Clouds (U-SMAC). II. A Far-UV Catalog of the Small Magellanic Cloud: Morphology and Kinematics of Young Stellar Population}",
      journal = {\aj},
         year = 2024,
        month = dec,
       volume = {168},
       number = {6},
          eid = {255},
        pages = {255},
          doi = {10.3847/1538-3881/ad7de2},
archivePrefix = {arXiv},
       eprint = {2409.13605},
 primaryClass = {astro-ph.GA},
       adsurl = {https://ui.adsabs.harvard.edu/abs/2024AJ....168..255H}
}

@ARTICLE{2026arXiv260306510S,
       author = {{Sabbi}, E. and {Meena}, B. and {Zeidler}, P. and {Bajaj}, V. and {Calzetti}, D. and {Eldridge}, J.~J. and {Facchini}, P. and {Linden}, S. and {Crowther}, P.~A. and {Adamo}, A. and {Bianchi}, L. and {Cignoni}, M. and {Elmegreen}, B.~G. and {Elmegreen}, D.~M. and {Gallagher}, III, J.~S. and {Gennaro}, M. and {Grebel}, E.~K. and {Klessen}, R.~S. and {Pasquali}, A. and {Smith}, L.~J. and {Wofford}, A.},
        title = "{Galaxy UV Legacy Project: Survey Description and First Insights Into NGC 4449 Recent History of Star Formation}",
      journal = {arXiv e-prints},
         year = 2026,
        month = mar,
          eid = {arXiv:2603.06510},
        pages = {arXiv:2603.06510},
          doi = {10.48550/arXiv.2603.06510},
archivePrefix = {arXiv},
       eprint = {2603.06510},
 primaryClass = {astro-ph.GA},
       adsurl = {https://ui.adsabs.harvard.edu/abs/2026arXiv260306510S}
}

@ARTICLE{2024ApJ...974..206W,
       author = {{Watts}, Chandan and {Das}, Mousumi and {Barway}, Sudhanshu},
        title = "{Anatomy of the Star Formation in a Tidally Disturbed Disk Galaxy: NGC 3718}",
      journal = {\apj},
         year = 2024,
        month = oct,
       volume = {974},
       number = {2},
          eid = {206},
        pages = {206},
          doi = {10.3847/1538-4357/ad738b},
archivePrefix = {arXiv},
       eprint = {2408.09898},
 primaryClass = {astro-ph.GA},
       adsurl = {https://ui.adsabs.harvard.edu/abs/2024ApJ...974..206W}
}

@article{Hao_2011,
doi = {10.1088/0004-637X/741/2/124},
url = {https://dx.doi.org/10.1088/0004-637X/741/2/124},
year = {2011},
month = {oct},
publisher = {The American Astronomical Society},
volume = {741},
number = {2},
pages = {124},
author = {Hao, Cai-Na and Kennicutt, Robert C. and Johnson, Benjamin D. and Calzetti, Daniela and Dale, Daniel A. and Moustakas, John},
title = {DUST-CORRECTED STAR FORMATION RATES OF GALAXIES. II. COMBINATIONS OF ULTRAVIOLET AND INFRARED TRACERS},
journal = {The Astrophysical Journal}
}

@ARTICLE{2016A&A...591A...6B,
       author = {{Boquien}, M. and {Kennicutt}, R. and {Calzetti}, D. and {Dale}, D. and {Galametz}, M. and {Sauvage}, M. and {Croxall}, K. and {Draine}, B. and {Kirkpatrick}, A. and {Kumari}, N. and {Hunt}, L. and {De Looze}, I. and {Pellegrini}, E. and {Rela{\~n}o}, M. and {Smith}, J. -D. and {Tabatabaei}, F.},
        title = "{Towards universal hybrid star formation rate estimators}",
      journal = {\aap},
         year = 2016,
        month = jun,
       volume = {591},
          eid = {A6},
        pages = {A6},
          doi = {10.1051/0004-6361/201527759},
archivePrefix = {arXiv},
       eprint = {1603.09340},
 primaryClass = {astro-ph.GA},
       adsurl = {https://ui.adsabs.harvard.edu/abs/2016A&A...591A...6B}
}

@ARTICLE{2025arXiv250808451C,
       author = {{Calzetti}, Daniela and {Kennicutt}, Robert C. and {Adamo}, Angela and {Sandstrom}, Karin and {Dale}, Daniel A. and {Elmegreen}, Bruce and {Gallagher}, John S. and {Gregg}, Benjamin and {Bajaj}, Varun and {Boker}, Torsten and {Bortolini}, Giacomo and {Boyer}, Martha and {Correnti}, Matteo and {De Looze}, Ilse and {Draine}, Bruce T. and {Duarte-Cabral}, Ana and {Faustino Vieira}, Helena and {Grasha}, Kathryn and {Hunt}, L.~K. and {Johnson}, Kelsey E. and {Klessen}, Ralf S. and {Krumholz}, Mark R. and {Lai}, Thomas S. -Y. and {Lapeer}, Drew and {Linden}, Sean T. and {Messa}, Matteo and {Ostlin}, Goeran and {Pedrini}, Alex and {Relano}, Monica and {Sabbi}, Elena and {Schinnerer}, Eva and {Skillman}, Evan and {Smith}, Linda J. and {Tosi}, Monica and {Walter}, Fabian and {Weinbeck}, Tony D.},
        title = "{Quantification of The Age Dependence of Mid-Infrared Star Formation Rate Indicators}",
      journal = {arXiv e-prints},
         year = 2025,
        month = aug,
          eid = {arXiv:2508.08451},
        pages = {arXiv:2508.08451},
          doi = {10.48550/arXiv.2508.08451},
archivePrefix = {arXiv},
       eprint = {2508.08451},
 primaryClass = {astro-ph.GA},
       adsurl = {https://ui.adsabs.harvard.edu/abs/2025arXiv250808451C}
}

@article{Hassani_2024,
doi = {10.3847/1538-4365/ad152c},
url = {https://dx.doi.org/10.3847/1538-4365/ad152c},
year = {2024},
month = {feb},
publisher = {The American Astronomical Society},
volume = {271},
number = {1},
pages = {2},
author = {Hassani, Hamid and Rosolowsky, Erik and Koch, Eric W. and Postma, Joseph and Nofech, Joseph and Corbould, Harrisen and Thilker, David and Leroy, Adam K. and Schinnerer, Eva and Belfiore, Francesco and Bigiel, Frank and Boquien, Médéric and Chevance, Mélanie and Dale, Daniel A. and Egorov, Oleg V. and Emsellem, Eric and Glover, Simon C. O. and Grasha, Kathryn and Groves, Brent and Henny, Kiana and Kim, Jaeyeon and Klessen, Ralf S. and Kreckel, Kathryn and Kruijssen, J. M. Diederik and Lee, Janice C. and Lopez, Laura A. and Neumann, Justus and Pan, Hsi-An and Sandstrom, Karin M. and Sarbadhicary, Sumit K. and Sun, Jiayi and Williams, Thomas G.},
title = {The PHANGS-AstroSat Atlas of Nearby Star-forming Galaxies},
journal = {The Astrophysical Journal Supplement Series}
}

@ARTICLE{Elmegreen_2014,
       author = {{Elmegreen}, Debra Meloy and {Elmegreen}, Bruce G. and {Adamo}, Angela and {Aloisi}, Alessandra and {Andrews}, Jennifer and {Annibali}, Francesca and {Bright}, Stacey N. and {Calzetti}, Daniela and {Cignoni}, Michele and {Evans}, Aaron S. and {Gallagher}, John S., III and {Gouliermis}, Dimitrios A. and {Grebel}, Eva K. and {Hunter}, Deidre A. and {Johnson}, Kelsey and {Kim}, Hwihyun and {Lee}, Janice and {Sabbi}, Elena and {Smith}, Linda J. and {Thilker}, David and {Tosi}, Monica and {Ubeda}, Leonardo},
        title = "{Hierarchical Star Formation in Nearby LEGUS Galaxies}",
      journal = {\apjl},
         year = 2014,
        month = may,
       volume = {787},
       number = {1},
          eid = {L15},
        pages = {L15},
          doi = {10.1088/2041-8205/787/1/L15},
archivePrefix = {arXiv},
       eprint = {1404.6001},
 primaryClass = {astro-ph.GA},
       adsurl = {https://ui.adsabs.harvard.edu/abs/2014ApJ...787L..15E}
}

@ARTICLE{Elmegreen_2006,
       author = {{Elmegreen}, Bruce G. and {Elmegreen}, Debra Meloy and {Chandar}, Rupali and {Whitmore}, Brad and {Regan}, Michael},
        title = "{Hierarchical Star Formation in the Spiral Galaxy NGC 628}",
      journal = {\apj},
         year = 2006,
        month = jun,
       volume = {644},
       number = {2},
        pages = {879-889},
          doi = {10.1086/503797},
archivePrefix = {arXiv},
       eprint = {astro-ph/0605523},
 primaryClass = {astro-ph},
       adsurl = {https://ui.adsabs.harvard.edu/abs/2006ApJ...644..879E}
}

@ARTICLE{2021MNRAS.507.5542M,
       author = {{Menon}, Shyam H. and {Grasha}, Kathryn and {Elmegreen}, Bruce G. and {Federrath}, Christoph and {Krumholz}, Mark R. and {Calzetti}, Daniela and {S{\'a}nchez}, N{\'e}stor and {Linden}, Sean T. and {Adamo}, Angela and {Messa}, Matteo and {Cook}, David O. and {Dale}, Daniel A. and {Grebel}, Eva K. and {Fumagalli}, Michele and {Sabbi}, Elena and {Johnson}, Kelsey E. and {Smith}, Linda J. and {Kennicutt}, Robert C.},
        title = "{The dependence of the hierarchical distribution of star clusters on galactic environment}",
      journal = {\mnras},
         year = 2021,
        month = nov,
       volume = {507},
       number = {4},
        pages = {5542-5566},
          doi = {10.1093/mnras/stab2413},
archivePrefix = {arXiv},
       eprint = {2108.04387},
 primaryClass = {astro-ph.GA},
       adsurl = {https://ui.adsabs.harvard.edu/abs/2021MNRAS.507.5542M}
}

@ARTICLE{Grasha_2017_Spatial,
       author = {{Grasha}, K. and {Calzetti}, D. and {Adamo}, A. and {Kim}, H. and {Elmegreen}, B.~G. and {Gouliermis}, D.~A. and {Dale}, D.~A. and {Fumagalli}, M. and {Grebel}, E.~K. and {Johnson}, K.~E. and {Kahre}, L. and {Kennicutt}, R.~C. and {Messa}, M. and {Pellerin}, A. and {Ryon}, J.~E. and {Smith}, L.~J. and {Shabani}, F. and {Thilker}, D. and {Ubeda}, L.},
        title = "{The Hierarchical Distribution of the Young Stellar Clusters in Six Local Star-forming Galaxies}",
      journal = {\apj},
         year = 2017,
        month = may,
       volume = {840},
       number = {2},
          eid = {113},
        pages = {113},
          doi = {10.3847/1538-4357/aa6f15},
archivePrefix = {arXiv},
       eprint = {1704.06321},
 primaryClass = {astro-ph.GA},
       adsurl = {https://ui.adsabs.harvard.edu/abs/2017ApJ...840..113G}
}

@ARTICLE{2018PASP..130g2001G,
       author = {{Gouliermis}, Dimitrios A.},
        title = "{Unbound Young Stellar Systems: Star Formation on the Loose}",
      journal = {\pasp},
         year = 2018,
        month = jul,
       volume = {130},
       number = {989},
        pages = {072001},
          doi = {10.1088/1538-3873/aac1fd},
archivePrefix = {arXiv},
       eprint = {1806.11541},
 primaryClass = {astro-ph.GA},
       adsurl = {https://ui.adsabs.harvard.edu/abs/2018PASP..130g2001G}
}

@ARTICLE{2025A&A...693A.188S,
       author = {{Shashank}, Gairola and {Subramanian}, Smitha and {Muraleedharan}, Sreedevi and {Menon}, Shyam H. and {Mondal}, Chayan and {Krishna}, Sriram and {Das}, Mousumi and {Subramaniam}, Annapurni},
        title = "{Tracing hierarchical star formation out to kiloparsec scales in nearby spiral galaxies with UVIT}",
      journal = {\aap},
         year = 2025,
        month = jan,
       volume = {693},
          eid = {A188},
        pages = {A188},
          doi = {10.1051/0004-6361/202451739},
archivePrefix = {arXiv},
       eprint = {2412.00872},
 primaryClass = {astro-ph.GA},
       adsurl = {https://ui.adsabs.harvard.edu/abs/2025A&A...693A.188S}
}

@ARTICLE{1991ApJ...379...52W,
       author = {{White}, Simon D.~M. and {Frenk}, Carlos S.},
        title = "{Galaxy Formation through Hierarchical Clustering}",
      journal = {\apj},
         year = 1991,
        month = sep,
       volume = {379},
        pages = {52},
          doi = {10.1086/170483},
       adsurl = {https://ui.adsabs.harvard.edu/abs/1991ApJ...379...52W}
}

@ARTICLE{1998MNRAS.295..319M,
       author = {{Mo}, H.~J. and {Mao}, Shude and {White}, Simon D.~M.},
        title = "{The formation of galactic discs}",
      journal = {\mnras},
         year = 1998,
        month = apr,
       volume = {295},
       number = {2},
        pages = {319-336},
          doi = {10.1046/j.1365-8711.1998.01227.x},
archivePrefix = {arXiv},
       eprint = {astro-ph/9707093},
 primaryClass = {astro-ph},
       adsurl = {https://ui.adsabs.harvard.edu/abs/1998MNRAS.295..319M}
}

@ARTICLE{2014ApJ...781...11E,
       author = {{Elmegreen}, Debra Meloy and {Elmegreen}, Bruce G.},
        title = "{The Onset of Spiral Structure in the Universe}",
      journal = {\apj},
         year = 2014,
        month = jan,
       volume = {781},
       number = {1},
          eid = {11},
        pages = {11},
          doi = {10.1088/0004-637X/781/1/11},
archivePrefix = {arXiv},
       eprint = {1312.2215},
 primaryClass = {astro-ph.GA},
       adsurl = {https://ui.adsabs.harvard.edu/abs/2014ApJ...781...11E}
}

@ARTICLE{2003AJ....125..525J,
       author = {{Jarrett}, T.~H. and {Chester}, T. and {Cutri}, R. and {Schneider}, S.~E. and {Huchra}, J.~P.},
        title = "{The 2MASS Large Galaxy Atlas}",
      journal = {\aj},
         year = 2003,
        month = feb,
       volume = {125},
       number = {2},
        pages = {525-554},
          doi = {10.1086/345794},
       adsurl = {https://ui.adsabs.harvard.edu/abs/2003AJ....125..525J}
}

@ARTICLE{2009ApJ...693.1821D,
       author = {{Dale}, D.~A. and {Smith}, J.~D.~T. and {Schlawin}, E.~A. and {Armus}, L. and {Buckalew}, B.~A. and {Cohen}, S.~A. and {Helou}, G. and {Jarrett}, T.~H. and {Johnson}, L.~C. and {Moustakas}, J. and {Murphy}, E.~J. and {Roussel}, H. and {Sheth}, K. and {Staudaher}, S. and {Bot}, C. and {Calzetti}, D. and {Engelbracht}, C.~W. and {Gordon}, K.~D. and {Hollenbach}, D.~J. and {Kennicutt}, R.~C. and {Malhotra}, S.},
        title = "{The Spitzer Infrared Nearby Galaxies Survey: A High-Resolution Spectroscopy Anthology}",
      journal = {\apj},
         year = 2009,
        month = mar,
       volume = {693},
       number = {2},
        pages = {1821-1834},
          doi = {10.1088/0004-637X/693/2/1821},
archivePrefix = {arXiv},
       eprint = {0811.4190},
 primaryClass = {astro-ph},
       adsurl = {https://ui.adsabs.harvard.edu/abs/2009ApJ...693.1821D}
}

@ARTICLE{2003PASP..115..928K,
       author = {{Kennicutt}, Jr., Robert C. and {Armus}, Lee and {Bendo}, George and {Calzetti}, Daniela and {Dale}, Daniel A. and {Draine}, Bruce T. and {Engelbracht}, Charles W. and {Gordon}, Karl D. and {Grauer}, Albert D. and {Helou}, George and {Hollenbach}, David J. and {Jarrett}, Thomas H. and {Kewley}, Lisa J. and {Leitherer}, Claus and {Li}, Aigen and {Malhotra}, Sangeeta and {Regan}, Michael W. and {Rieke}, George H. and {Rieke}, Marcia J. and {Roussel}, H{\'e}l{\`e}ne and {Smith}, John-David T. and {Thornley}, Michele D. and {Walter}, Fabian},
        title = "{SINGS: The SIRTF Nearby Galaxies Survey}",
      journal = {\pasp},
         year = 2003,
        month = aug,
       volume = {115},
       number = {810},
        pages = {928-952},
          doi = {10.1086/376941},
archivePrefix = {arXiv},
       eprint = {astro-ph/0305437},
 primaryClass = {astro-ph},
       adsurl = {https://ui.adsabs.harvard.edu/abs/2003PASP..115..928K}
}

@ARTICLE{1989ApJ...345..245C,
       author = {{Cardelli}, Jason A. and {Clayton}, Geoffrey C. and {Mathis}, John S.},
        title = "{The Relationship between Infrared, Optical, and Ultraviolet Extinction}",
      journal = {\apj},
         year = 1989,
        month = oct,
       volume = {345},
        pages = {245},
          doi = {10.1086/167900},
       adsurl = {https://ui.adsabs.harvard.edu/abs/1989ApJ...345..245C}
}

@INPROCEEDINGS{2012SPIE.8443E..1NK,
       author = {{Kumar}, Amit and {Ghosh}, S.~K. and {Hutchings}, J. and {Kamath}, P.~U. and {Kathiravan}, S. and {Mahesh}, P.~K. and {Murthy}, J. and {Nagbhushana}, S. and {Pati}, A.~K. and {Rao}, M.~N. and {Rao}, N.~K. and {Sriram}, S. and {Tandon}, S.~N.},
        title = "{Ultra Violet Imaging Telescope (UVIT) on ASTROSAT}",
    booktitle = {Space Telescopes and Instrumentation 2012: Ultraviolet to Gamma Ray},
         year = 2012,
       editor = {{Takahashi}, Tadayuki and {Murray}, Stephen S. and {den Herder}, Jan-Willem A.},
       series = {Society of Photo-Optical Instrumentation Engineers (SPIE) Conference Series},
       volume = {8443},
        month = sep,
          eid = {84431N},
        pages = {84431N},
          doi = {10.1117/12.924507},
archivePrefix = {arXiv},
       eprint = {1208.4670},
 primaryClass = {astro-ph.IM},
       adsurl = {https://ui.adsabs.harvard.edu/abs/2012SPIE.8443E..1NK}
}

@article{Tandon_2017,
doi = {10.3847/1538-3881/aa8451},
url = {https://dx.doi.org/10.3847/1538-3881/aa8451},
year = {2017},
month = {sep},
publisher = {The American Astronomical Society},
volume = {154},
number = {3},
pages = {128},
author = {Tandon, S. N. and Subramaniam, Annapurni and Girish, V. and Postma, J. and Sankarasubramanian, K. and Sriram, S. and Stalin, C. S. and Mondal, C. and Sahu, S. and Joseph, P. and Hutchings, J. and Ghosh, S. K. and Barve, I. V. and George, K. and Kamath, P. U. and Kathiravan, S. and Kumar, A. and Lancelot, J. P. and Leahy, D. and Mahesh, P. K. and Mohan, R. and Nagabhushana, S. and Pati, A. K. and Kameswara Rao, N. and Sreedhar, Y. H. and Sreekumar, P.},
title = {In-orbit Calibrations of the Ultraviolet Imaging Telescope},
journal = {The Astronomical Journal}
}

@article{Tandon_2020,
doi = {10.3847/1538-3881/ab72a3},
url = {https://dx.doi.org/10.3847/1538-3881/ab72a3},
year = {2020},
month = {mar},
publisher = {The American Astronomical Society},
volume = {159},
number = {4},
pages = {158},
author = {Tandon, S. N. and Postma, J. and Joseph, P. and Devaraj, A. and Subramaniam, A and Barve, I. V. and George, K. and Ghosh, S. K. and Girish, V. and Hutchings, J. B. and Kamath, P. U. and Kathiravan, S. and Kumar, A. and Lancelot, J. P. and Leahy, D. and Mahesh, P. K. and Mohan, R. and Nagabhushana, S. and Pati, A. K. and Rao, N. Kameswara and Sankarasubramanian, K. and Sriram, S. and Stalin, C. S.},
title = {Additional Calibration of the Ultraviolet Imaging Telescope on Board AstroSat},
journal = {The Astronomical Journal}
}

@ARTICLE{2017PASP..129k5002P,
       author = {{Postma}, Joseph E. and {Leahy}, Denis},
        title = "{CCDLAB: A Graphical User Interface FITS Image Data Reducer, Viewer, and Canadian UVIT Data Pipeline}",
      journal = {\pasp},
         year = 2017,
        month = nov,
       volume = {129},
       number = {981},
        pages = {115002},
          doi = {10.1088/1538-3873/aa8800},
       adsurl = {https://ui.adsabs.harvard.edu/abs/2017PASP..129k5002P}
}

@ARTICLE{2008ApJ...679.1338R,
       author = {{Rosolowsky}, E.~W. and {Pineda}, J.~E. and {Kauffmann}, J. and {Goodman}, A.~A.},
        title = "{Structural Analysis of Molecular Clouds: Dendrograms}",
      journal = {\apj},
         year = 2008,
        month = jun,
       volume = {679},
       number = {2},
        pages = {1338-1351},
          doi = {10.1086/587685},
archivePrefix = {arXiv},
       eprint = {0802.2944},
 primaryClass = {astro-ph},
       adsurl = {https://ui.adsabs.harvard.edu/abs/2008ApJ...679.1338R}
}

@ARTICLE{2005MNRAS.360.1413B,
       author = {{Burgarella}, D. and {Buat}, V. and {Iglesias-P{\'a}ramo}, J.},
        title = "{Star formation and dust attenuation properties in galaxies from a statistical ultraviolet-to-far-infrared analysis}",
      journal = {\mnras},
         year = 2005,
        month = jul,
       volume = {360},
       number = {4},
        pages = {1413-1425},
          doi = {10.1111/j.1365-2966.2005.09131.x},
archivePrefix = {arXiv},
       eprint = {astro-ph/0504434},
 primaryClass = {astro-ph},
       adsurl = {https://ui.adsabs.harvard.edu/abs/2005MNRAS.360.1413B}
}

@article{Buat_2005,
doi = {10.1086/423241},
url = {https://dx.doi.org/10.1086/423241},
year = {2005},
month = {jan},
publisher = {},
volume = {619},
number = {1},
pages = {L51},
author = {Buat, V. and Iglesias-Páramo, J. and Seibert, M. and Burgarella, D. and Charlot, S. and Martin, D. C. and Xu, C. K. and Heckman, T. M. and Boissier, S. and Boselli, A. and Barlow, T. and Bianchi, L. and Byun, Y.-I. and Donas, J. and Forster, K. and Friedman, P. G. and Jelinski, P. and Lee, Y.-W. and Madore, B. F. and Malina, R. and Milliard, B. and Morissey, P. and Neff, S. and Rich, M. and Schiminovitch, D. and Siegmund, O. and Small, T. and Szalay, A. S. and Welsh, B. and Wyder, T. K.},
title = {Dust Attenuation in the Nearby Universe: A Comparison between Galaxies Selected in the Ultraviolet and in the Far-Infrared},
journal = {The Astrophysical Journal}
}

@ARTICLE{1992ApJ...396L..69S,
       author = {{Sauvage}, Marc and {Thuan}, Trinh X.},
        title = "{On the Use of Far-Infrared Luminosity as a Star Formation Indicator in Galaxies}",
      journal = {\apjl},
         year = 1992,
        month = sep,
       volume = {396},
        pages = {L69},
          doi = {10.1086/186519},
       adsurl = {https://ui.adsabs.harvard.edu/abs/1992ApJ...396L..69S}
}

@ARTICLE{2011ApJ...737..103S,
       author = {{Schlafly}, Edward F. and {Finkbeiner}, Douglas P.},
        title = "{Measuring Reddening with Sloan Digital Sky Survey Stellar Spectra and Recalibrating SFD}",
      journal = {\apj},
         year = 2011,
        month = aug,
       volume = {737},
       number = {2},
          eid = {103},
        pages = {103},
          doi = {10.1088/0004-637X/737/2/103},
archivePrefix = {arXiv},
       eprint = {1012.4804},
 primaryClass = {astro-ph.GA},
       adsurl = {https://ui.adsabs.harvard.edu/abs/2011ApJ...737..103S}
}

@ARTICLE{2019A&A...622A.103B,
       author = {{Boquien}, M. and {Burgarella}, D. and {Roehlly}, Y. and {Buat}, V. and {Ciesla}, L. and {Corre}, D. and {Inoue}, A.~K. and {Salas}, H.},
        title = "{CIGALE: a python Code Investigating GALaxy Emission}",
      journal = {\aap},
         year = 2019,
        month = feb,
       volume = {622},
          eid = {A103},
        pages = {A103},
          doi = {10.1051/0004-6361/201834156},
archivePrefix = {arXiv},
       eprint = {1811.03094},
 primaryClass = {astro-ph.GA},
       adsurl = {https://ui.adsabs.harvard.edu/abs/2019A&A...622A.103B}
}

@ARTICLE{2008MNRAS.386.1157C,
       author = {{Cortese}, L. and {Boselli}, A. and {Franzetti}, P. and {Decarli}, R. and {Gavazzi}, G. and {Boissier}, S. and {Buat}, V.},
        title = "{Ultraviolet dust attenuation in star-forming galaxies - II. Calibrating the A(UV) versus L$_{TIR}$/L$_{UV}$ relation}",
      journal = {\mnras},
         year = 2008,
        month = may,
       volume = {386},
       number = {2},
        pages = {1157-1168},
          doi = {10.1111/j.1365-2966.2008.13118.x},
archivePrefix = {arXiv},
       eprint = {0802.3020},
 primaryClass = {astro-ph},
       adsurl = {https://ui.adsabs.harvard.edu/abs/2008MNRAS.386.1157C}
}

@article{Calzetti_2007,
doi = {10.1086/520082},
url = {https://dx.doi.org/10.1086/520082},
year = {2007},
month = {sep},
publisher = {},
volume = {666},
number = {2},
pages = {870},
author = {Calzetti, D. and Kennicutt, R. C. and Engelbracht, C. W. and Leitherer, C. and Draine, B. T. and Kewley, L. and Moustakas, J. and Sosey, M. and Dale, D. A. and Gordon, K. D. and Helou, G. X. and Hollenbach, D. J. and Armus, L. and Bendo, G. and Bot, C. and Buckalew, B. and Jarrett, T. and Li, A. and Meyer, M. and Murphy, E. J. and Prescott, M. and Regan, M. W. and Rieke, G. H. and Roussel, H. and Sheth, K. and Smith, J. D. T. and Thornley, M. D. and Walter, F.},
title = {The Calibration of Mid-Infrared Star Formation Rate Indicators*},
journal = {The Astrophysical Journal}
}

@ARTICLE{1999ApJS..123....3L,
       author = {{Leitherer}, Claus and {Schaerer}, Daniel and {Goldader}, Jeffrey D. and {Delgado}, Rosa M. Gonz{\'a}lez and {Robert}, Carmelle and {Kune}, Denis Foo and {de Mello}, Du{\'\i}lia F. and {Devost}, Daniel and {Heckman}, Timothy M.},
        title = "{Starburst99: Synthesis Models for Galaxies with Active Star Formation}",
      journal = {\apjs},
         year = 1999,
        month = jul,
       volume = {123},
       number = {1},
        pages = {3-40},
          doi = {10.1086/313233},
archivePrefix = {arXiv},
       eprint = {astro-ph/9902334},
 primaryClass = {astro-ph},
       adsurl = {https://ui.adsabs.harvard.edu/abs/1999ApJS..123....3L}
}

@ARTICLE{2001MNRAS.322..231K,
       author = {{Kroupa}, Pavel},
        title = "{On the variation of the initial mass function}",
      journal = {\mnras},
         year = 2001,
        month = apr,
       volume = {322},
       number = {2},
        pages = {231-246},
          doi = {10.1046/j.1365-8711.2001.04022.x},
archivePrefix = {arXiv},
       eprint = {astro-ph/0009005},
 primaryClass = {astro-ph},
       adsurl = {https://ui.adsabs.harvard.edu/abs/2001MNRAS.322..231K}
}

@ARTICLE{2013ApJ...771...62K,
       author = {{Kreckel}, Kathryn and {Groves}, Brent and {Schinnerer}, Eva and {Johnson}, Benjamin D. and {Aniano}, Gonzalo and {Calzetti}, Daniela and {Croxall}, Kevin V. and {Draine}, Bruce T. and {Gordon}, Karl D. and {Crocker}, Alison F. and {Dale}, Daniel A. and {Hunt}, Leslie K. and {Kennicutt}, Robert C. and {Meidt}, Sharon E. and {Smith}, J.~D.~T. and {Tabatabaei}, Fatemeh S.},
        title = "{Mapping Dust through Emission and Absorption in Nearby Galaxies}",
      journal = {\apj},
         year = 2013,
        month = jul,
       volume = {771},
       number = {1},
          eid = {62},
        pages = {62},
          doi = {10.1088/0004-637X/771/1/62},
archivePrefix = {arXiv},
       eprint = {1305.2923},
 primaryClass = {astro-ph.CO},
       adsurl = {https://ui.adsabs.harvard.edu/abs/2013ApJ...771...62K}
}

@ARTICLE{2023MNRAS.520.4902G,
       author = {{Groves}, B. and {Kreckel}, K. and {Santoro}, F. and {Belfiore}, F. and {Zavodnik}, E. and {Congiu}, E. and {Egorov}, O.~V. and {Emsellem}, E. and {Grasha}, K. and {Leroy}, A. and {Scheuermann}, F. and {Schinnerer}, E. and {Watkins}, E.~J. and {Barnes}, A.~T. and {Bigiel}, F. and {Dale}, D.~A. and {Glover}, S.~C.~O. and {Pessa}, I. and {Sanchez-Blazquez}, P. and {Williams}, T.~G.},
        title = "{The PHANGS-MUSE nebular catalogue}",
      journal = {\mnras},
         year = 2023,
        month = apr,
       volume = {520},
       number = {4},
        pages = {4902-4952},
          doi = {10.1093/mnras/stad114},
archivePrefix = {arXiv},
       eprint = {2301.03811},
 primaryClass = {astro-ph.GA},
       adsurl = {https://ui.adsabs.harvard.edu/abs/2023MNRAS.520.4902G}
}

@ARTICLE{2018MNRAS.481.1016G,
       author = {{Grasha}, K. and {Calzetti}, D. and {Bittle}, L. and {Johnson}, K.~E. and {Donovan Meyer}, J. and {Kennicutt}, R.~C. and {Elmegreen}, B.~G. and {Adamo}, A. and {Krumholz}, M.~R. and {Fumagalli}, M. and {Grebel}, E.~K. and {Gouliermis}, D.~A. and {Cook}, D.~O. and {Gallagher}, J.~S. and {Aloisi}, A. and {Dale}, D.~A. and {Linden}, S. and {Sacchi}, E. and {Thilker}, D.~A. and {Walterbos}, R.~A.~M. and {Messa}, M. and {Wofford}, A. and {Smith}, L.~J.},
        title = "{Connecting young star clusters to CO molecular gas in NGC 7793 with ALMA-LEGUS}",
      journal = {\mnras},
         year = 2018,
        month = nov,
       volume = {481},
       number = {1},
        pages = {1016-1027},
          doi = {10.1093/mnras/sty2154},
archivePrefix = {arXiv},
       eprint = {1808.02496},
 primaryClass = {astro-ph.GA},
       adsurl = {https://ui.adsabs.harvard.edu/abs/2018MNRAS.481.1016G}
}

@ARTICLE{2025ApJ...987...33M,
       author = {{Meena}, Beena and {Sabbi}, Elena and {Zeidler}, Peter and {Elmegreen}, Bruce G. and {Eldridge}, Jan J. and {Bajaj}, Varun and {Gennaro}, Mario and {Pasquali}, Anna and {Elmegreen}, Debra M. and {Klessen}, Ralf S. and {Smith}, Linda J. and {Bianchi}, Luciana and {Wofford}, Aida and {Facchini}, Pietro and {Gallagher}, John S. and {Calzetti}, Daniela and {Grebel}, Eva K. and {Adamo}, Angela and {(GULP)}},
        title = "{GULP. II. Hierarchical Distribution and Evolution of Young Stellar Structures in NGC 4449}",
      journal = {\apj},
         year = 2025,
        month = jul,
       volume = {987},
       number = {1},
          eid = {33},
        pages = {33},
          doi = {10.3847/1538-4357/add475},
archivePrefix = {arXiv},
       eprint = {2505.02890},
 primaryClass = {astro-ph.GA},
       adsurl = {https://ui.adsabs.harvard.edu/abs/2025ApJ...987...33M}
}

@software{larry_bradley_2024_10967176,
  author       = {Larry Bradley and
                  Brigitta Sip{\H o}cz and
                  Thomas Robitaille and
                  Erik Tollerud and
                  Z\`e Vin{\'{\i}}cius and
                  Christoph Deil and
                  Kyle Barbary and
                  Tom J Wilson and
                  Ivo Busko and
                  Axel Donath and
                  Hans Moritz G{\"u}nther and
                  Mihai Cara and
                  P. L. Lim and
                  Sebastian Me{\ss}linger and
                  Zach Burnett and
                  Simon Conseil and
                  Michael Droettboom and
                  Azalee Bostroem and
                  E. M. Bray and
                  Lars Andersen Bratholm and
                  William Jamieson and
                  Adam Ginsburg and
                  Geert Barentsen and
                  Matt Craig and
                  Sergio Pascual and
                  Shivangee Rathi and
                  Marshall Perrin and
                  Brett M. Morris and
                  Gabriel Perren},
  title        = {astropy/photutils: 1.12.0},
  month        = apr,
  year         = 2024,
  publisher    = {Zenodo},
  version      = {1.12.0},
  doi          = {10.5281/zenodo.10967176},
  url          = {https://doi.org/10.5281/zenodo.10967176}
}

@ARTICLE{2022JApA...43...80S,
       author = {{Subramaniam}, Annapurni},
        title = "{An overview of the proposed Indian spectroscopic and imaging space telescope}",
      journal = {Journal of Astrophysics and Astronomy},
         year = 2022,
        month = dec,
       volume = {43},
       number = {2},
          eid = {80},
        pages = {80},
          doi = {10.1007/s12036-022-09870-3},
archivePrefix = {arXiv},
       eprint = {2206.03771},
 primaryClass = {astro-ph.IM},
       adsurl = {https://ui.adsabs.harvard.edu/abs/2022JApA...43...80S}
}

@Article{Hunter07,
  Author    = {Hunter, J. D.},
  Title     = {Matplotlib: A 2D graphics environment},
  Journal   = {Computing in Science \& Engineering},
  Volume    = {9},
  Number    = {3},
  Pages     = {90--95},
  publisher = {IEEE COMPUTER SOC},
  doi       = {10.1109/MCSE.2007.55},
  year      = 2007
}

@book{python09,
 author = {Van Rossum, Guido and Drake, Fred L.},
 title = {Python 3 Reference Manual},
 year = {2009},
 isbn = {1441412697},
 publisher = {CreateSpace},
 address = {Scotts Valley, CA}
}

@ARTICLE{SciPy20,
  author  = {Virtanen, Pauli and Gommers, Ralf and Oliphant, Travis E. and
            Haberland, Matt and Reddy, Tyler and Cournapeau, David and
            Burovski, Evgeni and Peterson, Pearu and Weckesser, Warren and
            Bright, Jonathan and {van der Walt}, St{\'e}fan J. and
            Brett, Matthew and Wilson, Joshua and Millman, K. Jarrod and
            Mayorov, Nikolay and Nelson, Andrew R. J. and Jones, Eric and
            Kern, Robert and Larson, Eric and Carey, C J and
            Polat, {\.I}lhan and Feng, Yu and Moore, Eric W. and
            {VanderPlas}, Jake and Laxalde, Denis and Perktold, Josef and
            Cimrman, Robert and Henriksen, Ian and Quintero, E. A. and
            Harris, Charles R. and Archibald, Anne M. and
            Ribeiro, Ant{\^o}nio H. and Pedregosa, Fabian and
            {van Mulbregt}, Paul and {SciPy 1.0 Contributors}},
  title   = {{{SciPy} 1.0: Fundamental Algorithms for Scientific
            Computing in Python}},
  journal = {Nature Methods},
  year    = {2020},
  volume  = {17},
  pages   = {261--272},
  adsurl  = {https://rdcu.be/b08Wh},
  doi     = {10.1038/s41592-019-0686-2},
}

@article{JMLR:v12:pedregosa11a,
  author  = {Fabian Pedregosa and Ga{{\"e}}l Varoquaux and Alexandre Gramfort and Vincent Michel and Bertrand Thirion and Olivier Grisel and Mathieu Blondel and Peter Prettenhofer and Ron Weiss and Vincent Dubourg and Jake Vanderplas and Alexandre Passos and David Cournapeau and Matthieu Brucher and Matthieu Perrot and {{\'E}}douard Duchesnay},
  title   = {Scikit-learn: Machine Learning in Python},
  journal = {Journal of Machine Learning Research},
  year    = {2011},
  volume  = {12},
  number  = {85},
  pages   = {2825--2830},
  url     = {http://jmlr.org/papers/v12/pedregosa11a.html}
}

@INPROCEEDINGS{2012cidu.conf...47V,
       author = {{VanderPlas}, J. and {Connolly}, A.~J. and {Ivezic}, Z. and {Gray}, A.},
        title = "{Introduction to astroML: Machine learning for astrophysics}",
    booktitle = {Proceedings of Conference on Intelligent Data Understanding (CIDU},
         year = 2012,
        month = oct,
        pages = {47-54},
          doi = {10.1109/CIDU.2012.6382200},
archivePrefix = {arXiv},
       eprint = {1411.5039},
 primaryClass = {astro-ph.IM},
       adsurl = {https://ui.adsabs.harvard.edu/abs/2012cidu.conf...47V}
}

@ARTICLE{NumPy20,
  author  = {Harris, Charles R. and Millman, K. Jarrod and van der Walt, Stéfan J and Gommers, Ralf and Virtanen, Pauli and Cournapeau, David and Wieser, Eric and Taylor, Julian and Berg, Sebastian and Smith, Nathaniel J. and Kern, Robert and Picus, Matti and Hoyer, Stephan and van Kerkwijk, Marten H. and Brett, Matthew and Haldane, Allan and Fernández del Río, Jaime and Wiebe, Mark and Peterson, Pearu and Gérard-Marchant, Pierre and Sheppard, Kevin and Reddy, Tyler and Weckesser, Warren and Abbasi, Hameer and Gohlke, Christoph and Oliphant, Travis E.},
  title   = {Array programming with {NumPy}},
  journal = {Nature},
  year    = {2020},
  volume  = {585},
  pages   = {357–362},
  doi     = {10.1038/s41586-020-2649-2}
}

@ARTICLE{astropy_2018,
       author = {{Astropy Collaboration} and {Price-Whelan}, A.~M. and
         {Sip{\H{o}}cz}, B.~M. and {G{\"u}nther}, H.~M. and {Lim}, P.~L. and
         {Crawford}, S.~M. and {Conseil}, S. and {Shupe}, D.~L. and
         {Craig}, M.~W. and {Dencheva}, N. and {Ginsburg}, A. and {Vand
        erPlas}, J.~T. and {Bradley}, L.~D. and {P{\'e}rez-Su{\'a}rez}, D. and
         {de Val-Borro}, M. and {Aldcroft}, T.~L. and {Cruz}, K.~L. and
         {Robitaille}, T.~P. and {Tollerud}, E.~J. and {Ardelean}, C. and
         {Babej}, T. and {Bach}, Y.~P. and {Bachetti}, M. and {Bakanov}, A.~V. and
         {Bamford}, S.~P. and {Barentsen}, G. and {Barmby}, P. and
         {Baumbach}, A. and {Berry}, K.~L. and {Biscani}, F. and {Boquien}, M. and
         {Bostroem}, K.~A. and {Bouma}, L.~G. and {Brammer}, G.~B. and
         {Bray}, E.~M. and {Breytenbach}, H. and {Buddelmeijer}, H. and
         {Burke}, D.~J. and {Calderone}, G. and {Cano Rodr{\'\i}guez}, J.~L. and
         {Cara}, M. and {Cardoso}, J.~V.~M. and {Cheedella}, S. and {Copin}, Y. and
         {Corrales}, L. and {Crichton}, D. and {D'Avella}, D. and {Deil}, C. and
         {Depagne}, {\'E}. and {Dietrich}, J.~P. and {Donath}, A. and
         {Droettboom}, M. and {Earl}, N. and {Erben}, T. and {Fabbro}, S. and
         {Ferreira}, L.~A. and {Finethy}, T. and {Fox}, R.~T. and
         {Garrison}, L.~H. and {Gibbons}, S.~L.~J. and {Goldstein}, D.~A. and
         {Gommers}, R. and {Greco}, J.~P. and {Greenfield}, P. and
         {Groener}, A.~M. and {Grollier}, F. and {Hagen}, A. and {Hirst}, P. and
         {Homeier}, D. and {Horton}, A.~J. and {Hosseinzadeh}, G. and {Hu}, L. and
         {Hunkeler}, J.~S. and {Ivezi{\'c}}, {\v{Z}}. and {Jain}, A. and
         {Jenness}, T. and {Kanarek}, G. and {Kendrew}, S. and {Kern}, N.~S. and
         {Kerzendorf}, W.~E. and {Khvalko}, A. and {King}, J. and {Kirkby}, D. and
         {Kulkarni}, A.~M. and {Kumar}, A. and {Lee}, A. and {Lenz}, D. and
         {Littlefair}, S.~P. and {Ma}, Z. and {Macleod}, D.~M. and
         {Mastropietro}, M. and {McCully}, C. and {Montagnac}, S. and
         {Morris}, B.~M. and {Mueller}, M. and {Mumford}, S.~J. and {Muna}, D. and
         {Murphy}, N.~A. and {Nelson}, S. and {Nguyen}, G.~H. and
         {Ninan}, J.~P. and {N{\"o}the}, M. and {Ogaz}, S. and {Oh}, S. and
         {Parejko}, J.~K. and {Parley}, N. and {Pascual}, S. and {Patil}, R. and
         {Patil}, A.~A. and {Plunkett}, A.~L. and {Prochaska}, J.~X. and
         {Rastogi}, T. and {Reddy Janga}, V. and {Sabater}, J. and
         {Sakurikar}, P. and {Seifert}, M. and {Sherbert}, L.~E. and
         {Sherwood-Taylor}, H. and {Shih}, A.~Y. and {Sick}, J. and
         {Silbiger}, M.~T. and {Singanamalla}, S. and {Singer}, L.~P. and
         {Sladen}, P.~H. and {Sooley}, K.~A. and {Sornarajah}, S. and
         {Streicher}, O. and {Teuben}, P. and {Thomas}, S.~W. and
         {Tremblay}, G.~R. and {Turner}, J.~E.~H. and {Terr{\'o}n}, V. and
         {van Kerkwijk}, M.~H. and {de la Vega}, A. and {Watkins}, L.~L. and
         {Weaver}, B.~A. and {Whitmore}, J.~B. and {Woillez}, J. and
         {Zabalza}, V. and {Astropy Contributors}},
        title = "{The Astropy Project: Building an Open-science Project and Status of the v2.0 Core Package}",
      journal = {\aj},
         year = 2018,
        month = sep,
       volume = {156},
       number = {3},
          eid = {123},
        pages = {123},
          doi = {10.3847/1538-3881/aabc4f},
archivePrefix = {arXiv},
       eprint = {1801.02634},
 primaryClass = {astro-ph.IM},
       adsurl = {https://ui.adsabs.harvard.edu/abs/2018AJ....156..123A}
}

@ARTICLE{2010ApJ...712..858G,
       author = {{Gogarten}, Stephanie M. and {Dalcanton}, Julianne J. and {Williams}, Benjamin F. and {Ro{\v{s}}kar}, Rok and {Holtzman}, Jon and {Seth}, Anil C. and {Dolphin}, Andrew and {Weisz}, Daniel and {Cole}, Andrew and {Debattista}, Victor P. and {Gilbert}, Karoline M. and {Olsen}, Knut and {Skillman}, Evan and {de Jong}, Roelof S. and {Karachentsev}, Igor D. and {Quinn}, Thomas R.},
        title = "{The Advanced Camera for Surveys Nearby Galaxy Survey Treasury. V. Radial Star Formation History of NGC 300}",
      journal = {\apj},
         year = 2010,
        month = apr,
       volume = {712},
       number = {2},
        pages = {858-874},
          doi = {10.1088/0004-637X/712/2/858},
archivePrefix = {arXiv},
       eprint = {1002.1743},
 primaryClass = {astro-ph.GA},
       adsurl = {https://ui.adsabs.harvard.edu/abs/2010ApJ...712..858G}
}

@article{bianchi2005recent,
  title={Recent star formation in nearby galaxies from Galaxy Evolution Explorer imaging: M101 and M51},
  author={Bianchi, Luciana and Thilker, David A and Burgarella, Denis and Friedman, Peter G and Hoopes, Charles G and Boissier, Samuel and De Paz, Armando Gil and Barlow, Tom A and Byun, Yong-Ik and Donas, Jose and others},
  journal={The Astrophysical Journal},
  volume={619},
  number={1},
  pages={L71},
  year={2005},
  publisher={IOP Publishing}
}

@ARTICLE{2011ApJ...735...63L,
       author = {{Liu}, Guilin and {Koda}, Jin and {Calzetti}, Daniela and {Fukuhara}, Masayuki and {Momose}, Rieko},
        title = "{The Super-linear Slope of the Spatially Resolved Star Formation Law in NGC 3521 and NGC 5194 (M51a)}",
      journal = {\apj},
         year = 2011,
        month = jul,
       volume = {735},
       number = {1},
          eid = {63},
        pages = {63},
          doi = {10.1088/0004-637X/735/1/63},
archivePrefix = {arXiv},
       eprint = {1104.4122},
 primaryClass = {astro-ph.CO},
       adsurl = {https://ui.adsabs.harvard.edu/abs/2011ApJ...735...63L}
}

@ARTICLE{2022A&A...659A.191E,
       author = {{Emsellem}, Eric and {Schinnerer}, Eva and {Santoro}, Francesco and {Belfiore}, Francesco and {Pessa}, Ismael and {McElroy}, Rebecca and {Blanc}, Guillermo A. and {Congiu}, Enrico and {Groves}, Brent and {Ho}, I.-Ting and {Kreckel}, Kathryn and {Razza}, Alessandro and {Sanchez-Blazquez}, Patricia and {Egorov}, Oleg and {Faesi}, Chris and {Klessen}, Ralf S. and {Leroy}, Adam K. and {Meidt}, Sharon and {Querejeta}, Miguel and {Rosolowsky}, Erik and {Scheuermann}, Fabian and {Anand}, Gagandeep S. and {Barnes}, Ashley T. and {Be{\v{s}}li{\'c}}, Ivana and {Bigiel}, Frank and {Boquien}, M{\'e}d{\'e}ric and {Cao}, Yixian and {Chevance}, M{\'e}lanie and {Dale}, Daniel A. and {Eibensteiner}, Cosima and {Glover}, Simon C.~O. and {Grasha}, Kathryn and {Henshaw}, Jonathan D. and {Hughes}, Annie and {Koch}, Eric W. and {Kruijssen}, J.~M. Diederik and {Lee}, Janice and {Liu}, Daizhong and {Pan}, Hsi-An and {Pety}, J{\'e}r{\^o}me and {Saito}, Toshiki and {Sandstrom}, Karin M. and {Schruba}, Andreas and {Sun}, Jiayi and {Thilker}, David A. and {Usero}, Antonio and {Watkins}, Elizabeth J. and {Williams}, Thomas G.},
        title = "{The PHANGS-MUSE survey. Probing the chemo-dynamical evolution of disc galaxies}",
      journal = {\aap},
         year = 2022,
        month = mar,
       volume = {659},
          eid = {A191},
        pages = {A191},
          doi = {10.1051/0004-6361/202141727},
archivePrefix = {arXiv},
       eprint = {2110.03708},
 primaryClass = {astro-ph.GA},
       adsurl = {https://ui.adsabs.harvard.edu/abs/2022A&A...659A.191E}
}

@ARTICLE{2022A&A...658A.188S,
       author = {{Santoro}, Francesco and {Kreckel}, Kathryn and {Belfiore}, Francesco and {Groves}, Brent and {Congiu}, Enrico and {Thilker}, David A. and {Blanc}, Guillermo A. and {Schinnerer}, Eva and {Ho}, I.-Ting and {Kruijssen}, J.~M. Diederik and {Meidt}, Sharon and {Klessen}, Ralf S. and {Schruba}, Andreas and {Querejeta}, Miguel and {Pessa}, Ismael and {Chevance}, M{\'e}lanie and {Kim}, Jaeyeon and {Emsellem}, Eric and {McElroy}, Rebecca and {Barnes}, Ashley T. and {Bigiel}, Frank and {Boquien}, M{\'e}d{\'e}ric and {Dale}, Daniel A. and {Glover}, Simon C.~O. and {Grasha}, Kathryn and {Lee}, Janice and {Leroy}, Adam K. and {Pan}, Hsi-An and {Rosolowsky}, Erik and {Saito}, Toshiki and {Sanchez-Blazquez}, Patricia and {Watkins}, Elizabeth J. and {Williams}, Thomas G.},
        title = "{PHANGS-MUSE: The H II region luminosity function of local star-forming galaxies}",
      journal = {\aap},
         year = 2022,
        month = feb,
       volume = {658},
          eid = {A188},
        pages = {A188},
          doi = {10.1051/0004-6361/202141907},
archivePrefix = {arXiv},
       eprint = {2111.09362},
 primaryClass = {astro-ph.GA},
       adsurl = {https://ui.adsabs.harvard.edu/abs/2022A&A...658A.188S}
}

@ARTICLE{2011ApJ...743..137B,
       author = {{Barnes}, Kate L. and {van Zee}, Liese and {Skillman}, Evan D.},
        title = "{Star Formation in the Outer Disks of Spiral Galaxies: Ultraviolet and H{\ensuremath{\alpha}} Photometry}",
      journal = {\apj},
         year = 2011,
        month = dec,
       volume = {743},
       number = {2},
          eid = {137},
        pages = {137},
          doi = {10.1088/0004-637X/743/2/137},
       adsurl = {https://ui.adsabs.harvard.edu/abs/2011ApJ...743..137B}
}

@ARTICLE{2021MNRAS.502.5508P,
       author = {{Parikh}, Taniya and {Thomas}, Daniel and {Maraston}, Claudia and {Westfall}, Kyle B. and {Andrews}, Brett H. and {Boardman}, Nicholas Fraser and {Drory}, Niv and {Oyarzun}, Grecco},
        title = "{SDSS-IV MaNGA: radial gradients in stellar population properties of early-type and late-type galaxies}",
      journal = {\mnras},
         year = 2021,
        month = apr,
       volume = {502},
       number = {4},
        pages = {5508-5527},
          doi = {10.1093/mnras/stab449},
archivePrefix = {arXiv},
       eprint = {2102.06703},
 primaryClass = {astro-ph.GA},
       adsurl = {https://ui.adsabs.harvard.edu/abs/2021MNRAS.502.5508P}
}

@ARTICLE{1976MNRAS.176...31L,
       author = {{Larson}, R.~B.},
        title = "{Models for the formation of disc galaxies.}",
      journal = {\mnras},
         year = 1976,
        month = jul,
       volume = {176},
        pages = {31-52},
          doi = {10.1093/mnras/176.1.31},
       adsurl = {https://ui.adsabs.harvard.edu/abs/1976MNRAS.176...31L}
}

@ARTICLE{2006MNRAS.366..899N,
       author = {{Naab}, Thorsten and {Ostriker}, Jeremiah P.},
        title = "{A simple model for the evolution of disc galaxies: the Milky Way}",
      journal = {\mnras},
         year = 2006,
        month = mar,
       volume = {366},
       number = {3},
        pages = {899-917},
          doi = {10.1111/j.1365-2966.2005.09807.x},
archivePrefix = {arXiv},
       eprint = {astro-ph/0505594},
 primaryClass = {astro-ph},
       adsurl = {https://ui.adsabs.harvard.edu/abs/2006MNRAS.366..899N}
}

@ARTICLE{2017A&A...608A..27G,
       author = {{Garc{\'\i}a-Benito}, R. and {Gonz{\'a}lez Delgado}, R.~M. and {P{\'e}rez}, E. and {Cid Fernandes}, R. and {Cortijo-Ferrero}, C. and {L{\'o}pez Fern{\'a}ndez}, R. and {de Amorim}, A.~L. and {Lacerda}, E.~A.~D. and {Vale Asari}, N. and {S{\'a}nchez}, S.~F.},
        title = "{The spatially resolved star formation history of CALIFA galaxies. Cosmic time scales}",
      journal = {\aap},
         year = 2017,
        month = dec,
       volume = {608},
          eid = {A27},
        pages = {A27},
          doi = {10.1051/0004-6361/201731357},
archivePrefix = {arXiv},
       eprint = {1709.00413},
 primaryClass = {astro-ph.GA},
       adsurl = {https://ui.adsabs.harvard.edu/abs/2017A&A...608A..27G}
}

@ARTICLE{1964ApJ...140..646L,
       author = {{Lin}, C.~C. and {Shu}, Frank H.},
        title = "{On the Spiral Structure of Disk Galaxies.}",
      journal = {\apj},
         year = 1964,
        month = aug,
       volume = {140},
        pages = {646},
          doi = {10.1086/147955},
       adsurl = {https://ui.adsabs.harvard.edu/abs/1964ApJ...140..646L}
}

@ARTICLE{2010ApJ...725.1342D,
       author = {{Davidge}, T.~J.},
        title = "{Shaken, Not Stirred: The Disrupted Disk of the Starburst Galaxy NGC 253}",
      journal = {\apj},
         year = 2010,
        month = dec,
       volume = {725},
       number = {1},
        pages = {1342-1365},
          doi = {10.1088/0004-637X/725/1/1342},
archivePrefix = {arXiv},
       eprint = {1011.3006},
 primaryClass = {astro-ph.CO},
       adsurl = {https://ui.adsabs.harvard.edu/abs/2010ApJ...725.1342D}
}

@ARTICLE{2019ApJ...884...21K,
       author = {{Kim}, Duho and {Jansen}, Rolf A. and {Windhorst}, Rogier A. and {Cohen}, Seth H. and {McCabe}, Tyler J.},
        title = "{Analysis of the Spatially Resolved V-3.6 {\ensuremath{\mu}}m Colors and Dust Extinction in 257 Nearby NGC and IC Galaxies}",
      journal = {\apj},
         year = 2019,
        month = oct,
       volume = {884},
       number = {1},
          eid = {21},
        pages = {21},
          doi = {10.3847/1538-4357/ab385c},
archivePrefix = {arXiv},
       eprint = {1901.00565},
 primaryClass = {astro-ph.GA},
       adsurl = {https://ui.adsabs.harvard.edu/abs/2019ApJ...884...21K}
}

@ARTICLE{2024ApJ...977...20R,
       author = {{Riggs}, Claire L. and {Brooks}, Alyson M. and {Munshi}, Ferah and {Christensen}, Charlotte R. and {Cohen}, Roger E. and {Quinn}, Thomas R. and {Wadsley}, James},
        title = "{Testable Predictions of Outside-in Age Gradients in Dwarf Galaxies of All Types}",
      journal = {\apj},
         year = 2024,
        month = dec,
       volume = {977},
       number = {1},
          eid = {20},
        pages = {20},
          doi = {10.3847/1538-4357/ad8b1e},
archivePrefix = {arXiv},
       eprint = {2408.10379},
 primaryClass = {astro-ph.GA},
       adsurl = {https://ui.adsabs.harvard.edu/abs/2024ApJ...977...20R}
}

@ARTICLE{2019MNRAS.490.1186G,
       author = {{Graus}, Andrew S. and {Bullock}, James S. and {Fitts}, Alex and {Cooper}, Michael C. and {Boylan-Kolchin}, Michael and {Weisz}, Daniel R. and {Wetzel}, Andrew and {Feldmann}, Robert and {Faucher-Gigu{\`e}re}, Claude-Andr{\'e} and {Quataert}, Eliot and {Hopkins}, Philip F. and {Kere{\v{s}}}, Du{\v{s}}an},
        title = "{A predicted correlation between age gradient and star formation history in FIRE dwarf galaxies}",
      journal = {\mnras},
         year = 2019,
        month = nov,
       volume = {490},
       number = {1},
        pages = {1186-1201},
          doi = {10.1093/mnras/stz2649},
archivePrefix = {arXiv},
       eprint = {1901.05487},
 primaryClass = {astro-ph.GA},
       adsurl = {https://ui.adsabs.harvard.edu/abs/2019MNRAS.490.1186G}
}

@article{10.1117/1.JATIS.11.4.042202,
author = {Patrick C{\^o}t{\'e} and Tyrone Woods and John B. Hutchings and Jason D. Rhodes and Ruben Sanchez-Janssen and Alan D. Scott and John S. Pazder and Melissa Amenouche and Michael Balogh and Simon Blouin and Alain Cournoyer and Maria R. Drout and Nick Kuzmin and Katherine J. Mack and Laura Ferrarese and Wesley C. Fraser and Sarah  C.  Gallagher and Fr{\'e}d{\'e}ric Grandmont and Daryl Haggard and Paul Harrison and Vincent H{\'e}nault-Brunet and J. J. Kavelaars and Viraja Khatu and Joel C. Roediger and Jason Rowe and Marcin Sawicki and Jesper Skottfelt and Matt Taylor and Ludo van Waerbeke and Laurie Amen and Dhananjhay Bansal and Martin Bergeron and Toby Brown and Greg Burley and Hum Chand and Isaac Cheng and Ryan Cloutier and Nolan Dickson and Oleg Djazovski and Ivana Damjanov and James Doherty and Kyle Finner and Macarena Garc{\'i}a Del Valle Espinosa and Jennifer Glover and Ana I. G{\'o}mez de Castro and Or Graur and Tim Hardy and Michelle Kao and Denis A. Leahy and Deborah M. Lokhorst and A;ex Malz and Allison Man and Madeline Marshall and Sean McGee and Ryan McKenzie and Kai Michaud and Surhud S. More and David G. Morris and Patrick  W. Morris and Thibaud Moutard and Wasi Naqvi and Matt Nicholl and Ga{\"e}l Noirot and M.  S. Oey and Cyrielle Opitom and Samir Salim and Bryan Scott and Charles A. Shapiro and Daniel K. Stern and Annapurni Subramaniam and David Thilke and Ivan Wevers and Dmitry Vorobiev and L. Y. Aaron Yung and Fr{\'e}d{\'e}ric Zamkotsian and Suzanne Aigrain and Anahita Alavi and Martin A. Barstow and Peter Bartosik and Hadleigh Bluhm and Jo Bovy and Peter G. Cameron and Raymond G. Carlberg and Jessie L. Christiansen and Yuyang Chen and Paul Crowther and Kristen Dage and Aaron Dotter and Patrick Dufour and Jean Dupuis and Ben Dryer and Angaraj Duara and Gwendolyn M. Eadie and Marielle R. Eduardo and Vincente Estrada-Carpenter and Sebastien Fabbro and Andreas Faisst and Nicole M. Ford and Morgan Fraser and Boris G{\"a}nsicke and Shashikiran Ganesh and Poshak Gandhi and Melissa L. Graham and Rebecca Hamel and Martin Hellmich and John J. Hennessy and Kaitlyn Hessel and Jeremy Heyl and Catherine Heymans and Yashar Hezaveh and Renee Hlozek and Michael E. Hoenk and Andrew Holland and Eric Huff and Ian Hutchinson and Ikuru Iwata and April D. Jewell and Doug Johnstone and Maia Jones and Todd Jones and Dustin Lang and Jon Lapington and Justin Larivi{\`e}re and Cameron Lawlor-Forsyth and Denis Laurin and Charles Lee and Ronan Legin and Ting S. Li and Sungsoon Lim and Bethany Ludwig and Matt Kozun and Vivek M. and Robert Mann and Alan W. McConnachie and Evan McDonough and Stanimir Metchev and David R. Miller and Takashi Moriya and Cameron Morgan and Julio Navarro and Ya{\"e}l Naz{\'e} and Shouleh Nikzad and Vivek Oad and Nathalie Ouellette and Emily K. Pass and Will  J. Percival and Laurence Perreault Levasseur and Joe Postma and Nayyer Raza and Gordon T. Richards and Harvey Richer and Carmelle Robert and Erik Rosolowsky and John  J. Ruan and Sarah Rugheimer and Samar Safi-Harb and Kanak Saha and Vicky Scowcroft and Federico Sestito and Himanshu Sharma and James Sikora and Gregory  R.  Sivakoff and Thirupathi Sivarani and Patrick Smith and Warren Soh and Robert Sorba and Smitha Subramanian and Hossen Teimoorinia and Harry I. Teplitz and Shaylin Thadani and Shavon Thadani and Aaron Tohuvavohu and Kim A. Venn and Nicholas Vieira and Jeremy J. Webb and Paul Wiegert and Ryan Wierckx and Yanqin Wu and Jade Yeung and Sukyoung  K.  Yi},
title = {{The CASTOR mission}},
volume = {11},
journal = {Journal of Astronomical Telescopes, Instruments, and Systems},
number = {4},
publisher = {SPIE},
pages = {042202},
year = {2025},
doi = {10.1117/1.JATIS.11.4.042202},
URL = {https://doi.org/10.1117/1.JATIS.11.4.042202}
}

@ARTICLE{2008AJ....136.2782L,
       author = {{Leroy}, Adam K. and {Walter}, Fabian and {Brinks}, Elias and {Bigiel}, Frank and {de Blok}, W.~J.~G. and {Madore}, Barry and {Thornley}, M.~D.},
        title = "{The Star Formation Efficiency in Nearby Galaxies: Measuring Where Gas Forms Stars Effectively}",
      journal = {\aj},
         year = 2008,
        month = dec,
       volume = {136},
       number = {6},
        pages = {2782-2845},
          doi = {10.1088/0004-6256/136/6/2782},
archivePrefix = {arXiv},
       eprint = {0810.2556},
 primaryClass = {astro-ph},
       adsurl = {https://ui.adsabs.harvard.edu/abs/2008AJ....136.2782L}
}

@ARTICLE{2019MNRAS.490..467Z,
       author = {{Zheng}, Yong and {Putman}, Mary E. and {Emerick}, Andrew and {McQuinn}, Kristen B.~W. and {Werk}, Jessica K. and {Lockman}, Felix J. and {Oppenheimer}, Benjamin D. and {Fox}, Andrew J. and {Kirby}, Evan N. and {Burchett}, Joseph N.},
        title = "{Tentative detection of the circumgalactic medium of the isolated low-mass dwarf galaxy WLM}",
      journal = {\mnras},
         year = 2019,
        month = nov,
       volume = {490},
       number = {1},
        pages = {467-477},
          doi = {10.1093/mnras/stz2563},
archivePrefix = {arXiv},
       eprint = {1909.05407},
 primaryClass = {astro-ph.GA},
       adsurl = {https://ui.adsabs.harvard.edu/abs/2019MNRAS.490..467Z}
}

@ARTICLE{2022MNRAS.515.3270S,
       author = {{Smith}, Madison V. and {van Zee}, L. and {Dale}, D.~A. and {Hunter}, L.~C. and {Staudaher}, S. and {Wrock}, T.},
        title = "{A multiwavelength study of star formation in nearby galaxies: evidence for inside-out growth of the stellar disc}",
      journal = {\mnras},
         year = 2022,
        month = sep,
       volume = {515},
       number = {3},
        pages = {3270-3298},
          doi = {10.1093/mnras/stac1974},
       adsurl = {https://ui.adsabs.harvard.edu/abs/2022MNRAS.515.3270S}
}

@ARTICLE{2023ApJ...950...81N,
       author = {{Nandi}, Payel and {Stalin}, C.~S. and {Saikia}, D.~J. and {Muneer}, S. and {Mountrichas}, George and {Wylezalek}, Dominika and {Sagar}, R. and {Kissler-Patig}, Markus},
        title = "{Star Formation in the Dwarf Seyfert Galaxy NGC 4395: Evidence for Both AGN and SN Feedback?}",
      journal = {\apj},
         year = 2023,
        month = jun,
       volume = {950},
       number = {2},
          eid = {81},
        pages = {81},
          doi = {10.3847/1538-4357/accf1e},
archivePrefix = {arXiv},
       eprint = {2304.08986},
 primaryClass = {astro-ph.GA},
       adsurl = {https://ui.adsabs.harvard.edu/abs/2023ApJ...950...81N}
}

@ARTICLE{2015AJ....149....1Z,
       author = {{Zhou}, Zhi-Min and {Cao}, Chen and {Wu}, Hong},
        title = "{Star Formation Properties in Barred Galaxies. III. Statistical Study of Bar-Driven Secular Evolution Using a Sample of Nearby Barred Spirals}",
      journal = {\aj},
         year = 2015,
        month = jan,
       volume = {149},
       number = {1},
          eid = {1},
        pages = {1},
          doi = {10.1088/0004-6256/149/1/1},
archivePrefix = {arXiv},
       eprint = {1409.3045},
 primaryClass = {astro-ph.GA},
       adsurl = {https://ui.adsabs.harvard.edu/abs/2015AJ....149....1Z}
}

@ARTICLE{2021ApJS..257...43L,
       author = {{Leroy}, Adam K. and {Schinnerer}, Eva and {Hughes}, Annie and {Rosolowsky}, Erik and {Pety}, J{\'e}r{\^o}me and {Schruba}, Andreas and {Usero}, Antonio and {Blanc}, Guillermo A. and {Chevance}, M{\'e}lanie and {Emsellem}, Eric and {Faesi}, Christopher M. and {Herrera}, Cinthya N. and {Liu}, Daizhong and {Meidt}, Sharon E. and {Querejeta}, Miguel and {Saito}, Toshiki and {Sandstrom}, Karin M. and {Sun}, Jiayi and {Williams}, Thomas G. and {Anand}, Gagandeep S. and {Barnes}, Ashley T. and {Behrens}, Erica A. and {Belfiore}, Francesco and {Benincasa}, Samantha M. and {Be{\v{s}}li{\'c}}, Ivana and {Bigiel}, Frank and {Bolatto}, Alberto D. and {den Brok}, Jakob S. and {Cao}, Yixian and {Chandar}, Rupali and {Chastenet}, J{\'e}r{\'e}my and {Chiang}, I-Da and {Congiu}, Enrico and {Dale}, Daniel A. and {Deger}, Sinan and {Eibensteiner}, Cosima and {Egorov}, Oleg V. and {Garc{\'\i}a-Rodr{\'\i}guez}, Axel and {Glover}, Simon C.~O. and {Grasha}, Kathryn and {Henshaw}, Jonathan D. and {Ho}, I.-Ting and {Kepley}, Amanda A. and {Kim}, Jaeyeon and {Klessen}, Ralf S. and {Kreckel}, Kathryn and {Koch}, Eric W. and {Kruijssen}, J.~M. Diederik and {Larson}, Kirsten L. and {Lee}, Janice C. and {Lopez}, Laura A. and {Machado}, Josh and {Mayker}, Ness and {McElroy}, Rebecca and {Murphy}, Eric J. and {Ostriker}, Eve C. and {Pan}, Hsi-An and {Pessa}, Ismael and {Puschnig}, Johannes and {Razza}, Alessandro and {S{\'a}nchez-Bl{\'a}zquez}, Patricia and {Santoro}, Francesco and {Sardone}, Amy and {Scheuermann}, Fabian and {Sliwa}, Kazimierz and {Sormani}, Mattia C. and {Stuber}, Sophia K. and {Thilker}, David A. and {Turner}, Jordan A. and {Utomo}, Dyas and {Watkins}, Elizabeth J. and {Whitmore}, Bradley},
        title = "{PHANGS-ALMA: Arcsecond CO(2-1) Imaging of Nearby Star-forming Galaxies}",
      journal = {\apjs},
         year = 2021,
        month = dec,
       volume = {257},
       number = {2},
          eid = {43},
        pages = {43},
          doi = {10.3847/1538-4365/ac17f3},
archivePrefix = {arXiv},
       eprint = {2104.07739},
 primaryClass = {astro-ph.GA},
       adsurl = {https://ui.adsabs.harvard.edu/abs/2021ApJS..257...43L}
}

@ARTICLE{2007A&A...462..933C,
       author = {{Chy{\.z}y}, K.~T. and {Bomans}, D.~J. and {Krause}, M. and {Beck}, R. and {Soida}, M. and {Urbanik}, M.},
        title = "{Magnetic fields and ionized gas in nearby late type galaxies}",
      journal = {\aap},
         year = 2007,
        month = feb,
       volume = {462},
       number = {3},
        pages = {933-941},
          doi = {10.1051/0004-6361:20065932},
archivePrefix = {arXiv},
       eprint = {astro-ph/0611316},
 primaryClass = {astro-ph},
       adsurl = {https://ui.adsabs.harvard.edu/abs/2007A&A...462..933C}
}

@ARTICLE{2019A&A...621A..51H,
       author = {{Hunt}, L.~K. and {De Looze}, I. and {Boquien}, M. and {Nikutta}, R. and {Rossi}, A. and {Bianchi}, S. and {Dale}, D.~A. and {Granato}, G.~L. and {Kennicutt}, R.~C. and {Silva}, L. and {Ciesla}, L. and {Rela{\~n}o}, M. and {Viaene}, S. and {Brandl}, B. and {Calzetti}, D. and {Croxall}, K.~V. and {Draine}, B.~T. and {Galametz}, M. and {Gordon}, K.~D. and {Groves}, B.~A. and {Helou}, G. and {Herrera-Camus}, R. and {Hinz}, J.~L. and {Koda}, J. and {Salim}, S. and {Sandstrom}, K.~M. and {Smith}, J.~D. and {Wilson}, C.~D. and {Zibetti}, S.},
        title = "{Comprehensive comparison of models for spectral energy distributions from 0.1 {\ensuremath{\mu}}m to 1 mm of nearby star-forming galaxies}",
      journal = {\aap},
         year = 2019,
        month = jan,
       volume = {621},
          eid = {A51},
        pages = {A51},
          doi = {10.1051/0004-6361/201834212},
archivePrefix = {arXiv},
       eprint = {1809.04088},
 primaryClass = {astro-ph.GA},
       adsurl = {https://ui.adsabs.harvard.edu/abs/2019A&A...621A..51H}
}

@ARTICLE{1997MNRAS.290...15T,
       author = {{Thean}, A.~H.~C. and {Mundell}, C.~G. and {Pedlar}, A. and {Nicholson}, R.~A.},
        title = "{A neutral hydrogen study of the Seyfert galaxy NGC 5033}",
      journal = {\mnras},
         year = 1997,
        month = sep,
       volume = {290},
       number = {1},
        pages = {15-24},
          doi = {10.1093/mnras/290.1.15},
       adsurl = {https://ui.adsabs.harvard.edu/abs/1997MNRAS.290...15T}
}

@ARTICLE{2019MNRAS.488.3826B,
       author = {{Bresolin}, Fabio},
        title = "{Metallicity gradients in small and nearby spiral galaxies}",
      journal = {\mnras},
         year = 2019,
        month = sep,
       volume = {488},
       number = {3},
        pages = {3826-3843},
          doi = {10.1093/mnras/stz1947},
archivePrefix = {arXiv},
       eprint = {1907.05071},
 primaryClass = {astro-ph.GA},
       adsurl = {https://ui.adsabs.harvard.edu/abs/2019MNRAS.488.3826B}
}

@ARTICLE{2004ApJ...613..898T,
       author = {{Tremonti}, Christy A. and {Heckman}, Timothy M. and {Kauffmann}, Guinevere and {Brinchmann}, Jarle and {Charlot}, St{\'e}phane and {White}, Simon D.~M. and {Seibert}, Mark and {Peng}, Eric W. and {Schlegel}, David J. and {Uomoto}, Alan and {Fukugita}, Masataka and {Brinkmann}, Jon},
        title = "{The Origin of the Mass-Metallicity Relation: Insights from 53,000 Star-forming Galaxies in the Sloan Digital Sky Survey}",
      journal = {\apj},
         year = 2004,
        month = oct,
       volume = {613},
       number = {2},
        pages = {898-913},
          doi = {10.1086/423264},
archivePrefix = {arXiv},
       eprint = {astro-ph/0405537},
 primaryClass = {astro-ph},
       adsurl = {https://ui.adsabs.harvard.edu/abs/2004ApJ...613..898T}
}

@ARTICLE{2019ApJ...886...28Q,
       author = {{Qin}, Jianbo and {Zheng}, Xian Zhong and {Wuyts}, Stijn and {Pan}, Zhizheng and {Ren}, Jian},
        title = "{Understanding the Discrepancy between IRX and Balmer Decrement in Tracing Galaxy Dust Attenuation}",
      journal = {\apj},
         year = 2019,
        month = nov,
       volume = {886},
       number = {1},
          eid = {28},
        pages = {28},
          doi = {10.3847/1538-4357/ab4a04},
archivePrefix = {arXiv},
       eprint = {1909.13505},
 primaryClass = {astro-ph.GA},
       adsurl = {https://ui.adsabs.harvard.edu/abs/2019ApJ...886...28Q}
}

@ARTICLE{2019PASJ...71....8K,
       author = {{Koyama}, Yusei and {Shimakawa}, Rhythm and {Yamamura}, Issei and {Kodama}, Tadayuki and {Hayashi}, Masao},
        title = "{On the different levels of dust attenuation to nebular and stellar light in star-forming galaxies}",
      journal = {\pasj},
         year = 2019,
        month = jan,
       volume = {71},
       number = {1},
          eid = {8},
        pages = {8},
          doi = {10.1093/pasj/psy113},
archivePrefix = {arXiv},
       eprint = {1809.03715},
 primaryClass = {astro-ph.GA},
       adsurl = {https://ui.adsabs.harvard.edu/abs/2019PASJ...71....8K}
}

\appendix


\section{Astrodendro sensitivity analysis}\ 
\label{apdx_Astrodendro_sensitivity}

To address the robustness of SFC detections as a function of variation in astrodendro parameters, we performed the following sensitivity analysis. NGC 7793 was selected for this due to its proximity and higher exposure data. We separately varied the detection threshold (min\_value), contrast (min\_delta) and number of pixels (min\_npix) by a few times and tabulated the number of leaves and SFCs with photometric uncertainties smaller than 0.10 magnitude (see Table \ref{table_dendrogram_variation}). Additionally, we also varied all three parameters simultaneously by a factor of two. We observed a small loss (0-25\%) in the surviving number of SFCs as higher values of astrodendro parameters are used. We conclude that the surviving number of SFCs is not very sensitive to the variation in the astrodendro parameters; at least by a factor of two.

\begin{figure}[H]
    \hfill
    \includegraphics[width=0.45\textwidth]{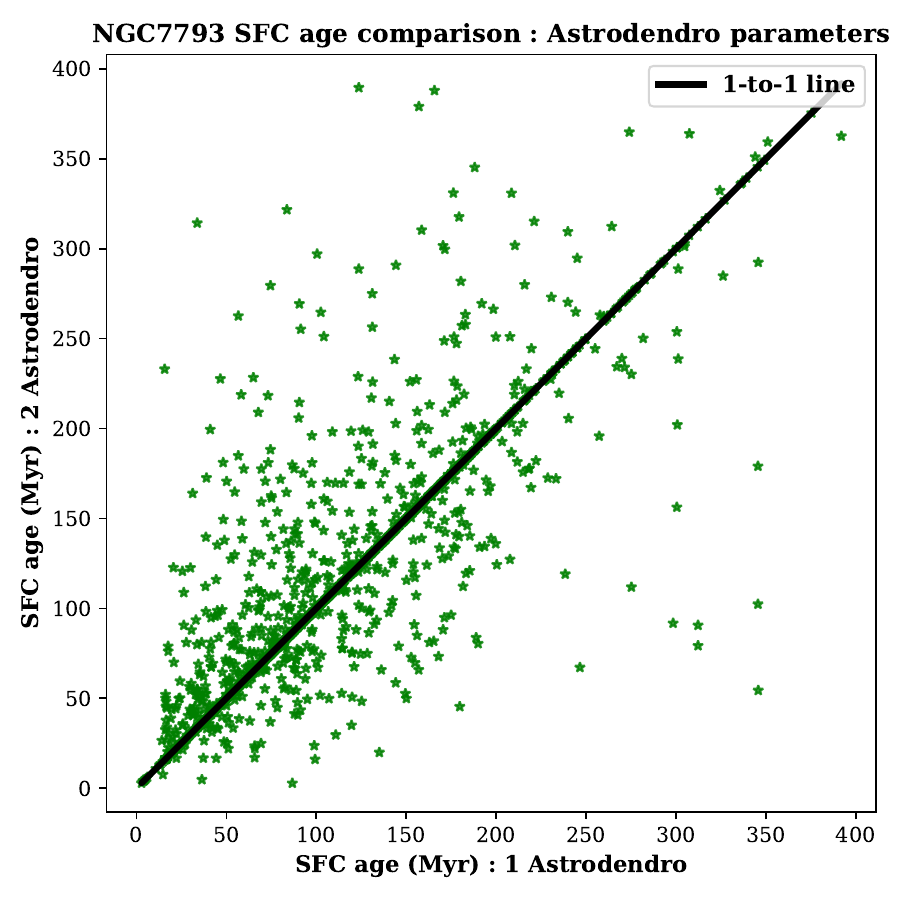}  
    \hfill
    \includegraphics[width=0.45\textwidth]{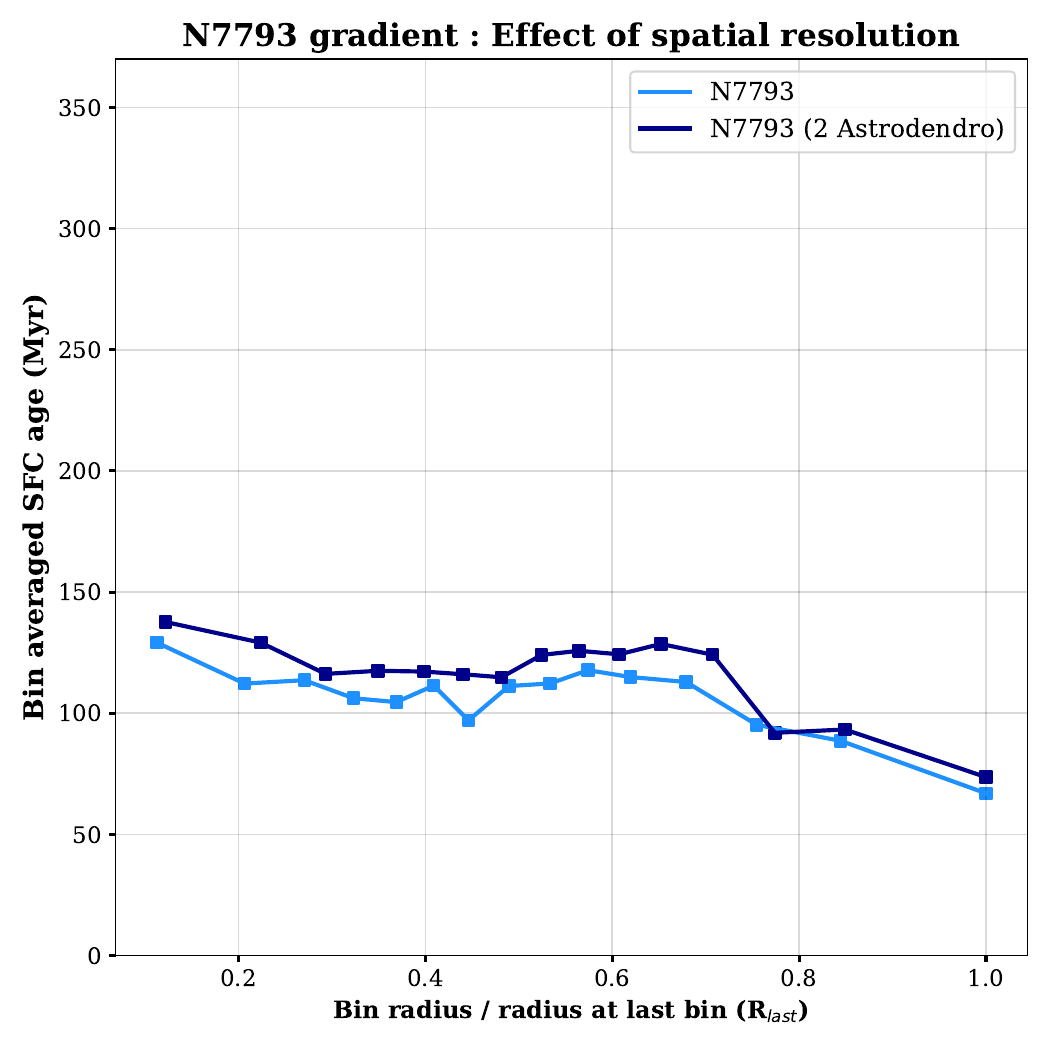}  
    \hfill
    \caption{Left : Comparison of SFC ages in NGC 7793, measured at (1bg + 3 sigma, 1 sigma, 11 px) (x-axis) and (2 bg + 6 sigma, 2 sigma, 22 px) astrodendro parameters. A good one-to-one correlation is observed. Right : Radial SFC age gradient for NGC 7793 created using the (1bg + 3 sigma, 1 sigma, 11 px) (light blue) and (2 bg + 6 sigma, 2 sigma, 22 px) (dark blue) astrodendro parameter choices. The shape of the radial age gradient is preserved,  with slightly older ages measured for higher astrodendro parameters.}
  \label{fig_astrodendro_sensitivity}
\end{figure}

Next, we visually inspected the spatial age map created at the (1bg + 3 sigma, 1 sigma, 11 px) astrodendro parameters with the map created at (2 bg + 6 sigma, 2 sigma, 22 px) parameters. A strong match in the location of young and old SFCs was observed. A strong one-to-one correlation in the SFC ages was also observed (see Figure \ref{fig_astrodendro_sensitivity} (left)). Additionally, the shape of the radial age gradient was preserved, with slightly older ages measured for higher astrodendro parameters (see Figure \ref{fig_astrodendro_sensitivity} (right)). This demonstrates that the main conclusions of our paper are robust to reasonable variations in the astrodendro parameters. Similar conclusions were drawn in \cite{Mondal_2021} regarding the effect of varying astrodendro parameters on the resulting number of SFCs and their properties.

\begin{table*}
\centering
\caption{Summary of the astrodendro parameters used, corresponding number of identified leaves and the number of SFCs surviving the 0.10 magnitude error cut in our sensitivity analysis.}
\label{table_dendrogram_variation}
\begin{tabular}{lcccccc}
\hline
Category & min\_value & min\_delta & min\_npix & Number of leaves & Number of SFCs \\\hline

Used in the paper & $1\mathrm{bg}+3\sigma$ & $1\sigma$ & 11 & 8177 & 1775 \\\hline

min\_value variation & $1\mathrm{bg}+5\sigma$ & $1\sigma$ & 11 & 6186 & 1759 \\
& $1\mathrm{bg}+10\sigma$ & $1\sigma$ & 11 & 3765 & 1664 \\
& $1\mathrm{bg}+20\sigma$ & $1\sigma$ & 11 & 2061 & 1404 \\\hline

min\_delta variation& $1\mathrm{bg}+3\sigma$ & $3\sigma$ & 11 & 7964 & 1780 \\
& $1\mathrm{bg}+3\sigma$ & $5\sigma$ & 11 & 6621 & 1798 \\
& $1\mathrm{bg}+3\sigma$ & $10\sigma$ & 11 & 3184 & 1924 \\\hline

min\_npix variation& $1\mathrm{bg}+3\sigma$ & $1\sigma$ & 22 & 3939 & 1661 \\
& $1\mathrm{bg}+3\sigma$ & $1\sigma$ & 33 & 2599 & 1444 \\
& $1\mathrm{bg}+3\sigma$ & $1\sigma$ & 44 & 1920 & 1262 \\\hline

All parameters varied & $2\mathrm{bg}+6\sigma$ & $2\sigma$ & 22 & 2577 & 1587 \\\hline
\end{tabular}
\tablecomments{Counter intuitively, when the min\_delta value is increased, the number of surviving SFCs increases. This is because the detected leaf sizes increase and so, more SFCs can survive the 0.10 magnitude error cut.}
\end{table*}


\section{Effect of adopting more than one metallicity value in SB99 to account for radial metallicity gradients}\ 
\label{apdx_effect_of_metallicity}

As the existence of negative metallicity gradients is well established in disc galaxies, the metallicity adopted to simulate the SFCs in SB99 should ideally vary with galactocentric radius. However, the metallicity choices available in SB99 are quite sparse, with only Z = 0.001, 0.004, 0.008, 0.02 and 0.04 as options. Consider NGC 0628 and NGC 1566, both of which have stellar masses greater than 10$^{10}$ $M_{\odot}$ and therefore, we adopted a metallicity value of Z = 0.02 for these galaxies. \cite{2022A&A...658A.188S} measured the metallicity gradients to be 12 + (O/H) = 8.53 $-$ 0.014 R$_{gal}$ for NGC 0628 and 12 + (O/H) = 8.62 $-$ 0.012 R$_{gal}$ for NGC 1566, in their Figure A.19. Here, R$_{gal}$ is the galactocentric radius. Now consider NGC 7793, for which \cite{2021A&A...650A.103D} measured median 12 + (O/H) = 8.37. The stellar mass of NGC 7793 is $\times$ 10$^{9}$ $M_{\odot}$ and we adopted Z = 0.008 for this galaxy (similar to \citep{Mondal_2021}). The sparse metallicity grid available in SB99 meant that for NGC 7793, we could only choose Z = 0.008. Using the metallicity gradient relations above, the 12 + (O/H) value of NGC 0628 and NGC 1566 will fall to 8.37 (equivalent to Z = 0.008, i.e NGC 7793's metallicity) at R$_{gal}$ = 11.4 kpc and 21.4 kpc. These radii are comparable to the R$_{25}$ of the galaxies i.e. 15 kpc for NGC 0628 and 21 kpc for NGC 1566. This demonstrates that the observed metallicity gradients are not steep enough to justify the use of two different SB99 metallicity values in the inner and outer regions of a galaxy. Therefore, our choice of adopting a single metallicity value for each galaxy is reasonable. Even in past literature, when star clusters are characterized using spectral energy distribution fitting, only one metallicity value per galaxy is used \citep{2017ApJ...841..131A, 2021MNRAS.502.1366T, 2021MNRAS.507.5542M, Linden_2022}. In such works, the main focus is on breaking the age-extinction degeneracy and not on constraining the metallicity effectively

\begin{figure}[H]
    \includegraphics[width=0.42\textwidth]{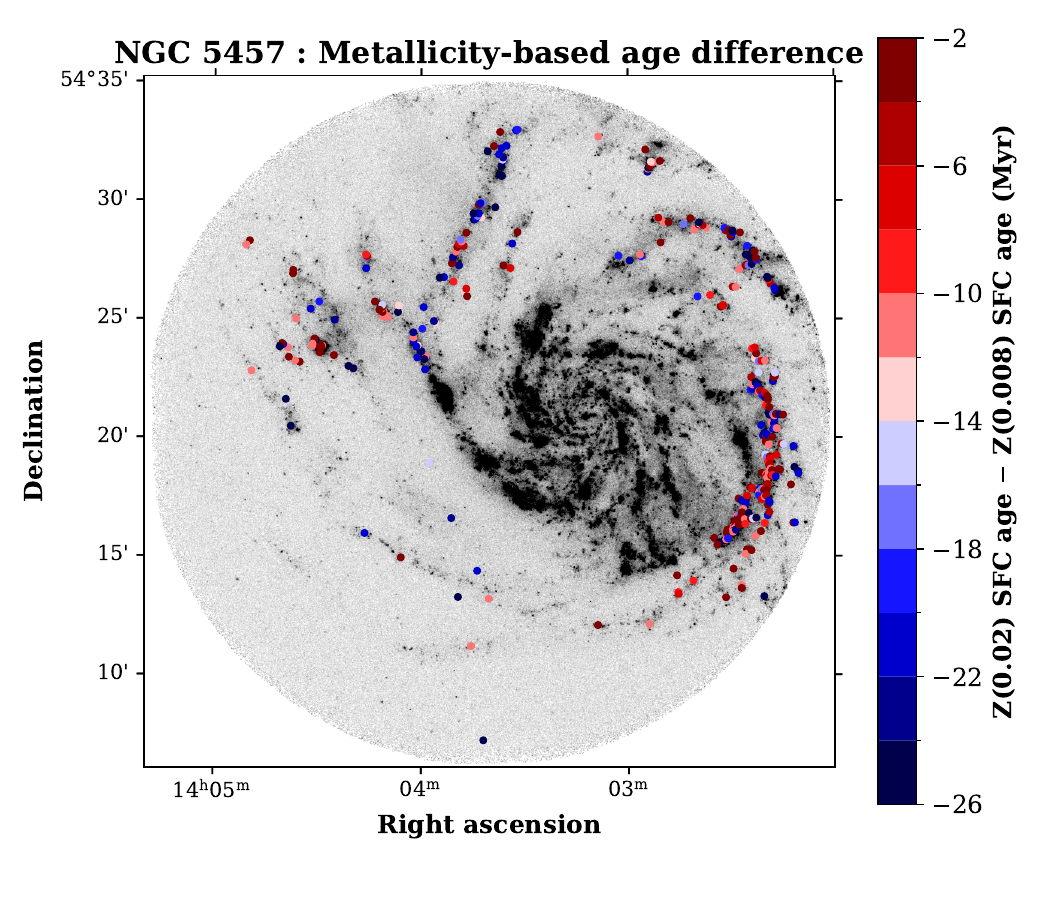}  
    \hfill
    \includegraphics[width=0.42\textwidth]{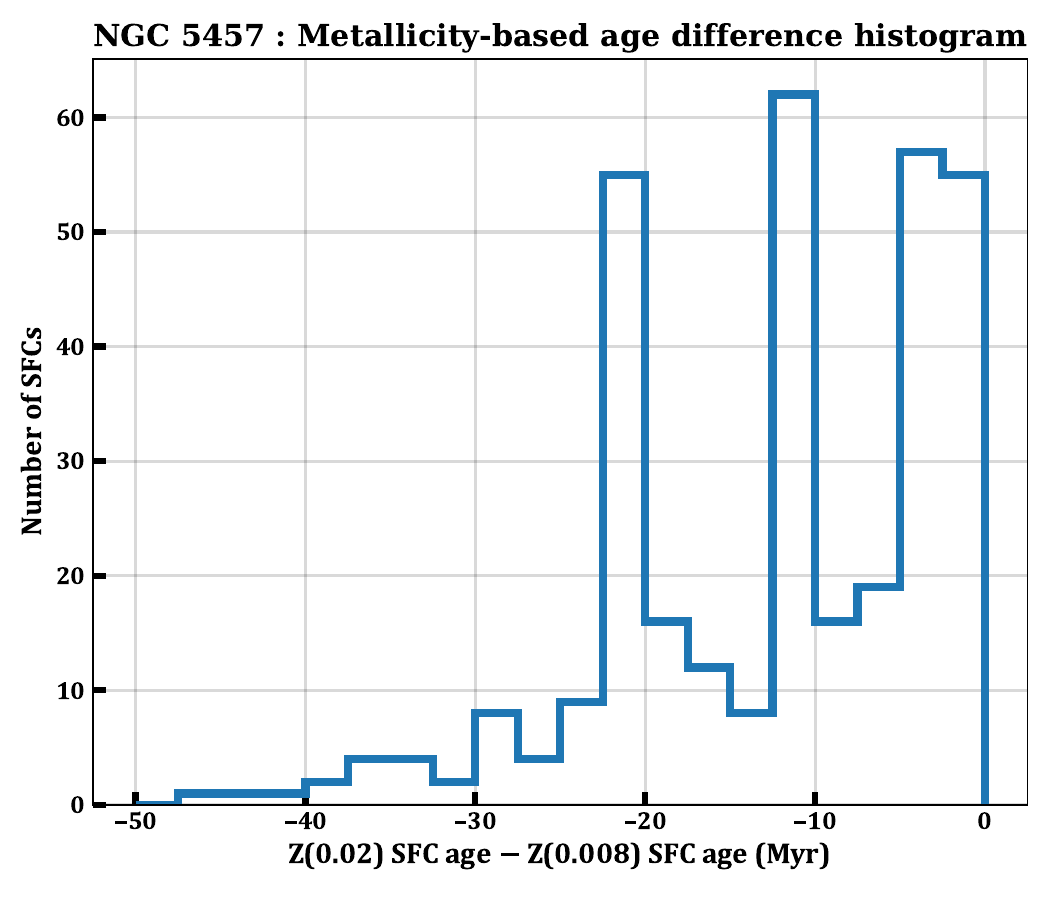}  
    \caption{\textbf{Left : Spatial age difference map, and right : age difference histogram for SFCs in NGC 5457, assuming Z = 0.02 in the inner parts (galactocentric radius $<$ R$_{25}$/2) and Z = 0.008 in the outer parts (galactocentric radius $>$ R$_{25}$/2) of the galaxy. Age-differences in the range of 0-25 Myr are observed.}}
  \label{fig_effect_of_metallicity}
\end{figure}

However, to constrain the maximum variation possible in SFC ages with two different metallicity values being used in a galaxy, we performed the following test. It is well established that using lower metallicity for a simulated star cluster (in SB99) leads to bluer colors. In our tests, the UV CMD demonstrated this effect, when simulated evolutionary tracks for solar metallicity (Z = 0.02, solid lines) and sub-solar metallicity (Z = 0.008, dashed lines) were plotted together. As a result, for a given FUV-NUV color, the derived SFC age should be older at a lower, sub-Solar metallicity. Next, we assumed Z = 0.02 in the inner parts (galactocentric radius $<$ R$_{25}$/2) of NGC 5457 and Z = 0.008 in the outer parts. We created spatial age difference maps and histograms with respect to the SFC ages measured assuming a single metallicity value of Z = 0.02 across the entire galaxy (the ages used in the paper). This way only the outer SFCs are expected to exhibit age differences. The age difference map and histogram suggest that lower metallicity leads to older SFC ages within an age range of $\sim$0-25 Myr (see Figure \ref{fig_effect_of_metallicity}).

\section{SFC catalog table}\ 
\label{apdx_catalog_table}
In Table \ref{table_SFC_catalog}, we present a brief description of the properties of the first SFC for each of our 17 galaxies. The full SFC catalog consists of $\sim$25000 SFC within our 17 galaxies.

\begin{sidewaystable*}
\caption{Description of the physical properties of the $\sim$25000 SFC characterized within our 17 galaxies.}
\resizebox{0.995\textwidth}{!}{%
\centering
\begin{tabular}{ccccccccccccccccccc}
\hline
Galaxy & index & area & x & y & RA & DEC & magF & magF$_{err}$ & magN & magN$_{err}$ & F\_N & F\_N$_{err}$ & $A_V$ & magF$_{corr}$ & magN$_{corr}$ & F\_N$_{corr}$ & Age & Age$_{err}$ \\
 &  & (px) &  &  & (deg) & (deg) & (mag) & (mag) & (mag) & (mag) & (mag) & (mag) & (mag) & (mag) & (mag) & (mag) & (Myr) & (Myr)\\
(1) & (2) & (3) & (4) & (5) & (6) & (7) & (8) & (9) & (10) & (11) & (12) & (13) & (14) & (15) & (16) & (17) & (18) & (19)\\\hline
N0253	& 9 & 50 & 717.97	& 235.69	& 12.0008408	& -25.1647814	& 21.02	& 0.05	& 22.70	& 0.07	& -1.67	& 0.09	& 0.44	& 19.75	& 21.49	& -1.73	& 1	& 3.68 \\\hline
N0300	& 19489	& 51	& 1222.98	& 1490.43	& 13.8064014	& -37.6938989	& 17.56	& 0.01	& 18.06	& 0.00	& -0.49	& 0.01	& 0.16	& 17.03	& 17.56	& -0.53	& 1	& 0.65\\\hline
N0628	& 52	& 34	& 1691.48	& 1044.44	& 24.1500971	& 15.8364181 & 22.35 & 0.13 & 22.79 & 0.24	& -0.43	& 0.27	& 0.69	& 20.05	& 20.93 & -0.87	& 1	& 8.44\\\hline
N1512	& 16	& 15	& 1025.59	& 205.14	& 61.0349597 &	-43.2103999	& 22.76	& 0.15	& 23.38	& 0.11	& -0.61	& 0.19	& 0.47	& 21.44 & 22.11 & -0.67	& 1	& 7.58\\\hline
N1566	& 3479	& 12	& 687.39	& 909.39	& 65.0444367	& -54.9389384	& 22.40	& 0.14	& 22.81	& 0.17	& -0.40	& 0.22	& 0.52	& 20.90	& 21.66	& -0.72	& 1	& 6.67\\\hline
N2403	& 6770	& 46	& 2267.05	& 1658.14	& 114.0823638	& 65.6190926	& 16.68	& 0.01	& 17.37	& 0.00	& -0.68	& 0.01	& 0.25	& 15.70	& 16.47 & -0.75	& 1	& 0.60 \\\hline
N2903	& 4831	& 44	& 1063.66	& 1398.08	& 143.0592276	& 21.4778557	& 22.68	& 0.16	& 22.69	& 0.17	& -0.01	& 0.23	& 0.98	& 19.80	& 20.44 & -0.62	& 1.45	& 6.82\\\hline
N3486	& 394	& 26	& 893.93	& 565.56	& 165.0738159	& 28.9906197	& 22.28	& 0.17	& 22.83	& 0.21	& -0.55	& 0.27	& 0.30	& 21.30	& 22.07	& -0.75	& 1	& 9.55\\\hline
N4228	& 27	& 73	& 755.57	& 267.70	& 183.9339123   & 36.4000494	& 21.11	& 0.05	& 21.82	& 0.24	& -0.70	& 0.25	& 0.11	& 20.60	& 21.25	& -0.62	& 1	& 13.08\\\hline
N4236	& 713	& 16	& 1290.46	& 1367.40	& 184.0784856	& 69.5126831	& 21.94	& 0.10	& 22.21	& 0.11	& -0.26	& 0.15	& 0.88	& 19.50	& 20.26 & -0.80	& 1	& 5.44\\\hline
N4395	& 10880	& 12	& 957.91	& 1570.63	& 186.4872514	& 33.5039902	& 21.37	& 0.03	& 21.84	& 0.15	& -0.47	& 0.16	& 0.14	& 20.90	& 21.44	& -0.58	& 2.38	& 5.63\\\hline
N5033	& 479	& 20	& 811.16	& 667.41	& 198.3917000	& 36.6324485	& 22.59	& 0.16	& 23.13	& 0.21	& -0.54	& 0.26	& 0.29	& 21.70	& 22.44	& -0.73	& 1	& 7.74\\\hline
N5194	& 1654	& 49	& 838.21	& 961.66	& 202.5311662	& 47.2227261	& 21.98	& 0.15	& 21.57	& 0.17	& 0.41	& 0.23	& 1.54	& 17.60	& 18.11	& -0.53	& 2.19 & 6.69\\\hline
N5457	& 14264	& 88	& 2902.07	& 3032.20	& 210.7585709	& 54.2370707	& 20.76	& 0.06	& 20.75	& 0.065	& 0.01	& 0.09	& 1.22	& 17.4	& 18.11	& -0.70 & 1	& 2.68\\\hline
N7793	& 12718	& 425	& 1034.33	& 1224.31	& 359.4528215	& -32.6170667	& 16.37	& 0.01	& 16.67	& 0.01	& -0.29	& 0.01	& 0.09	& 16.0	& 16.30	& -0.32	& 2.83	& 0.68\\\hline
Holmber II & 99	& 42	& 724.36	& 343.48	& 124.7976815	& 70.7728837	& 22.86	& 0.08	& 23.38	& 0.08	& -0.51	& 0.12	& 0.046	& 22.52	& 23.06	& -0.53	& 1	& 5.57\\\hline	
WLM	& 112	& 37	& 502.41	& 530.94	& 0.5160534	& -15.3949499	& 22.73	& 0.09	& 23.27	& 0.18	& -0.54	& 0.20	& 0.07	& 22.30	& 22.90	& -0.64	& 2.14	& 7.74\\\hline	
\end{tabular}
}
\tablecomments{(1) Galaxy name, (2) index, (3) area (in pixel units, where 1px = 0.416\arcsec~for UVIT) , (4) x-coordinate in the UVIT image, (5) y-coordinate in the UVIT image, (6) right ascension, (7) declination, (8) observed FUV magnitude, (9) FUV magnitude error, (10) observed NUV magnitude, (11) NUV magnitude error, (12) observed FUV$-$NUV color, (13) FUV$-$NUV color error, (14) $A_V$ value corresponding to the SFC derived using the MB16 method (refer Section \ref{subsec_attenuation}), (15) attenuation corrected FUV magnitude, (16) attenuation corrected NUV magnitude, (17) attenuation corrected FUV$-$NUV color, (18) derived age and the (19) associated age error for each SFC respectively, in our catalog of $\sim$25000 SFCs characterized in our 17 galaxies. We present the properties of the first SFC for each of our 17 galaxies in this table. This full SFC catalog table can be accessed with the online version of this paper.} 
\label{table_SFC_catalog}
\end{sidewaystable*}


\section{Radius distribution of the SFCs}
\label{apdx_radius_distribution}
\begin{figure}[H]
    \centering
    \includegraphics[width=0.55\textwidth]{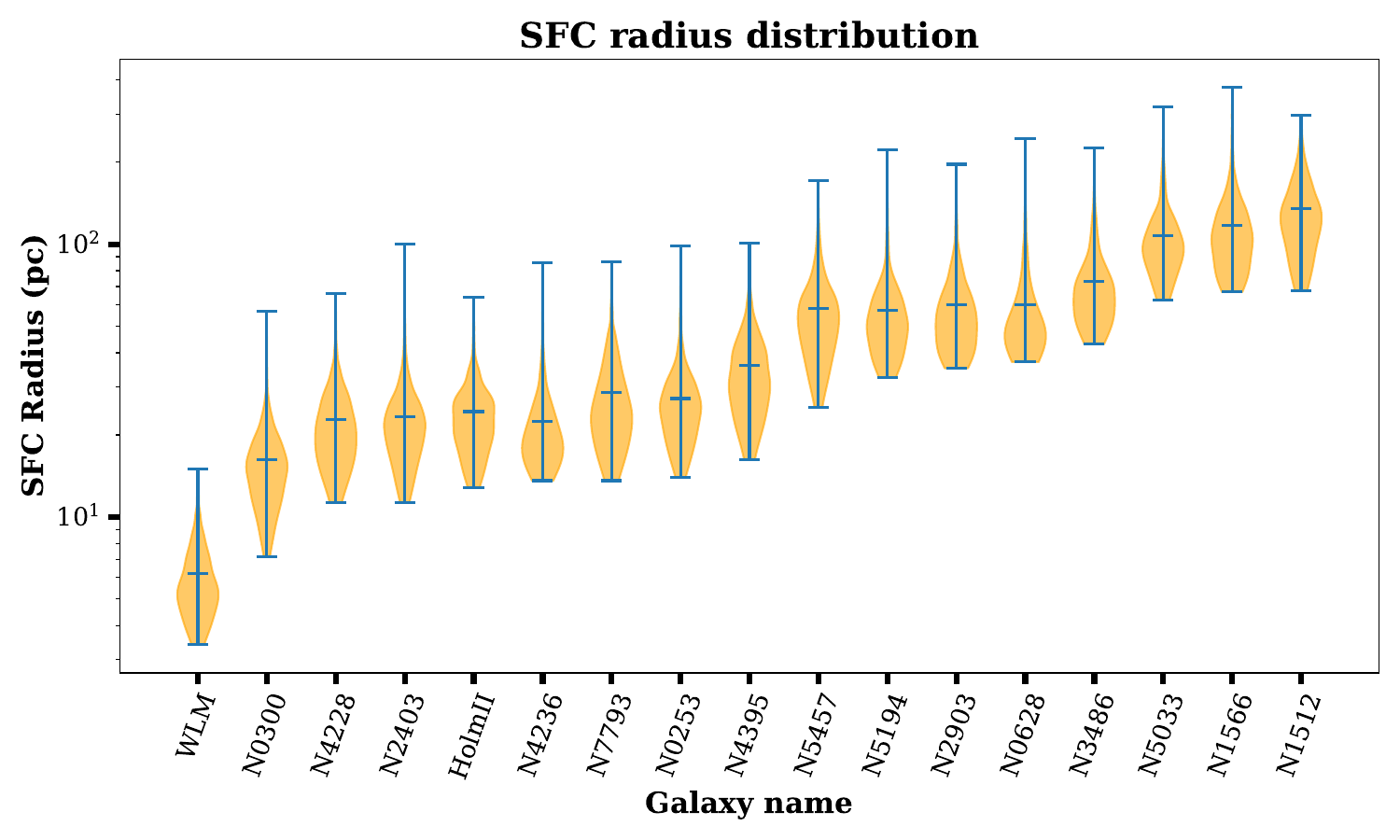}  
    \caption{Violin plots for the radius distribution of the observed SFCs in all 17 galaxies, arranged in the increasing order of their distances. The central, top and bottom bars represent the median, maximum and minimum SFC radius values for the galaxy, respectively. As expected, we detect larger SFC in galaxies that are located further away and vice versa.}
  \label{fig_radius_distribution}
\end{figure}

Figure \ref{fig_radius_distribution} presents the radius distribution of the characterized SFCs in our 17 galaxies. 

\section{$A_V$ maps of galaxies at the SFC positions}
\label{apdx_AVmaps}
\begin{figure*}[hbt!]
      \centering
		\includegraphics[width=0.24\linewidth]{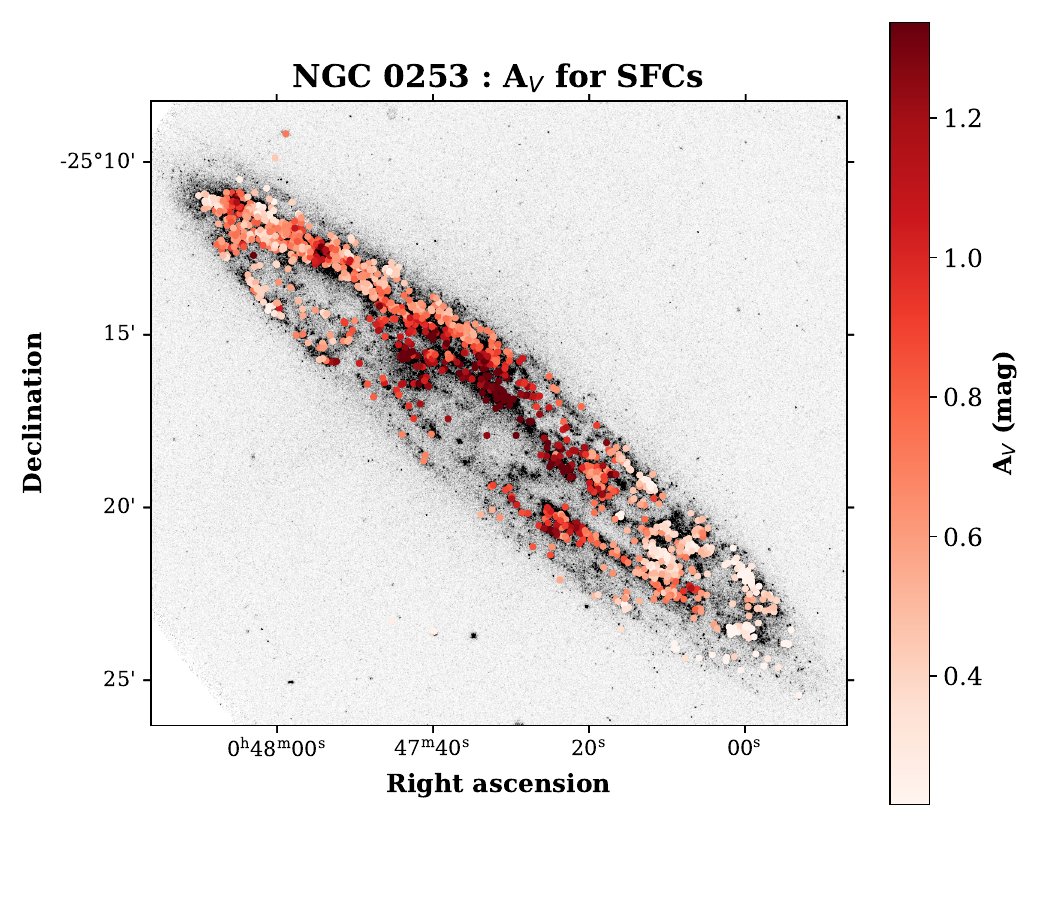}
		\includegraphics[width=0.24\linewidth]{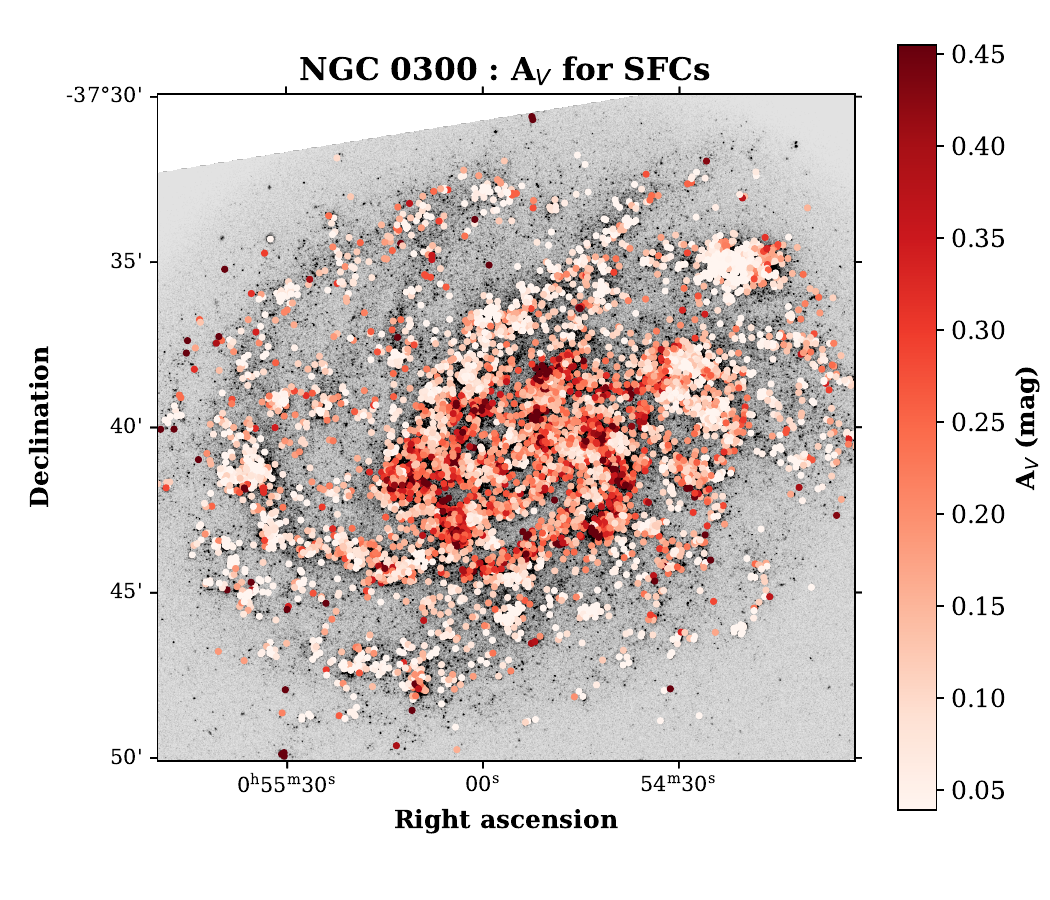}
		\includegraphics[width=0.24\linewidth]{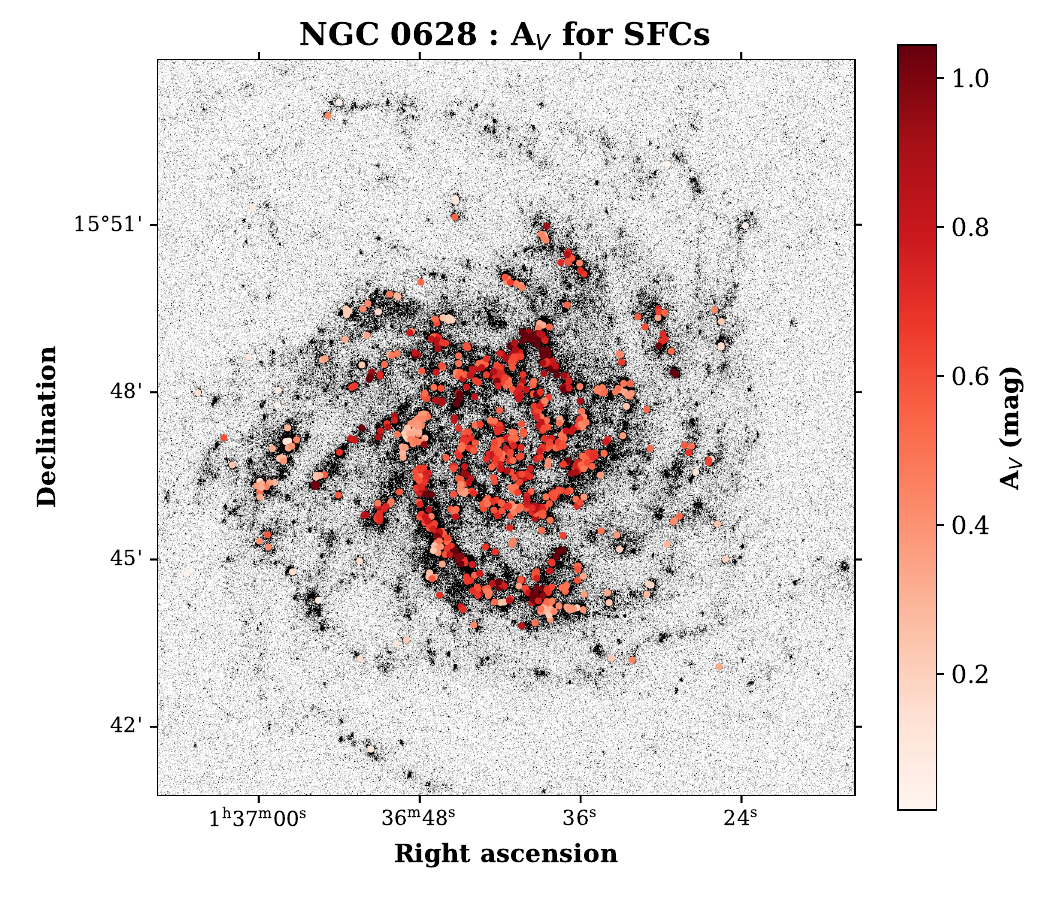}
		\includegraphics[width=0.24\linewidth]{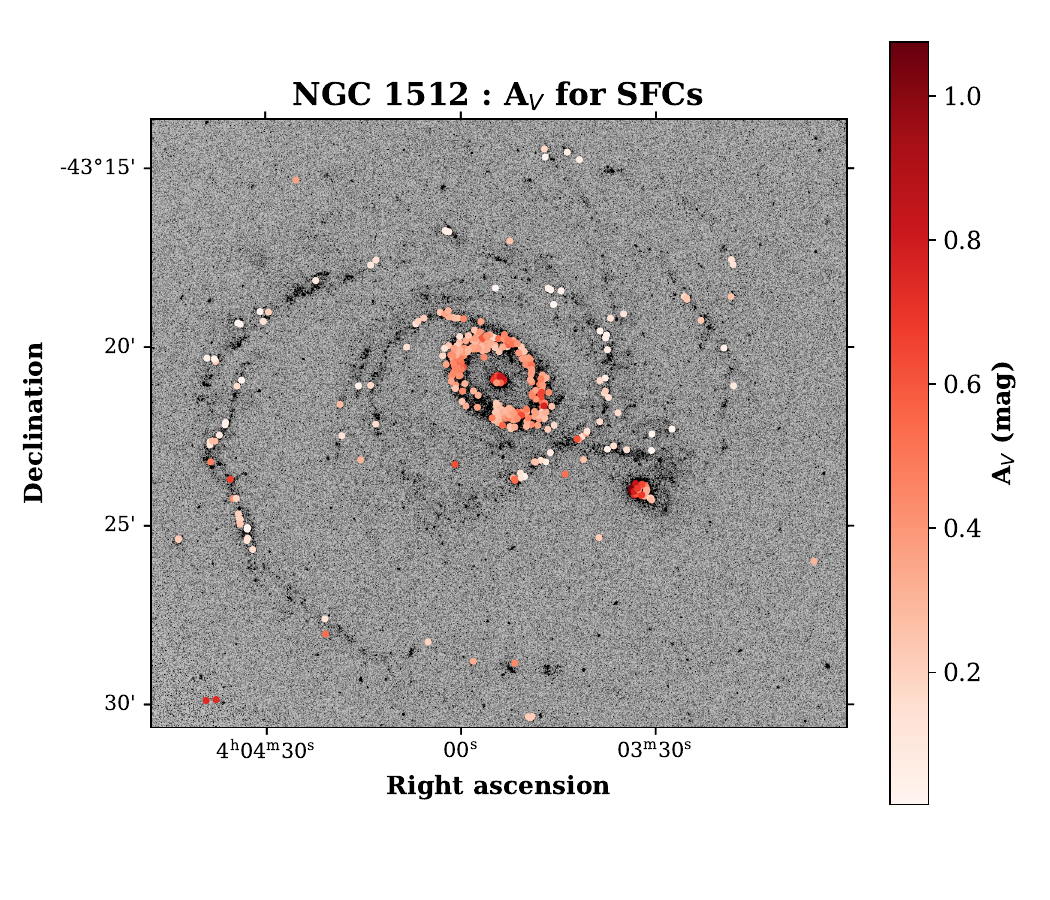}
	\vfill	
	\vfill	
		\includegraphics[width=0.24\linewidth]{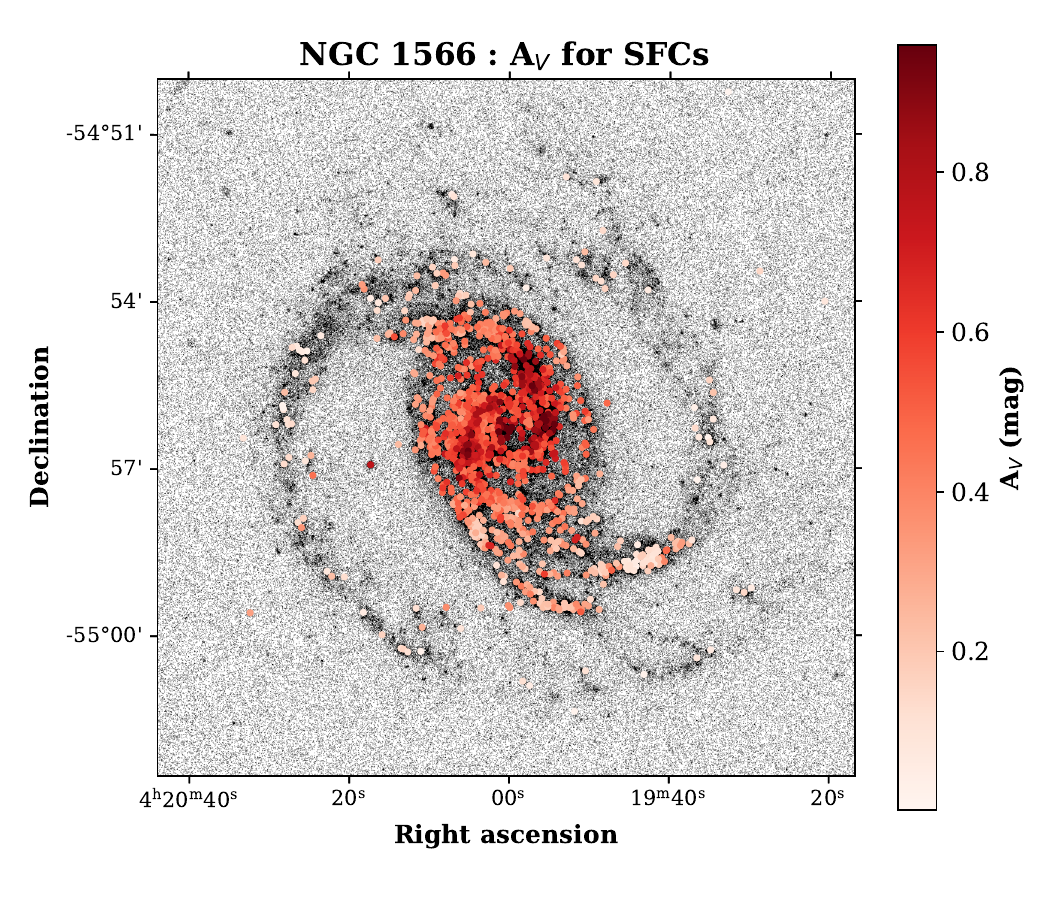}
		\includegraphics[width=0.24\linewidth]{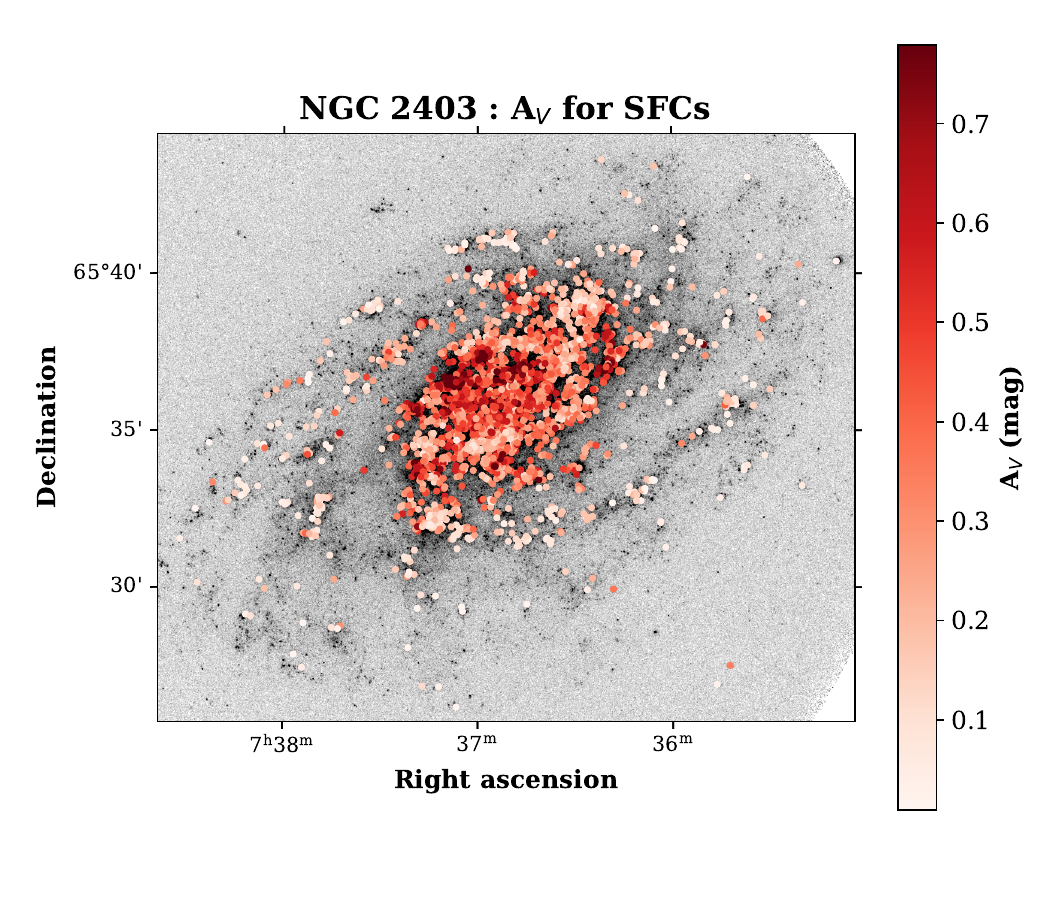}
		\includegraphics[width=0.24\linewidth]{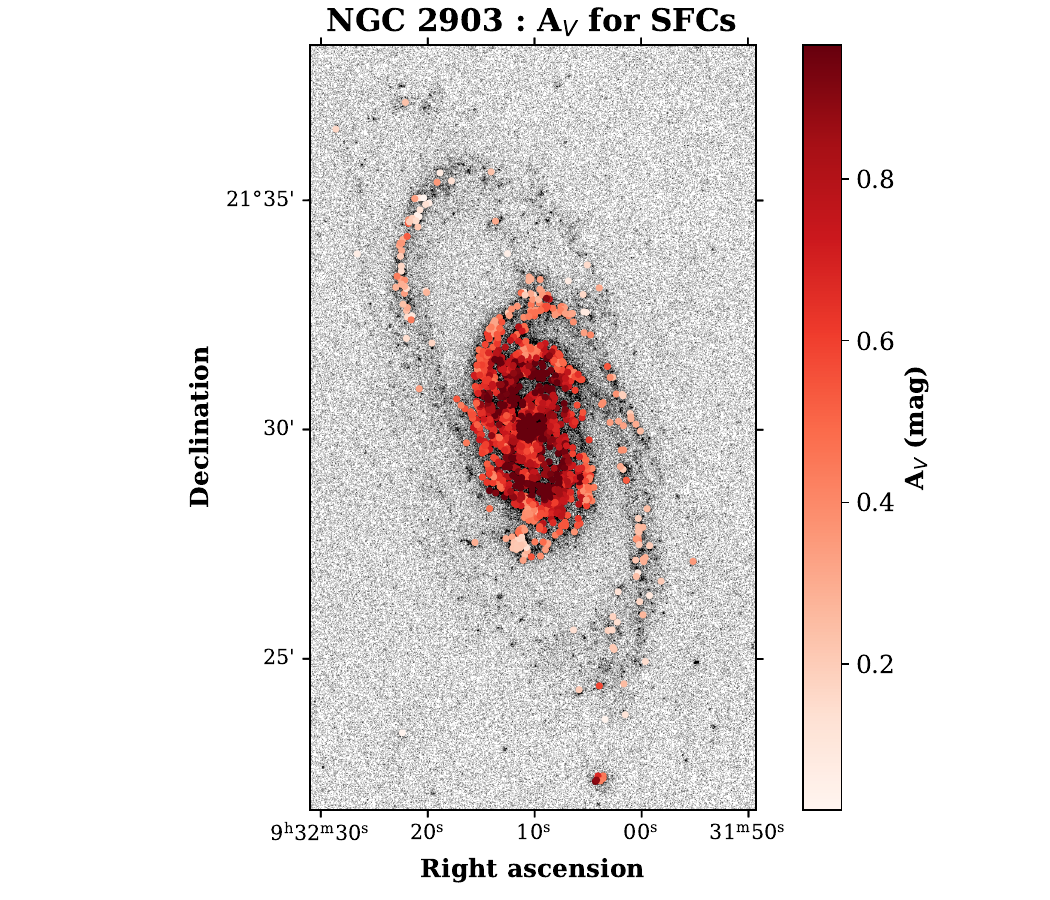}
		\includegraphics[width=0.24\linewidth]{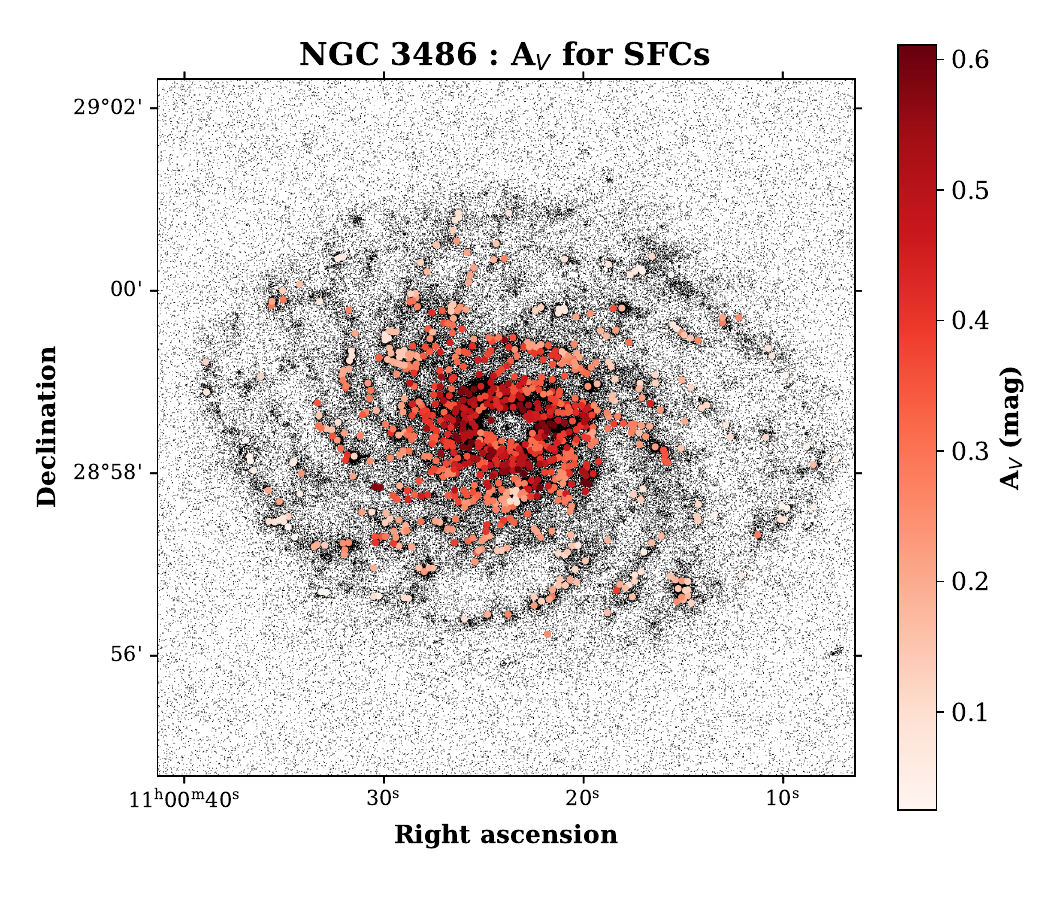}
    \vfill
    \vfill
		\includegraphics[width=0.24\linewidth]{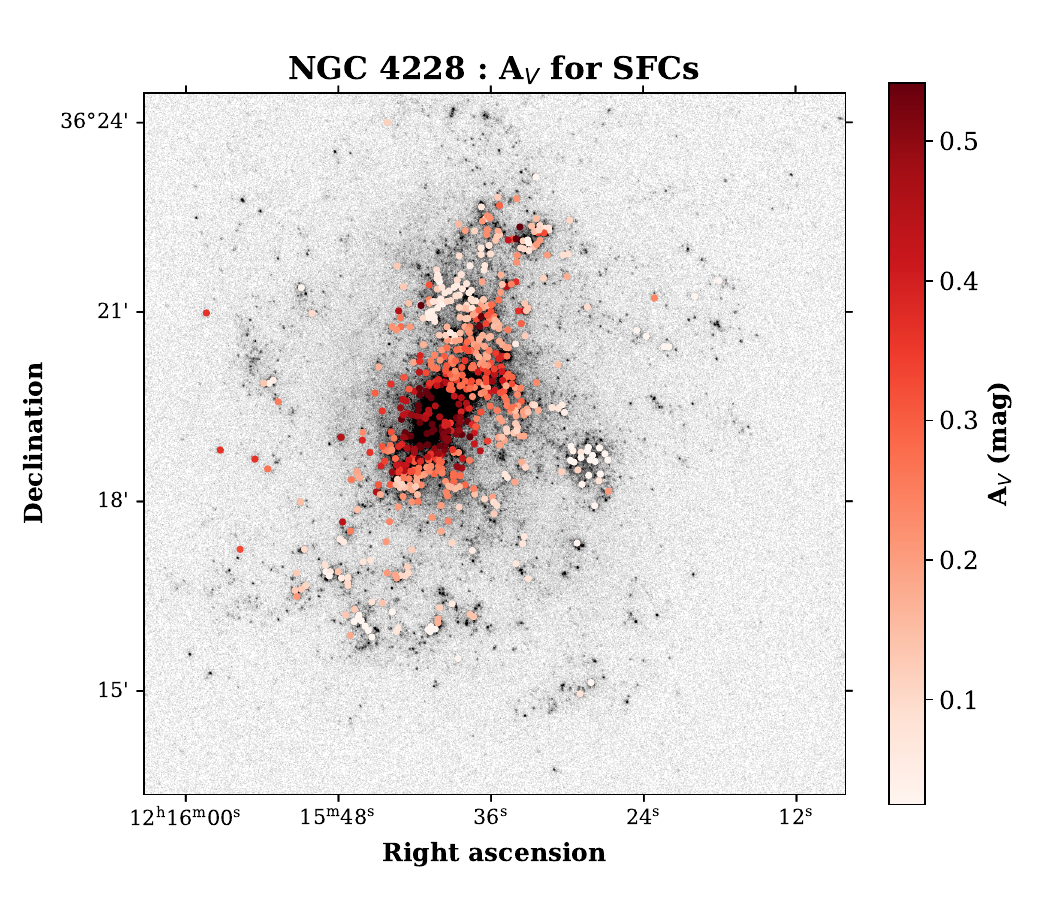}
        \includegraphics[width=0.24\linewidth]{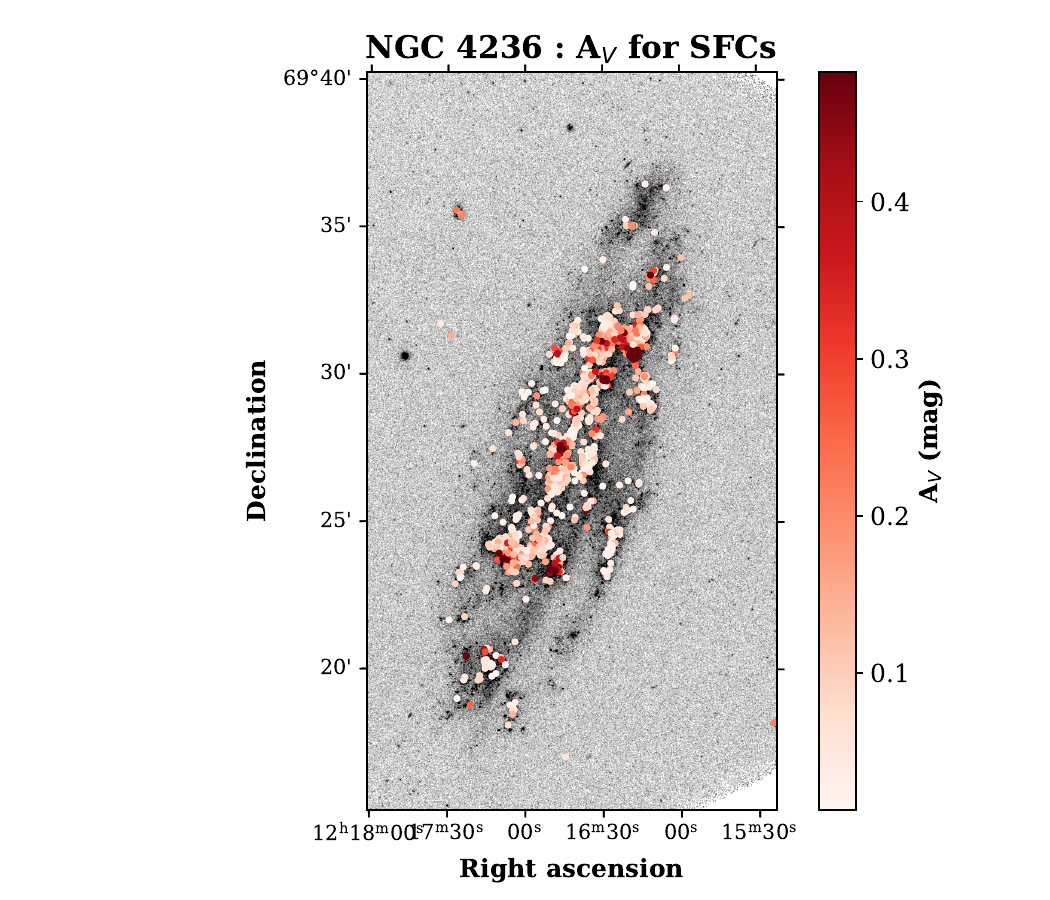}
        \includegraphics[width=0.24\linewidth]{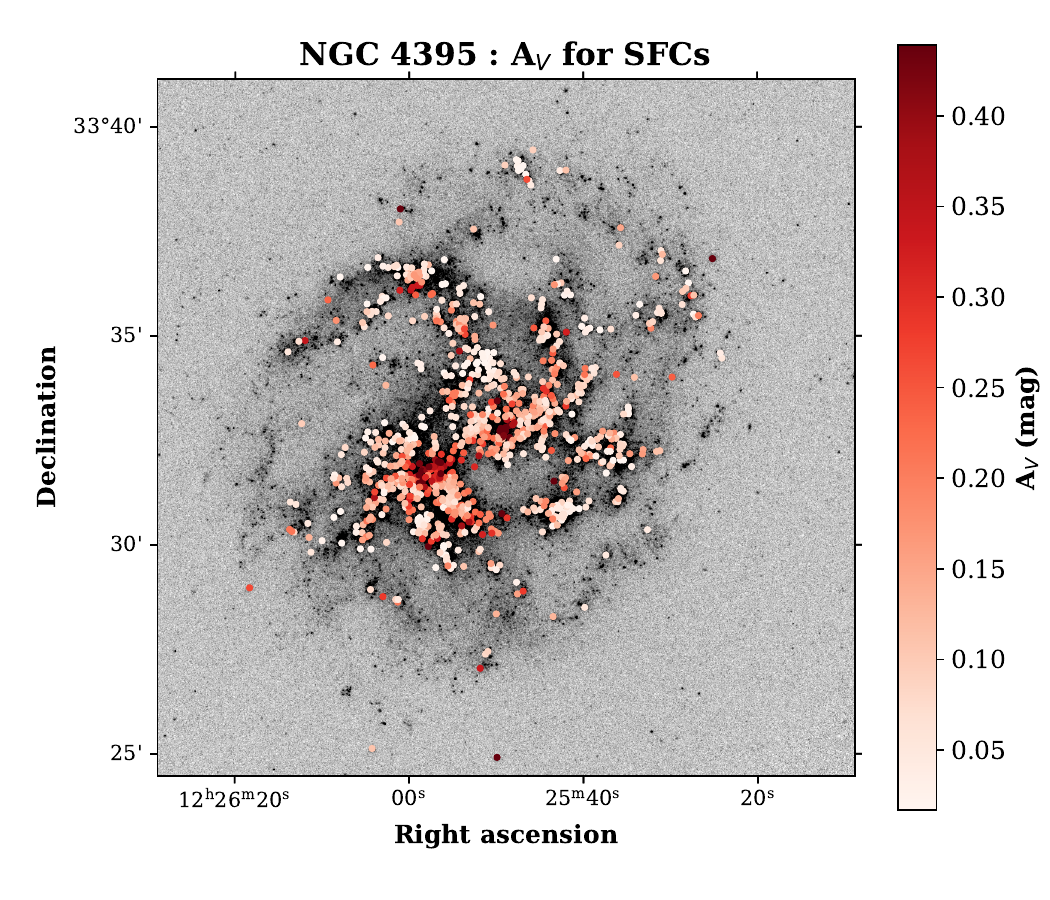}
		\includegraphics[width=0.24\linewidth]{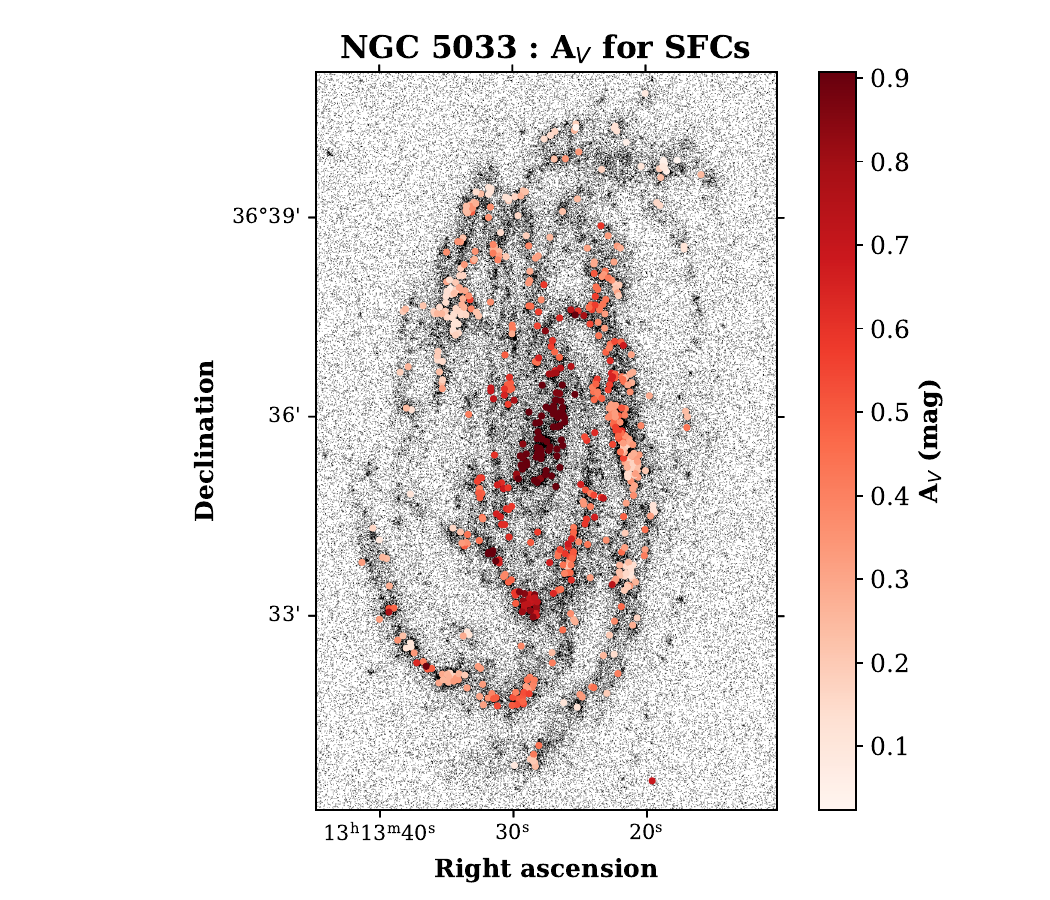}
    \vfill
    \vfill
  	    \includegraphics[width=0.24\linewidth]{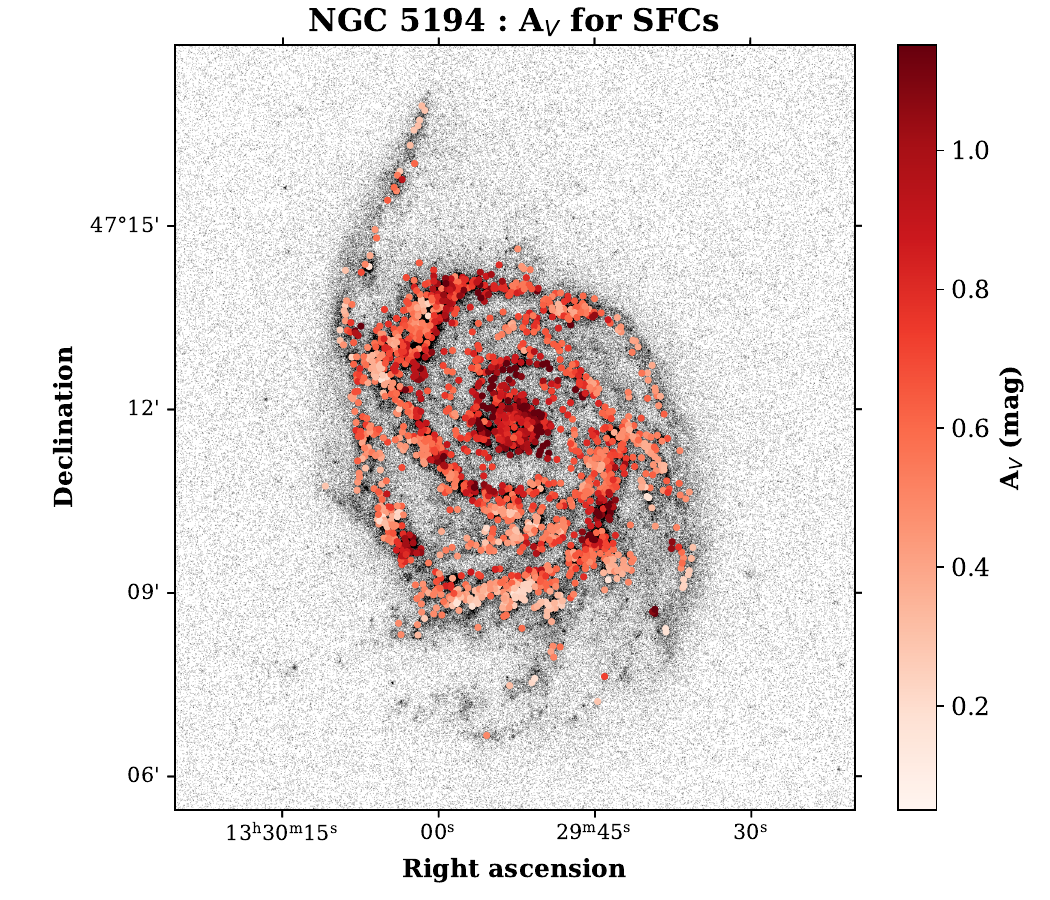}
        \includegraphics[width=0.24\linewidth]{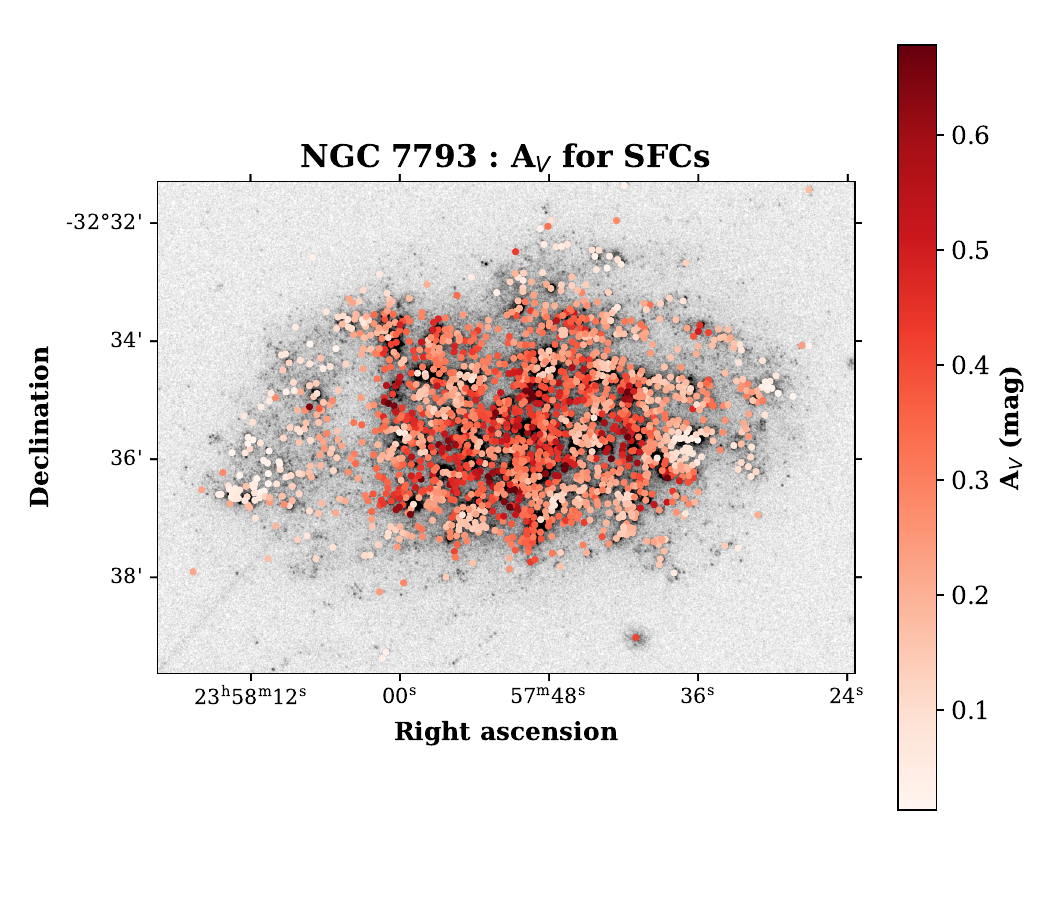}
		\includegraphics[width=0.24\linewidth]{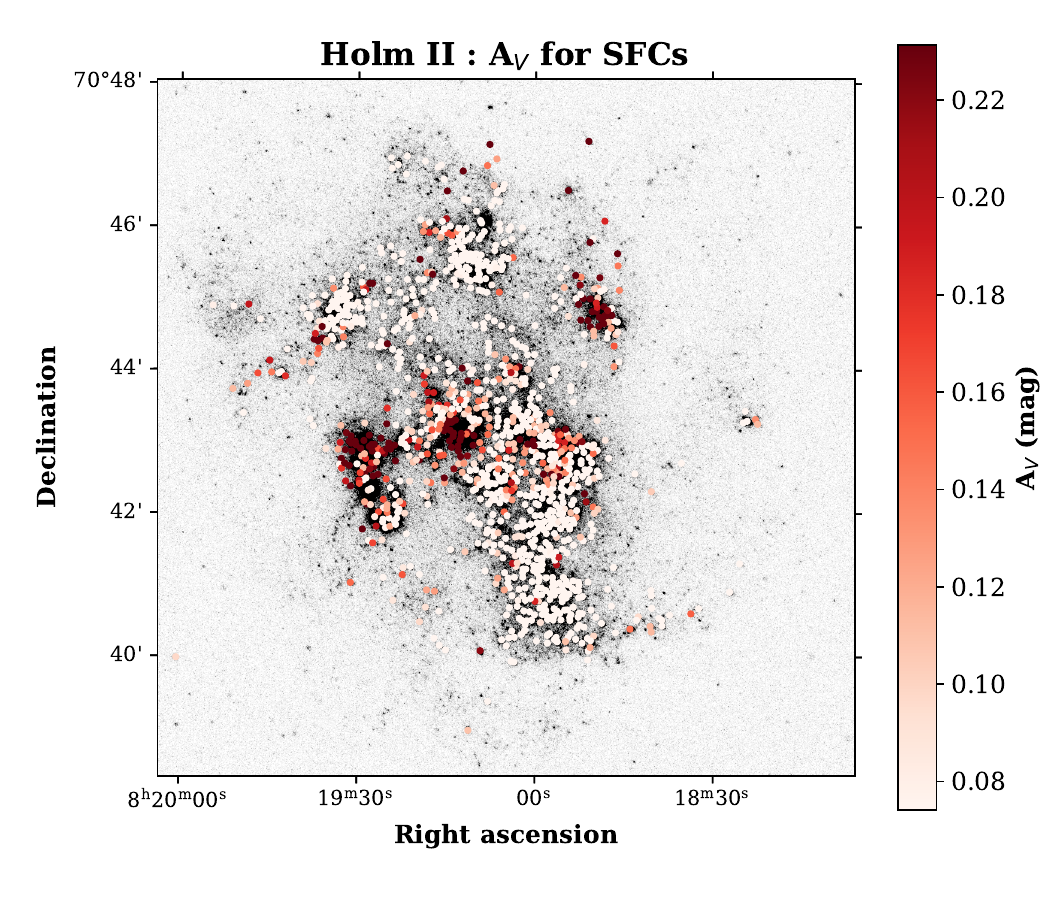}
		\includegraphics[width=0.24\linewidth]{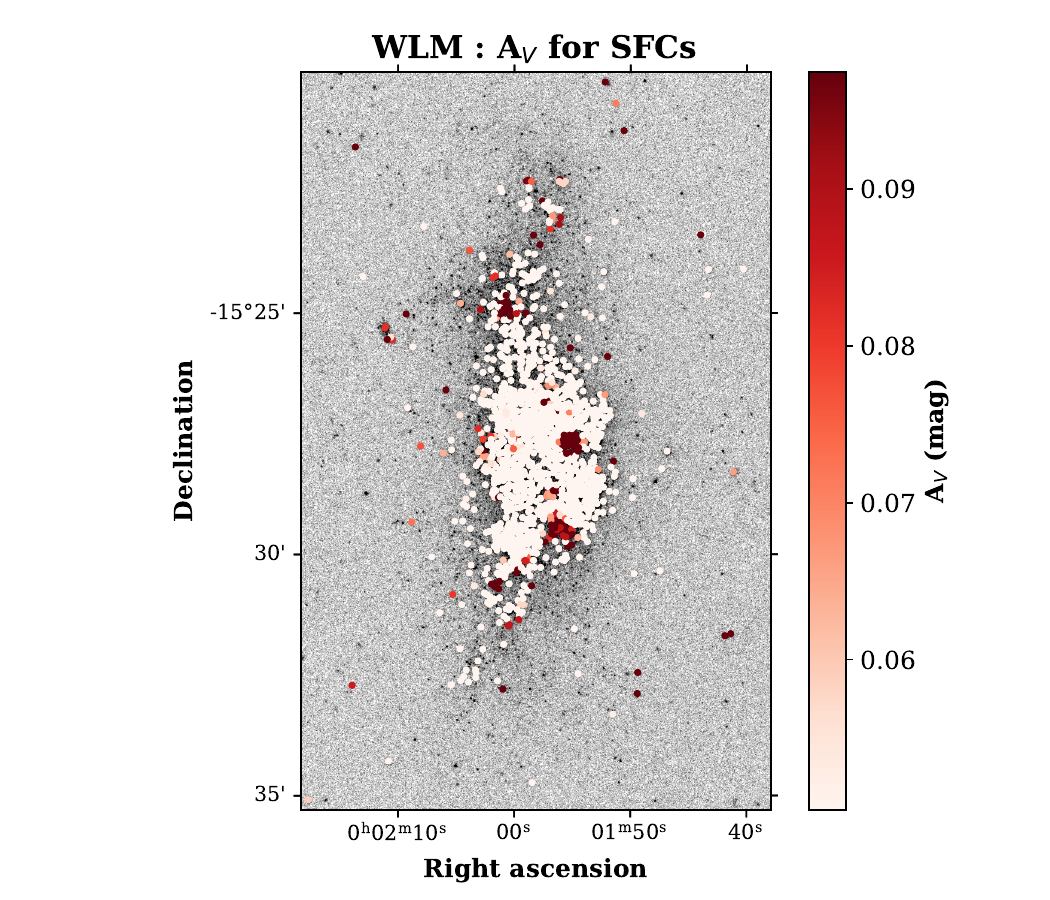}
    \caption{$A_{V}$ maps of the 16 out of 17 galaxies at the SFC positions, barring NGC 5457 for which see Figure \ref{fig_AV_5457}}
  \label{fig_AV_maps}
\end{figure*}

In Figure \ref{fig_AV_maps}, we present the $A_V$ maps corresponding to the positions of the characterized SFCs within our galaxies. The full $A_V$ tables with much larger coverage (see Table \ref{table_AV_values}) used to create these dust attenuation maps with 6\arcsec~resolution rectangular grids can be accessed with the online version of this paper. 

\begin{table} 
\caption{The $A_V$ tables used to create the dust attenuation maps used in this paper}
\centering
\begin{tabular}{cccccc}
\hline
Galaxy & Index & R.A. & Dec. & $k_{24\mu}$ & $A_V$ \\
(1) & (2) & (3) & (4) & (5) & (6) \\\hline
N0253 & 2 & 11.7047086	& -25.4197496	& 3.91	& 0.25 \\
.. & .. & .. & .. & .. & ..\\\hline
\end{tabular}
\tablecomments{(1) Galaxy name, (2) Index of the 6\arcsec~rectangular grid, (3) right ascension, (4) declination, (5) $k_{24\mu}$ value (refer Section \ref{subsec_attenuation}) (6) $A_V$ value measured for the particular bin.}
\label{table_AV_values}
\end{table}


\section{Mathematical expressions used for dust attenuation correction}
\label{apdx_formulae}

\begin{enumerate}
\item The conversion of FUV and NUV source counts into the AB magnitude and the spectral flux units is performed according to the calibrations provided in \cite{Tandon_2017, Tandon_2020}.
   
\item 2MASS J-band magnitudes are defined in the Vega system and need to be converted to AB magnitude system in our analysis; In the J-band fits files accessed from NED, pixel values are written in counts.   

\begin{equation}
magJ_{AB} = magJ_{Vega} + 0.91\ =  ZP_{J} + 2.5 * log_{10}(counts_{J}) + 0.91
\end{equation}

\item In the process of $A_{FUV}$ estimation, MIPS 24$\mu$ and FUV luminosities are used. The pixel values of MIPS 24$\mu$ fits files taken from NED are written in mega-Jansky per steradian (MJy/Sr). The FUV fits files are written in source counts. The conversion of both these data between Jansky (mJy) and ergs/second unit is achieved as follows.

\begin{equation}
FUV\\ (in\\\ mJy) = k_1 * cps_{FUV} 
\end{equation}
where k\_{1} = 0.227 for F148W, 0.282 for F154W and 0.419 for F169M filter. (derived using the Tandon et al. (2017) calibrations.
\begin{equation}
24\mu\\ (in\\\ mJy) = 0.529*counts_{24\mu} 
\end{equation}
Finally, the conversion of FUV and 24$\mu$ spectral energy densities into luminosity units and their subsequent use in the derivation of $A_{FUV}$ is performed as, 
\begin{equation}
A_{FUV} = 2.5*log_{10} [ 1 +  \frac {k_{i} * 24\mu\ (mJy)} {ratio * FUV ( mJy)}]\
\end{equation}  

where ki = the \cite{2016A&A...591A...6B}'s equivalent factor for the \cite{Hao_2011}'s 3.89 values; ratio = 162 for F148W, 156 for F154W and 142 for F169M filter and is simply the ratio of central wavelength of MIPS 24$\mu$ and the respective FUV filter.

\end{enumerate}


\section{Effect of variation in A$_V$ within a 6\arcsec~ resolution element on SFC ages}
\label{apdx_AV_resolution_mismatch}

To quantify the variation in A$_V$ within a resolution element owing to the mismatch in the angular resolution of our attenuation maps (6\arcsec~) and the UVIT's detection scale (1.5\arcsec~), we selected NGC 0628 as our test case. NGC 0628 is the nearest galaxy for which MUSE-based A$_V$ maps at $\sim$0.92\arcsec~ resolution \citep{2022A&A...658A.188S} are available and it is assumed that the MUSE-based A$_V$ maps have comparable resolution to the UVIT's (1.5\arcsec~).  

\begin{figure}[t]
    \includegraphics[width=0.31\textwidth]{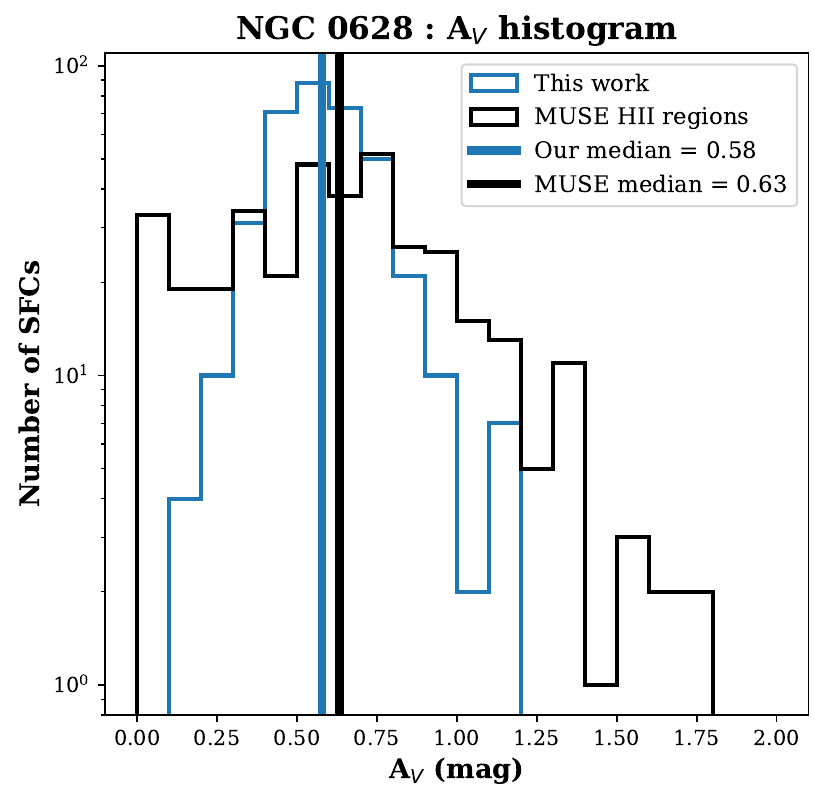}  
    \hfill
    \includegraphics[width=0.31\textwidth]{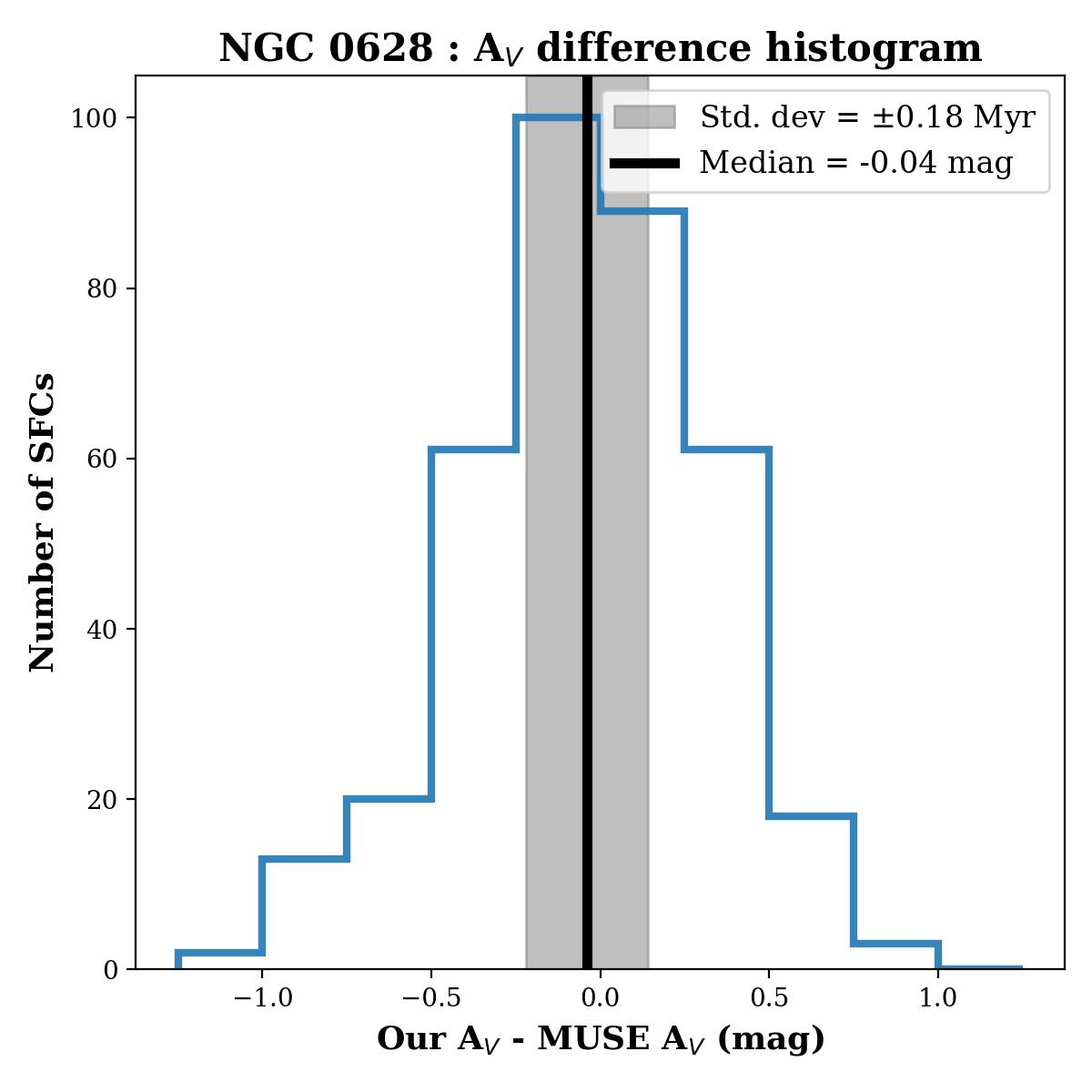}
    \hfill
    \includegraphics[width=0.36\textwidth]{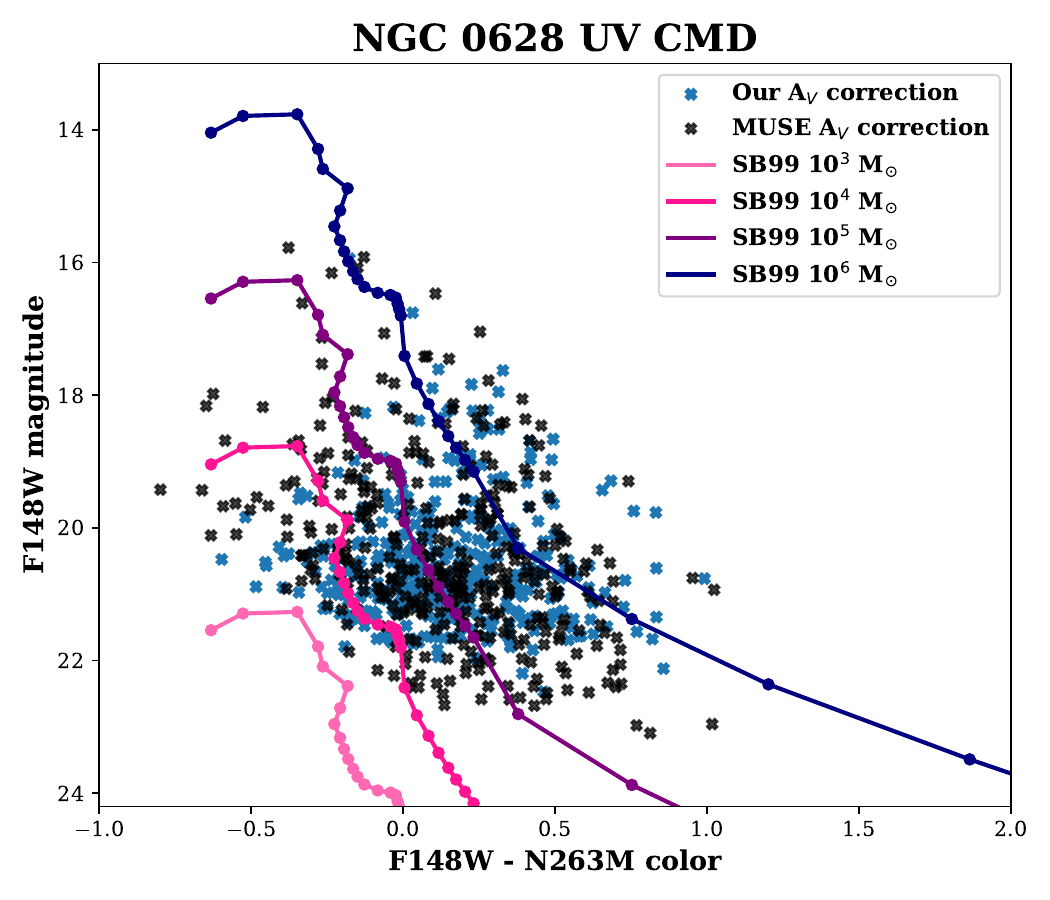}
    \vfill
    \hfill
    \includegraphics[width=0.47\textwidth]{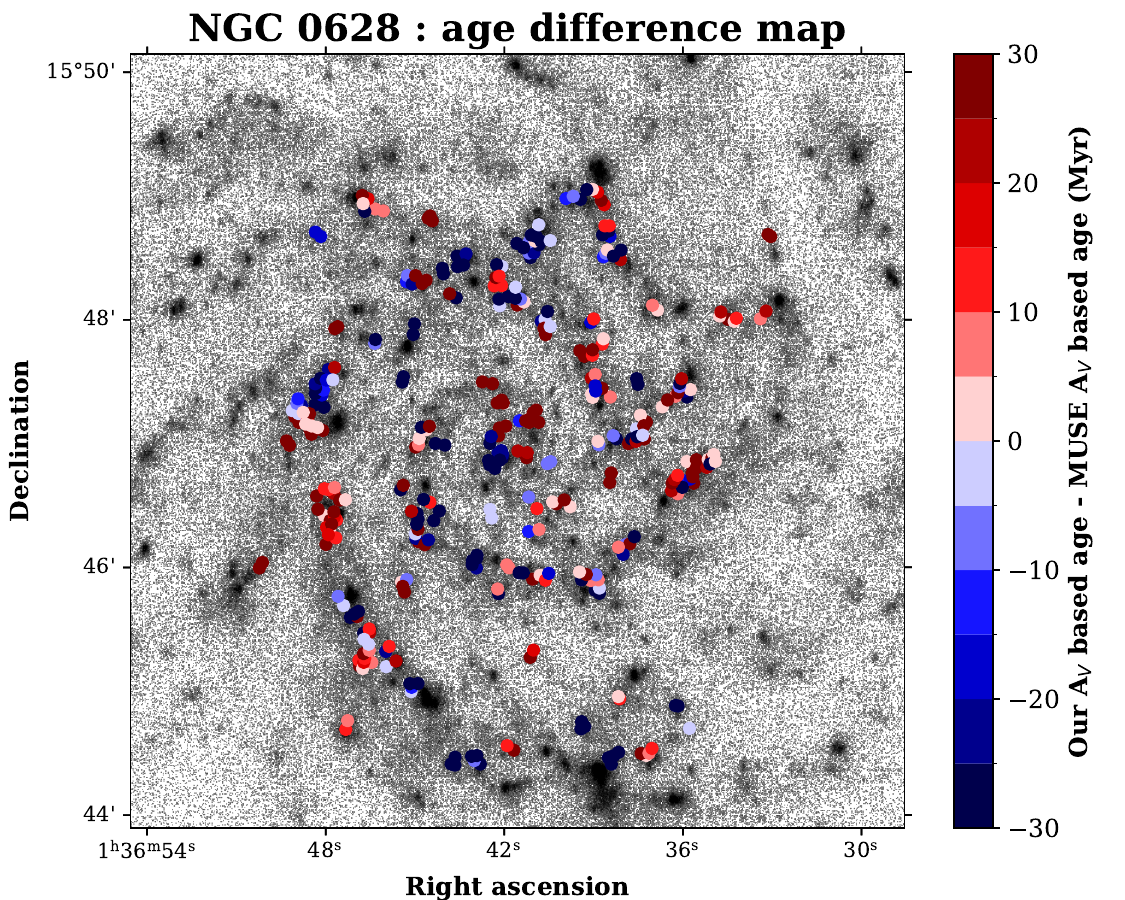}
    \hfill
    \includegraphics[width=0.39\textwidth]{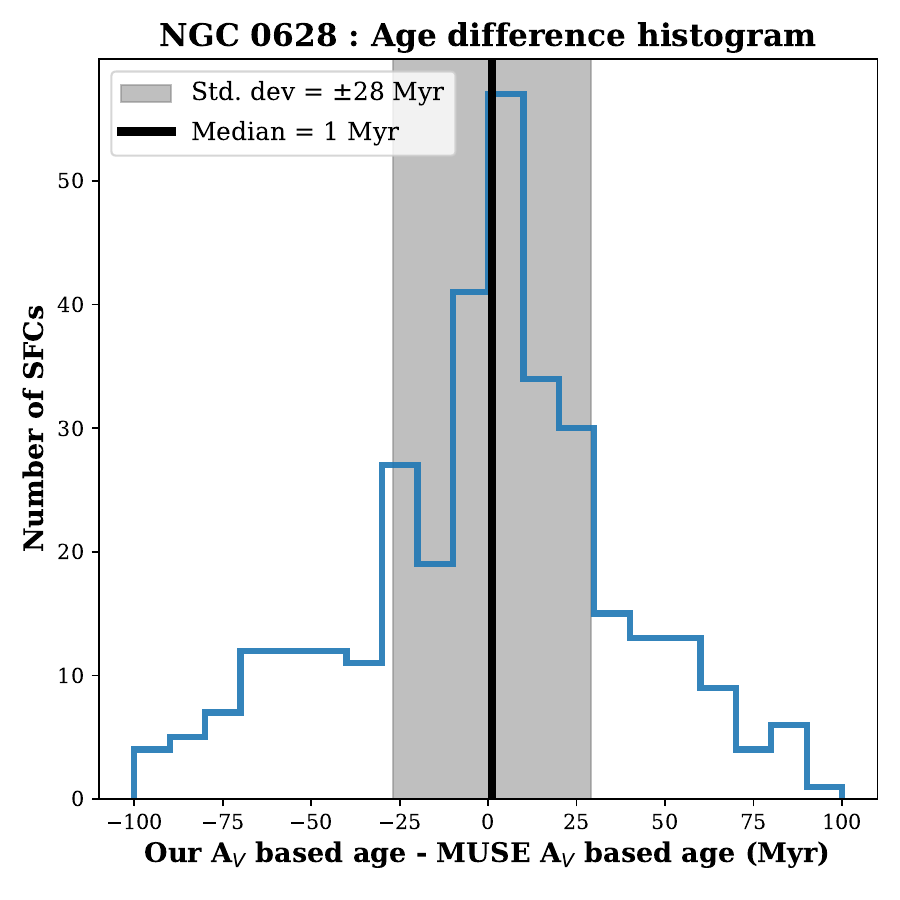}  
    \hfill
    \hfill
    \caption{Top left : Histogram of our 6\arcsec~resolution A$_V$ values and MUSE-based A$_V$ values for the MUSE-covered region of NGC 0628. MUSE-based A$_V$ values span a slightly wider range, whereas the extreme values at both the lower and upper end of A$_V$ are likely smoothed out in our A$_V$ maps. Top middle : A$_V$ difference histogram showing that over 85\% of the difference values lie between -0.5 to 0.5 magnitude. Top right : UV CMD for the 367 SFCs in NGC 0628 after correcting them using our A$_V$ values and MUSE-based A$_V$ values. Bottom left : Spatial age-difference map and; Bottom right : Age-difference histogram for these SFCs reveals that most of the age differences lie between $\pm$ 30 Myr, which is within the associated SFC age errors.}
  \label{fig_AV_within_resolution}
\end{figure}

We have 882 SFCs in the full disc of NGC 0628, out of which $\sim$450 SFCs lie within the MUSE-covered area of the galaxy. $\sim$2200 HII regions have been identified in this area of NGC 0628 by \cite{2022A&A...658A.188S}. First, we cross-matched the SFC positions with our A$_V$ map and identified 148 6\arcsec~ bins in which more than one SFCs are present. In total, 367 SFCs are located within these 148 6\arcsec~ bins. Next, we cross-matched these 367 SFC positions with the MUSE HII regions and measured the higher resolution MUSE-based A$_V$ values.
  
This allows us to measure the A$_V$ variation within a resolution element, using the 367 SFCs as probes. Firstly, we observed that our A$_V$ values and MUSE-based A$_V$ values span a similar distribution with comparable median values (see Figure \ref{fig_AV_within_resolution}). However, it is noteworthy that MUSE-based A$_V$ measurements include extreme values at both the lower end (close to 0) and the upper end (between 1.20 to 1.80) of A$_V$. Owing to our coarser 6\arcsec~ resolution, such values are missing in our A$_V$ maps and are likely smoothed out (similar to the histograms presented in Figure \ref{fig_AV_comparison_0628} and \ref{fig_AV_comparison_1566}). The A$_V$ difference histogram roughly spans between -1.0 to +1.0 magnitude, with over 85\% values lying between -0.5 to 0.5 magnitude. The A$_V$ difference median is -0.04 magnitude and standard deviation is $\pm$0.18 magnitude.

In order to understand how the resolution-dependent A$_V$ variation affects the SFC ages, we corrected these 367 SFCs using our A$_V$ values (used in this paper) and MUSE-based A$_V$ values. The UV CMD thus created indicates small variations in the FUV-NUV color. However, this variation is not systematic toward either bluer or redder colors. Next, we created an age-difference map and histogram for the SFCs, after correcting these with our A$_V$ values and MUSE-based A$_V$ values. Approximately 60\% of the SFCs exhibit age variations within 30 Myr, whereas the median and standard deviation of the age difference distribution is measured to be 1 Myr and $\pm$28 Myr. These age differences are largely within the magnitude error, and the corresponding age errors associated with the SFCs. This analysis allows us to conclude that the resolution mismatch between the UVIT and our own A$_V$ map leads to no systematic shifts in the SFC ages. Using a higher resolution A$_V$ map may lead to more accurate dust attenuation, however its impact on the derived ages and the overall age demographics may not be particularly significant.


\section{Effect of variable spatial resolution across the galaxy sample on the age demographic}
\label{apdx_robust_against_resolution}
\begin{figure*}[hbt!]
  	    \includegraphics[width=0.32\linewidth]{N2403_age_map.pdf}
	\hfill
		\includegraphics[width=0.32\linewidth]{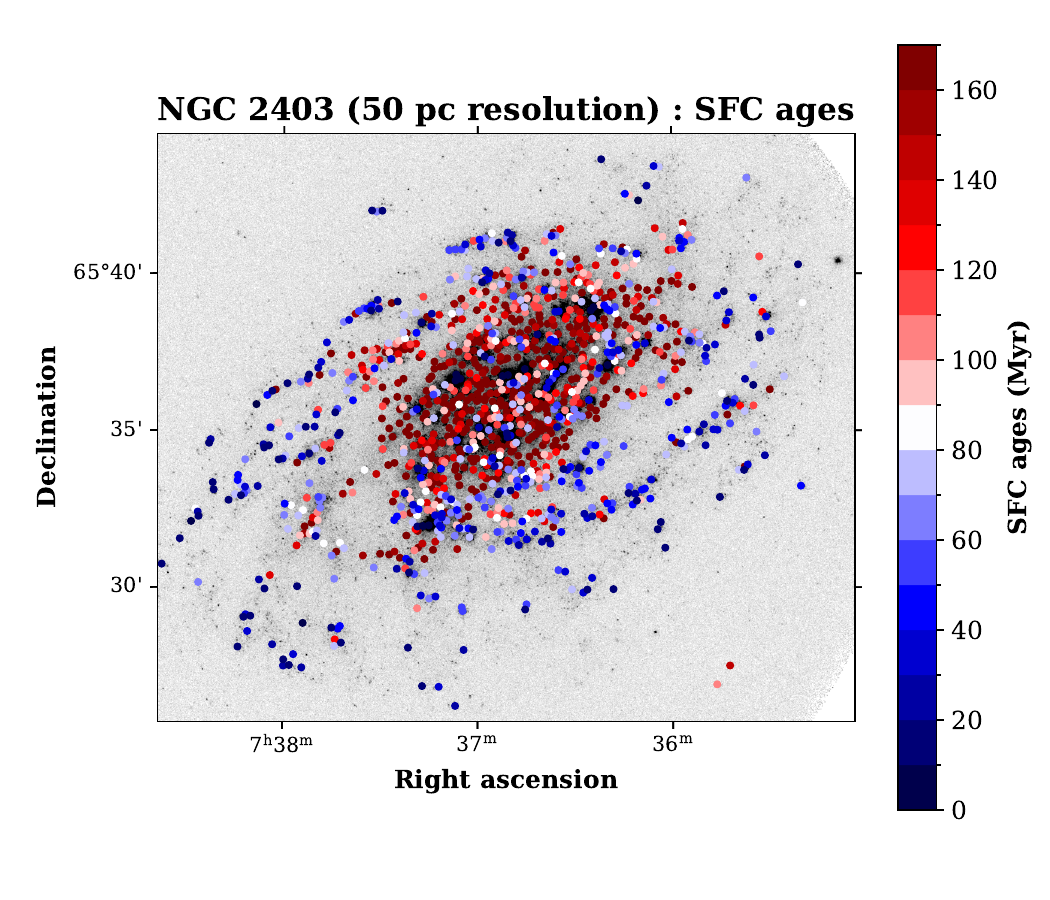}
    \hfill
         \includegraphics[width=0.32\linewidth]{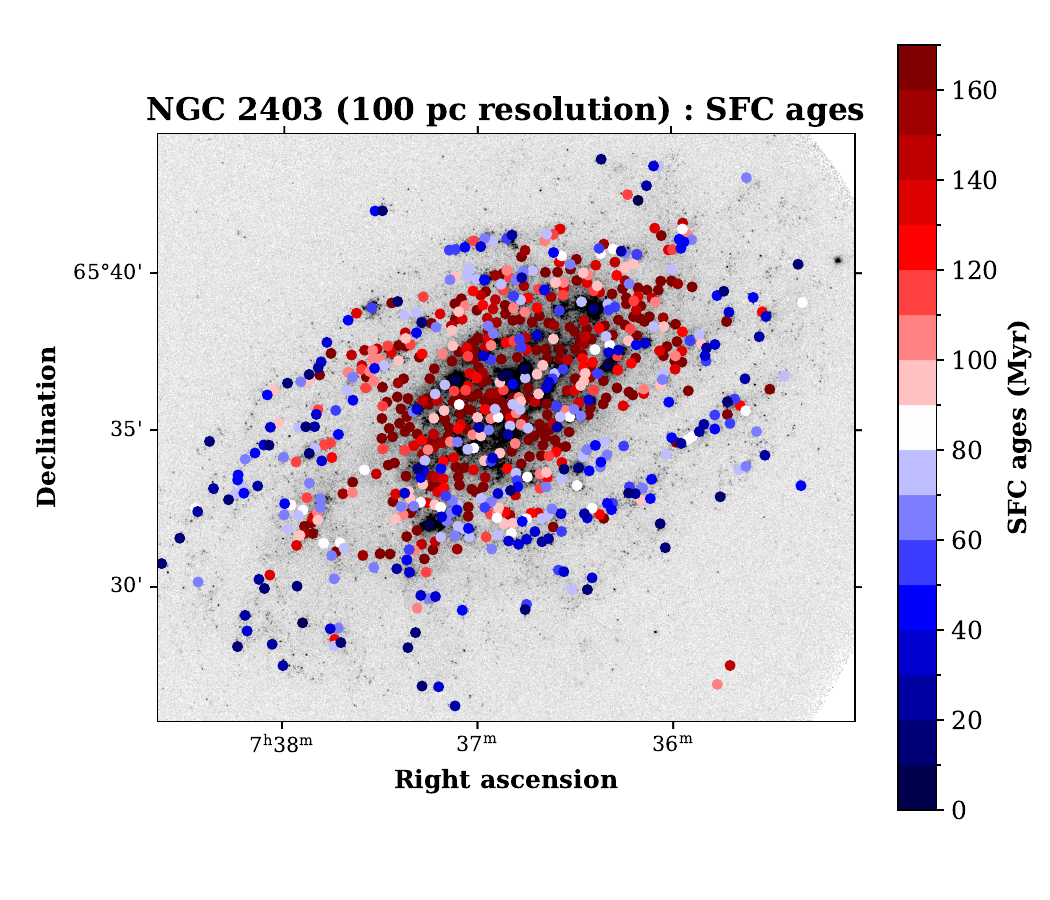}
	\vfill
		\includegraphics[width=0.32\linewidth]{N7793_age_map.pdf}
    \hfill
         \includegraphics[width=0.32\linewidth]{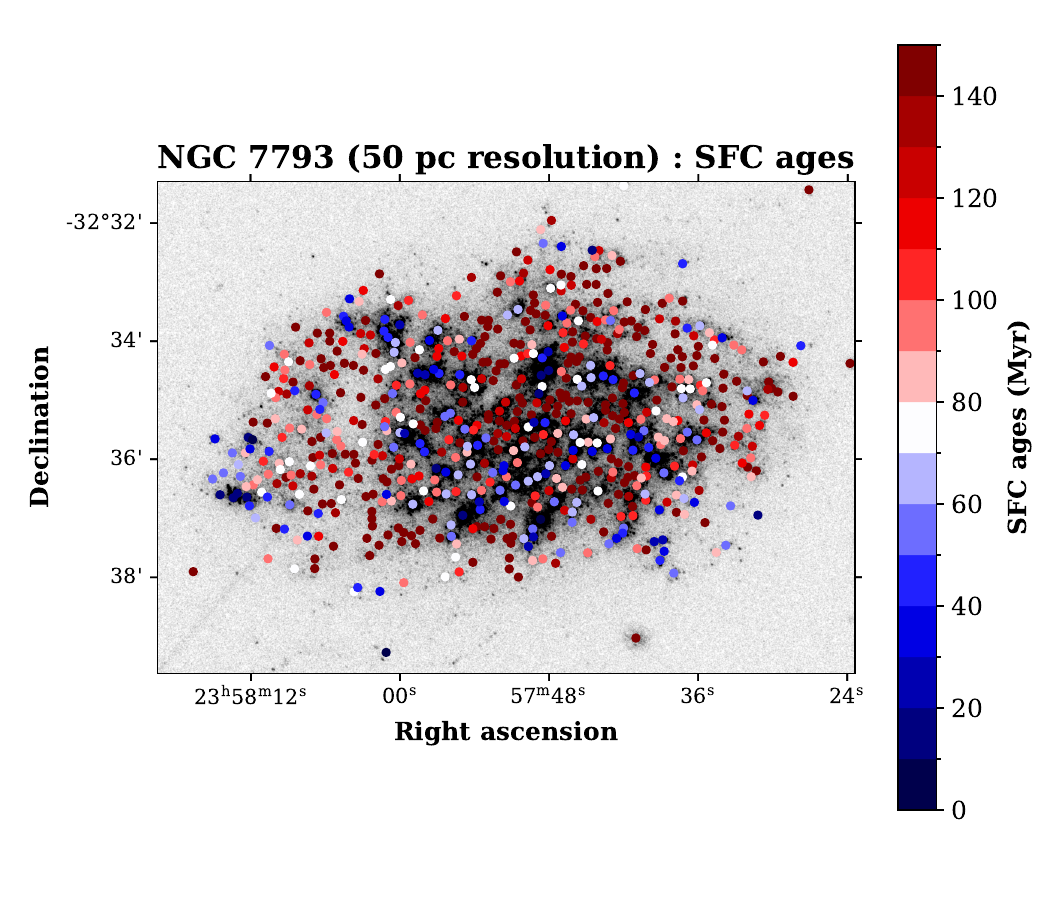}
	\hfill
		\includegraphics[width=0.32\linewidth]{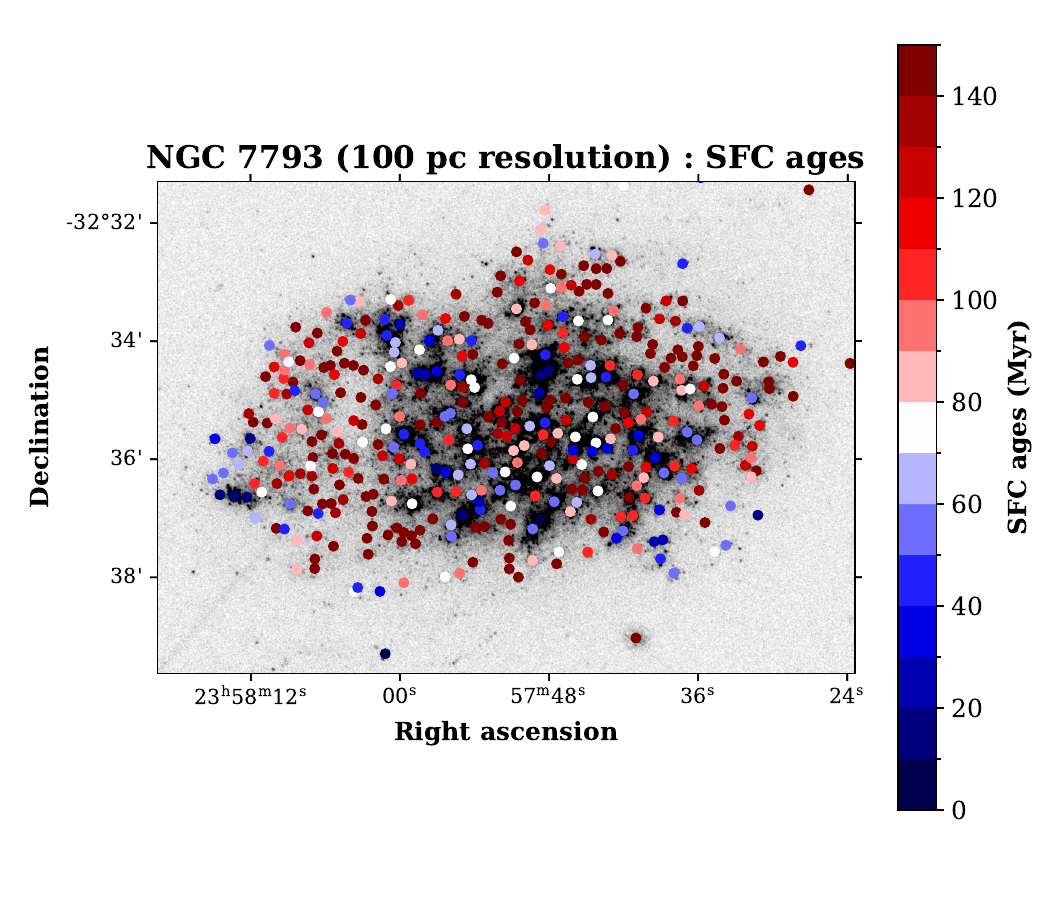}
    \vfill
    \hfill
         \includegraphics[width=0.40\linewidth]{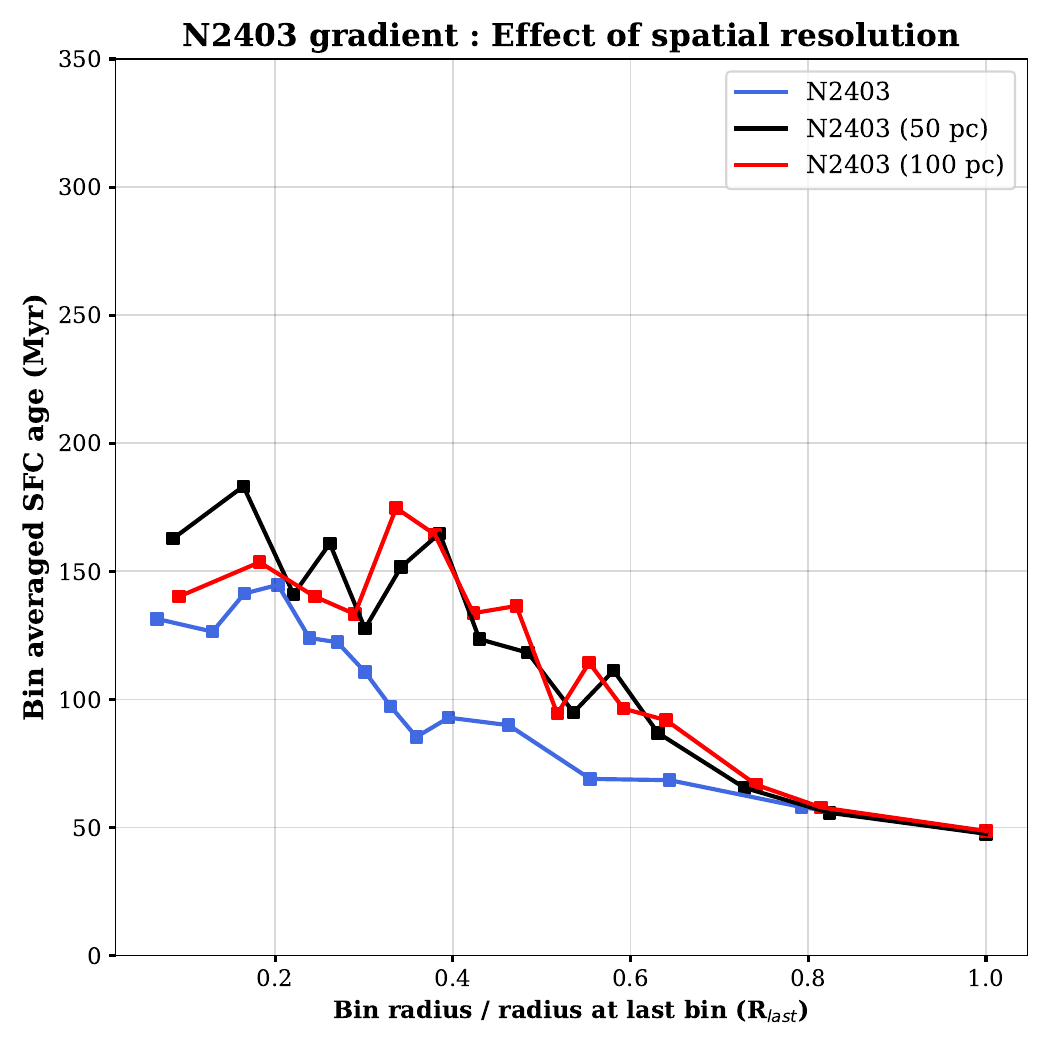}
    \hfill
         \includegraphics[width=0.40\linewidth]{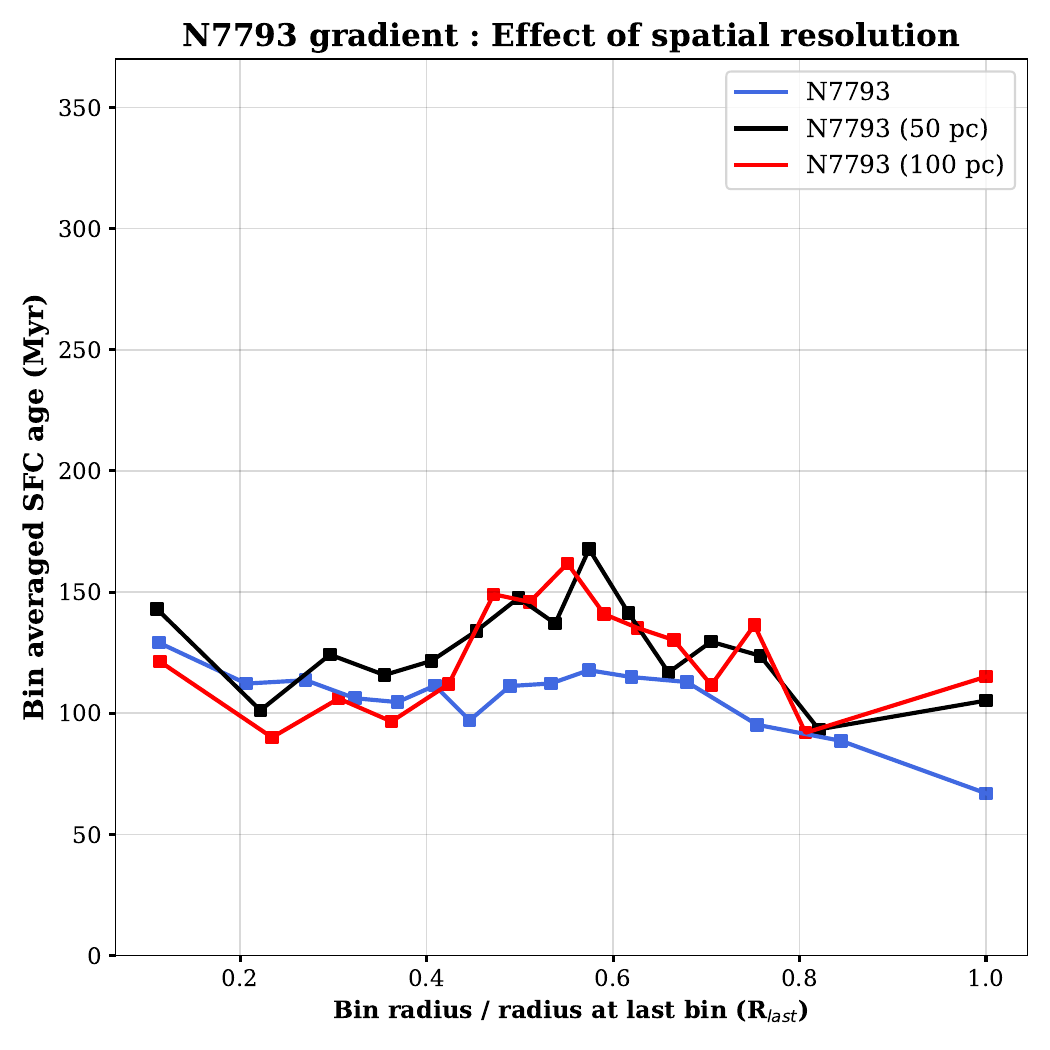}
    \hfill
    \hfill
    \caption{Top and middle row; from left to right : Spatial age maps of SFCs with $<$0.10 magnitude error cuts in NGC 2403 (top row) and NGC 7793 (middle row), created at native resolution (same as Figures \ref{fig_age_maps} and \ref{fig_age_maps_contd}), 50 pc resolution and 100 pc resolution. A reasonable match in the location of young and old SFCs within the galaxies is observed. Bottom row : Radial SFC Age gradients in NGC 2403 (left) and NGC 7793 (right), measured using the native resolution (blue points), 50 pc resolution (black points) and 100 pc resolution (red points) SFC age maps. The negative slope and shape of the gradients is mostly preserved, irrespective of the spatial resolution.}
  \label{fig_robust_against_resolution}
\end{figure*}

In order to establish that our age trends are not driven by the variation in physical resolution across our galaxy sample (between 6 pc to 137 pc), we tested the robustness of our age demographics and gradients. We chose NGC 7793 and NGC 2403 for this test due to their proximity, large exposure times, and an abundance of SFCs with better than 10 sigma detection (i.e. 0.10 magnitude error cut). The UVIT resolution corresponds to a native physical scale of $\sim$19-24 pc for these galaxies. We degraded the galaxy images to 50 pc and 100 pc resolution and characterized the SFCs with less than 0.10 magnitude error (equivalent to 10$\sigma$ detection) at these resolution values. We then compared the native resolution, 50 pc and 100 pc resolution age maps and radial gradients. The visual inspection of age maps (see Figure \ref{fig_robust_against_resolution}) reveals a reasonable match in the location of young and old SFCs within galaxies. In terms of absolute ages, it can be observed that poorer resolution images lead to systematically older SFC ages, which is further supported by the observed age gradients (The gradient shifts upwards in the y-axis as poorer resolution images are used). This happens due to the contamination caused by underlying older disc stars, as the SFC flux is integrated over a larger area \citep{Calzetti_2025}. However, the negative slope and shape of the gradient are mostly preserved, considering the age uncertainties of ($\sim$23 Myrs for 0.10 magnitude error) associated with our SFCs.

\end{document}